\documentclass[12pt]{article}
\pdfoutput=1
\usepackage{putex}
\usepackage{mathtools}
\usepackage{mathrsfs}
\usepackage{graphicx}
\usepackage{caption}
\usepackage{subcaption}
\usepackage{epstopdf}
\usepackage{enumerate}
\usepackage{cite}
\usepackage{youngtab}
\usepackage{tensor}
\usepackage{slashed}
\usepackage[aligntableaux=center]{ytableau}
\usepackage{url}
\usepackage{bbold}

\usepackage{enumitem} 
\setlist[itemize]{leftmargin=*}
\setlist[enumerate]{leftmargin=*}

\numberwithin{equation}{section}
\allowdisplaybreaks 

\newcommand{\abs}[1]{\left\lvert #1 \right\rvert}

\newcommand {\be} {\begin {equation}}
\newcommand {\ee} {\end {equation}}

\newcommand {\bes} {\begin {equation*}}
\newcommand {\ees} {\end {equation*}}

\newcommand{\es}[2] {\begin{equation} \label{#1} \begin{split} #2 \end{split} \end{equation}}

\newcommand{\Z}{\mathbb{Z}}

\newcommand{\R}{\mathbb{R}}
\newcommand{\C}{\mathbb{C}}

\def\Tr{\mop{Tr}}

\newcommand{\beq}{\begin{equation}}
\newcommand{\eeq}{\end{equation}}

\def\MD#1{}

\def\<{\langle}
\def\>{\rangle}

\newcommand{\bR}{\ensuremath{\mathbb{R}}}

\newcommand{\cB}{\ensuremath{\mathcal{B}}}

\newcommand{\cD}{\ensuremath{\mathcal{D}}}

\newcommand{\cH}{\ensuremath{\mathcal{H}}}
\newcommand{\cI}{\ensuremath{\mathcal{I}}}

\newcommand{\cL}{\ensuremath{\mathcal{L}}}
\newcommand{\cM}{\ensuremath{\mathcal{M}}}
\newcommand{\cN}{\ensuremath{\mathcal{N}}}
\newcommand{\cO}{\ensuremath{\mathcal{O}}}

\newcommand{\cQ}{\ensuremath{\mathcal{Q}}}
\newcommand{\cR}{\ensuremath{\mathcal{R}}}
\newcommand{\cS}{\ensuremath{\mathcal{S}}}

\newcommand{\cV}{\ensuremath{\mathcal{V}}}
\newcommand{\cW}{\ensuremath{\mathcal{W}}}
\newcommand{\cX}{\ensuremath{\mathcal{X}}}
\newcommand{\cY}{\ensuremath{\mathcal{Y}}}
\newcommand{\cZ}{\ensuremath{\mathcal{Z}}}

\newcommand{\pD}{\mathscr{D}}

\newcommand{\ed}{\,.}
\newcommand{\ec}{\,,}
\newcommand{\ecq}{\ec\quad}

\newcommand{\tpsi}{\widetilde{\psi}}
\newcommand{\tq}{\widetilde{q}}
\newcommand{\tQ}{\widetilde{Q}}

\newcommand{\tF}{\widetilde{F}}
\newcommand{\vphi}{\varphi}
\DeclareMathOperator{\trace}{Tr}

\begin{document}

\preprint{CALT-TH 2017-064\\ PUPT-2547 \\ WIS/05/17-Dec-DPPA}

\institution{Caltech}{Walter Burke Institute for Theoretical Physics, California Institute of Technology, \cr Pasadena, CA 91125, USA}
\institution{PU}{Department of Physics, Princeton University, Princeton, NJ 08544, USA}
\institution{Weizmann}{Department of Particle Physics and Astrophysics, Weizmann Institute of Science, \cr Rehovot 76100, Israel}

\title{Coulomb Branch Operators and \linebreak Mirror Symmetry in Three Dimensions}

\authors{Mykola Dedushenko,\worksat{\Caltech} Yale Fan,\worksat{\PU} Silviu S.~Pufu,\worksat{\PU} and Ran Yacoby\worksat{\Weizmann}}

\abstract{

We develop new techniques for computing exact correlation functions of a class of local operators, including certain monopole operators, in three-dimensional $\mathcal{N} = 4$ abelian gauge theories that have superconformal infrared limits.  These operators are position-dependent linear combinations of Coulomb branch operators.  They form a one-dimensional topological sector that encodes a deformation quantization of the Coulomb branch chiral ring, and their correlation functions completely fix the ($n\leq 3$)-point functions of all half-BPS Coulomb branch operators.  Using these results, we provide new derivations of the conformal dimension of half-BPS monopole operators as well as new and detailed tests of mirror symmetry.  Our main approach involves supersymmetric localization on a hemisphere $HS^3$ with half-BPS boundary conditions, where operator insertions within the hemisphere are represented by certain shift operators acting on the $HS^3$ wavefunction.  By gluing a pair of such wavefunctions, we obtain correlators on $S^3$ with an arbitrary number of operator insertions.  Finally, we show that our results can be recovered by dimensionally reducing the Schur index of 4D $\mathcal{N} = 2$ theories decorated by BPS 't~Hooft-Wilson loops.

}

\date{}

\maketitle

\tableofcontents
\setlength{\unitlength}{1mm}

\newpage
\section{Introduction}
\label{INTRO}

${\cal N} = 4$ supersymmetry in three dimensions provides a rich middle ground between the availability of calculable supersymmetry-protected observables and nontrivial dynamics.  As an example that will be relevant to us, ${\cal N} = 4$ gauge theories with matter hypermultiplets exhibit an infrared duality known as mirror symmetry  \cite{Intriligator:1996ex}, under which the Higgs and Coulomb branches of the vacuum moduli space of a given theory are mapped to the Coulomb and Higgs branches of the other.  In particular, the half-BPS operators that acquire expectation values when the theory is taken to the Higgs/Coulomb branch, henceforth referred to as Higgs/Coulomb branch operators (HBOs/CBOs), are mapped to the CBOs/HBOs of the mirror dual theory.  The duality is nontrivial for several reasons:  while the Higgs branch is protected by a non-renormalization theorem and can simply be fixed classically from the UV Lagrangian \cite{Aharony:1997bx}, the Coulomb branch generically receives quantum corrections;  the duality exchanges certain order operators and disorder operators; and non-abelian flavor symmetries visible in one theory may be accidental in the mirror dual.  At the same time, ${\cal N} = 4$ supersymmetry allows for various calculations of protected observables that led to the discovery of the duality and  to various tests thereof, such as the match between the infrared metrics of the Coulomb and Higgs branches \cite{deBoer:1996ck}, scaling dimensions of monopole operators \cite{Borokhov:2002cg}, various curved-space partition functions \cite{Kapustin:2010xq, Krattenthaler:2011da, Closset:2016arn}, expectation values of loop operators \cite{Kapustin:2012iw,Drukker:2012sr}, and the Hilbert series \cite{Cremonesi:2013lqa}.

Our goal in the present paper is to provide new insights into the mirror symmetry duality and, more generally, into 3D ${\cal N} = 4$ QFTs, by developing new techniques for calculating correlation functions of certain CBOs that include monopole operators.  These techniques are related to the observation of \cite{Beem:2016cbd,Chester:2014mea} that all ${\cal N} = 4$ superconformal field theories (SCFTs) contain two one-dimensional topological sectors, one associated with the Higgs branch and one associated with the Coulomb branch.  These sectors are described abstractly as consisting of the cohomology classes with respect to a pair of nilpotent supercharges, and each cohomology class can be represented by a position-dependent linear combination of HBOs/CBOs that can be inserted anywhere along a line.  For the Higgs branch case, it was shown in \cite{Dedushenko:2016jxl} that the 1D sector has a Lagrangian description that can be obtained by supersymmetric localization and that gives a simple way of computing all correlation functions of the 1D Higgs branch theory.  The objective of this work is to provide an explicit description of the Coulomb branch topological sector.  Having explicit descriptions of both the Higgs and Coulomb branch 1D sectors allows for more explicit tests of mirror symmetry, including a precise mapping between all half-BPS operators of the two theories.

For simplicity, in this work, we focus only on abelian ${\cal N} = 4$ gauge theories.\footnote{In fact, our results can easily be generalized to theories with both ordinary and twisted multiplets coupled through BF terms, first studied in \cite{Brooks:1994nn}.}  Any abelian $\cN=4$ gauge theory has a known mirror dual, which is also abelian. The fundamental abelian mirror duality, proven in \cite{Borokhov:2002cg}, states that the IR limit of $\cN=4$ SQED with one flavor coincides with a free (twisted) hypermultiplet.  All other abelian mirror pairs can be formally deduced from the fundamental one by gauging global symmetries \cite{Kapustin:1999ha}. 

Compared to the Higgs branch 1D theory described in \cite{Dedushenko:2016jxl}, the description of the Coulomb branch theory is more complicated because it involves monopole operators.  Monopole operators in 3D gauge theories are local disorder operators, meaning that they cannot be expressed as polynomials in the classical fields. Instead, their insertion in the path integral is realized by assigning boundary conditions for the fields near the insertion point. Specifically, a monopole operator is defined by letting the gauge field approach the singular configuration of an abelian Dirac monopole at a point.  Calculations involving monopole operators are notoriously difficult, even in perturbation theory. Following \cite{Borokhov:2002ib}, the IR conformal dimensions of monopole operators have been estimated for various non-supersymmetric theories using the $1/N_f$ expansion \cite{Murthy:1989ps, Metlitski:2008dw,Pufu:2013eda,Pufu:2013vpa,Dyer:2013fja,Dyer:2015zha}, the $(4-\epsilon)$-expansion \cite{Chester:2015wao}, and the conformal bootstrap \cite{Chester:2016wrc}. In supersymmetric theories, one can also construct BPS monopole operators by assigning additional singular boundary conditions for some of the scalars in the vector multiplet. For such BPS monopoles, some nonperturbative results are known: for instance, in $\cN=4$ theories, their exact conformal dimension was determined in \cite{Borokhov:2002cg,Gaiotto:2008ak,Benna:2009xd,Bashkirov:2010kz}.\footnote{The exact results mentioned above are valid for ``good'' or ``ugly'' theories, to use the terminology of \cite{Gaiotto:2008ak}. We will only consider such theories in this paper.} The correlation functions that we calculate in this paper provide additional nonperturbative results involving BPS monopole operators.

The Coulomb branch 1D theory whose description we will derive encodes information on the geometry of the quantum-corrected Coulomb branch.  The Coulomb branch is constrained by supersymmetry to be a (singular) hyperk\"ahler manifold which, with respect to a fixed complex structure, can be viewed as a complex symplectic manifold whose holomorphic symplectic structure endows its coordinate ring with Poisson brackets.\footnote{The description of the Coulomb branch as a complex symplectic manifold is not sufficient to reconstruct its hyperk\"ahler metric. It would be interesting to understand whether, and how, information on this metric is encoded in the SCFT.}  The holomorphic coordinate ring of the Coulomb branch, which describes it as a complex variety, is believed to coincide with the ring of chiral CBOs\@.  As explained in \cite{Beem:2016cbd}, the OPE of the 1D Coulomb branch theory provides a deformation quantization of the Poisson algebra associated with the chiral ring.

In brief, we obtain an explicit description of the Coulomb branch 1D theory as follows.  First, we stereographically map the ${\cal N} = 4$ theory from $\R^3$ to $S^3$.  While the 1D theory is defined on a straight line in $\R^3$, after the mapping to $S^3$, it is defined on a great circle.  Ideally, we would like to perform supersymmetric localization on $S^3$ with respect to a judiciously chosen supercharge such that the 3D theory localizes to a theory on the great circle (this is how the description of the 1D Higgs branch theory was obtained in \cite{Dedushenko:2016jxl}).  Unfortunately, it is challenging to calculate functional determinants in the presence of an arbitrary number of disorder operators inserted along the great circle.  To circumvent this problem, we develop another approach in which we cut the $S^3$ into two hemispheres $HS^3$ glued along an $S^2$ that intersects the great circle at two points, and then calculate the $HS^3$ wavefunction.  Because we can add a localizing term on $S^2$, it is sufficient to evaluate the $HS^3$ wavefunction along a finite-dimensional locus in field space.  For every insertion within the hemisphere, we derive a corresponding operator acting on the $HS^3$ wavefunction.  As we will explain, gluing two hemisphere wavefunctions allows us to compute arbitrary correlators of the 1D theory.

We hope that the methods presented in this paper can be generalized and applied also to non-abelian $\cN=4$ theories.  In these theories, both the Coulomb branch geometry and mirror symmetry are less understood than in the abelian case. In particular, the mirror duals of non-abelian theories are not always known, and the Coulomb branch metric can no longer be simply computed due to nonperturbative effects that are absent in abelian theories. A general picture for the Coulomb branch geometry was recently proposed in \cite{Bullimore:2015lsa}, and it should be possible to verify it rigorously using correlators of CBOs (there have also been a number of papers on Coulomb branches of 3D $\cN=4$ theories in the mathematical literature \cite{Nakajima:2015txa, Braverman:2016wma, Braverman:2016pwk, Braverman:2017ofm, Nakajima:2017bdt}). Furthermore, correlators of CBOs and HBOs could shed light on non-abelian mirror symmetry, because this duality maps these two classes of operators to each other.  We hope to report on progress in answering these interesting questions in the near future.

The remainder of this section contains a technical overview of our approach and a summary of our results.  The rest of the paper is organized as follows.  In Section~\ref{prelim}, we introduce in detail the theories that we study and their 1D topological sectors.  In Section~\ref{S3loc}, we perform supersymmetric localization on $S^3$ with monopole-antimonopole insertions at opposite points on the sphere.  In Section~\ref{SGLUE}, we perform supersymmetric localization on a hemisphere and on its boundary and explain how to glue two hemisphere wavefunctions.  In Section~\ref{multipleinsertions}, we explain how to compute correlators in the 1D theory with multiple operator insertions.  In Section~\ref{applications}, we discuss, as applications of our results, a derivation of the chiral ring relations, and we provide several new tests of mirror symmetry.  Several technical details are relegated to the appendices.

\subsection{Technical Overview}

Let us now describe the general logic behind our computation, which closely follows that of \cite{Dedushenko:2016jxl}. Consider an $\cN=4$ theory with gauge group $G$ and a hypermultiplet transforming in a (generally reducible) unitary representation $\cR$ of $G$. The theory could also be deformed by real masses and FI parameters, which, for simplicity, we set to zero until further notice. The above information determines an $\cN=4$ preserving Lagrangian $\cL_{\bR^3}$ on $\bR^3$ and another Lagrangian $\cL_{S^3}$ on an $S^3$ with radius $r$,  both of which coincide when $r\to\infty$. Furthermore, the theories on $\bR^3$ and $S^3$ have the same IR limit, and we will consider examples in which it is a nontrivial SCFT.\footnote{The limit $g_{\rm YM},r\to\infty$ on $S^3$ is identical to the flat space IR SCFT. Instead, taking $g_{\rm YM}\to\infty$ at fixed $r$ leads to an SCFT on $S^3$ whose correlators are equivalent to those of the IR SCFT on $\bR^3$, by a conformal map from $S^3$ to $\bR^3$. One subtlety in this procedure, first noted in \cite{Gerchkovitz:2016gxx}, is that on $S^3$, there can be mixing between operators of different conformal dimensions, though this mixing can always be resolved.} From our point of view, the advantage of working on $S^3$ is that $\cL_{S^3}$ preserves certain supercharges $\cQ^C$ and $\cQ^H$, which are only symmetries of the flat space theory at the IR fixed point. The attractive property of $\cQ^C$ (or $\cQ^H$) is that its cohomology contains local operators which have nontrivial correlation functions, and which form a subset of the full family of CBOs (or HBOs).\footnote{This cohomology is distinct from the chiral ring, as will be explained later.
} It follows that the correlators of these $\cQ^C$-closed ($\cQ^H$-closed) operators, which are known as {\it twisted} CBOs (HBOs), could possibly be computed using supersymmetric localization of the path integral on $S^3$ with respect to $\cQ^C$ ($\cQ^H$). Indeed, the problem of localizing with respect to $\cQ^H$ was fully solved in \cite{Dedushenko:2016jxl}, thus making correlators of twisted HBOs calculable.

In this work, we are interested in correlators of twisted CBOs, which can be described abstractly as follows. First, each CBO is a Lorentz scalar transforming in a spin-$j$ irrep of an $SU(2)$ R-symmetry, such that in the IR SCFT, it is a superconformal primary of dimension $\Delta=j$.\footnote{Strictly speaking, the RG flow on $S^3$ only preserves a $U(1)$ subgroup of the $SU(2)$ R-symmetry mentioned above. Nevertheless, it is useful (and possible) to group CBOs into $SU(2)$ irreps also along the flow, even if it only becomes a true symmetry in the IR. \label{su2breaking}} Each twisted CBO is given by a certain position-dependent linear combination of the $SU(2)$ R-symmetry components of a CBO, and is restricted to lie on the great circle fixed by the $S^3$ isometry generated by $\cQ^C$. Furthermore, at each point on this circle, the twisted CBOs are chiral with respect to a distinct $\cN=2$ subalgebra. More details will be given in Section \ref{prelim}. Restricting our 3D theories to the cohomology of $\cQ^C$, therefore, results in some 1D field theory on a circle whose local operators can be identified with cohomology classes of twisted CBOs, which, in turn, are in one-to-one correspondence with Coulomb branch chiral ring operators. 

The above 1D theory provides a significant simplification of the original 3D problem of computing correlators of CBOs, due to the following properties. First, the IR two- and three-point functions of twisted CBOs in the 1D theory are sufficient to fix the corresponding correlators of CBOs in the full 3D SCFT, simply because a two- or three-point function of Lorentz scalar primary operators is fixed by conformal invariance up to an overall constant (see, e.g., Section 6.4 of \cite{Dedushenko:2016jxl}). 
Moreover, it turns out that the 1D theory is topological in the sense that its correlators are independent of the relative separation between insertions, but can depend on their order on the circle. We will refer to this theory as the Coulomb branch 1D topological quantum field theory (TQFT). The topological correlators could in principle be functions of dimensionless parameters along the flow. Because we set all the real masses and FI terms to zero, the only remaining dimensionless parameter is $g_{\rm YM}^2 r$. However, the 1D theory is independent of $g_{\rm YM}$ (and therefore of $g_{\rm YM}^2 r$) because, as shown in \cite{Dedushenko:2016jxl}, the Yang-Mills action is $\cQ^C$-exact. It follows that the correlators of twisted CBOs are RG-invariant and can be identified, all along the flow, with those of the IR SCFT. The same results also hold for twisted HBOs, whose associated 1D TQFT is obtained by passing to the cohomology of $\cQ^H$. The above properties of the 1D TQFTs turn them into a powerful framework to study correlators of half-BPS operators in $\cN=4$ theories.

The observation that some BPS operators in $d$-dimensional theories with eight supercharges admit a lower-dimensional description was made for SCFTs in \cite{Beem:2013sza}. Earlier works achieved an analogous suppression of non-compact spacetime directions in four dimensions via the Omega-background: see \cite{Nekrasov:2002qd, Nekrasov:2009rc, Nekrasov:2010ka} for the original discussion. In both approaches, equivariance plays an important role, though the precise relation between them has not yet been worked out. It is believed that in four dimensions, the SCFT approach of \cite{Beem:2013sza} corresponds to a new type of Omega-deformation. In three dimensions, on the other hand, the Omega-deformation and the associated quantizations of moduli spaces, first discussed in \cite{Yagi:2014toa, Bullimore:2015lsa, Bullimore:2016hdc}, are most likely directly related to quantization in the SCFT picture. 

Following the work of \cite{Beem:2013sza}, the 1D TQFTs associated with 3D $\cN=4$ SCFTs were studied in detail in \cite{Chester:2014mea,Beem:2016cbd}. It was shown in \cite{Chester:2014mea,Beem:2016cbd} that conformal bootstrap arguments can be used to fix the 1D TQFT in some simple examples, though doing this for general 3D $\cN=4$ SCFTs proved to be difficult. Finally, the fact that the 1D TQFTs can also be defined along $\cN=4$ RG flows on $S^3$, as we just reviewed, was discovered in \cite{Dedushenko:2016jxl}. This fact allows for the use of supersymmetric localization to calculate correlators in the 1D TQFTs for 3D $\cN=4$ theories described in the UV by a Lagrangian. Moreover, it follows that the 1D theory is also defined along relevant deformations of the theory on $S^3$ by real masses and FI parameters. The correlators of twisted CBOs are in general sensitive to these deformations, providing nonperturbatively calculable examples of correlators along RG flows.\footnote{The topological invariance of the Coulomb (Higgs) branch 1D theory is lost upon turning on FI (real mass) parameters. However, the resulting position dependence of correlators turns out to be very simple.} 

We develop three complementary approaches to computing correlators of twisted CBOs.  In Section \ref{S3loc}, we use localization on $S^3$ in an $SO(3)$-symmetric background created by a monopole-antimonopole pair to compute correlators involving two twisted monopole CBOs and an arbitrary number of non-defect twisted CBOs.  In Sections \ref{SGLUE} and \ref{multipleinsertions}, we explain how to vastly generalize these results by localizing on a hemisphere $HS^3$ with half-BPS boundary conditions, which allows for insertions of twisted CBOs anywhere along a great semicircle.  These insertions are conveniently described by certain operators acting on the $HS^3$ wavefunction.  Pairs of such wavefunctions can then be glued along their $S^2$ boundary to reproduce the $S^3$ partition function with an arbitrary number of twisted CBOs.  In Section \ref{multipleinsertions}, we further show how to interpret our results as a dimensional reduction of the Schur index of 4D $\mathcal{N}=2$ theories enriched by BPS 't Hooft-Wilson loops.  

\subsection{Summary of Results}

Let us now summarize our results and fix our notation. We consider $\cN=4$ theories with gauge group $G=U(1)^r$ and $N_h\geq r$ hypermultiplets of gauge charges $\vec{q}_I=(q^1_I,\ldots,q^r_I)\in\mathbb{Z}^r$ with $I=1,\ldots,N_h$. Viewing $q$ as an $N_h\times r$ matrix, we demand that $\mathrm{rank}(q) = r$ to avoid having $U(1)$ subgroups of $G$ with no charged matter. The theory has flavor symmetry $G_H\times G_C$ where $G_H$ acts on the hypermultiplet, while $G_C$ generally emerges in the IR and acts on the Coulomb branch. Only a maximal torus of $G_C$ is manifest in the UV as a ``topological symmetry'' $U(1)^r$ acting on monopole operators and generated by currents $j_T$ constructed from the field strength as $j_T \sim \ast F$. 

\begin{figure}[t!]
	\centering
	\includegraphics[width=0.8\textwidth]{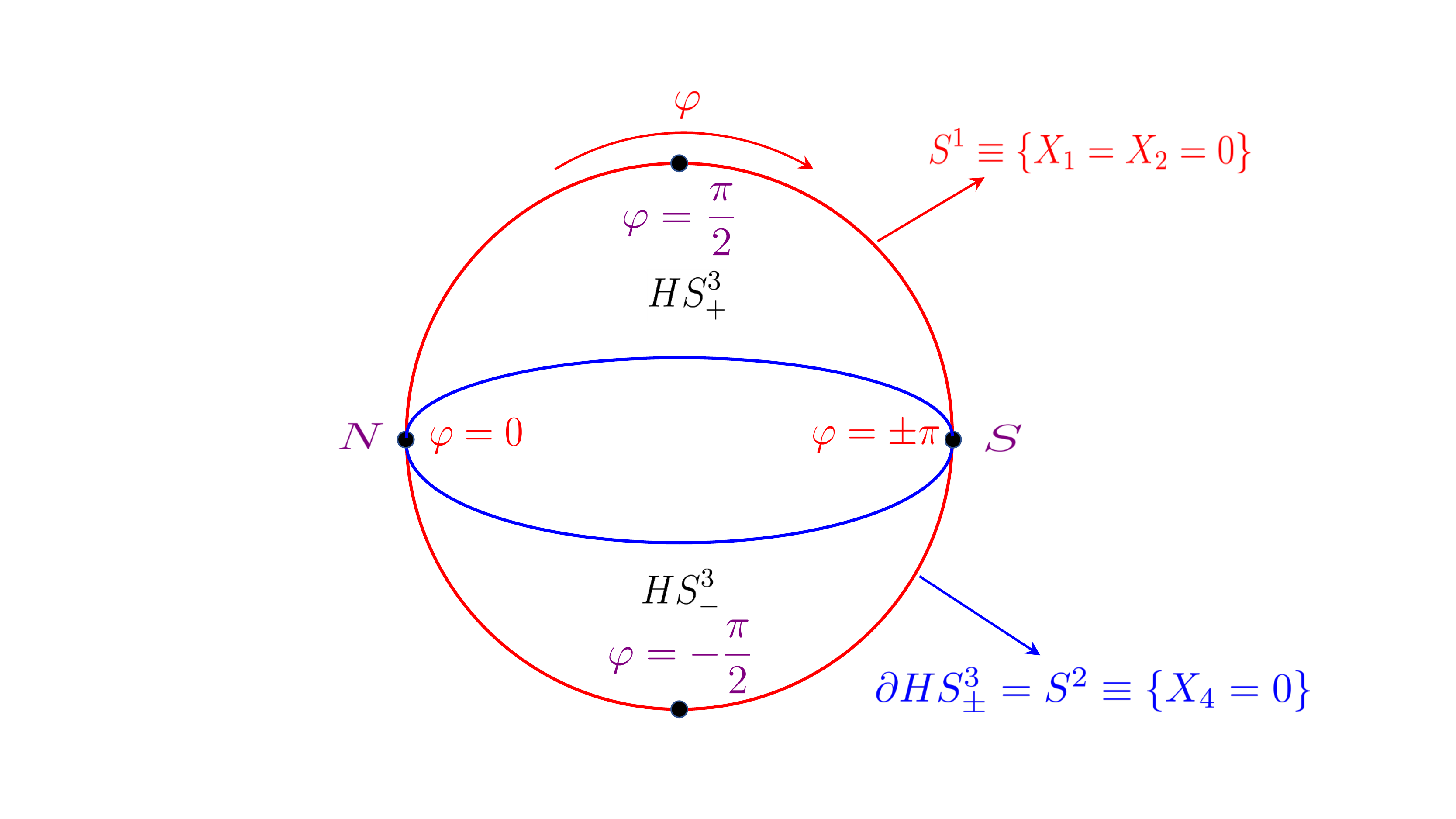}
	\caption{\label{fig:sphere} A schematic 2D representation of $S^3$ given by $X_1^2+X_2^2+X_3^2+X_4^2=r^2$. The 1D TQFT lives on the $S^1$ defined by $X_1=X_2=0$ (red) and parametrized by the angle $\varphi$. The $S^3$ can be cut into two hemispheres $HS^3_{\pm}\cong B^3$ whose boundary forms an $S^2=\partial HS^3_{\pm}$ (blue circle) defined by $X_4=0$. The 1D TQFT circle intersects this $S^2$ at two points identified with its North ($N$) and South ($S$) poles.}
\end{figure}

Let the 1D theory live on a great circle parametrized by $\varphi$ (see Figure \ref{fig:sphere}). The $\cQ^C$-closed twisted CBOs are constructed from products of bare twisted monopole operators $\cM^{\vec{b}}(\varphi)$, labeled by their $G_C$ charge $\vec{b}\in\Gamma_m\subset\mathbb{R}^r$ where $\Gamma_m$ is the monopole charge lattice determined by Dirac quantization, as well as twisted vector multiplet scalars $\vec{\Phi}(\varphi) = (\Phi^1(\varphi),\ldots,\Phi^r(\varphi))$ corresponding to each $U(1)$ factor of $G$. 
As we will see in Section \ref{prelim}, $\vec{\Phi}$ is a position-dependent linear combination of the three real vector multiplet scalars, while $\cM^{\vec{b}}$ can be described as a particular $\cQ^C$-invariant background for the vector multiplet fields, which inserts the appropriate Dirac monopole singularity. These singular backgrounds are described in detail in Appendix \ref{moremonopoles}.

In Section \ref{multipleinsertions}, we present a matrix model expression for a correlator with $n$ insertions of twisted CBOs  $\cO^{(k)}(\varphi_k)$, where $k=1,\ldots,n$. To describe this expression, it is useful to think of $S^3$ as a union of two hemispheres $HS^3_{\pm}\cong B^3$ joined along their $S^2$ boundary, as depicted in Figure \ref{fig:sphere}. The 1D TQFT circle intersects the boundary $S^2$ at its North and South poles labeled, respectively, by $N$ and $S$ in Figure \ref{fig:sphere}. Under this decomposition, the path integral on $S^3$ can be thought of as an inner product (more accurately, a bilinear form) composing the wavefunctions of $HS^3_+$ and $HS^3_-$. Moreover, in this language, the insertions of twisted CBOs can be represented as certain shift operators acting on the hemisphere wavefunctions. 

Explicitly, consider the case in which the $\cO^{(k)}(\varphi_k)$ are all inserted along the semicircle inside the upper hemisphere $HS^3_+$ ($0<\varphi<\pi$) in the order $0<\varphi_1<\varphi_2<\cdots<\varphi_n<\pi$. There is no loss of generality in inserting all operators in $HS^3_+$ because the 1D TQFT is topological, so only the order of the insertions is important. Our analysis then implies that this correlator can be computed in terms of an ordinary $r$-fold integral given by
\begin{align}
\langle \cO^{(1)}(\varphi_1)\cdots \cO^{(n)}(\varphi_n)\rangle_{S^3} &= 
\frac{1}{Z_{S^3}}\sum_{\vec{B}\in\Gamma_m}\int_{\R^r} [d\vec\sigma]_{\vec B}\, \Psi_-(\vec{\sigma},\vec{B}) \widehat{\cO}^{(1)}_{N} \cdots \widehat{\cO}^{(n)}_{N} \Psi_+(\vec{\sigma},\vec{B}) \ed \label{1dcorGenS}
\end{align}
Let us now unpack the notation in \eqref{1dcorGenS}:
\begin{itemize}
	\item The $\Psi_{\pm}(\vec{\sigma},\vec{B})$ represent wavefunctions defined by the path integral on the hemispheres $HS^3_{\pm}\cong B^3$ evaluated with certain half-BPS boundary conditions on $\partial HS^3_{\pm}\cong S^2$. We will show in Section \ref{SGLUE} that these boundary conditions are parametrized by constants $\vec{\sigma}\in\mathbb{R}^r$ and by the monopole charge $\vec{B}\in \Gamma_m$. In particular, the vacuum wavefunctions $\Psi_{\pm}(\vec{\sigma},\vec{B})$, which have zero monopole charge, are given by\footnote{In general, the above correlator can be written as $(1/Z_{S^3})\sum_{\vec B}\int [d\vec\sigma]_{\vec B} \Psi_1(\vec{\sigma},\vec{B}) \Psi_2(\vec{\sigma},\vec{B})$, where $\Psi_1$ and $\Psi_2$ are hemisphere wavefunctions with arbitrary insertions.  In \eqref{1dcorGenS}, we represent insertions by shift operators acting only on the (empty) upper hemisphere wavefunction, in which case the sum over $\vec B$ collapses to the $\vec{B}=0$ term.}
	\begin{align}
	\Psi_{\pm}(\vec{\sigma},\vec{B}) = \delta_{\vec{B},\vec{0}} \prod_{I=1}^{N_h}\frac{1}{\sqrt{2\pi}}\Gamma\left(\frac{1}{2}-i \vec{q}_I\cdot\vec{\sigma} \right)\ed \label{Psipm}
	\end{align}
	The variables $\vec{\sigma}$ arise from localization of scalars in the vector multiplet. 
	
	\item In \eqref{1dcorGenS}, each of the twisted CBOs $\cO^{(k)}$ is represented by a certain shift operator, denoted by $\widehat{\cO}^{(k)}_{N}$, acting on the $HS^3_+$ wavefunction $\Psi_+(\vec{\sigma},\vec{B})$. The label $N$ on the $\widehat{\cO}^{(k)}_{N}$ implies that it represents an insertion of $\cO^{(k)}$ through the North pole of $\partial HS^
	3_{\pm}=S^2$, labeled by $N$ in Figure \ref{fig:sphere}. The order in which the shift operators $\widehat{\cO}^{(k)}_{N}$ act on $\Psi_+$ represents the order of insertions on the semicircle. There is a second set of shift operators $\widehat{\cO}^{(k)}_{S}$ representing insertions through the South pole (labeled by $S$ in Figure \ref{fig:sphere}), such that the same correlator \eqref{1dcorGenS} is given by
	\begin{align}
	\langle \cO^{(1)}(\varphi_1)\cdots \cO^{(n)}(\varphi_n)\rangle_{S^3} &=
	\frac{1}{Z_{S^3}}\sum_{\vec{B}\in\Gamma_m}\int_{\R^r} [d\vec{\sigma}]_{\vec B}\,\Psi_-(\vec{\sigma},\vec{B}) \widehat{\cO}^{(n)}_{S} \cdots \widehat{\cO}^{(1)}_{S} \Psi_+(\vec{\sigma},\vec{B}) \ed \label{1dcorGenN}
	\end{align}
	The order in which the $S$ operators act on $\Psi_+$ also represents the order of insertions on the semicircle, but in the opposite direction. The shift operators corresponding to the bare twisted monopoles $\cM^{\vec{b}}(\varphi)$ and the vector multiplet scalars $\vec{\Phi}(\varphi)$ are written explicitly in \eqref{MN}, \eqref{MS}, and \eqref{hatPhiNS}, respectively. It is important that the shift operators do not depend on the insertion point. This must be the case because the correlators are topological and depend only on the order of the insertions, which is reflected in the nontrivial commutation relations between the shift operators.
	
	\item The $HS^3_{\pm}$ wavefunctions can be glued into a partition function on $S^3$ with the measure as in \eqref{1dcorGenS}, where $[d\vec{\sigma}]_{\vec{B}}$ is given explicitly by 
	\begin{align}
	[d\vec{\sigma}]_{\vec{B}}&= \mu(\vec{\sigma}, \vec{B})\, d^r\sigma\ec\cr
	\mu(\vec{\sigma}, \vec{B}) &= \prod_{I=1}^{N_h}(-1)^{\frac{|\vec{q}_I\cdot\vec{B}|-\vec{q}_I\cdot\vec{B}}{2}}\frac{\Gamma\left(\frac{1+|\vec{q}_I\cdot\vec{B}|}{2}+i\vec{q}_I\cdot\vec{\sigma}\right)}{\Gamma\left(\frac{1+|\vec{q}_I\cdot\vec{B}|}{2}-i\vec{q}_I\cdot\vec{\sigma}\right)} \ed \label{dsig}
	\end{align}
	This measure is simply the $S^2$ partition function of $N_h$ chiral multiplets in a 2D $\cN=(2,2)$ theory, coupled to $U(1)^r$ vector multiplets with magnetic charge $\vec{B}$ \cite{Benini:2016qnm}. We have normalized the correlators \eqref{1dcorGenS} by the $S^3$ partition function $Z_{S^3}$, such that $\langle 1\rangle_{S^3}=1$. 		
	\item The above expressions can be generalized straightforwardly to include deformations by real masses and FI parameters. This will be described in Section \ref{massFI}.
\end{itemize}
The above description of correlators of twisted CBOs in terms of hemispheres and shift operators, while derived using localization in 3D, was inspired by computations of Schur indices with line defects in 4D $\cN=2$ theories \cite{Dimofte:2011py,Gang:2012yr,Cordova:2016uwk}.\footnote{In turn, the interpretation of loop operator insertions on $S^3\times S^1$ as shift operators acting on half-indices in \cite{Dimofte:2011py,Gang:2012yr,Cordova:2016uwk} was inspired by earlier works \cite{Alday:2009fs, Drukker:2009id, Gomis:2010kv}, where loop operator insertions on $S^4$ were also understood as shift operators acting on the $HS^4$ wavefunction, as derived via localization in \cite{Gomis:2011pf}.} In fact, as we show in Section \ref{multipleinsertions}, these problems are closely related. The defect Schur index can be computed by a path integral on $S^3\times S^1$ with 't~Hooft-Wilson loops wrapping the $S^1$. To preserve supersymmetry, the defects should be inserted at points along a great circle in $S^3$. As we will show, upon dimensional reduction of the 4D index along $S^1$, the line defects become twisted CBOs in the 3D dimensionally reduced theory. The above expressions for correlators of twisted CBOs can all be derived from the 4D defect Schur index, providing a strong consistency check of our results.

\section{Preliminaries} \label{prelim}

In this section, we set the stage for the problems that we study in the rest of the paper. We start by reviewing the construction of $\cN=4$ supersymmetric Lagrangians using vector multiplets and hypermultiplets on $S^3$. We then describe a BPS sector of these theories that is captured by a 1D theory, focusing on the case of the Coulomb branch.  Finally, we give a careful definition of BPS monopole operators, which are of primary interest in this paper, and explain some of their properties.

In this section, we try to be maximally general and define everything for non-abelian gauge theories. However, the actual localization computations in the rest of the paper will be performed only for abelian theories.

\subsection{$\cN=4$ Theories on $S^3$}

The theories that we analyze in this paper are Lagrangian 3D $\cN=4$ gauge theories. We start by giving a short review of their structure and summarizing our conventions, referring the reader to \cite{Dedushenko:2016jxl} for more details. 

\subsubsection{Supersymmetry Algebra}

$\cN=4$ supersymmetry on $S^3$ is based on the superalgebra $\mathfrak{su}(2|1)_\ell \oplus \mathfrak{su}(2|1)_r$ or a central extension thereof. Its even subalgebra contains the $\mathfrak{su}(2)_\ell\oplus \mathfrak{su}(2)_r$ isometries of $S^3$, whose generators we denote by $J^{(\ell)}_{\alpha\beta}$ and $J^{(r)}_{\alpha\beta}$, as well as the R-symmetry subalgebra $\mathfrak{u}(1)_\ell\oplus \mathfrak{u}(1)_r$ generated by $R_\ell$ and $R_r$. The odd generators are denoted by $\cQ_{\alpha}^{(\ell_{\pm})}$ and $\cQ_{\alpha}^{(r_{\pm})}$.\footnote{Above, $\alpha,\beta,\ldots=1,2$ are spinor indices. They can be raised and lowered from the left with the anti-symmetric symbols $\varepsilon_{\alpha\beta}$, $\varepsilon^{\alpha\beta}$, where $\varepsilon_{21}=\varepsilon^{12}=1$. The same raising/lowering convention will also be used for the fundamental indices of $\mathfrak{su}(2)$ R-symmetries.  See Appendix \ref{conventions} for a full list of our conventions.} The algebra obeyed by $J^{(\ell)}_{\alpha\beta}$, $R_\ell$, $\cQ_{\alpha}^{(\ell_{\pm})}$ is
\begin{alignat}{3}
[J^{(\ell)}_i, J^{(\ell)}_j]  &= i \epsilon_{ijk} J^{(\ell)}_k \ecq
&[J^{(\ell)}_{\alpha \beta}, \cQ^{(\ell_{\pm})}_{\gamma}] &= \frac{1}{2}\left(\varepsilon_{\alpha\gamma} \cQ^{(\ell_{\pm})}_{\beta} + \varepsilon_{\beta \gamma } \cQ^{(\ell_{\pm})}_{\alpha} \right) \ec\\
[R_{\ell}, \cQ_{\alpha}^{(\ell_{\pm})}] &= \pm \cQ_{\alpha}^{(\ell_{\pm})} \ecq &\{\cQ^{(\ell_+)}_{\alpha}, \cQ^{(\ell_-)}_{\beta}\} &= -\frac{4i}{r}\left(J_{\alpha\beta}^{(\ell)} + \frac{1}{2} \varepsilon_{\alpha\beta}R_\ell\right)\ec \label{Commutators}
\end{alignat}
where we have set
\begin{align}
J_{\alpha \beta}^{(\ell)} \equiv \begin{pmatrix}
-(J_1^{(\ell)} + i J_2^{(\ell)}) & J_3^{(\ell)} \\
J_3^{(\ell)} & J_1^{(\ell)} - i J_2^{(\ell)}
\end{pmatrix} \ed
\end{align}
The generators of $\mathfrak{su}(2|1)_r$ obey the same relations with $\ell \to r$.

The generators $J_i^\ell$ and $J_i^r$ act by Lie derivatives $\cL_{v_i^{\ell}}$ and $\cL_{v_i^r}$ with respect to the left- and right-invariant vector fields $v_i^{\ell}$ and $v_i^r$ on $S^3$. The generators $J_3^\ell$ and $J_3^r$ will often be important to us, and their corresponding vector fields are given by
\begin{equation}
v_3^\ell=-\frac{i}2(\partial_\tau +\partial_\varphi)\ec \quad v_3^r = -\frac{i}2 (\partial_\tau - \partial_\varphi) \ed
\end{equation}
Above, we have used coordinates that exhibit $S^3$ as a $U(1)$ fibration over a disk $D^2$ with the fiber shrinking at the boundary, which will be useful in the remainder of the paper (see Appendix \ref{coordinates} for details). Explicitly, let us embed $S^3$ in $\mathbb{R}^4$ as
\begin{equation}
X_1^2 + X_2^2 + X_3^2 + X_4^2 = r^2
\end{equation}
and parametrize the $X_i$ by
\begin{equation}
X_1 + iX_2 = r \cos\theta e^{i\tau}\ec \quad X_3 + iX_4 = r\sin\theta e^{i\varphi} \ec \label{roundvars}
\end{equation}
where $0\leq \theta\leq \frac{\pi}{2}$ and $-\pi\leq \varphi,\tau\leq \pi$. In these coordinates, $\sin\theta e^{i\varphi}$ parametrizes the unit disk, and $e^{i\tau}$ the $U(1)$ fiber. We also sometimes use the notation 
\begin{equation}
P_\tau = -(J_3^\ell + J_3^r)\ec \quad P_\varphi = -J_3^\ell + J_3^r
\end{equation}
to denote the $\tau$ and $\varphi$ rotation isometries of $S^3$.

It is convenient to think of $\mathfrak{su}(2|1)_\ell \oplus \mathfrak{su}(2|1)_r$ as a subalgebra of the 3D $\cN=4$ superconformal algebra $\mathfrak{osp}(4|4)$, whose R-symmetry subalgebra is $\mathfrak{so}(4)\cong \mathfrak{su}(2)_H\oplus\mathfrak{su}(2)_C$. 
This embedding is parametrized by the choice of the $\mathfrak{u}(1)_\ell \oplus \mathfrak{u}(1)_r$ subalgebra of $\mathfrak{su}(2)_H\oplus \mathfrak{su}(2)_C$, which is specified by the Cartan elements
\begin{equation}
h_a{}^b\in \mathfrak{su}(2)_H,\quad \overline{h}^{\dot a}{}_{\dot b}\in \mathfrak{su}(2)_C \ec
\end{equation}
where $a,b,\ldots = 1,2$ ($\dot{a},\dot{b},\ldots=1,2$) label the fundamental irrep of $\mathfrak{su}(2)_H$ ($\mathfrak{su}(2)_C$).  Here, $h_a{}^b$ and $\overline{h}^{\dot a}{}_{\dot b}$ are traceless Hermitian matrices satisfying $h_a{}^c h_c{}^b = \delta_a{}^b$ and $\bar{h}^{\dot{a}}{}_{\dot{c}}\bar{h}^{\dot{c}}{}_{\dot{b}} = \delta^{\dot{a}}{}_{\dot{b}}$.
They determine a relation between the generators $R_\ell$, $R_r$ of $\mathfrak{u}(1)_\ell\oplus \mathfrak{u}(1)_r$ and the generators $R_a{}^b$, $\overline{R}^{\dot a}{}_{\dot b}$ of $\mathfrak{su}(2)_H\oplus \mathfrak{su}(2)_C$:
\begin{equation}
\frac12 (R_\ell + R_r)  =\frac12 h_a{}^bR_b{}^a\equiv R_H,\quad \frac12 (R_\ell - R_r) = \frac12 \overline{h}^{\dot a}{}_{\dot b}\overline{R}^{\dot b}{}_{\dot a}\equiv R_C \ed
\end{equation}
The superconformal symmetries of $\mathfrak{osp}(4|4)$ are parametrized by conformal Killing spinors $\xi_{\alpha a\dot a}$ satisfying the conformal Killing spinor equations on $S^3$:
\begin{equation}
\label{killing1}
\nabla_\mu \xi_{a\dot a}=\gamma_\mu\xi_{a\dot a}' \ecq \nabla_\mu \xi_{a\dot a}'=-\frac{1}{4r^2}\gamma_\mu\xi_{a\dot a} \ec
\end{equation}
where $\gamma_\mu$ are curved-space gamma matrices and $r$ is the radius of $S^3$ (the first equation implies the second via $\gamma^\mu\nabla_\mu\xi_{a\dot{a}}' = -\frac{1}{8}R\xi_{a\dot{a}}$ where $R_{S^3} = 6/r^2$). Those that correspond to supersymmetries within the subalgebra $\mathfrak{su}(2|1)_\ell \oplus \mathfrak{su}(2|1)_r$ satisfy the additional condition
\begin{equation}
\label{killing2}
\xi_{a\dot a}'=\frac{i}{2r}h_a{}^b \xi_{b\dot b}\overline{h}^{\dot b}{}_{\dot a} \ed
\end{equation}
To conform with previous works, we use the convention that
\begin{equation}
\label{Cartan_choice}
h_a{}^b=-\sigma^2,\quad \overline{h}^{\dot a}{}_{\dot b}=-\sigma^3 \ed
\end{equation}
Different choices of $h, \bar{h}$ are related by conjugation with $SU(2)_H\times SU(2)_C$ and, as will be explained shortly, determine which components in the triplets of FI and mass parameters can be present on the sphere: $\zeta = h_a{}^b(\zeta_\text{flat})_b{}^a$ and $m = \bar{h}^{\dot a}{}_{\dot b}(m_\text{flat})^{\dot b}{}_{\dot a}$. In Appendix \ref{SUGRA_BG}, we describe how the $\mathfrak{su}(2|1)_{\ell}\oplus\mathfrak{su}(2|1)_r$ algebra is obtained from the rigid limit of off-shell 3D $\cN=4$ conformal supergravity, following the philosophy of \cite{Festuccia:2011ws}. The latter point of view elucidates the origin of the matrices $h$ and $\bar{h}$ as background values for scalar fields within a certain 3D Kaluza-Klein supergravity multiplet.

\subsubsection{Lagrangians}

The supersymmetry algebra just described acts in Lagrangian theories constructed from a vector multiplet $\cV$ and a hypermultiplet $\cH$. The vector multiplet transforms in the adjoint representation of the gauge group $G$ and has components
\begin{align}
\cV = (A_{\mu}, \lambda_{\alpha a\dot a}, \Phi_{\dot{a}\dot b}, D_{ab})\ec \label{Vmul}
\end{align}
consisting of the gauge field $A_{\mu}$, gaugino $\lambda_{\alpha a\dot a}$, and scalars $\Phi_{\dot{a}\dot b}=\Phi_{\dot{b}\dot a}$ and $D_{ab}=D_{ba}$, which transform in the trivial, $(\mathbf{2},\mathbf{2})$, $(\mathbf{1},\mathbf{3})$, and $(\mathbf{3},\mathbf{1})$ irreps of the $\mathfrak{su}(2)_H\oplus\mathfrak{su}(2)_C$ R-symmetry, respectively. The hypermultiplet transforms in some unitary representation $\cR$ of $G$ and has components
\begin{align}
\cH = (q_a, \tq_a, \psi_{\alpha\dot a}, \tpsi_{\alpha\dot a}) \label{Hmul}
\end{align}
where $q_a, \tq_a$ are scalars transforming as $(\mathbf{2},\mathbf{1})$ under the R-symmetry and as $\cR, \overline{\cR}$ under $G$, respectively, while $\psi_{\alpha\dot a}, \tpsi_{\alpha\dot a}$ are their fermionic superpartners and transform as $(\mathbf{1},\mathbf{2})$ under the R-symmetry. 
The SUSY transformations of $\cV$ and $\cH$ are collected in Appendix \ref{susytrans}.

The action for $\cH$ coupled to $\cV$ is 
\begin{align}
S_{\text{hyper}}[\cH,\cV] &= \int d^3x \sqrt{g} \bigg[ \cD^{\mu}\tq^{a} \cD_{\mu} q_a - i\tpsi^{\dot{a}}\slashed{D}\psi_{\dot{a}} + \frac{3}{4r^2} \tq^{a} q_a + i \tq^{a} D_a{}^b q_b  \notag\\
&\hspace{3 cm}- \frac{1}{2}\tq^a \Phi^{\dot{a}\dot{b}}\Phi_{\dot{a}\dot{b}}q_a - i\tpsi^{\dot{a}}\Phi_{\dot{a}}{}^{\dot{b}}\psi_{\dot{b}} +i\left( \tq^ a\lambda_a{}^{\dot{b}}\psi_{\dot{b}} + \tpsi^{\dot{a}}\lambda^b{}_{\dot{a}}q_b\right)\bigg] \ec \label{Shyper}
\end{align}
which actually preserves the full superconformal symmetry $\mathfrak{osp}(4|4)$. The super Yang-Mills action preserves only the $\mathfrak{su}(2|1)_\ell \oplus \mathfrak{su}(2|1)_r$ subalgebra and is given by
\begin{align}
S_{\text{YM}}[\cV] &= \frac{1}{g^2_\text{YM}}\int d^3x\sqrt{g}\trace\bigg(F^{\mu\nu}F_{\mu\nu}  - \cD^{\mu}\Phi^{\dot{c}\dot d}\cD_{\mu}\Phi_{\dot{c}\dot d} +i\lambda^{a\dot a}\slashed{\cD}\lambda_{a\dot a} - D^{cd}D_{cd}  -i \lambda^{a\dot a}[\lambda_a{}^{\dot b},\Phi_{\dot{a}\dot b}] \notag\\
&\hspace{-0.25 cm}- \frac{1}{4}[\Phi^{\dot a}{}_{\dot b}, \Phi^{\dot c}{}_{\dot d}][\Phi^{\dot b}{}_{\dot a}, \Phi^{\dot d}{}_{\dot c}]- \frac{1}{2r}h^{ab}\bar{h}^{\dot{a}\dot b}\lambda_{a\dot a}\lambda_{b\dot b} + \frac{1}{r}(h_a{}^bD_b{}^a)(\bar{h}^{\dot a}{}_{\dot b}\Phi^{\dot b}{}_{\dot a}) -\frac{1}{r^2}\Phi^{\dot{c}\dot{d}}\Phi_{\dot{c}\dot d}\bigg) \ed \label{SYM}
\end{align}
The theory \eqref{Shyper} has flavor symmetry group $G_H\times G_C$, whose Cartan subalgebra we denote by $\mathfrak{t}_H\oplus\mathfrak{t}_C$. The factor $G_H$ acts on the hypermultiplets, while $G_C\cong U(1)^\text{$\# U(1)$'s in $G$}$ contains the topological $U(1)$ symmetries that act on monopole operators.\footnote{$G_C$ may be enhanced to a non-abelian group in the IR.} It is possible to couple the theory to a supersymmetric background twisted vector multiplet in $\mathfrak{t}_C$, which on $S^3$ leads to a single FI parameter $\zeta$ for every $U(1)$ factor of the gauge group (as opposed to an $\mathfrak{su}(2)_H$ triplet on $\mathbb{R}^3$). The corresponding FI action is given by
\begin{align}
S_{\text{FI}}[\cV] &= i\sum_{I=1}^{\dim(\mathfrak{t}_C)}\zeta_I\int d^3x\sqrt{g}\left(h_a{}^b (D^{(I)})_b{}^a - \frac{1}{r} \bar{h}^{\dot a}{}_{\dot b}(\Phi^{(I)})^{\dot b}{}_{\dot a} \right) \ec\label{FIAction}
\end{align}
where $D^{(I)}_{ab}$ and $\Phi^{(I)}_{\dot{a}\dot b}$ are the scalars in the vector multiplet gauging the $I^\textrm{th}$ $U(1)$ factor of $G$.
Similarly, one can introduce real masses for the hypermultiplets by turning on background vector multiplets $\cV_{\rm b.g.}$ in $\mathfrak{t}_H$. In order to preserve supersymmetry, all the components of $\cV_{\text{b.g.}}$ are set to zero except for
\es{PhiDBack}{
	\widehat{m}=-\frac{1}{2}\bar{h}^{\dot a}{}_{\dot b}(\Phi_{\text{b.g.}})^{\dot b}{}_{\dot a} = \frac{r}{2}h_a{}^b(D_{\text{b.g.}})_b{}^a \ed
}
In particular, on $S^3$, there is a single real mass parameter for every generator in $\mathfrak{t}_H$ (as opposed to an $\mathfrak{su}(2)_C$ triplet on $\mathbb{R}^3$). In the presence of nonzero real mass and FI parameters, the $\mathfrak{su}(2|1)_\ell \oplus \mathfrak{su}(2|1)_r$ algebra is centrally extended by charges $Z_\ell$ and $Z_r$ for the respective factors of the superalgebra. The central charges are related to the mass/FI parameters by
\begin{equation}
\label{CentrExt}
\frac1{r} (Z_\ell + Z_r)=i\widehat{m}\in i\mathfrak{t}_H,\quad \frac1{r}(Z_\ell - Z_r)=i\widehat{\zeta}\in i\mathfrak{t}_C \ed
\end{equation}
A more detailed description of the superalgebras can be found in \cite{Dedushenko:2016jxl}.

Finally, let us specify the contour of integration in the path integral. Because we work in Euclidean signature, the fermionic fields do not obey any reality conditions, while the bosonic fields satisfy
\begin{equation}
\label{RealqAPhD}
q^\dagger_a = \tq^a, \quad A_\mu^\dagger = A_\mu, \quad \Phi_{\dot a\dot b}^\dagger = -\Phi^{\dot a\dot b}, \quad D_{ab}^\dagger = -D^{ab},
\end{equation}
where the Hermitian conjugate is taken in the corresponding representation.

\subsubsection{Abelian Gauge Theories}
\label{abeliantheories}

In the bulk of the paper, we will focus exclusively on abelian gauge theories.  Specifically, we will consider a $G=U(1)^r$ gauge theory coupled to $N_h$ hypermultiplets with gauge charges $\vec{q}_I=(q_I^1,\dots, q_I^r)\in \Z^r$, where $I=1,\dots, N_h$. The maximal tori of the global symmetry algebras in this case are given by $\mathfrak{t}_H\cong U(1)^{N_h-r}$ and $\mathfrak{t}_C\cong U(1)^r$. The hypermultiplets transform under $G_H$ with weights $\vec{Q}_I = (Q_I^1,\ldots,Q_I^{N_h-r})\in\mathbb{Z}^{N_h-r}$, while monopole operators transform under the topological symmetry $G_C$ with charges  $\vec{b}\in\Gamma_m\subset\mathbb{R}^r$. The monopole charge lattice $\Gamma_m\subset\mathbb{R}^r$ is defined through Dirac quantization by the constraints $\vec{q}\cdot \vec{b}\in \Z$ where $\vec{q}$ ranges over all gauge charges allowed in the theory. We assume throughout this paper that charges have been normalized such that $\Gamma_m = \Z^r$.

\subsection{Twisted Operators and the 1D Theory}

Supersymmetric field theories with eight supercharges in various dimensions have subsectors of operators which can be described by lower dimensional theories. Our 3D $\cN=4$ theories are among those that have such sectors, which, moreover, turn out to furnish certain 1D theories. This fact was originally noticed for SCFTs in \cite{Beem:2013sza}, further developed in \cite{Chester:2014mea,Beem:2016cbd}, and extended to non-conformal $\cN=4$ theories on $S^3$ in \cite{Dedushenko:2016jxl}.

Following \cite{Dedushenko:2016jxl}, we consider two pairs of supercharges within $\mathfrak{su}(2|1)_\ell\oplus\mathfrak{su}(2|1)_r$.\footnote{The embedding of these supercharges inside $\mathfrak{osp}(4|4)$ is given in Appendix \ref{susytrans}.} 
Those associated with the Higgs branch are
\begin{equation}
\cQ^H_1=\cQ_1^{(\ell +)}+\cQ_1^{(r-)}\ec \quad \cQ_2^H=\cQ_2^{(\ell-)}+\cQ_2^{(r+)} \ec \label{QH12} 
\end{equation}
and those associated with the Coulomb branch are
\begin{equation}
\cQ^C_1=\cQ_1^{(\ell +)}+\cQ_1^{(r+)}\ec \quad \cQ^C_2=\cQ_2^{(\ell-)}+\cQ_2^{(r-)} \ed \label{QC12}
\end{equation}
Each of these four supercharges is nilpotent.  There exists a 1D theory associated with co\-homology classes of $\cQ^H_{1,2}$ and another associated with those of $\cQ^C_{1,2}$. To see this, let us focus on the (equivariant) cohomology of $\cQ_\beta^H=\cQ_1^H + \beta \cQ_2^H$ or $\cQ_\beta^C=\cQ_1^C + \beta \cQ_2^C$ acting on local operators, for an arbitrary constant $\beta\neq 0$. Because of the relations
\begin{align}
(\cQ^H_\beta)^2 &= \frac{4i\beta}{r}(P_\tau + R_C + ir\widehat{\zeta}) \ec \label{QHsqr}\\
(\cQ^C_\beta)^2 &= \frac{4i\beta}{r}(P_\tau + R_H + ir\widehat{m}) \ec \label{QCsqr}
\end{align}
local operators in the cohomology of $\cQ_{\beta}^H$ or $\cQ^C_{\beta}$ must be annihilated by the right-hand side of \eqref{QHsqr} or \eqref{QCsqr},  respectively. This implies that local operators can only be inserted at the fixed points of the $P_\tau$ isometry, which form a great circle parametrized by $\varphi$ at $\theta=\pi/2$, where the $\tau$-circle shrinks (see \eqref{roundvars}).\footnote{It also follows from \eqref{QHsqr} (or \eqref{QCsqr}) that the spins and R-charges of $\cQ^H_{\beta}$- (or $\cQ^C_{\beta}$-) closed operators should be related. However, this constraint turns out to be trivial because all these operators turn out to be Lorentz scalars transforming trivially under  $\mathfrak{su}(2)_C$ (or $\mathfrak{su}(2)_H$).}  In flat space, $P_\tau$ is the rotation that fixes the line along which operators are inserted.

Another important property emphasized in \cite{Dedushenko:2016jxl} is that 
\begin{align}
\{\cQ_\beta^H, \dots\} &= P_\varphi + R_H + ir\widehat{m}\ec \label{cohomologousH} \\
\{\cQ_\beta^C, \dots\} &= P_\varphi + R_C + ir\widehat{\zeta}\ec \label{cohomologousC}
\end{align}
which leads to the definitions of twisted translations:
\begin{align}
\widehat{P}^H_\varphi &= P_\varphi + R_H\ec\label{HtwistedTR}\\
\widehat{P}^C_\varphi &= P_\varphi + R_C\ed\label{twistedTR}
\end{align}
The twisted translations $\widehat{P}^H_\varphi$ (or $\widehat{P}^C_\varphi$) are $\cQ^H_{\beta}$- (or $\cQ^C_{\beta}$-) closed, and can therefore be used to translate cohomology classes along the great $\varphi$-circle. The cohomology classes of $\cQ^H_{\beta}$ and $\cQ^C_{\beta}$ therefore form two distinct 1D theories. Furthermore, when $\widehat{m}=0$ (or $\widehat{\zeta}=0$), the twisted translation $\widehat{P}^H_{\varphi}$ (or $\widehat{P}^C_{\varphi}$) is exact under $\cQ^H_{\beta}$ (or $\cQ^C_{\beta}$). The twisted-translated cohomology classes then become independent of the position $\varphi$ along the circle. In such a situation, the cohomology classes furnish a 1D TQFT, meaning that their OPE is independent of the separation between operators, but can depend on their ordering along the circle. This OPE therefore determines an associative but non-commutative product, which can be thought of as a star product on some variety. 

The operators in the cohomology are most easily classified at the superconformal point, where the symmetry is enhanced to $\mathfrak{osp}(4|4)$. In this case, one finds that for every fixed insertion point $\varphi$, the operators in the cohomology of $\cQ_\beta^H$ and $\cQ_\beta^C$ are in the Higgs and Coulomb branch chiral rings, respectively, with respect to some $\cN=2$ superconformal subalgebra of $\mathfrak{osp}(4|4)$.\footnote{In particular, the star product in the 1D TQFT then yields a deformation quantization of the chiral ring, which describes the Higgs or Coulomb branch of the moduli space of the theory as a complex variety; this point of view was advocated in \cite{Beem:2016cbd}.} Indeed, for SCFTs, we have the algebraic relations
\begin{align}
\{\cQ_1^H,\cQ_1^{H\dagger}\}&=\{\cQ_2^H,\cQ_2^{H\dagger}\}=8(D-R_1{}^1)\ec \label{QQdagH}\\
\{\cQ_1^C,\cQ_1^{C\dagger}\}&=\{\cQ_2^C,\cQ_2^{C\dagger}\}=\textstyle 8(D-\frac12(\bar{R}_{\dot1}{}^{\dot2} + \bar{R}_{\dot2}{}^{\dot1}))\ec \label{QQdagC}
\end{align}
where $D$ is the generator of dilatations. The relation \eqref{QQdagH}, together with the state-operator map (which yields an inner product, hence a notion of adjoint in radial quantization) and the standard Hodge theory reasoning (which exhibits a unique harmonic representative of each cohomology class), implies that representatives of the cohomology of $\cQ_\beta^H$, when inserted at the origin, satisfy
\begin{equation}D=R_1{}^1\ed \end{equation}
Such operators belong to the Higgs branch chiral ring. They are the $\mathfrak{su}(2)_H$ highest-weight components of HBOs $H_{a_1\cdots a_n}$, which are half-BPS superconformal primaries transforming in the spin-$\frac{n}{2}$ irrep of $\mathfrak{su}(2)_H$, and are Lorentz scalars of dimension $\Delta=\frac{n}{2}$. Similarly, \eqref{QQdagC} implies that the representatives of $\cQ_\beta^C$ cohomology at the origin satisfy
\begin{equation}D=\frac12(\bar{R}_{\dot1}{}^{\dot2} + \bar{R}_{\dot2}{}^{\dot1})\ec \end{equation} 
which is the defining relation of Coulomb branch chiral ring operators for the appropriate choice of $\mathfrak{u}(1)_C\subset \mathfrak{su}(2)_C$. They are the $\mathfrak{su}(2)_C$ highest-weight components of CBOs $C_{\dot{a}_1\cdots \dot{a}_m}$, which have the same quantum numbers as HBOs with $\mathfrak{su}(2)_H$ interchanged with $\mathfrak{su}(2)_C$.  

To define the operators in the cohomology away from the origin, one simply applies the appropriate twisted translation \eqref{HtwistedTR} or \eqref{twistedTR}. For the HBOs $H_{a_1\cdots a_n}$, the corresponding twisted-translated operator is given by
\begin{equation}
\label{twistedH}
H(\varphi) = H_{a_1\cdots a_n}u^{a_1}\cdots u^{a_n}\ec \quad u^a=\left( \begin{matrix}
\cos\frac{\varphi}{2}\cr 
\sin\frac{\varphi}{2}
\end{matrix} \right) \ed
\end{equation}
For the CBOs $C_{\dot a_1\cdots \dot a_m}$, the corresponding twisted-translated operator is given by
\begin{equation}
\label{twistedC}
C(\varphi) = C_{\dot a_1\cdots \dot a_n}v^{\dot a_1}\cdots v^{\dot a_m}\ec \quad v^{\dot a}=\left( \begin{matrix}
\frac1{\sqrt{2}}e^{i\varphi/2}\cr 
\frac1{\sqrt{2}}e^{-i\varphi/2}
\end{matrix} \right) \ed	
\end{equation}
In \eqref{twistedH} and \eqref{twistedC}, it is understood that the operators are restricted to the $\theta=\frac{\pi}{2}$ circle. The reason that $u^a$ and $v^{\dot a}$ are different in \eqref{twistedH} and \eqref{twistedC} is that in defining the $\mathfrak{su}(2|1)_{\ell}\oplus\mathfrak{su}(2|1)_r$ algebra on $S^3$, we chose different Cartan elements \eqref{Cartan_choice} for $\mathfrak{su}(2)_H$ and for $\mathfrak{su}(2)_C$. Because the translation in \eqref{HtwistedTR} (or \eqref{twistedTR}) is accompanied by an R-symmetry rotation, the twisted operators \eqref{twistedH}, \eqref{twistedC} at $\varphi=0$ and $\varphi\ne 0$ are both in chiral rings, but with respect to distinct Cartan elements of $\mathfrak{su}(2)_H$ (or $\mathfrak{su}(2)_C$). This twist allows us to go beyond the chiral ring data. In particular, cohomology classes at different points $\varphi$ are not mutually chiral, and may thus have nontrivial SCFT correlators.

Above, we have formally classified operators in the cohomology within SCFTs. In practice, for what follows, we need a definition of such operators along RG flows on $S^3$, where only $\mathfrak{su}(2|1)_{\ell}\oplus\mathfrak{su}(2|1)_r \subset \mathfrak{osp}(4|4)$ is preserved. Some of the properties mentioned above for HBOs, CBOs, and their twisted analogs then become imprecise, and we would like to clarify some possible confusions. In particular, along the flow, the $\mathfrak{su}(2)_{H,C}$ symmetries are broken to their $\mathfrak{u}(1)_{H,C}$ Cartans. The operators $H_{a_1\cdots a_n}$ and $C_{\dot{a}_1\cdots \dot{a}_n}$ are generally still present, but their different $a_i,\dot{a}_i=1,2$ components are no longer related by $\mathfrak{su}(2)_{H,C}$, and their correlators therefore need not respect these symmetries away from the fixed point. However, the twisted operators \eqref{twistedH} and \eqref{twistedC} are still in the cohomology, and this notion is well-defined along the flow. For example, the components $q_1$ and $q_2$ of the hypermultiplet scalars need not be related by $\mathfrak{su}(2)_H$ along the flow. Nevertheless, they are still well-defined operators, and the twisted operator $Q(\varphi) = \cos\frac{\varphi}{2}q_1(\varphi) + \sin\frac{\varphi}{2}q_2(\varphi)$ is still in $\cQ^H_{\beta}$-cohomology. 
Furthermore, we stress that $H(\varphi)$ and $C(\varphi)$ are not chiral with respect to any $\cN=2$ subalgebra of the $\mathfrak{su}(2|1)_{\ell}\oplus\mathfrak{su}(2|1)_r$ symmetry preserved along the flow; they become chiral with respect to certain such subalgebras of $\mathfrak{osp}(4|4)$, which is only realized at the fixed point. Nevertheless, it can be checked by inspection that they are half-BPS under $\mathfrak{su}(2|1)_{\ell}\oplus\mathfrak{su}(2|1)_r$.\footnote{In particular, $Q(\varphi=\pi/2) = (q_1(\pi/2)+q_2(\pi/2))/\sqrt{2}$ is invariant not only under the $\smash{\cQ^H_{1,2}}$ in \eqref{QH12}, but also under $\cQ^{(\ell+)}_2-\cQ^{(r-)}_2$ and $\cQ^{(\ell-)}_1-\cQ^{(r+)}_1$, as can be checked by using the explicit SUSY transformations \eqref{qpsivar}. Similarly, the twisted CBOs $C(\varphi=\pi/2)\vphantom{\cQ^{(\ell)}_1}$ that will be constructed explicitly for our theories in the following sections can be checked to be invariant under $\cQ^C_{1,2}$ as well as $\cQ^{(\ell-)}_1-\cQ^{(r-)}_1$ and $\cQ^{(\ell+)}_2-\cQ^{(r+)}_2$. \label{halfBPStip}}

\subsection{Coulomb Branch Operators}\label{CoulOps}

In \cite{Dedushenko:2016jxl}, the Higgs branch case was studied in detail, and all twisted HBOs were constructed from the hypermultiplet scalars. Our focus here is on the Coulomb branch, so let us first understand what the corresponding observables are.

Twisted CBOs are observables in the cohomology of $\cQ_\beta^C$. If we try to construct them from local fields, we find that there is only one such operator:
\begin{equation}
\Phi(\varphi) = \Phi_{\dot a\dot b}(\varphi)v^{\dot a} v^{\dot b}\biggr|_{\theta=\frac{\pi}{2}}.
\end{equation}
However, it is well-known that a complete picture of the Coulomb branch must also include monopole operators. 
Let us first summarize the prescription for inserting these operators, before providing a more detailed explanation. A twisted-translated monopole operator inserted at the point $p$ with coordinate $\varphi$ along the great circle is defined via the following prescription:
\begin{itemize}
	\item Pick a monopole charge $b$. For $G=U(1)$, $b\in\mathbb{Z}$. For $G=U(1)^r$, $b$ belongs to a lattice $\Gamma_m\subset \R^r$ of magnetic charges allowed by Dirac quantization. For non-abelian semisimple $G$, it is a cocharacter $b: U(1) \to G$, and we use the same letter $b$ to denote the image of $1$ at the level of maps of Lie algebras: $\R \to \mathfrak{g}$, $1\mapsto b$. 
	\item Near the insertion point $p$, impose the singularities
	\begin{equation}
	*F \sim b\frac{y_\mu dy^\mu}{|y|^3}\,, \qquad \Phi_{\dot1\dot1}= -(\Phi_{\dot 2\dot 2})^\dagger\sim -\frac{b}{2|y|} e^{-i\varphi}\,, \qquad \Phi_{\dot1\dot2} \sim 0\,,
	\label{MonSing}
	\end{equation}
	where the notation ``$\sim$'' means ``$=$ up to regular terms'' and $y^{\mu}$, $\mu = 1, 2, 3$, are local Euclidean coordinates centered at $p$ (i.e., Riemann normal coordinates).
	\item Further restrict the space of fields by requiring that all vector multiplet fields commute with $b$ at the insertion point, which we write formally as:\footnote{Because $[b,b]=0$, the regular part of the vector multiplet commutes with $b$ at $p$ by itself. 
	}
	\begin{equation}
	[\cV,b]\big|_p=0.
	\end{equation}
	\item Restrict gauge transformations at $p$ to a subgroup $G_b\subset G$ preserving $b$. In other words, allow only gauge transformations by $g(x)$ such that
	\begin{equation}
	g(p)b g(p)^{-1}=b\ed
	\end{equation}
	\item The actions \eqref{Shyper}, \eqref{SYM}, \eqref{FIAction} must be modified by certain boundary terms near the insertion at $p$. Namely, we cut out a ball $U_p(\epsilon)$ of radius $\epsilon$ at $p$ and modify the action as
	\begin{align}
	S^{(\rm mon.)} = \lim_{\epsilon\to 0}\left[\int_{S^3\setminus U_p(\epsilon)} \cL - \int_{\partial U_p(\epsilon)}\Sigma\right] \ec
	\end{align}
	where $\cL$ is viewed as a top form and $\Sigma$ will be referred to as the ``monopole counterterm.'' Without $\Sigma$, the action can diverge in the monopole background, and may also not preserve the right amount of supersymmetry. While the boundary terms $\Sigma$ do not seem to leave any imprint on our calculations, it is important that there exists a choice of $\Sigma$ such that the modification $S^{(\rm mon.)}_{\rm YM}$ of $S_{\rm YM}$ in \eqref{SYM} is $\cQ^C_{\beta}$-exact, because we will use it as a localizing term.
\end{itemize}
In the remainder of this section, we provide additional details regarding the above definition, including discussions of the monopole counterterm and of subtleties in defining the normalization of monopole operators via the path integral, which may be skipped at first reading. In particular, the singular part of the twisted monopole operator background \eqref{MonSing} will be derived from the results of \cite{Borokhov:2002cg} on half-BPS monopole operators. This background can alternatively be viewed as a solution to the $\cQ^C_{\beta}$ BPS equations, with a Dirac monopole singularity $\ast F \sim b \frac{y_{\mu}dy^{\mu}}{|y|^3}$. These solutions, which also involve fixing the regular parts in \eqref{MonSing}, will be classified in Section \ref{S3loc} and Appendix \ref{moremonopoles}.

\subsection{Remarks on Monopoles}\label{monoRemarks}

Monopoles introduce point-like sources of magnetic flux and are characterized, in the case of $U(1)$ gauge group, by a number $b$---their magnetic charge. They are analogs of 't~Hooft lines in 4D theories, and in Kaluza-Klein (KK) reduction from 4D to 3D, monopole operators correspond to 't~Hooft lines (worldlines of 4D magnetic monopoles) winding the KK circle. At the location of the 3D monopole operator, the gauge field strength is prescribed to have a singularity of the form $(*F)_\mu \sim b \frac{y_\mu}{|y|^3}$. In the path integral formulation, we are instructed to integrate over field configurations with such a fixed singularity. For non-abelian gauge group $G$, we simply embed the $U(1)$ monopole in $G$ as a GNO monopole whose charge is given by a cocharacter
\begin{equation}
b: U(1) \to G.
\end{equation}
Note that the topological charge of a monopole (corresponding to the conserved topological current) is labeled by $\pi_1(G)$, while its GNO charges are labeled by cocharacters of $G$, modulo gauge and Weyl symmetries \cite{Goddard:1976qe}.  Unless $G = U(1)$, in which case topological and GNO charge coincide, each topological class contains infinitely many GNO monopoles.  For instance, when $G = U(N)$, the topological charge is the sum of the GNO charges.

There exists a supersymmetric version of the monopole operator that is of particular relevance to us. In \cite{Borokhov:2002cg}, such observables were defined for theories with $\cN=2$ supersymmetry as well as in the $\cN=4$ context. In the $\cN=2$ case, they were constructed as half-BPS operators sitting in the lowest component of the short multiplet, and therefore contributing to the chiral ring. The half-BPS property requires that, in addition to the gauge field being singular, the real scalar in the $\cN=2$ vector multiplet diverge as $\frac{b}{2|y|}$ near the monopole.\footnote{This follows from the vanishing of the SUSY variation of the gaugino.}

More precisely, if the monopole charge is given by a cocharacter $b: U(1) \to G$, then at the level of Lie algebras, there is a map $\R \to \mathfrak{g}$, and we denote the image of $1$ by the same letter $b$. Denoting the real scalar in the $\cN=2$ vector multiplet by $\chi$ (we only need it in this paragraph, so this notation is by all means temporary), the singularity is prescribed to be:
\begin{equation}
\ast F = b\frac{y_\mu dy^\mu}{|y|^3} + \ast F^{\rm reg},\qquad \chi = b\frac{1}{2|y|} + \chi^{\rm reg},
\end{equation}
while the rest of the fields are regular. Consistency also implies that the monopole operator slightly breaks the gauge group: at the location of the monopole, the gauge transformations are restricted to lie in $G_b$, where $G_b\subset G$ is the centralizer of $b$. This also means that $F^{\rm reg}$ and $\chi^{\rm reg}$, as well as the gauginos (that is, all fields in the vector multiplet), commute with $b$ at the location of the monopole.

Extending this definition to the $\cN=4$ case is straightforward, as long as we still impose that the operator be an element of the chiral ring. Indeed, the definition of $\cN=4$ Higgs and Coulomb branch chiral rings involves picking an $\cN=2$ subalgebra and considering operators that are chiral with respect to this subalgebra. This choice is equivalent to choosing a Cartan subalgebra in the $\mathfrak{su}(2)_H\oplus \mathfrak{su}(2)_C$ R-symmetry of the $\cN=4$ theory. 

In particular, the choice of $U(1)_C\subset SU(2)_C$ is parametrized by $SU(2)_C/U(1)_C = \C P^1_C$, which is discussed extensively in Section \ref{phaseAmb}. This same choice tells us which components of the triplet of scalars $\Phi_{\dot a\dot b}$ belong to the $\cN=2$ chiral multiplet, and which component is part of the $\cN=2$ vector multiplet. Let us parametrize points of this $\C P^1$ by $\alpha,\psi$, and pick a local section of the Hopf fibration as:
\begin{equation}
\label{cp1Rsym}
v=\left(\begin{matrix}
\cos\frac{\alpha}{2} e^{i\psi/2}\\
\sin\frac{\alpha}{2} e^{-i\psi/2}
\end{matrix}\right)\, .
\end{equation}
We refer to this vector $v$ as the R-symmetry polarization. This $v$ is acted on by $SU(2)_C$ in the fundamental representation, and $U(1)_C$ simply multiplies it by a phase. This means that it is the highest-weight vector with respect to the choice of $U(1)_C$. For any operator in the spin-$\frac{n}{2}$ representation of $SU(2)_C$ written as a symmetric tensor with $n$ indices $M_{\dot a_1\dots \dot a_n}$, the highest-weight component is then given (up to an arbitrary phase) by
\begin{equation}
M(v)=M_{\dot a_1\dots \dot a_n}v^{\dot a_1}\dots v^{\dot a_n}.
\end{equation}
This component has the maximal $R_C$-charge $n/2$, as measured by the generator of $U(1)_C$. It is this component that contributes to the chiral ring if the multiplet is short. In particular, for $\Phi_{\dot a\dot b}$, the component
\begin{equation}
\Phi(v)=\Phi_{\dot a\dot b}v^{\dot a} v^{\dot b}
\end{equation}
is in the chiral ring: it is the complex scalar in the $\cN=2$ chiral multiplet. This implies that, according to the definition of the $\cN=2$ half-BPS monopole operator, this component should remain regular near the insertion point $p$ of the monopole:
\begin{equation}
\Phi_{\dot a\dot b}v^{\dot a} v^{\dot b}\big|_p\sim 0.
\end{equation}
Note that $\Phi(v)$ has $U(1)_C$-charge (weight) $+1$. Acting with lowering operators of $SU(2)_C$, one can obtain the component of weight zero (the $\cN=2$ vector multiplet scalar) and the component of weight $-1$ (the antichiral conjugate of $\Phi(v)$). Only the component of weight zero is required to blow up like $\frac{b}{2|y|}$ near the monopole. This translates into the following boundary conditions defining the chiral component of the BPS $\cN=4$ monopole:
\begin{align}
\label{N4Monopole}
*F &\sim b \frac{y_\mu dy^\mu}{|y|^3} \ecq\quad \Phi_{\dot{a}\dot b}\sim \Phi_{\dot{a}\dot b}^{(v)}\ec 
\end{align}
where $\Phi_{\dot a\dot b}^{(v)}$ denotes the $v$-dependent singular part of $\Phi_{\dot a\dot b}$, given by
\begin{align}
\Phi_{\dot1\dot2}^{(v)} = \frac{b}{2|y|}\cos\alpha\ecq\quad  \Phi_{\dot1\dot1}^{(v)}=-(\Phi_{\dot2\dot2}^{(v)})^\ast= -\frac{b}{2|y|}\sin\alpha\, e^{-i\psi}\ed 
\label{gensing}
\end{align}
Again, the regular parts of these fields should commute with $b$ at $y=0$: the gauge group is broken to $G_b$ at the location of the monopole. 

The reason we have kept $v$ general should be clear by now: we want to define twisted-translated monopoles, and for that, we should know how to construct different R-symmetry components. Comparing \eqref{twistedC} with \eqref{cp1Rsym}, we see that for twisted-translated operators, the R-symmetry polarization vector has $\alpha=\pi/2$ and $\psi=\varphi$.  The resulting singularity \eqref{gensing} is precisely as announced in \eqref{MonSing}.

To further determine the normalization of monopole operators requires careful study of the path integral measure in the presence of monopole singularities. We will be able to avoid this subtle issue by finding alternative ways to fix the normalization in Sections \ref{S3loc} and \ref{SGLUE}.

\subsubsection*{An Observation}

Notice one curious feature. The monopole operator, written as a symmetric tensor $M_{\dot a_1\dots \dot a_n}$, transforms in the spin-$\frac{n}{2}$ representation of $\mathfrak{su}(2)_C$. Acting on its chiral component $M(v)=M_{\dot a_1\dots \dot a_n}v^{\dot a_1}\dots v^{\dot a_n}$ with an R-symmetry transformation $U\in SU(2)_C$, we obtain
\begin{equation}
U M(v) U^{-1} = (U_{\dot a_1}{}^{\dot b_1}\dots U_{\dot a_n}{}^{\dot b_n}M_{\dot b_1\dots\dot b_n})v^{\dot a_1}\dots v^{\dot a_n}=M(\widetilde{v})
\end{equation}
where:
\begin{equation}
\widetilde{v}^{\dot a}=U_{\dot b}{}^{\dot a}v^{\dot b}.
\end{equation}
In other words, the action of $U$ on $M(v)$ produces a different chiral component of $M$ characterized by the R-symmetry polarization $\widetilde{v}$. Notice that our definition of the monopole is such that $\Phi_{\dot a\dot b}v^{\dot a}v^{\dot b}$ remains regular. For the singular part of $\Phi_{\dot a\dot b}$ called $\Phi_{\dot a\dot b}^{(v)}$, we simply have $\Phi^{(v)}_{\dot a\dot b}v^{\dot a}v^{\dot b}=0$. To build the chiral component along the R-symmetry polarization vector $\widetilde{v}$, we should also have $\Phi^{(\widetilde{v})}_{\dot a\dot b}\widetilde{v}^{\dot a}\widetilde{v}^{\dot b}=0$. Therefore, we claim that
\begin{equation}
\Phi_{\dot a\dot b}^{(\widetilde{v})}=(U^{-1})_{\dot a}{}^{\dot c}(U^{-1})_{\dot b}{}^{\dot d}\Phi_{\dot c\dot d}^{(v)}.
\end{equation}
What this observation illustrates is that acting with $U$ on a monopole operator is equivalent to acting with $U^{-1}$ on the corresponding boundary condition. In fact, this is quite a general observation about defect operators, whose detailed derivation is given in Appendix \ref{DefectsSymm}.

\subsubsection{The Monopole Counterterm}
\label{boundaryaction}

The last ingredient needed to have a complete and well-defined notion of ``monopole operator'' is the monopole counterterm. Already in the non-supersymmetric case, merely imposing $\ast F \sim b \frac{y_\mu dy^\mu}{|y|^3}$ makes the Yang-Mills action infinite, with the divergent piece given by $\frac{8\pi\, {\rm Tr}\, b^2}{\epsilon g_{\text{YM}}^2}$. In this case, simply accompanying each monopole insertion by a factor of $\exp\big(\frac{8\pi {\rm Tr}\, b^2}{\epsilon g_{\text{YM}}^2}\big)$ suffices, as it cancels the divergence and makes the action at least na\"ively well-defined in the $\epsilon\to 0$ limit.

The problem is slightly more complicated for BPS monopoles. One reason is that the divergent part of the action receives another contribution from the singular boundary condition for the scalar. Another reason is that, even if the supersymmetry equations hold,\footnote{The equations $\delta_\text{SUSY}(\text{fermions})=0$ were used in \cite{Borokhov:2002cg} to argue that the vector multiplet scalar should also be singular near the monopole.} the presence of the singularity might break too much SUSY in the following way. Our prescription for evaluating the action involves cutting out balls of radius $\epsilon$ around the monopole insertions (followed by subtracting divergent pieces and taking $\epsilon\to 0$). Since the SUSY variation of the Lagrangian is actually a total derivative, not just zero, this can generate boundary terms in the SUSY variation. These boundary terms might not vanish in the $\epsilon\to 0$ limit, thus breaking SUSY.

The resolution of this problem is to include a proper boundary counterterm which will cancel not only divergences in the $\epsilon\to 0$ limit, but also SUSY-breaking terms. The choice of such a counterterm is not unique: we can always add a term which remains finite in the $\epsilon\to 0$ limit and whose SUSY variation vanishes in this limit. 

A very natural and convenient boundary counterterm is constructed as follows. First of all, we note that only the Yang-Mills action becomes divergent and requires a boundary counterterm, while the hypermultiplet action and the FI term both remain finite and supersymmetric in the presence of monopoles. We know from \cite{Dedushenko:2016jxl} that the Yang-Mills action is $\cQ_\beta^C$-exact. For the Lagrangian, this means that
\begin{equation}
\cL_\text{YM} = \{\cQ_\beta^C,\Psi\} + d\Sigma,
\end{equation}
where $\Psi$ is some fermionic operator. We will simply use this $\Sigma$ to construct the boundary correction. Namely, every monopole insertion should be accompanied by a term
\begin{equation}
-\int_{\partial U(\epsilon)} \Sigma
\label{moncounterterm}
\end{equation}
in the action, where $U(\epsilon)$ is a ball of radius $\epsilon$ around the monopole insertion point. With such a choice, the Yang-Mills action plus boundary counterterms are written together as:
\begin{equation}
\label{renAction}
S_{{\rm YM}}^{{\rm ren.}} = \int_{S^3} \{\cQ_\beta^C,\Psi\},
\end{equation}
regardless of how many monopoles we have inserted.

The action \eqref{renAction} is now manifestly supersymmetric because, as it turns out, $(\cQ_\beta^C)^2$ annihilates $\Psi$. Moreover, it is finite in the presence of BPS monopole insertions simply because $\int_{S^3} \{\cQ_\beta^C,\Psi\}$ vanishes on solutions to the BPS equations.  This ensures the cancellation of the leading $\frac1{\epsilon}$ divergence. (If it were not canceled, it would be present even for the action evaluated on the BPS solution.) A possible subleading $\log\epsilon$ divergence is absent, as can be checked by inspecting each term of the classical action---this is actually ensured by the fact that $B$ commutes with all the vector multiplet fields at the insertion point. So what remains is a finite action, just as we wanted.\footnote{The reader might be wondering how it is possible that in \cite{Borokhov:2002cg}, supersymmetry implied a relation between the singularities for $F_{\mu\nu}$ and for $\Phi_{\dot a\dot b}$, while here, supersymmetry holds without additional conditions. The answer is that even though the action \eqref{renAction} is manifestly supersymmetric, in order for it to stay finite, we still need to impose the same relation between the singularities of $F_{\mu\nu}$ and of $\Phi_{\dot a\dot b}$.}

The proper monopole counterterm $\Sigma$ as defined above is explicitly constructed in Appendix \ref{CounterSigma}.

\subsection{Remarks on Normalization} 
\label{normamb}
\subsubsection{Phase Ambiguity of Chiral Operators}\label{phaseAmb}

Suppose we are given an HBO $H_{a_1\dots a_n}$ whose $\mathfrak{su}(2)_H$ R-charge equals its conformal dimension, or a CBO $C_{\dot{a}_1\dots\dot{a}_m}$ with the analogous property. The highest weight component of $H_{a_1\dots a_n}$ or $C_{\dot{a}_1\dots\dot{a}_m}$ will then give an element of the corresponding chiral ring: it lives in the bottom of the chiral multiplet in the $\cN=2$ decomposition of the corresponding $\cN=4$ multiplet. To define the highest weight vector, we need to pick maximal tori $U(1)_H\subset SU(2)_H$ and $U(1)_C\subset SU(2)_C$. These choices are parametrized by $\C P^1_H= SU(2)_H/U(1)_H$ and $\C P^1_C = SU(2)_C/U(1)_C$, the twistor spheres of the Higgs and Coulomb branches (which are hyperk\"ahler cones). However, a point of the twistor sphere only determines the chiral operator up to a phase. In the following few paragraphs, we explain this freedom for the Higgs branch case. The Coulomb branch case is completely analogous and can be obtained by replacing the label $H$ by $C$ in what follows.

Suppose we are given a point of $\C P^1_H$ with homogeneous coordinates $(x:y)$. We can pick a point of the tautological bundle $\cO(-1)$ that belongs to the fiber above $(x:y)$, say $u=(\begin{array}{cc}x& y \end{array})^T$. Na\"ively, the corresponding chiral operator is
\begin{equation}
H(u) = H_{a_1\dots a_n}u^{a_1}\dots u^{a_n},
\end{equation}
since this object is the highest weight component of $H$. However, $u$ is only defined up to an overall $\C^*$ scaling: thus this definition is not unique. In fact, $H(u)$ gives a polynomial function of degree $n$ on the total space of $\cO(-1)$ with values in operators, or equivalently, an operator-valued holomorphic section of $\cO(n)$ over $\C P^1_H$.\footnote{This has been noted for the Coulomb branch chiral operators in, e.g., \cite{Borokhov:2002cg}.} Alternatively, we can pick $u$ to be normalized as $u^\dagger u=1$. Then it parametrizes points of $S^3_{\rm Hopf}$, the total space of the Hopf fibration. $H(u)$ becomes an operator-valued function on $S^3_{\rm Hopf}$, i.e., for each point of $S^3_{\rm Hopf}$, there is a unique and unambiguous choice for the chiral operator $H(u)$.

This suggests that we cannot identify chiral operators for each point of $\C P^1_H$ globally: to do that, one would have to pick a global section of the Hopf fibration and plug it into $H(u)$, but such a section simply does not exist. So at best, we can do so locally on $\C P^1_H$, say if we remove a point from it. Even in this case, for each point of $\C P^1_H$, $H(u)$ is only defined up to a phase, since we still have to pick a local section of the Hopf fibration. So, to emphasize, the definition of $H(u)$ for a point of $\C P^1_H$ involves a phase ambiguity and requires making an arbitrary choice. The Coulomb branch version of this story is exactly the same.

This phase ambiguity is rather innocent in the Higgs branch case, since all Higgs branch operators are constructed from fields in the Lagrangian. Then for each $u$, we have a direct definition of the operator $H(u)$, and there is no real need to talk about points of $\C P^1_H$. The Coulomb branch case is more involved, as we will leave the normalization of the path integral measure undetermined, in addition to making a non-unique choice for the monopole counterterm. Therefore, our path integral definition of the monopole operator only encodes the point of $\C P^1_C$, any possible additional data being ignored. Thus the phase is not manifestly fixed, and we will have to use some other reasoning to pin down the normalization of monopole operators.

\subsubsection{Subtleties with Antiperiodicity} \label{periSubtle} 

In our analysis, we have not needed to directly confront the fact that $H(u)$ or $C(v)$ cannot be written globally on $\C P^1_H$ or $\C P^1_C$. Indeed, we are mostly interested in twisted-translated operators, and such operators have $u$ and $v$ as in \eqref{twistedH} and \eqref{twistedC}, which are only defined on great circles of $\C P^1_H$ and $\C P^1_C$. Clearly, we can trivialize the Hopf bundle if we restrict it to a circle on the base. However, due to the definition of twisted translations, we are forced to consider sections that are antiperiodic on this circle. Indeed, both $u$ and $v$ from \eqref{twistedH}, \eqref{twistedC} are antiperiodic under $\varphi \to \varphi + 2\pi$. Therefore, the periodicity of $H(u)$ or $C(v)$ depends on the sign of $(-1)^n$ or $(-1)^m$: twisted translations give antiperiodic operators on the circle for half-integral R-spins.\footnote{In the language of \cite{Borokhov:2002cg}, this sign arises due to the Berry phase: parallel transport in $\cO(n)$ along the great circle of $\C P^1$ results in a holonomy $(-1)^n$.}  The occurrence of antiperiodic observables on $S^1$ is of course familiar from the study of twisted HBOs in \cite{Chester:2014mea,Dedushenko:2016jxl}. Here, we have simply emphasized the similar origin of these antiperiodicities in both the Higgs and Coulomb cases.

If we have some twisted-translated observable on a circle $\cO(\varphi)$ that happens to be antiperiodic, then we should take extra care in defining its sign. This is directly related to the phase ambiguity of general chiral operators discussed in the previous subsection. Once we pick $u$ and $v$ as in \eqref{twistedH} and \eqref{twistedC}, we fix the phase ambiguity almost completely, except for operators of half-integral R-charge, whose sign remains undefined. Such observables are only single-valued on the double cover of $S^1$. We deal with this ambiguity by inserting a ``branch point'' somewhere on the circle. Then we choose to insert all observables away from the branch point, and if we ever have to move an observable past the branch point, it should pick up an extra sign of $(-1)^n$ in the Higgs branch case or of $(-1)^m$ in the Coulomb branch case (here, $n/2$ is an $\mathfrak{su}(2)_H$ spin and $m/2$ is an $\mathfrak{su}(2)_C$ spin). In the presence of such a branch point, all observables become single-valued. 

For each observable, we pick its sign at $\varphi=0$, and then apply twisted translations to extend the definition to the rest of the circle (away from the branch point). This procedure is trivial in the Higgs branch case: because all Higgs branch operators are constructed from the hypermultiplet scalars $q_a$, and these are both single-valued and canonically normalized, the sign choice is simply a choice for the value $\cos\varphi/2\big|_{\varphi=0}=1$ (as opposed to $-1$, which would also be valid since $\cos\varphi/2$ is only defined up to a sign on the circle).\footnote{The choice of this sign at $\varphi=0$ does not affect physical answers due to R-symmetry: every nonvanishing correlator has total R-charge zero, so flipping the signs of all observables of half-integer R-charge does not change the answer.} The sign choice is less trivial in the Coulomb branch case because, as we have already mentioned, the disorder-type definition of a monopole does not come with any canonical normalization. We will use a different consideration to fix the phase, and in particular, the sign.

In Section \ref{S3loc}, we will fix the sign by comparing with the two-point function in the SCFT on $\R^3$. According to \cite{Dedushenko:2016jxl, Beem:2016cbd, Chester:2014mea}, twisted-translated operators are inserted along the $x_3$-axis, and we choose the normalization such that the two-point function of a monopole at $x_3>0$ and an antimonopole at $x_3<0$ is positive. Identifying $\R^3$ with $S^3$ via stereographic projection such that $\varphi=0$, $\theta=\pi/2$ maps to the origin allows us to pin down the signs as in \eqref{mon2}. With such an identification, the $x_3$-axis maps to the interval $-\pi<\varphi<\pi$ of the great circle, implying that the branch point is located at $\varphi=\pm\pi$. Had we chosen to perform stereographic projection with $\varphi=\pm\pi$, $\theta=\pi/2$ taken as the origin, but with the same normalization in $\R^3$, we would have obtained a sign differing by $(-1)^B$ where $B$ is the monopole charge. The interval $(0,2\pi)$ would then have mapped to the $x_3$-axis, resulting in a branch point at $\varphi=0$. So we see that the choice of branch point is correlated with choice of the sign in \eqref{mon2}. Our convention is to always put the branch point at $\varphi=\pm\pi$.

From the point of view of the discussion in Section \ref{SGLUE}, this sign will be slightly more obscure. There, we cut the sphere into two equal halves and then glue the hemisphere wavefunctions together. It turns out that the two hemispheres give precisely equal contributions, so the sign should be contained entirely in what we refer to as the ``gluing measure'' $\mu(\sigma, B)$. In accordance with the rest of the paper, we assume that the branch point is at $\varphi=\pm\pi$. Then, under stereographic projection, the upper hemisphere corresponds to the upper half-space $x_3>0$ while the lower hemisphere corresponds to $x_3<0$. Putting the branch point at $\varphi=0$ instead (which is the only possibility other than $\varphi=\pm\pi$ consistent with cutting and gluing, as other locations would break the symmetry between the upper and lower hemispheres) would correspond to swapping these identifications, and would need to be accompanied by a sign in the gluing measure for consistency. This can be achieved by simply replacing $\mu(\sigma,B) \to \mu(\sigma, -B)$.

We can give one more argument to demonstrate that our method of fixing the signs is correct. Suppose we have a monopole at $\varphi=\pi/2$, an antimonopole at $\varphi=-\pi/2$, and a branch point at $\varphi=\pm\pi$. Let us perform a twisted translation by $+\pi$ while simultaneously moving the branch point by $+\pi$. The two-point function will remain the same, simply because the correlator can only depend on the distance between the observables, and no operator crosses the branch point in this process. We end up with a monopole at $-\pi/2$, an antimonopole at $+\pi/2$, and a branch point at $2\pi$ (or, equivalently, at $0$). Next, we switch the monopole with the antimonopole, so that we end up with the initial configuration for the operator insertions, except that now the branch point is at $\varphi=0$. This swap of monopole with antimonopole produces exactly the sign difference explained in the previous paragraphs, as we will see from our results.

\section{Localization on $S^3$}\label{S3loc}

We now perform supersymmetric localization of abelian $\cN=4$ theories on $S^3$ with respect to the supercharge $\cQ_\beta^C$. As described in the previous section, the cohomology of $\cQ_\beta^C$ includes twisted-translated monopole operators that can be inserted anywhere along a great circle of $S^3$. 
In what follows, we will derive a matrix model expression for correlators containing such a monopole, a corresponding antimonopole, and arbitrary additional insertions of twisted-translated operators constructed from the vector multiplet scalars.

\subsection{BPS Equations and Their Solutions}

Let us start by describing the vector multiplet BPS equations $\delta_{\xi_\beta^C}\lambda_{a\dot b}=0$, where the SUSY transformation rule is given in \eqref{lamvar} and $\xi_\beta^C$ is the Killing spinor corresponding to $\cQ_\beta^C$.\footnote{In the stereographic frame, we have
	\begin{align}
	(\xi_\beta^C)_{1a \dot b}=\frac{e^{\Omega/2}}{2r}\begin{pmatrix}
	c + \beta d & c - \beta \bar{d} \\ i(c - \beta d) & i(c + \beta\bar{d})
	\end{pmatrix} \ecq (\xi_\beta^C)_{2a \dot b}=\frac{e^{\Omega/2}}{2r}\begin{pmatrix}
	d + \beta \bar{c} & -\bar{d} + \beta \bar{c} \\ i(d - \beta \bar{c}) & -i(\bar{d}+\beta\bar{c}) 
	\end{pmatrix}
	\label{xiCbeta}
	\end{align}
	where $e^{\Omega} = (1+x^2/4r^2)^{-1}$, $c= i x_1 +x_2$, $d=2r- i x_3$, and $x_i$ are the standard stereographic coordinates on $S^3$ (see Appendix \ref{coordinates}).} The results are most simply expressed in terms of the fields
\begin{align}
\Phi_r \equiv \Re( R e^{i\varphi}\Phi_{\dot 1\dot 1}) \ecq \Phi_i \equiv \Im( R e^{i\varphi}\Phi_{\dot 1\dot 1}) \ec \label{PHir}
\end{align}
where $R=\sin\theta\in[0,1]$ and the coordinates $(\theta,\varphi,\tau)$ were defined in \eqref{roundvars}. Note that $\Phi_{\dot 1\dot 1}$ is regular at $R=0$, as there are no insertions there, implying that $\Phi_{r,i}$ in \eqref{PHir} satisfy
\begin{align}
\lim_{R\to 0} \Phi_{r,i} = 0 \ed \label{PHirLim}
\end{align}
In terms of \eqref{PHir}, the BPS equations can be summarized as
\begin{gather}
D_{12}=\Re(D_{11}) = 0 \ecq \Im(D_{11}) = \textstyle -\frac{1}{r}\Phi_{\dot{1}\dot 2} \ec \label{Deqs}\\
\partial_{\mu}\Phi_{\dot{1}\dot 2} = \partial_{\tau}\Phi_i = \partial_{\tau}\Phi_r = 0 \ec \label{DPH}\\
R \partial_R\Phi_i + \partial_{\varphi}\Phi_r = 0 \ec \label{PHeq1}\\
R(1-R^2)\partial_R\Phi_r - \partial_{\varphi}\Phi_i =0 \ec \label{PHeq2}\\
F_{\mu\nu} = \varepsilon_{\mu\nu\rho}\partial^{\rho}\Phi_r \ed \label{BogoEq}
\end{gather}
Note that \eqref{DPH} implies that the vector multiplet scalars are independent of $\tau$ on the BPS locus. Together with \eqref{Deqs} and \eqref{BogoEq}, it follows that all of the vector multiplet fields are $\tau$-independent. This is, of course, also an immediate consequence of \eqref{QCsqr}. The BPS field configurations can therefore be viewed as functions on the disk parametrized by $(R,\varphi)$. Clearly, the remaining content of the first two sets of equations \eqref{Deqs}, \eqref{DPH} is that $\Phi_{\dot 1\dot 2}$ is a constant, in terms of which $D_{ab}$ is determined. In what follows, we will study the remaining equations \eqref{PHeq1}--\eqref{BogoEq}.

\subsubsection{Non-Singular Solutions}
\label{monsol}

Let us first review the non-singular solutions to \eqref{PHeq1}--\eqref{BogoEq}, which were already described in \cite{Dedushenko:2016jxl} from a slightly different point of view. Equation \eqref{BogoEq} is the Bogomolny equation on $S^3$. Its only regular solutions have $A_{\mu}=0$ and $\Phi_r$ constant. Equations \eqref{PHeq1} and \eqref{PHeq2} then  imply that $\Phi_i$ is also a constant. As argued around \eqref{PHirLim}, these constants must vanish to avoid having a singularity at $R=0$.  Therefore, $\Phi_{i,r}=0$. To summarize, the non-singular BPS locus is given by
\begin{align}
\cV_{\text{loc}} = \{A_{\mu}^{\text{loc}}, \lambda_{a\dot b}^{\text{loc}}, \Phi_{\dot{a}\dot b}^{\text{loc}}, D_{ab}^{\text{loc}}\} \ec \label{VecLoc1}
\end{align}
where
\begin{align}
\Phi_{\dot{1}\dot 2}^{\text{loc}} = irD_{11}^{\text{loc}} = irD_{22}^{\text{loc}} = \frac{1}{r}\sigma \ecq A_{\mu}^{\text{loc}} = D_{12}^{\text{loc}}=\Phi_{\dot{1}\dot 1}^{\text{loc}} = \Phi_{\dot{2}\dot 2}^{\text{loc}} = \lambda_{a\dot b}^{\text{loc}} = 0 \ec \label{VecLoc2}
\end{align}
and for a $U(1)^r$ gauge group, $\sigma\in\mathbb{R}^{r}$ is a constant $r$-vector. Note that the non-singular $\cQ_\beta^C$ BPS locus \eqref{VecLoc2} coincides with the saddle points of the $\cN=4$ Yang-Mills action \cite{Kapustin:2009kz,Hama:2010av}. Indeed, as shown in \cite{Dedushenko:2016jxl}, the Yang-Mills action is $\cQ_\beta^C$-exact. It can therefore be used as a localizing term, so that the path integral reduces to a sum over its saddles. 

The cohomology of $\cQ_\beta^C$ includes local operators constructed from the vector multiplet scalars $\Phi_{\dot a\dot b}$. As shown in \cite{Dedushenko:2016jxl}, and as we now review, these operators evaluate to polynomials in $\sigma$ on the BPS locus \eqref{VecLoc2}. According to the prescription \eqref{twistedC}, gauge-invariant polynomials in 
\begin{align}
\Phi(\varphi) = \Phi_{\dot a\dot b}v^{\dot a}v^{\dot b}\biggr|_{R=1} = \Phi_{\dot 1\dot 2} + i \Phi_i\biggr|_{R=1}  \label{HatPhiDef}
\end{align}
are $\cQ_\beta^C$-closed. This fact can be readily checked using the SUSY variations given in \eqref{phivar}. Plugging in \eqref{VecLoc2}, we see that in the absence of defect operators, \eqref{HatPhiDef} localizes to 
\begin{align}
\Phi(\varphi) \rightarrow \Phi_{\rm loc}(\varphi) =\frac{1}{r}\sigma \ed \label{HatPhiLoc}
\end{align}
As we will see later, insertions of monopole operators modify the RHS of \eqref{HatPhiLoc}, since they lead to a nontrivial background for $\Phi_i$.

\subsubsection{The Two-Monopole Background}
\label{2monopolebackgronud}

The BPS equations \eqref{PHeq1}--\eqref{BogoEq} also admit singular solutions describing insertions of twisted-translated monopole operators. In Appendix \ref{moremonopoles}, we explicitly construct these solutions for any number of insertions of such operators at $R=1$. As shown there, the solution is uniquely determined by the values of $\Phi_i(R,\varphi)$ at the boundary of the disk ($R=1$), where it must be a piecewise constant periodic function of $\varphi$. In particular, for $n$ insertions, it takes the form
\begin{align}
\Phi_i(R=1,\varphi) = -\frac{1}{2r}\sum_{k=1}^nb_k\left[\sgn\left(\varphi-\varphi_k\right) + \frac{\varphi_k}{\pi}\right] \ec \label{PHiBdy}
\end{align} 
where $b_k\in\Gamma_m$ is the charge of the $k^\textrm{th}$ monopole, $\varphi_k$ is its angular position at $R=1$, and $\sum_{k=1}^nb_k=0$ because the total charge on $S^3$ must vanish.\footnote{The sign function in \eqref{PHiBdy} is defined for $\varphi\in(-\pi,\pi]$. For other values of $\varphi$, it should be replaced by $\sgn\left(\cos\frac{\varphi}{2}\sin\frac{\varphi-\varphi_k}{2}\right)$.} The solutions for general configurations of monopole operators are given in terms of complicated expansions, such as \eqref{PHiExp}, which are difficult to use in explicit localization computations. Instead, we will work with a simple background corresponding to the insertion of two monopole operators. As we will see in Section \ref{multipleinsertions}, this is sufficient to construct arbitrary correlators with $n>2$ insertions of monopole operators.

Let us now describe the two-monopole background. 
Consider a monopole of charge $b$ at $\varphi=\frac{\pi}{2}$ and one of charge $-b$ at $\varphi=-\frac{\pi}{2}$. In this case,
\begin{align}
\Phi_i(R=1,\varphi) = \frac{b}{2r}\sgn\cos\varphi \ed \label{PHiBdy2}
\end{align}
The (unique) solution to \eqref{PHeq1}--\eqref{BogoEq} with boundary condition \eqref{PHiBdy2} is given by
\begin{align}
\Phi_r &= -\frac{b}{2r}\frac{R\sin\varphi}{\sqrt{1-R^2\sin^2\varphi}} \ec \label{PHr2Mon}\\
\Phi_i &= \frac{b}{2r}\frac{R\cos\varphi}{\sqrt{1-R^2\sin^2\varphi}} \ec\label{PHi2Mon}\\
A^{\pm} &= \frac{b}{2} \left(\frac{R\cos\varphi}{\sqrt{1-R^2\sin^2\varphi}} \pm 1\right) d\tau \ec \label{A2Mon}
\end{align}
where $A^-$ is defined only in the patch $\{0\leq R <1\}\cup\{R=1,-\frac{\pi}{2}<\varphi<\frac{\pi}{2}\}$, while $A^+$ is defined in  $\{0\leq R<1\}\cup\{R=1,-\pi<\varphi<-\frac{\pi}{2}\cup\frac{\pi}{2}<\varphi<\pi\}$.

The background \eqref{PHr2Mon}--\eqref{A2Mon} can be rewritten in a more familiar form by passing to spherical coordinates $\eta,\psi\in[0,\pi]$ and $\tau\in (-\pi,\pi]$, defined as
\begin{alignat}{3}
X_1 &= r\sin\eta\sin\psi\cos\tau \ecq &X_2 &= r\sin\eta\sin\psi\sin\tau \ec\notag\\
X_3 &= -r\sin\eta\cos\psi \ecq &X_4 &= r\cos\eta \ec \label{sphericalvars}
\end{alignat}
and in which the metric is given by
\begin{align}
ds^2 &= r^2\left(d\eta^2 + \sin^2\eta\, ds^2_{S^2}\right) \ecq ds^2_{S^2} = d\psi^2 + \sin^2\psi\, d\tau^2 \ed
\end{align}
In the coordinates \eqref{sphericalvars}, the monopole is inserted at $\eta=0$ and the antimonopole at $\eta=\pi$. In particular, \eqref{PHr2Mon}--\eqref{A2Mon} become
\begin{align}
\Phi_{\dot{1}\dot 1} &= -(\Phi_{\dot{2}\dot 2})^* = \frac{i b}{2r\sin\eta}\ec \label{PHspherical}\\
A^{\pm} &= -\frac{b}{2}\left(\cos\psi\mp 1\right)d\tau \ed \label{Aspherical}
\end{align}
The background \eqref{PHspherical}, \eqref{Aspherical} is stereographically projected to a half-BPS monopole operator of charge $b$ inserted at the origin of $\mathbb{R}^3$. Indeed, one can check that it preserves the supercharges $\cQ_1^{(\ell\pm)}\pm \cQ_1^{(r\pm)}$ and $\cQ_2^{(\ell\pm)}\mp \cQ_2^{(r\pm)}$, a fact that will become useful in Section \ref{SGLUE}. In what follows, we will compute correlation functions with two twisted-translated monopole operators by using the solution \eqref{PHspherical} and \eqref{Aspherical}.

\subsection{Localization of Correlators with Monopoles}

Let us now discuss some general aspects of our localization problem.\footnote{For an introduction to supersymmetric localization, we refer the reader to \cite{Pestun:2016zxk} and references therein.} We wish to calculate correlators of $\cQ_\beta^C$-closed local operators. These operators include the monopole operators described above as singular supersymmetric backgrounds, as well as polynomials in the twisted-translated vector multiplet scalars $\Phi(\varphi)$ defined in \eqref{HatPhiDef}. They are all inserted along the great circle at $R=1$, which is parametrized by the angle $\varphi$.\footnote{The cohomology of $\cQ_\beta^C$ also includes BPS vortex loops wrapping the $R=1$ circle, as well as BPS Wilson loops wrapping $\tau$-circles. The former line operators were first described in the context of localization of 3D $\cN=2$ theories in \cite{Kapustin:2012iw,Drukker:2012sr}, and for $\cN=4$ theories in \cite{Assel:2015oxa}. We will not discuss them in this paper.} The path integral expressions of such correlators are given by
\begin{align}
\langle \cM^{b_1}(\varphi_1)\cdots \cM^{b_n}(\varphi_n) \cdots \rangle = \frac{1}{Z_{S^3}}\int \pD\cH \pD\widetilde{\cV}\,  e^{-S_{\text{YM}}[\cV_{\rm b.g.}+\widetilde{\cV}] -S_{\text{hyper}}[\cH,\cV_{\rm b.g.}+\widetilde{\cV}]} ({\cdots})\ec \label{cordefgen}
\end{align}
where the $\cM^{b_k}(\varphi_k)$ denote charge-$b_k$ twisted-translated monopole operators. On the RHS of \eqref{cordefgen}, $Z_{S^3}$ is the $S^3$ partition function, while $\cV_{\text{b.g.}}$ denotes the monopole background described in the previous subsection and in Appendix \ref{moremonopoles}; the fluctuations around it are denoted by $\widetilde{\cV}$. The final ellipses $({\cdots})$ in \eqref{cordefgen} represent arbitrary additional insertions of $\Phi(\varphi)$ polynomials at different points on the great circle. The Yang-Mills and hypermultiplet actions $S_{\text{YM}}$ and $S_{\text{hyper}}$ were defined in \eqref{SYM} and \eqref{Shyper}. We will assume that these actions also contain appropriate boundary terms (the ``monopole counterterms'') at the positions of the defects,  as discussed in Section \ref{boundaryaction}, though their explicit form will not be needed.  

Localizing the path integral \eqref{cordefgen} over $\widetilde{\cV}$ for abelian theories is very simple. By taking into account the counterterm required to define the insertions of twisted-translated monopole operators, it was argued in Section \ref{boundaryaction} that the Yang-Mills action is $\cQ_\beta^C$-exact (and closed). It can therefore be used as a localizing term. Because the gauge group is abelian, this action is quadratic, and in fact completely independent of the background $\cV_{\rm b.g.}$. The localization locus for $\widetilde{\cV}$ is therefore identical to the one written in \eqref{VecLoc2}, which was derived assuming $\cV_{\rm b.g.}=0$. The Yang-Mills action vanishes on the localization locus. Moreover, the one-loop determinant of fluctuations around it is known to be equal to $1$ (see, \cite{Kapustin:2009kz,Hama:2010av}). 

We conclude that for a $U(1)^r$ gauge group, \eqref{cordefgen} localizes to
\begin{align}
\langle \cM^{b_1}(\varphi_1)\cdots \cM^{b_n}(\varphi_n) \cdots \rangle &= \frac{1}{Z_{S^3}}\int\left(\prod_{i=1}^rd\sigma_i\right)Z(\sigma,b_1,\ldots,b_n)({\cdots})  \ec \label{corLocDef}\\
Z(\sigma,b_1,\ldots,b_n) &\equiv \int \pD\cH\,   e^{-S_{\text{hyper}}[\cH,\cV_{\rm b.g.}+\widetilde{\cV}_{\text{loc}}]} \ed \label{ZsigDef}
\end{align}
In \eqref{corLocDef} and \eqref{ZsigDef}, $\widetilde{\cV}_{\text{loc}}$ is the same as $\cV_{\text{loc}}$ in \eqref{VecLoc2}, depending only on the $r$ real constants $\sigma_i$, and the $({\cdots})$ denote additional insertions of localized $\Phi(\varphi)$ polynomials. Note that in the presence of monopoles, insertions of $\Phi(\varphi)$ do not quite localize to $\sigma$ as in \eqref{HatPhiLoc}. Instead, using \eqref{HatPhiDef}, we have:
\begin{align}
\Phi(\varphi) \to \Phi_{\rm loc}(\varphi) =  \frac{1}{r}\sigma + i \Phi_i(R=1,\varphi) \ec\label{HatPhiMon}
\end{align}
with $\Phi_i(R=1,\varphi)$ given in \eqref{PHiBdy}.
All that is left is to calculate $Z(\sigma,b_1,\ldots,b_n)$ in \eqref{ZsigDef}. Note that in principle, it should be possible to evaluate the path integral in \eqref{ZsigDef} explicitly, even without localization, because $S_{\text{hyper}}$ is quadratic in $\cH$. We now carry out this step for $n=2$ insertions of twisted-translated monopole operators. 

\subsubsection{Two Monopole Insertions}

To evaluate the localization formula for correlators of a twisted-translated monopole operator $\cM^b(\varphi)$, an antimonopole $\cM^{-b}(\varphi)$, and additional insertions of polynomials in the twisted-translated vector multiplet scalars $\Phi(\varphi)$ requires calculating the hypermultiplet path integral \eqref{ZsigDef} around the singular background given in \eqref{PHspherical}, \eqref{Aspherical}, which corresponds to inserting $\cM^b(\varphi)$ at $\varphi=\frac{\pi}{2}$ and $\cM^{-b}(\varphi)$ at $\varphi=-\frac{\pi}{2}$. Because correlators of twisted-translated operators are topological, there is no loss of generality in fixing the insertion points in this way. Note that by using \eqref{PHiBdy2}, we find that in the two-monopole configuration, $\Phi(\varphi)$ localizes to:
\begin{align}
\Phi_{\rm loc}(\varphi)  =\frac{1}{r}\left(\sigma + i\frac{b}{2}\sgn\cos\varphi\right) \ed\label{HatPhi2Mon}
\end{align}
Let us now describe the computation of $Z(\sigma,b)\equiv Z(\sigma,b,-b)$. Because $S_{\text{hyper}}$ is quadratic, the $\cH$ path integral in \eqref{corLocDef} is given by the ratio of one-loop determinants
\begin{align}
Z(\sigma,b)\equiv \int \pD\cH\,   e^{-S_{\text{hyper}}[\cH,\cV_{\rm b.g.}+\widetilde{\cV}_{\text{loc}}]} = \frac{\det \Lambda^{-1}\cD_f}{\det \Lambda^{-2}\cD_b} \ec \label{Zsig}
\end{align}
where $\cD_b$ and $\cD_f$ are differential operators appearing in the bosonic and fermionic quadratic pieces of $S_{\text{hyper}}$, respectively. These differential operators depend explicitly on $\sigma$ and on the monopole background \eqref{PHspherical}, \eqref{Aspherical}. As we show in Appendix \ref{1loopdetails}, they can be diagonalized explicitly by expanding their eigenfunctions in monopole spherical harmonics on the $S^2$ parametrized by $\psi$ and $\tau$. In \eqref{Zsig}, we have introduced an arbitrary scale $\Lambda$ on dimensional grounds. It should be thought of as a UV scale necessary for a proper definition of the path integral, and will be removed at the end of the computation by a renormalization of the monopole operators.

Let us first summarize the results of Appendix \ref{1loopdetails} for ${\rm SQED}_1$, in which $b\in\mathbb{Z}$.  In this case, the spectrum of $\cD_b$ is given by
\begin{align}
\lambda_{\pm,n}^b = \frac{1}{r^2}\left(n+\frac{1+|b|}{2}\pm i\sigma\right)\left(n+\frac{3+|b|}{2}\mp i \sigma\right) \ecq n=0,1,\ldots \ec \label{DbSpec}
\end{align}
with degeneracies $d^b_{\pm,n} =(n+1)(n+|b|+1)$. The spectrum of $\cD_f$ is given by
\begin{align}
\lambda_{\pm,n}^f &= \frac{1}{r}\left[\pm\left(n+\frac{3+|b|}{2}\right)+i\sigma\right] \ecq n=0, 1,\ldots \ec \label{DfSpec} \\
\lambda_{\pm,n}^{f,0} &= \frac{1}{r}\left[\pm\left(n+\frac{1+|b|}{2}\right)+i\sigma\right] \ecq n=0, 1,\ldots \ec \label{Df0Spec}
\end{align}
with corresponding degeneracies $d_{\pm,n}^f = 2(n+|b|+2)(n+1)$ and $d_{\pm,n}^{f,0}=|b|$.\footnote{In \eqref{DfSpec} and \eqref{Df0Spec}, we actually quote the spectrum of $\tilde{\cD}_f$ defined in \eqref{tDF}. Its determinant is equal to that of $\cD_f$. The fermionic eigenvalues $\lambda_{\pm,n}^{f,0}$ in \eqref{Df0Spec} arise from zero modes of the Dirac operator on $S^2$.} Using the above spectrum, and the fact that $d^b_{\pm,n-1} - d^f_{\pm,n-1} + d^b_{\pm,n} - d^{f,0}_{\pm,n} = 1$, we can write the real part of the $S^3$ free energy as
\es{ReF}{
	\Re F = \sum_{n=0}^{\infty}\log\left|\frac{n+\frac{1+|b|}{2}+i\sigma}{\Lambda r} \right|^2 \,.
}
To evaluate this sum, let us define 
\es{fDef}{
	f(s) &= \sum_{n=0}^{\infty} \left[ \frac{(\Lambda r)^s}{\left(n+\frac{1+|b|}{2}+ i\sigma\right)^s} + \frac{(\Lambda r)^s}{\left(n+\frac{1+|b|}{2}- i\sigma\right)^s} \right] \,.
}
This function is related to $\Re F$ in \eqref{ReF} by
\begin{align}
\Re{F} = \frac{df}{ds}\biggr|_{s=0} \ed\label{f2F}
\end{align}
Moreover, the infinite sum defining $f(s)$ in \eqref{fDef} is convergent for large enough $s$, and can be analytically continued to small $s$ using the Hurwitz zeta function $\zeta(s,q) = \sum_{n=0}^{\infty}\frac{1}{(n+q)^s}$:
\begin{align}
f(s)&=-(\Lambda r)^s\left[\zeta\left( s, \frac{1+|B|}{2}+i\sigma\right)+\zeta\left( s, \frac{1+|B|}{2}-i\sigma\right)\right]\ed \label{fs}
\end{align}
Plugging \eqref{fs} into \eqref{f2F}, and using $\zeta(0,q)=\frac{1}{2}-q$ and $\frac{d\zeta(s,q)}{ds}\Big|_{s=0}=\log\frac{\Gamma(q)}{\sqrt{2\pi}}$, results in
\begin{align}
\Re F = |b|\log (\Lambda r) - \log\left|\frac{\Gamma\left(\frac{1+|b|}{2}+i\sigma\right)}{\sqrt{2\pi}}\right|^2 \ed
\end{align}
We conclude that for ${\rm SQED}_1$, the absolute value of \eqref{Zsig} is given by
\begin{align}
|Z(\sigma,b)| = (\Lambda r)^{-|b|}\frac{1}{2\pi}\Gamma\left(\frac{1+|b|}{2}-i\sigma\right)\Gamma\left(\frac{1+|b|}{2}+i\sigma\right) \ed \label{detRatioTmp}
\end{align}
As a check of \eqref{detRatioTmp}, we find that
\begin{align}
Z_{S^3, \sigma}\equiv |Z(\sigma,b=0)| = \frac{1}{2\cosh\pi\sigma} \ec
\end{align}
which is the correct $S^3$ partition function of a free hypermultiplet coupled to a real mass $m=\sigma$. To complete the calculation, the overall phase of $Z(\sigma,b)$ still needs to be determined. We have not been able to compute this phase rigorously, but we postulate that the full answer takes the form
\begin{align}
Z(\sigma,b) = (-1)^{\frac{|b|-b}{2}}\frac{1}{2\pi (\Lambda r)^{\abs{b}}}\Gamma\left(\frac{1+|b|}{2}-i\sigma\right)\Gamma\left(\frac{1+|b|}{2}+i\sigma\right) \ed \label{detRatio}
\end{align}
The overall sign in \eqref{detRatio} will be explained momentarily. 

First, we note that according to \eqref{corLocDef}, integrating $Z(\sigma,b)$ over $\sigma$ gives us the twisted monopole two-point function in ${\rm SQED}_1$. In particular,
\begin{align}
\langle \cM^b(\pi/2) \cM^{-b}(-\pi/2)\rangle_\text{$S^3$, SQED$_1$} = \frac{1}{Z_{S^3}}\int_{-\infty}^{\infty}d\sigma\, Z(\sigma,b) = (-1)^{\frac{|b|-b}{2}}\frac{(|b|)!}{(2\Lambda r)^{|b|}} \ed \label{2pntTmp}
\end{align}
The IR limit is obtained by renormalizing the monopole operators as $\cM^b\to \Lambda^{\frac{|b|}{2}}\cM^b$ and sending $\Lambda \to\infty$, while keeping $r$ fixed. From the power of $ r$ in \eqref{2pntTmp}, it follows that the dimension of a charge-$b$ monopole CBO in ${\rm SQED}_1$ is given by $\Delta_{\cM^b} = |b|/2$. This is a new derivation of the dimensions of the half-BPS monopole operators of ${\rm SQED}_1$, which were first obtained in \cite{Borokhov:2002cg}.\footnote{In particular, $\Delta_{\cM^{b=1}}=1/2$, so the IR limit of ${\rm SQED}_1$ is the theory of a free twisted hypermultiplet.}  Note that while the classical dimensions of hypermultiplet fields are the same as their dimensions in the IR SCFT, the dimensions and R-charges of monopole operators are inherently quantum: they cannot be read off from the action, and they are related to proper regularization of the path integral.  Supersymmetry requires that the dimensions of monopoles induced by quantum effects coincide with their IR R-charges.

The sign of $(-1)^{\frac{|b|-b}{2}}$ in \eqref{detRatio} can now be understood as follows. As shown in \cite{Chester:2014mea,Beem:2016cbd,Dedushenko:2016jxl}, the two-point function of a twisted operator $\cO(\varphi)$ corresponding to a CBO of dimension $\Delta$ has position dependence
\begin{align}
\langle \cO(\varphi_1)\cO(\varphi_2)\rangle_{S^3} = c \sgn(\varphi_1-\varphi_2)^{2\Delta}
\end{align}
for some constant $c$. 
It follows that 
\begin{align}
\langle \cM^b(\varphi_1) \cM^{-b}(\varphi_2)\rangle_\text{$S^3$, SQED$_1$} &= (\sgn(b)\sgn(\varphi_1-\varphi_2) )^{|b|} h(|b|)  \notag\\
&= (-1)^{\frac{|b|-b}{2}} (\sgn(\varphi_1-\varphi_2) )^{|b|} h(|b|) \label{mon2}
\end{align}
where, crucially, the factor of $\sgn(b)^{|b|}=(-1)^{\frac{|b|-b}{2}}$ accounts for the permutation symmetry $\varphi_1\leftrightarrow\varphi_2$, $b\leftrightarrow-b$, and $h$ depends only on $|b|$. In our calculation of the two-point function \eqref{mon2}, we fixed $\varphi_1=\pi/2$ and $\varphi_2=-\pi/2$, but this still leaves us with the $b$-dependent prefactor $(-1)^{\frac{|b|-b}{2}}$.\footnote{Note that replacing $(-1)^{(|b|-b)/2}\to(-1)^{(|b|+b)/2}$ in \eqref{detRatio} would also lead to an expression satisfying the desired properties of twisted monopole two-point functions.  We will see in Section \ref{matchsection} that the choice in \eqref{detRatio} is the one consistent with our conventions.} 
From the point of view of the determinant calculation of this section, the origin of this sign is quite mysterious because the spectrum is symmetric under $b\leftrightarrow -b$. Nevertheless, the above argument strongly suggests that it should be included in the final answer (see Section \ref{normamb} for further remarks). In Section \ref{SGLUE}, we will provide an alternative derivation of the $(-1)^{\frac{|b|-b}{2}}$ factor.

It is straightforward to generalize \eqref{detRatio} to abelian theories with $G=U(1)^r$ and $N_h$ hypermultiplets, as defined in Section \ref{abeliantheories}. 
For these theories, we have 
\begin{align}
Z(\vec{\sigma},\vec{b}) = \prod_{I=1}^{N_h} \frac{(-1)^{\frac{|\vec{q}_I\cdot \vec{b}|-\vec{q}_I\cdot \vec{b}}{2}}}{2\pi(\Lambda r)^{|\vec{q}_I\cdot \vec{b}|}}\Gamma\bigg(\frac{1+|\vec{q}_I\cdot \vec{b}|}{2}-i\vec{q}_I\cdot \vec{\sigma}\bigg)\Gamma\bigg(\frac{1+|\vec{q}_I\cdot \vec{b}|}{2}+i\vec{q}_I\cdot \vec{\sigma}\bigg) \label{detRatioGen}
\end{align}
(recall that $\vec{q}_I\in\mathbb{Z}^r$ is the vector of gauge charges of the $I^\textrm{th}$ hypermultiplet). From \eqref{detRatioGen}, one can read off the BPS monopole operator dimensions to be $\Delta_{\vec{b}} = \sum_{I=1}^{N_h}\frac{|\vec{q}_I\cdot \vec{b}|}{2}$, which is indeed the correct answer.

To summarize, we have shown that arbitrary correlators involving two twisted-translated monopole operators can be calculated by solving the matrix model
\begin{align}
\langle \cM^{\vec b}(\pi/2) \cM^{-\vec b}(-\pi/2) \cdots \rangle = \frac{1}{Z_{S^3}}\int\left(\prod_{i=1}^rd\sigma_i\right) Z(\vec{\sigma},\vec{b}) ({\cdots}) \ec \label{2moncor}
\end{align}
where $Z(\vec{\sigma},\vec{b})$ is given in \eqref{detRatioGen} and the $({\cdots})$ are some polynomials in $\Phi(\varphi)$, which in the monopole background localizes to \eqref{HatPhi2Mon}. We will discuss applications of the formula \eqref{2moncor} in Section \ref{applications}. Before doing so, we will show that the product over $\Gamma\big(\frac{1+|\vec{q}_I\cdot \vec{b}|}{2}-i\vec{q}_I\cdot \vec{\sigma}\big)$ in \eqref{detRatioGen} can be viewed as the partition function on a hemisphere with $\vec{b}$ units of flux threading its boundary $S^2$. The full expression \eqref{detRatioGen} can be viewed as the result of ``gluing'' two such partition functions. This point of view will lead to a simple generalization of \eqref{2moncor} to correlators with an arbitrary number of insertions of twisted-translated monopole operators.

\section{Localization on $HS^3$ and $\partial HS^3$}\label{SGLUE} 

A very useful representation of correlators of twisted CBOs, powerful enough to facilitate computations with an arbitrary number of monopole insertions, can be obtained by cutting $S^3$ into two hemispheres $HS^3$ along the equatorial $S^2$ that is orthogonal to the great circle where the 1D theory lives. The path integral on $HS^3$ then generates a state at the boundary $\partial HS^3= S^2$, and insertions of twisted CBOs can be represented by certain differential operators acting on this state. Gluing two hemispheres back together then allows one to recover the full $S^3$ answer.\footnote{We thank Davide Gaiotto for sharing the idea to use this approach.} As we will see, the boundary states (with insertions) in our case are $\cQ^C_\beta$-closed. It follows that the gluing of two such $\cQ_\beta^C$-closed states depends only on their cohomology classes.\footnote{In fact, we are going to compose a $\smash{\cQ^C_\beta}$-closed vector $|\Psi_+\rangle$ ($\smash{\cQ^C_\beta}|\Psi_+\rangle=0$) with a $\smash{\cQ^C_\beta}$-(co)closed covector $\langle\Psi_-|$ ($\langle\Psi_-|\smash{\cQ^C_\beta}=0$). This composition indeed descends to a composition on cohomology.} We will not, in practice, describe these cohomology classes: rather, we will utilize a slightly different philosophy, outlined in the next paragraph. 

Our strategy for gluing can be summarized as follows. Gluing two hemispheres along their common boundary is represented by a path integral on $S^2$, which we refer to as the ``gluing theory.'' This integral is taken over the space of boundary conditions corresponding to a fixed polarization on the phase space of the bulk theory. 
As will be explained, for our particular choice of supersymmetric polarization, the gluing theory itself preserves 2D $\cN=(2,2)$ supersymmetry on $S^2$. Applying supersymmetric localization to the gluing theory then reduces the infinite-dimensional functional integration at the boundary $S^2$ to a finite-dimensional integral over the space of half-BPS boundary conditions. 
In what follows, we will describe this technique, derive the gluing formula \eqref{Glue_F} via boundary localization, and derive the hemisphere partition function (or wavefunction) via localization on $HS^3$.

\subsection{Cutting and Gluing}\label{CutGlue}

The cutting and gluing axiom is one of the most fundamental properties of any local quantum field theory. The essence of cutting is that under a decomposition of a spacetime manifold into two components, the QFT dynamics as described by the path integral will generate physical states at the boundaries. The two boundary components in this decomposition have opposite orientations, so that one component supports a state living in some Hilbert space $\cH$, whereas the other supports a state living in the dual $\cH^\vee$. The gluing property refers to the opposite procedure: if we have two identical boundary components of opposite orientation, they support states in $\cH$ and $\cH^\vee$, respectively, and we can glue the spacetime manifold along these boundary components simply by composing the corresponding states.

In the context of supersymmetric boundary conditions and domain walls, the gluing procedure has appeared in various forms throughout the literature, a few examples being \cite{Drukker:2010jp, Gaiotto:2014gha, Beem:2012mb, Pasquetti:2016dyl, Bullimore:2014nla, Hori:2013ika, Cabo-Bizet:2016ars, Gava:2016oep, LeFloch:2017lbt, Dimofte:2017tpi}. In some of these works, concrete expressions for gluing are derived with the aid of heuristic arguments---for example, in \cite{Gava:2016oep}, where the need for a more illuminating derivation was emphasized. Here, we describe such a first-principles derivation for 3D $\cN=4$ theories, explaining the proper framework and relevant concepts along the way. A more detailed exposition of the gluing procedure and related symplectic geometry is presented in \cite{Dedushenko:2018aox, Dedushenko:2018tgx}.

In this problem, it is natural to start with a Hamiltonian formalism. Indeed, close to a boundary component $C\subset \partial M$, the manifold looks like a cylinder $C\times \R$. In the Hamiltonian description, $\R$ plays the role of time and the space of fields on $C$ is the configuration space. The bosonic fields and their time derivatives become, respectively, bosonic ``positions'' and ``momenta,'' while half of the fermionic fields become fermionic ``positions'' and the other half become fermionic ``momenta.'' There is a canonical Poisson bracket defined on the fields. This describes the phase space of the model, which is of course infinite-dimensional, unless we work with quantum mechanics (a 1D QFT).

To describe a boundary state, one has to choose what is called a  ``polarization''\cite{Woodhouse:1992de}: roughly, to pick one half of the phase space coordinates that Poisson-commute with each other and declare them to be ``position coordinates.'' States can then be defined as functionals of these position coordinates. The simplest situation occurs in quantum mechanics, where the phase space is $\R^{2n}$ parametrized by $p_i, q^i$, $i=1\dots n$, with the canonical Poisson bracket. Then the standard choice is to define states as square-integrable functions of $q^i$. In the path integral formulation, the action corresponding to this choice of polarization is written as
\begin{equation}
\label{actionSq}
S_q=\int_0^T (p_i \dot q^i - H) dt\ec
\end{equation}
with $H$ being the Hamiltonian. The boundary conditions are allowed to fix $q^i$ at the boundary, leading to the path integral formula for states in the ``position representation.'' For example, one can write $\psi(x)=\langle q=x|e^{-i HT}|q=0\rangle$ using the path integral as
\begin{align}
\psi(x) = \int_{\substack{q(T)=x\ec\\ q(0)=0}}\pD q\,\pD p\, e^{i\int_{0}^T(p_i \dot{q}^i - H(p,q))dt}\ed
\end{align}
In an alternative polarization, one could choose to fix the momenta $p_i$ at the boundary: this is commonly referred to as the ``momentum representation.'' It is known that for these boundary conditions to work, one has to write the action as:
\begin{equation}
S_p = \int_0^T (-q^i \dot p_i - H) dt = S_q - p_i q^i \big|_{t=T} + p_i q^i\big|_{t=0}\equiv  S_q - p_iq^i\big|_{t=0}^{t=T}\ed
\end{equation}
This differs from the action $S_q$ that was appropriate for the position picture by the boundary terms $-p_i q^i\big|_0^T$. We could also choose to fix coordinates for the first $k$ degrees of freedom and momenta for the remaining $n-k$ degrees of freedom. Then the proper boundary terms would be $-\sum_{i=k+1}^n p_i q^i \big|_0^T$.

One of the reasons that the boundary terms show up is to make the variational problem well-defined, i.e., to ensure that there are no boundary corrections to the equations of motion.\footnote{Note: this is a different perspective from the one adopted in some literature on supersymmetric boundary conditions, where boundary terms in the equations of motion are used to \emph{derive} boundary conditions, e.g., in \cite{Dimofte:2017tpi}. From that perspective, one would start with the action \eqref{actionSq} \emph{without} any boundary conditions and conclude that boundary equations of motion enforce $p(0)=p(T)=0$. This gives a single boundary condition, as opposed to a family of boundary conditions parametrized by $q$. We need the latter perspective, in which $p_i\delta q^i\big|$ vanishes because of $\delta q^i\big|=0$, not because of $p_i\big|=0$, to be able to describe boundary states.} For example, the variation of the position picture action $S_q$ is
\begin{equation}
\delta S_q = \int dt\left[\delta p_i(\dot{q}^i - \partial H/\partial p_i) - \delta q^i(\dot{p}_i + \partial H/\partial q^i)\right] + p_i \delta q^i\big|_0^T\ed
\end{equation}
Generically, the Hamiltonian equations of motion follow from the above variation if $\delta q^i$ vanish at the boundary, so that the positions $q^i$ take fixed values thereon. If this is not the case and we are considering more general boundary conditions, then we are forced to include boundary terms $F_{1,2}$ such that $p_i \delta q^i +\delta F_1 \big|_{t=0}=0$ and $p_i \delta q^i +\delta F_2 \big|_{t=T}=0$. The case of general boundary conditions given by Lagrangian submanifolds, and in particular the question of how to construct boundary terms in that case, is studied in \cite{Dedushenko:2018aox}.

In the upcoming subsections, we will use the fact that if the theory has a symmetry that preserves the polarization, then this symmetry is induced in the gluing path integral \cite{Dedushenko:2018aox}. For us, the relevant symmetry will be supersymmetry. What does it mean that a symmetry preserves the polarization? If we choose to fix the positions $q^i$ at the boundary, it simply means that the symmetry transforms a $q^i={\rm const.}$ submanifold into $q^i = \widetilde{\rm const.}$, where $\widetilde{\rm const.}$ are some other constants.

Let us illustrate this statement for the simplest example of a position-based polarization, in which the wavefunctions depend only on $q^i$. Suppose that a theory has a symmetry whose generating function is
\begin{equation}
Y = \sum_i c^i p_i + a(q)\ec
\end{equation}
where $c^i$ are constants. The corresponding Hamiltonian vector field, $X_Y=\sum_i c^i \frac{\partial}{\partial q^i} - \frac{\partial a(q)}{\partial q^i}\frac{\partial}{\partial p_i}$, obviously preserves the position-based polarization: every subspace $q^i={\rm const.}$ is transformed into another subspace of the same type. 
Suppose that $\psi_1$ and $\psi_2$ are states annihilated by the symmetry generated by $Y$, i.e.,
\begin{equation}
-i\sum_j c^j \frac{\partial}{\partial q^j}\psi_{1,2} + a(q)\psi_{1,2}=0\ec
\end{equation}
or in infinitesimal form,
\begin{equation}
\psi_{1,2}(q^i + \epsilon c^i) = e^{-i\epsilon a(q)}\psi_{1,2}\ed
\end{equation}
Then, clearly, the following holds:
\begin{equation}
\psi_1^*(q^i+\epsilon c^i)\psi_2(q^i+\epsilon c^i)=\psi_1^*(q^i)\psi_2(q^i)\ed
\end{equation}
The symmetry $Y$ induces a transformation $q^i \mapsto q^i + \epsilon c^i$ on the positions, and the product $\psi_1^*(q)\psi_2(q)$ is invariant with respect to it. This means that the integral performing the gluing,
\begin{equation}
\int d^nq\, \psi_1^*(q)\psi_2(q)\ec
\end{equation}
has a symmetry $q^i \to q^i + \epsilon c^i$.  We can say that $|\psi_2\rangle\in\cH$, whereupon $\psi_1$ determines an element of the dual space:
\begin{equation}
\langle\psi_1| \in\cH^\vee,\quad \langle\psi_1|\psi_2\rangle = \int d^nq\, \langle\psi_1|q\rangle \langle q|\psi_2\rangle,\quad \langle\psi_1|q\rangle = \psi_1^*(q),\quad \langle q|\psi_2\rangle = \psi_2(q)\ed
\end{equation}
This formulation is very natural: two copies of the boundary, with opposite orientations, support the Hilbert space $\cH$ and its dual $\cH^\vee$, with $\langle q|\psi_2\rangle$ and $\langle\psi_1|q\rangle$ representing their elements, respectively. The complex conjugation comes into play only if we use the Hilbert space structure on $\cH$ to relate it with $\cH^\vee$.

The above quantum mechanics example is a model of what is going to happen in our 3D theory: the symmetry $Y$ will be replaced by supersymmetry and the boundary states will be supersymmetric, as will the boundary path integral performing the gluing. This will allow for the use of supersymmetric localization to simplify the gluing.

\subsection{Supersymmetric Cutting and Gluing of Hemispheres}

Upon cutting $S^3$ into two hemispheres along the equatorial $S^2$, the isometry group $SO(4)$ is broken down to the isometry group $SO(3)$ of $S^2$. Correspondingly, the $\cN=4$ superalgebra $\mathfrak{su}(2|1)_\ell\oplus \mathfrak{su}(2|1)_r$ is broken as well. The maximal subalgebra that can remain unbroken is $\mathfrak{su}(2|1)$, which is the $\cN=(2,2)$ superalgebra on $S^2$. As is well-known, the latter comes in two versions, $\mathfrak{su}(2|1)_A$ and $\mathfrak{su}(2|1)_B$, related by 2D mirror symmetry \cite{Doroud:2013pka}. Correspondingly, we can impose two types of boundary conditions on an empty hemisphere, preserving either $\mathfrak{su}(2|1)_A$ or $\mathfrak{su}(2|1)_B$.\footnote{In the language of \cite{Chung:2016pgt}, these two types of boundary conditions are both called ``A-type,'' while their ``B-type'' preserves $(0,4)$ SUSY and has no counterpart in our story.} To see how this works in relation to 3D mirror symmetry, consider an outer automorphism $\mathfrak{a}$ of $\mathfrak{su}(2|1)_\ell\oplus \mathfrak{su}(2|1)_r$ that acts trivially on all generators, except:
\begin{align}
\mathfrak{a}(R_r)&=-R_r\ec\cr
\mathfrak{a}(\cQ_\alpha^{(r\pm)})&=\cQ_\alpha^{(r\mp)}\ed
\end{align}
This is the automorphism underlying 3D mirror symmetry: in particular, it switches $R_H$ and $R_C$. Up to conjugation, one can identify the $\mathfrak{su}(2|1)_A$ subalgebra as $\operatorname{diag}\left[\mathfrak{su}(2|1)_\ell\oplus \mathfrak{su}(2|1)_r\right]$. Then, up to conjugation, the $\mathfrak{su}(2|1)_B$ subalgebra is $\operatorname{diag}\left[\mathfrak{su}(2|1)_\ell\oplus \mathfrak{a}(\mathfrak{su}(2|1)_r)\right]$. We observe that $\cQ^C_\alpha\in\mathfrak{su}(2|1)_A$ and $\cQ^H_\alpha\in\mathfrak{su}(2|1)_B$. Furthermore, insertions of twisted CBOs at the tip of the hemisphere preserve $\mathfrak{su}(2|1)_A$, while similar insertions of twisted HBOs preserve $\mathfrak{su}(2|1)_B$ (see Footnote \ref{halfBPStip}). This implies that in this paper, we need only preserve $\mathfrak{su}(2|1)_A$ at the boundary, as $\mathfrak{su}(2|1)_B$ would be relevant for the mirror Higgs branch story. For this reason, we drop the subscript $A$ in what follows and simply write $\mathfrak{su}(2|1)$. 

In our conventions, the diagonal subalgebra $\diag\left[\mathfrak{su}(2|1)_\ell\oplus \mathfrak{su}(2|1)_r\right]$ preserves the great $S^2$ located at $\varphi=\pm\pi/2$. We choose to perform a cut along a different great $S^2$ located at $\varphi=0$ and $\varphi=\pm\pi$. 
Correspondingly, we will denote by $HS^3_+$ the hemisphere with $0<\varphi<\pi$, and by $HS^3_-$ the one with $-\pi<\varphi<0$. 
The $\mathfrak{su}(2|1)$ preserved by this cut is conjugate to $\diag\left[\mathfrak{su}(2|1)_\ell\oplus \mathfrak{su}(2|1)_r\right]$. More explicitly, in terms of the $\mathfrak{su}(2|1)_\ell \oplus \mathfrak{su}(2|1)_r$ supercharges $\cQ^{(\ell_{\pm})}_{\alpha}$ and $\cQ^{(r_{\pm})}_{\alpha}$, the $\mathfrak{su}(2|1)$ subalgebra preserved on our $HS^3$ is generated by
\begin{align}
Q_1^+ \equiv \cQ_1^{(\ell+)} + \cQ_1^{(r+)},\quad Q_2^+ \equiv \cQ_2^{(\ell+)} - \cQ_2^{(r+)} \ec\label{Qp22}\\ 
Q_1^- \equiv \cQ_1^{(\ell-)} - \cQ_1^{(r-)},\quad Q_2^- \equiv \cQ_2^{(\ell-)} + \cQ_2^{(r-)} \ed\label{Qm22}
\end{align}
Most importantly, our Coulomb branch supercharges \eqref{QC12} are part of this algebra, and identified as $\cQ_1^C = Q^+_1$ and $\cQ_2^C = Q^-_2$.

In the presence of real masses $\widehat{m}$, it is the central extension of $\mathfrak{su}(2|1)$ that becomes relevant. Indeed, by \eqref{CentrExt}, the central charge entering $\diag\left[\mathfrak{su}(2|1)_\ell\oplus \mathfrak{su}(2|1)_r\right]$ is $ir\widehat{m}$. This fact is not changed by conjugation, so the central extension appearing in our $\mathfrak{su}(2|1)$  always corresponds to mass deformations. The central transformations generated by FI terms, on the other hand, are not symmetries on $HS^3$. Those transformations multiply monopoles by a phase proportional to their charge, and because the total charge need not vanish on $HS^3$, they are not symmetries there. Note that the twisted translation \eqref{twistedTR} is also not a symmetry on $HS^3$, and only becomes one on the full $S^3$.

In what follows, we will first discuss how to include insertions of twisted CBOs on $HS^3$ in an $\mathfrak{su}(2|1)$-invariant way. We will then describe the phase space of our theories close to the $S^2$ boundary, and show that there is an $\mathfrak{su}(2|1)$-preserving polarization in the sense described in the previous subsection. 

\subsubsection{Operator Insertions and $\mathfrak{su}(2|1)$}

The path integral on an empty $HS^3$ generates a state at the boundary $S^2$ which is invariant under all supersymmetries, and in particular under $\mathfrak{su}(2|1)$. Moreover, the tip of $HS^3$ is a fixed point of the $SO(3)$ isometry, so the latter is not broken by insertions of scalar local operators there. In fact, it turns out that the full $\mathfrak{su}(2|1)$ symmetry is preserved by insertions of twisted CBOs at the tip of $HS^3$ (see Footnote \ref{halfBPStip} and the previous subsection). On the other hand, insertions of twisted-translated operators along the great semicircle of $HS^3$ away from the tip generally break the $\mathfrak{su}(2|1)$ symmetry. However, by performing a simple step before cutting $S^3$ into two halves, we can reduce calculations involving generic insertions to those involving only $\mathfrak{su}(2|1)$-invariant insertions, without any loss of generality.

Indeed, owing to \eqref{cohomologousH}--\eqref{cohomologousC} and \eqref{HtwistedTR}--\eqref{twistedTR}, we know exactly how correlators of twisted-translated operators on $S^3$ depend on their insertion points along the great circle. Specifically, suppose that $\cO_1(\varphi_1), \dots, \cO_n(\varphi_n)$ are twisted-translated operators inserted at points $\varphi_1<\dots <\varphi_n$, and suppose that they carry monopole charges $b_1, \dots, b_n$. From \eqref{cohomologousH}--\eqref{cohomologousC} and \eqref{HtwistedTR}--\eqref{twistedTR} (which state that $\widehat{P}^H_\varphi$ and $\widehat{P}^C_\varphi$ are cohomologous to $-ir\widehat{m}$ and $-ir\widehat{\zeta}$, respectively), we deduce that their correlation function on $S^3$ has position dependence 
\begin{equation}
\langle \cO_1(\varphi_1)\dots \cO_n(\varphi_n)\rangle \propto \exp\left(-r\sum_{k=1}^n \zeta b_k \varphi_k \right)\ec
\end{equation}
where $\zeta$ is an FI parameter (if the gauge group contains multiple $U(1)$ factors, then $\zeta$ and $b_k$ are vectors, and they are dotted into each other in the expression above). In particular, for vanishing FI parameters, the correlator has no position dependence at all, as long as we keep the ordering of operators unchanged. 

Now suppose that we cut $S^3$ along the equator at $\varphi=0$ and $\pm\pi$. Some insertions (say, $\cO_1,\ldots, \cO_k$) will end up on the hemisphere $HS^3_-$, while the others end up on $HS^3_+$. Let us move all operators to the tip of their corresponding hemisphere. Using the OPE, we define
\begin{align}
&\lim_{\substack{\varphi_i\to \pi/2 \\ \varphi_{k+1}<\cdots<\varphi_n}}\, \cO_{k+1}(\varphi_{k+1})\,\cO_{k+2}(\varphi_{k+2})\,\dots\, \cO_n(\varphi_n) = \cO_+(\pi/2) + \{\cQ^C,\dots\}\ec\cr
&\lim_{\substack{\varphi_i\to -\pi/2 \\ \varphi_1<\cdots<\varphi_k}}\, \cO_1(\varphi_1)\,\cO_2(\varphi_2)\,\dots\, \cO_k(\varphi_k) = \cO_-(-\pi/2) + \{\cQ^C,\dots\}\ec
\end{align}
where $\cO_{\pm}$ are some twisted CBOs. Then the full correlation function on $S^3$ is simply
\begin{equation}
\langle \cO_1(\varphi_1)\dots \cO_n(\varphi_n)\rangle=\langle \cO_+(\pi/2) \cO_-(-\pi/2) \rangle \times \exp\left(-r\sum_{k=1}^n \zeta b_k \varphi_k \right)\ed
\end{equation}
Now we can safely cut $S^3$ into two halves, with $\cO_{\pm}$ inserted at the tip of $HS^3_{\pm}$. These configurations generate $\mathfrak{su}(2|1)$-invariant states $\Psi_{\pm}$ at the boundaries of $HS^3_{\pm}$. The use of the OPE above is a bit formal, as we do not know it \emph{a priori}. In Section \ref{multipleinsertions}, we will see how it can nevertheless be determined only from knowing how to glue $HS^3$ wavefunctions with insertions at their tips.

\subsubsection{The Phase Space}

To apply the canonical formalism, we start by describing the phase space for the theory on $S^2\times \R$. Note that close to the equator, $S^3$ looks like $S^2\times \R$.  Hence there is no need to separately study actions on $S^2\times \R$, as all relevant information can be read off from the action on $S^3$. In other words, the Hilbert space of states on $S^2$ does not depend on which three-manifold this $S^2$ bounds: it could be $HS^3$, a half-cylinder $S^2\times \R_+$, or anything else.  The role of the bulk is merely to prepare a certain state at the boundary.

Let $\partial_\perp$ denote the derivative along the unit normal to $S^2$. In the canonical formalism, $\partial_\perp$ is thought of as the ``time derivative.'' On $S^2\times \R$, we have $\partial_\perp=\frac{\partial}{\partial x^0}$ with $x^0$ being the coordinate on $\R$. On $HS^3_+$ ($0<\varphi<\pi$), $\partial_\perp$ is given by
\begin{equation}
\partial_\perp = -\frac{\sgn(\cos \varphi)}{r\sin\theta}\frac{\partial}{\partial\varphi} \ec
\end{equation}
while on $HS^3_-$ ($-\pi<\varphi<0$), it is given by
\begin{equation}
\partial_\perp = +\frac{\sgn(\cos \varphi)}{r\sin\theta}\frac{\partial}{\partial\varphi}\ed
\end{equation}
With respect to $\partial_\perp$, the momenta canonically conjugate to $q_a$ and $\tq^a$ are, respectively, $p^a=\cD_\perp \tq^a$ and $\widetilde{p}_a=\cD_\perp q_a$. The corresponding Poisson brackets are\footnote{Strictly speaking, $\partial_\perp \widetilde{q}^a$ here is not really a derivative but merely a symbol standing for $p^a + i A_\perp \widetilde{q}^a$, and similarly for $\partial_\perp q_a$. This distinction will not be important anywhere in this paper.}
\begin{align}
\{q_a(x), \partial_\perp \tq^b(y) \}_P &= \delta_a^b\, \mathbb{1}_\cR \delta_{S^2}(x-y)\ecq\quad
\{\tq^a(x), \partial_\perp q_b(y) \}_P = \delta^a_b\, \mathbb{1}_\cR \delta_{S^2}(x-y)\ec 
\end{align}
where $\mathbb{1}_\cR$ is the identity matrix in the representation $\cR$, and $\delta_{S^2}(x-y)$ is a delta-function on $S^2$. Similarly, the Poisson brackets for $\Phi_{\dot a\dot b}$ are
\begin{align}
\{\Phi_{\dot a\dot b}(x), \partial_\perp \Phi^{\dot c\dot d}(y)\}_P &= -\frac{g_\text{YM}^2}4\left(\delta_{\dot a}^{\dot c} \delta_{\dot b}^{\dot d} + \delta_{\dot a}^{\dot d} \delta_{\dot b}^{\dot c} \right)\mathbb{1}_{\rm Adj}\delta_{S^2}(x-y)\ed
\end{align}
The auxiliary fields $D_{ab}$ are eliminated in the canonical formalism because the action does not include their derivatives. There are many equivalent ways to understand this. For example, they could simply be integrated out before quantizing the theory. Alternatively, recall that the phase space can be interpreted as the space of solutions to the classical equations of motion modulo gauge equivalences. The classical equations for $D_{ab}$ are algebraic and can be used to express $D_{ab}$ in terms of the other fields. Finally, we could apply Dirac's procedure by introducing conjugate momenta $\Pi_D^{ab}$ for $D_{ab}$ with the Poisson bracket 
\begin{equation}
\{D_{ab}(x),\Pi_D^{cd}(y)\}_P=\frac12 \left(\delta_{a}^{c} \delta_{b}^{d} + \delta_{a}^{d} \delta_{b}^{c}\right)\mathbb{1}_{\rm Adj}\delta_{S^2}(x-y)\ec
\end{equation}
which satisfy the constraint $\Pi_D^{ab}=0$. This induces a secondary constraint putting $D_{ab}$ on shell:
\begin{equation}
\label{onshD}
D^A_{ab}=- \frac{ig_\text{YM}^2}{2}\tq_{(a} T^A q_{b)} - \frac{1}{2r} h_{ab} (\bar{h}^{\dot a}{}_{\dot b} \Phi^{A, \dot b}{}_{\dot a})-\frac{i}2 g_\text{YM}^2 \zeta(T^A) h_{ab},
\end{equation}
where $T^A$ ($A=1\dots \dim G$) denote the generators of $G$ in the representation $\cR$ and $\zeta(T^A)$ denote possible FI terms that can only be present for those $T^A$ corresponding to $U(1)$ factors of $G$. Again, auxiliary fields are eliminated, the physical subspace being constructed as the solutions to \eqref{onshD} and $\Pi_D^{ab}=0$ (modded out by gauge symmetries). Note that because of \eqref{onshD}, $D^A_{ab}$ has nontrivial Poisson brackets with other fields on the physical subspace:
\begin{align}
\label{Dpoiss}
\{ D^A_{ab}(x), \partial_\perp \tq^c(y) \}_P &= -\frac{i g_\text{YM}^2}{2} \delta_{(b}^c\, \tq_{a)} T^A \delta_{S^2}(x-y) \ec\cr
\{ D^A_{ab}(x), \partial_\perp q^c(y) \}_P &= \frac{i g_\text{YM}^2}{2} \delta_{(a}^c\, T^A q_{b)} \delta_{S^2}(x-y) \ec\\
\{D^A_{ab}(x), \partial_\perp \Phi^{B, \dot{a}\dot{b}}(y)\}_P &= - \frac{g_\text{YM}^2}{4r}h_{ab} \bar{h}^{\dot{a}\dot{b}}\delta^{AB} \delta_{S^2}(x-y)\ec\notag
\end{align}
where we have left the representation label $\cR$ on hypermultiplet scalars implicit.

As usual, it is useful to keep in mind all equivalent descriptions of the phase space at once. In particular, we will often have $D_{ab}$ present in our equations, alluding to the latter description. On the other hand, the definition of the phase space as the space of solutions to the classical EOMs allows us to be cavalier about closing SUSY off shell: when we act with SUSY in the phase space, we simply transform one classical solution into another, so we are completely free to use the equations of motion.

Proceeding with our description of the phase space, the remaining bosonic fields are gauge fields. We denote the component of $A_\mu$ along the $\R$ direction in $S^2\times \R$ by $A_0$ and the components along the $S^2$ directions by $A_i$.  The canonical formalism complements them by conjugate momenta $\pi^0$ and $\pi^i$, as well as the constraints
\begin{equation}
\label{gauge_phase}
A_0 = \pi^0 =0\ecq\quad \cD^i A_i = \cD_i\pi^i=0\ed
\end{equation}
The canonical Poisson bracket $\{A_\mu(x), \pi^\nu(y)\}_P=\delta_\mu^\nu \delta_{S^2}(x-y)$ induces a Poisson bracket on the constraint subspace.\footnote{Note that $A_i$ can be interpreted as a gauge field on $S^2$, and the constraint $\cD^i A_i=0$ as a gauge-fixing condition. If we choose a position-based polarization and describe wavefunctions as functionals of $A_i$ on a subspace determined by $\cD^i A_i=0$, then we can alternatively relax this constraint and say that wavefunctions for the gauge field are simply gauge-invariant functionals of $A_i$.} On $HS^3$, we will identify $A_\perp$ with $A_0$ where, as before, $\perp$ denotes the direction normal to $S^2$. In this situation, it will be convenient to interpret the constraint $A_\perp=0$ as a partial gauge-fixing on $HS^3$.

Finally, let us turn to the fermions. In the canonical formalism, half of them become ``positions'' and the other half their conjugate momenta. This is simply because the action for fermions is of first order in derivatives. The Poisson brackets turn out to be\footnote{A na\"ive application of the canonical formalism would not give a factor of $1/2$ in the second equation of \eqref{ferm_P}: to obtain this coefficient, one must properly account for the second class constraints and construct the Dirac bracket on the constraint surface.}
\begin{align}
\label{ferm_P}
\{\psi_{\alpha\dot a}(x), \tpsi^{\beta\dot b}(y)\}_P&=i\delta_{\dot a}^{\dot b}\left(\gamma_\perp\right)_\alpha{}^\beta \delta_{S^2}(x-y)\ec\cr
\{\lambda_{\alpha a \dot a}(x), \lambda_{\beta b\dot b}(y)\}_P&=-\frac{i}{2} \varepsilon_{ab}\varepsilon_{\dot a\dot b}\left(\gamma_\perp\right)_{\alpha\beta} \delta_{S^2}(x-y)\ec
\end{align}
where $\gamma_\perp$ is the component of $\gamma_\mu$ along the unit vector field normal to $S^2$. In particular, $\gamma_\perp = -\frac{\sgn(\cos\varphi)}{r\sin\theta}\gamma_\varphi\big|_{\text{boundary}}=-\sigma_3$.

\subsubsection{The $\mathfrak{su}(2|1)$-Invariant Polarization} 

We would now like to describe a proper choice of splitting of the phase space variables of our theory, such that half of them define an $\mathfrak{su}(2|1)$-invariant polarization. In other words, we want to find field combinations that form $\mathfrak{su}(2|1)$ multiplets, in addition to Poisson-commuting with each other at the boundary.\footnote{Poisson commutativity would hold everywhere if we were working on $S^2\times \R$, but on $HS^3$, it only needs to hold at the boundary.} Fixing such field combinations on $S^2$ will provide us with the appropriate family of boundary conditions, inducing 2D $\cN=(2,2)$ supersymmetry in the gluing theory and allowing for localization of the gluing path integral.

Our strategy is to start with the combinations of scalars (familiar from \cite{Dedushenko:2016jxl})
\begin{align}
q_\pm &= q_1 \pm i q_2\ecq\quad \tq_\pm = \tq_1 \pm i \tq_2\ed\label{qpm}
\end{align}
Under $\mathfrak{su}(2|1)$ SUSY transformations (restricted to the boundary), the combinations \eqref{qpm} transform into the boundary fermions
\begin{align}
\chi&=(\psi_{\dot 1}-\sigma_3 \psi_{\dot 2})\big|\ec \quad \widetilde{\chi}=i(\tpsi_{\dot 1}+\sigma_3 \tpsi_{\dot 2})\big|\ec\label{xi}\\
\widetilde{\rho}&=(\tpsi_{\dot 1}-\sigma_3 \tpsi_{\dot 2})\big|\ec\quad \rho=-i(\psi_{\dot 1}+\sigma_3\psi_{\dot 2})\big| \ec\label{rho}
\end{align}
where the notation $X\big|$ denotes the restriction of $X$ to the boundary $S^2$.
The only nonvanishing Poisson brackets between the fermions in \eqref{xi} and \eqref{rho} are given by
\begin{align}
\{\chi_\alpha(x),\widetilde\rho_\beta(y)\}_P=\{\widetilde\chi_\alpha(x),\rho_\beta(y)\}_P=2i\varepsilon_{\alpha\beta} \delta_{S^2}(x-y)\ec
\end{align}
suggesting that, e.g., $\chi$ and $\widetilde\chi$ could be good candidates for the ``positions'' of the hypermultiplet fermions (and indeed they are, as we will see momentarily). Further acting on $\chi$ and $\widetilde{\chi}$ with supersymmetry generates entire 2D $\cN=(2,2)$ multiplets that Poisson-commute.

Let us summarize the results of this lengthy calculation. We identify a 2D $\cN=(2,2)$ chiral multiplet $\Phi^{(2d)}$ and a vector multiplet $V^{(2d)}$, whose components we denote by
\begin{align}
\Phi^{(2d)} &= (\phi, \widetilde{\phi}, \chi_{\alpha}, \widetilde{\chi}_{\alpha}, f, \widetilde{f}) \ec \label{PH2d}\\
V^{(2d)} &= (a, \lambda_{\alpha}, \tilde{\lambda}_{\alpha}, s_1, s_2, D^{(2d)}) \ed \label{V2d}
\end{align}
In \eqref{PH2d}, the scalars $\phi$ and $\widetilde{\phi}$ are complex conjugates, $\chi_{\alpha}$ and $\widetilde{\chi}_{\alpha}$ are their fermionic superpartners defined in terms of the bulk fields in \eqref{xi}, and $f$ and $\widetilde{f}$ are the complex conjugate auxiliary fields of the 2D chiral multiplet. In \eqref{V2d}, $a$ denotes the 2D gauge field on $S^2$, $\lambda$ and $\widetilde{\lambda}$ are the gauginos, $s_{1,2}$ are real scalar fields, and $D^{(2d)}$ is the auxiliary scalar in the 2D vector multiplet.

Apart from $\chi$ and $\widetilde{\chi}$, which are already written in \eqref{xi}, the other components of $\Phi^{(2d)}$ in \eqref{PH2d} are identified with boundary values of bulk fields as
\begin{align}
\phi &= q_+\big|  \ecq\quad f = \left(-\cD_\perp q_- - \frac{\Phi_{\dot1\dot1} - \Phi_{\dot2\dot2}}{2}q_-\right)\bigg| \ec \label{phif}
\end{align}
where $q_+$ was defined in \eqref{qpm} and the conjugate components $\widetilde{\phi} = \phi^\ast$ and $\widetilde{f} = f^\ast$ can be found using the reality conditions \eqref{RealqAPhD} satisfied by the bulk fields.

The components of $V^{(2d)}$ can be written in terms of the bulk fields as
\begin{align}
a &= A_{||}\big| \ec \label{A2d} \\
\lambda &= -\frac{1}2\left( \lambda_{1\dot2} - i\lambda_{2\dot2} +\sigma_3(\lambda_{1\dot1}-i\lambda_{2\dot1}) \right)\big|\ec \\
\widetilde{\lambda} &= -\frac{1}2\left( \lambda_{1\dot2} + i\lambda_{2\dot2} -\sigma_3(\lambda_{1\dot1}+i\lambda_{2\dot1}) \right)\big| \ec \\
s_1 &= \frac{\Phi_{\dot1\dot1}+\Phi_{\dot2\dot2}}{2i}\bigg|\ecq \quad s_2 = -\Phi_{\dot1\dot2}\big| \ec \\
D_{2d} &= \left[-\frac{\Phi_{\dot1\dot2}}{r} +\frac{i}2 \left(D^\text{on-shell}_{11}+D^\text{on-shell}_{22}\right)  + i\cD_\perp\left(\frac{\Phi_{\dot1\dot1}-\Phi_{\dot2\dot2}}{2}\right)\right]\bigg| \ed \label{D2d}
\end{align}
In \eqref{A2d}, we have defined $A_{\parallel} \equiv A_{\theta}d\theta+A_{\tau}d\tau$. The $D_{ab}^\text{on-shell}$ appearing in \eqref{D2d} denote the on-shell values of the auxiliary fields given in \eqref{onshD}. This traces back to the fact that in the description of the phase space, $D_{ab}$ takes its on-shell value.\footnote{Using \eqref{onshD}, the equation for $D_{2d}$ can alternatively be written as $D^A_{2d} =\frac{g_\text{YM}^2}{8}(\tq_+ T^A q_- + \tq_- T^A q_+) - \frac{i}2 g^2\zeta(T^A) + i\cD_\perp \left(\frac{\Phi^A_{\dot1\dot1}-\Phi^A_{\dot2\dot2}}{2}\right)\Big|$.} Finally, in addition to \eqref{xi}, \eqref{phif}, and \eqref{A2d}--\eqref{D2d}, which fix the 2D multiplets $\Phi^{(2d)}$ and $V^{(2d)}$ at the boundary, we impose the boundary condition
\begin{equation}
A_\perp\big|=0 \ed \label{Aperp}
\end{equation}
The condition \eqref{Aperp} should be interpreted as a partial gauge-fixing on $HS^3$. The necessity of imposing this condition follows from the description of the phase space for gauge fields in \eqref{gauge_phase}.

It is trivial to verify that the field combinations defined in \eqref{xi}, \eqref{phif}, \eqref{A2d}--\eqref{D2d}, and \eqref{Aperp} form a maximal subset of the phase space variables that all Poisson-commute at the boundary $S^2$.\footnote{When checking this, one should keep in mind that $D_{ab}$ has nonzero Poisson brackets with some other fields, as in \eqref{Dpoiss}.} This means that fixing them on $S^2$ is a consistent boundary condition for the path integral on $HS^3$. Moreover, one can check that under the $\mathfrak{su}(2|1)$ transformations restricted to $S^2$, the combinations \eqref{xi}, \eqref{phif}, and \eqref{A2d}--\eqref{D2d} indeed transform as 2D $\cN=(2,2)$ chiral and vector multiplets, respectively. These transformations, as well as further details on the boundary $\mathfrak{su}(2|1)$ SUSY variations, are summarized in Appendix \ref{susytrans}.\footnote{In particular, when applying the SUSY variations in our formalism, one should impose the equations of motion because the 3D $\cN=4$ algebra does not close off shell on the hypermultiplet. Furthermore, SUSY breaks the gauge-fixing condition $A_{\perp}\big|=0$, so this must be compensated for by a gauge transformation.}

For completeness, let us also describe the boundary terms that one must add to the action to guarantee that the variational problem is well-defined with the above boundary conditions. To do so, we introduce another set of fermionic variables
\begin{align}
\omega&=\frac{1}2 (\lambda_{1\dot2} - i\lambda_{2\dot2} -\sigma_3(\lambda_{1\dot1}-i\lambda_{2\dot1}))\ec\cr
\widetilde{\omega}&=\frac12 (\lambda_{1\dot2}+i\lambda_{2\dot2} + \sigma_3(\lambda_{1\dot1}+i\lambda_{2\dot1}))\ed
\end{align}
These are canonically conjugate to $\lambda$ and $\widetilde\lambda$, so the only nonzero Poisson brackets are
\begin{equation}
\{\lambda_\alpha(x),\widetilde{\omega}_\beta(y)\}_P=\{\widetilde{\lambda}_\alpha(x),\omega_\beta(y)\}_P=\frac12 \varepsilon_{\alpha\beta}\delta_{S^2}(x-y)\ed
\end{equation}
The proper boundary term can then be written as follows:
\begin{gather}
S_{\partial}= \frac{i}2\int_{S^2} d^2x \left(\tq_+ (\cD_\perp q_- + \frac1{2}(\Phi_{\dot1\dot1}-\Phi_{\dot2\dot2})q_-) + q_-(\cD_\perp\tq_+ + \frac{1}{2}(\Phi_{\dot1\dot1}-\Phi_{\dot2\dot2})\tq_+)+ \widetilde{\chi}^\alpha\rho_\alpha\right)\notag\\
{}+\frac{1}{g_\text{YM}^2}\int_{S^2} d^2x\left( (\Phi^{\dot1\dot1}-\Phi^{\dot2\dot2})\cD_\perp(\Phi_{\dot1\dot1}-\Phi_{\dot2\dot2}) +  \lambda^\alpha\widetilde{\omega}_\alpha + \widetilde{\lambda}^\alpha\omega_\alpha)\right)\ed \label{b_term}
\end{gather}
The boundary term \eqref{b_term} can be constructed along the lines of the discussion in Section \ref{CutGlue}. Adding it to the action ensures that the path integral on the upper hemisphere $HS^3_+$ produces a boundary state written in our polarization. While \eqref{b_term} is needed for consistency, we will see shortly that it vanishes on the localization locus of the $HS^3$ path integral, and being a term in the classical action, it does not contribute in the localization computation.

If we denote the boundary conditions collectively by
\begin{equation}
\mathscr{B} = (\Phi^{(2d)}, V^{(2d)})\ec
\end{equation}
then the state $|\Psi_+\rangle$ generated by the $HS^3_+$ path integral with these boundary conditions (and the gauge fixing \eqref{Aperp}) is represented by a functional of $\mathscr{B}$:
\begin{equation}
\Psi_+[\mathscr{B}]=\langle\mathscr{B}|\Psi_+\rangle\ed
\end{equation}
The dual state $\langle\Psi_-|$ generated by the $HS^3_-$ path integral with the same boundary conditions can be written as 
\begin{equation}
\Psi_-[\mathscr{B}]=\langle\Psi_-|\mathscr{B}\rangle \ed
\end{equation}
Gluing these states is tantamount to computing the path integral 
\begin{equation}
\label{susy2dPI}
\int \pD \mathscr{B}\, \langle\Psi_-|\mathscr{B}\rangle \langle\mathscr{B}|\Psi_+\rangle\ec
\end{equation}
which has $\mathfrak{su}(2|1)$ supersymmetry due to the $\mathfrak{su}(2|1)$-invariance of the polarization. The computation of this path integral will be performed in the following subsection.

\subsection{Boundary Localization and the Gluing Formula}

With the answer \eqref{susy2dPI} in hand, all we must do is localize it. Localization of $\cN=(2,2)$ theories on $S^2$ was studied in \cite{Benini:2012ui,Doroud:2012xw,Gomis:2012wy} and reviewed in \cite{Benini:2016qnm}. We will simply borrow these results, mostly following \cite{Doroud:2012xw}. Notice that the supercharge used in \cite{Doroud:2012xw} for localization is
\begin{equation}
Q_1^+ + Q_2^- = \cQ_1^C + \cQ^C_2 \ec
\end{equation}
which is precisely our $\cQ^C_\beta$ at $\beta=1$. This fact implies that as long as we use our boundary conditions, we do not really need the full $\mathfrak{su}(2|1)$ symmetry to localize the gluing theory. It is enough to have only $\cQ^C_{1,2}$ preserved, and this gives us the freedom to move twisted CBOs along the great semicircle of $HS^3$ as well as to include certain nonlocal observables. For simplicity, we will not exploit this freedom in what follows: we will simply restrict our attention to insertions of twisted CBOs at the tip of $HS^3$.

The results of \cite{Doroud:2012xw} come in two forms: those corresponding to Coulomb branch and to Higgs branch localization. The one relevant to us is the former. On the localization locus, all the 2D fermions vanish and the bosons take the following values:
\begin{gather}
\label{locloc2d}
a=\pm\frac{B}{2} (\sin\theta- 1)\, d\tau\ec\quad D_{2d}=0\ec\quad s_1=\frac{B}{2r}\ec\quad s_2=-\frac{\sigma}{r}={\rm const.}\ec\cr
\phi=0\ec\quad f=0\ed
\end{gather}
In \eqref{locloc2d}, $B\in\mathfrak{t}$ is the magnetic charge, where $\mathfrak{t}$ is the Cartan of the gauge algebra $\mathfrak{g}$ and $\sigma\in\mathfrak{g}$ is the Coulomb branch parameter (which can be further restricted to $\mathfrak{t}$ at the cost of a Vandermonde determinant).\footnote{In \eqref{locloc2d}, we took $B$ to have the opposite sign as compared to \cite{Doroud:2012xw}.  The reason is that the boundary conditions with $B$ as in \eqref{locloc2d} correspond to a monopole of charge $B$ inserted at the tip of $HS^3$. This can be checked by taking the background solution \eqref{A2Mon} and restricting it to $\varphi=0$. Thus to account for the orientation of $S^2$, in borrowing any results from \cite{Doroud:2012xw}, one has to replace $B\to -B$.} The signs in the expression for the 2D gauge field $a$ correspond to its values on different patches of $S^2$: in each of the two patches, $\theta$ takes values in $[0,\pi/2]$, with $\theta=\pi/2$ corresponding to the North and South poles of $S^2$ as in Figure \ref{fig:sphere} and $\theta=0$ being the equator of $S^2$, along which the patches are sewed.

It now follows from supersymmetric localization on $S^2$ that to compute \eqref{susy2dPI}, it suffices to evaluate the functionals $\Psi_\pm[\mathscr{B}]$ on the localization locus \eqref{locloc2d}: we denote this restriction by $\Psi_\pm(\sigma, B)\equiv\Psi_\pm[\mathscr{B}]\big|_{\eqref{locloc2d}}$. Furthermore, we must include the one-loop determinant from the localization on $S^2$.\footnote{Note that there is no contribution from a 2D classical action evaluated on the localization locus \eqref{locloc2d}, simply because the gluing theory does not have such an action.} This one-loop determinant plays the role of a ``gluing measure,'' which we denote by $\mu(\sigma, B)$. To summarize, the full $S^3$ answer can be written as
\begin{equation}
\label{Glue_F}
\sum_{B\in \Gamma_m} \int_{\mathfrak{t}}\, d\sigma\, \mu(\sigma, B)\,\Psi_-(\sigma, B) \Psi_+(\sigma, B)
\end{equation}
where $\Gamma_m$ is the lattice of magnetic charges allowed by the Dirac quantization condition. The gluing formula \eqref{Glue_F} holds in all 3D $\cN=4$ gauge theories, including non-abelian ones. 

In this paper, we are concerned only with the abelian theories described in Section \ref{abeliantheories}. For those theories, the one-loop determinant $\mu(\sigma,B)$ appearing in \eqref{Glue_F} only receives contributions from the 2D chiral multiplets, and is given by
\begin{equation}
\mu(\vec{\sigma}, \vec{B})=\prod_{I=1}^{N_h} (-1)^{\frac{|\vec{q}_I\cdot \vec{B}|-\vec{q}_I\cdot \vec{B}}{2}} (\Lambda r)^{-2i \vec{q}_I\cdot \vec{\sigma}} \frac{\Gamma\Big(\frac{1+\abs{\vec{q}_I\cdot \vec{B}}}{2} + i\vec{q}_I\cdot \vec{\sigma}\Big)}{\Gamma\Big( \frac{1+\abs{\vec{q}_I\cdot \vec{B}}}{2} - i\vec{q}_I\cdot \vec{\sigma} \Big)}\ed
\label{mudef}
\end{equation}
Note that the dependence of \eqref{mudef} on the UV cutoff $\Lambda$ simply exhibits the one-loop exact logarithmic running of the 2D FI term.\footnote{We have omitted an extra factor of $(\Lambda r)^{-1}$ in the formula for the gluing measure because it cancels with factors of $\sqrt{\Lambda r}$ arising from the determinant of the vector multiplet on the hemisphere.} 

With the localized boundary conditions \eqref{locloc2d}, the boundary correction \eqref{b_term} simplifies to:
\begin{equation}
\label{b_term_simple}
S_\partial = -\int_{S^2} d^2x \left(\zeta(T^A) (\Phi_{\dot1\dot1}^A - \Phi_{\dot2\dot2}^A) + \frac{i}4 \tq_+ (\Phi_{\dot1\dot1}-\Phi_{\dot2\dot2})q_-  \right)\ed
\end{equation}
Aside from ensuring that the gluing procedure is consistent, this boundary action plays another important role: its SUSY variation cancels the boundary terms \eqref{hyperBNDRY} and \eqref{vectorBNDRY} generated by the SUSY variation of the bulk action. This follows from the general formalism of Section \ref{CutGlue}, as explained in more detail in \cite{Dedushenko:2018aox}, and can also be checked by an explicit computation. Hence the total action, with \eqref{b_term_simple} included, preserves the required four supercharges on $HS^3$ that form $\cN=(2,2)$ supersymmetry at the boundary.

\subsubsection{The Monopole $HS^3$ Wavefunction}

The remaining pieces of our solution are the hemisphere wavefunctions $\Psi_{\pm}(\vec{\sigma},\vec{B})$ entering the gluing formula \eqref{Glue_F}. They are both determined by a path integral on $HS^3$ with the boundary conditions \eqref{xi}, \eqref{phif}, \eqref{A2d}--\eqref{D2d}, and \eqref{Aperp}, restricted to the localization locus \eqref{locloc2d}. We now compute $\Psi_{+}(\vec{\sigma},\vec{B})$ for vanishing FI parameters and in the presence of a charge-$\vec{b}$ twisted monopole operator $\cM^{\vec{b}}(\varphi)$ at the tip of $HS^3_+$; the result will be denoted by $\Psi_{+}(\vec{\sigma},\vec{B};\cM^{\vec{b}})$.

We can argue that the wavefunction $\Psi_{+}(\vec{\sigma},\vec{B};\cM^{\vec{b}})$ will be equal to $\Psi_{-}(\vec{\sigma},\vec{B};\cM^{-\vec{b}})$, i.e., the $HS^3_-$ wavefunction with an insertion of an oppositely charged twisted monopole $\cM^{-\vec{b}}$ at its tip. We therefore need not compute $\Psi_+$ and $\Psi_-$ separately. One way to understand this fact is to consider the background \eqref{PHspherical}, \eqref{Aspherical} of Section \ref{S3loc}, which represents the insertion of $\cM^{\vec{b}}$ at the tip of $HS^3_+$ and of $\cM^{-\vec{b}}$ at the tip of $HS^3_-$ (see also Section \ref{sec:conj}). This background is invariant under the coordinate change $\eta\to \pi-\eta$, which exchanges the upper and lower hemispheres. Therefore, the path integrals that generate $\Psi_{+}(\vec{\sigma},\vec{B};\cM^{\vec{b}})$ and $\Psi_{-}(\vec{\sigma},\vec{B};\cM^{-\vec{b}})$ are the same, and these two wavefunctions are equal:
\begin{align}
\Psi_{+}(\vec{\sigma},\vec{B};\cM^{\vec{b}}) = \Psi_{-}(\vec{\sigma},\vec{B};\cM^{-\vec{b}}) \ed\label{psipm}
\end{align}
Moreover, by evaluating \eqref{PHspherical}, \eqref{Aspherical} on $S^2$ (i.e., setting $\eta=\frac{\pi}{2}$), one sees that this background is compatible with the 2D localization locus \eqref{locloc2d} of the gluing theory precisely when $\vec{B}=\vec{b}$. In particular, this implies that $\Psi_{\pm}(\vec{\sigma},\vec{B};\cM^{\vec{b}})=0$ if $\vec{B}\neq\vec{b}$:\footnote{The latter fact does not change if we have some dressed monopole $\cO^{\vec{b}}$ instead of $\cM^{\vec{b}}$ at the tip, because insertions of order operators do not change the value of the background on $S^2$.} 
\begin{align}
\Psi_{+}(\vec{\sigma},\vec{B};\cM^{\vec{b}}) = \delta_{\vec{B},\vec{b}} Z_{HS^3}(\vec{\sigma},\vec{b}) \label{psipb}
\end{align}
where $Z_{HS^3}$ is the $HS^3$ partition function in the twisted monopole background, with boundary conditions specified in the previous subsection and $\vec{B}=\vec{b}$. 

Since the boundary conditions determined in \eqref{locloc2d} are half-BPS, we can apply supersymmetric localization on $HS^3$ to compute $Z_{HS^3}(\vec{\sigma},\vec{b})$. With such boundary conditions, the BPS equations on $HS^3_\pm$ have the same solutions as on $S^3$, described in Section \ref{2monopolebackgronud} (restricted to the corresponding hemisphere). In particular, the boundary correction \eqref{b_term_simple} vanishes on the localization locus.\footnote{This is simply because \eqref{b_term_simple} is proportional to $\Phi_{\dot1\dot1}-\Phi_{\dot2\dot2}$, which is zero on $S^2$ when evaluated on the bulk localization locus.} Being part of the classical action, \eqref{b_term_simple} therefore leaves no imprint on the localization computation, in a similar manner to the monopole counterterm.

The boundary conditions for fluctuations of the hypermultiplet fields around the BPS locus simplify to 
\begin{equation}
q_+| = 0, \quad \partial_\perp q_-| = 0, \quad (\psi_{\dot{1}}  - \sigma_3\psi_{\dot{2}})| = 0, \quad (\tilde{\psi}_{\dot{1}} + \sigma_3\tilde{\psi}_{\dot{2}})| = 0 \ed
\label{boundaryconditions}
\end{equation}
As was the case on $S^3$, the hypermultiplet path integral on $HS^3$ is given by the ratio of determinants \eqref{Zsig}. Now, however, the modes of the differential operators appearing in \eqref{Zsig} must be truncated according to \eqref{boundaryconditions}. Recall that in abelian theories, the vector multiplet contribution is trivial, so the partition function is fully accounted for by the hypermultiplet one-loop determinant.

Let us summarize the results of the calculation of this determinant for SQED$_1$, leaving the details to Appendix \ref{hemisphere-details}.  Assuming \eqref{boundaryconditions}, the bosonic eigenvalues are
\begin{equation}
\lambda_{B, N}^\pm = \frac{1}{r^2} \left(N  + \frac{1 + \abs{b}}{2}\pm i\sigma\right)\left(N  + \frac{3 + \abs{b}}{2}\mp i\sigma\right) \ecq N = 0, 1, \ldots \ec
\end{equation}
and have degeneracies
\begin{equation}
d_{B, N}^\pm = \frac{(N + 1)(N + 1\mp 1)}{2} + \frac{\abs{b}}{2} \times \begin{cases} N + 1\mp 1 & \text{($N$ even)}\ec \\ N + 1 & \text{($N$ odd)}\ed \end{cases} 
\label{bosdegen}
\end{equation}
The fermionic eigenvalues are
\begin{equation}
\lambda_{F, N}^\pm = \frac{1}{r} \left[ \pm\left(N  + \frac{1 + \abs{b}}{2} \right) + i\sigma \right] \ecq N = 0, 1, \ldots \ec
\end{equation}
and have degeneracies
\begin{equation}
d_{F, N}^\pm = N(N + 1) + \abs{b}\times \begin{cases} N + 1/2\pm 1/2 & \text{($N$ even)} \ec \\ N + 1/2\mp 1/2 & \text{($N$ odd)}\ed \end{cases}
\label{ferdegen}
\end{equation} 
The $HS^3$ free energy can then be written as
\begin{align}
F_{HS^3} &= \sum_{N=0}^\infty \Bigg[ (d_{B, N}^+ + d_{B, N-1}^- - d_{F, N}^-)\log \frac{N + \frac{1 + \abs{b}}{2}+ i\sigma}{\Lambda r} \notag \\
&\hspace{2.5 cm} + (d_{B, N}^- + d_{B, N-1}^+ - d_{F, N}^+) \log \frac{N  + \frac{1 + \abs{b}}{2}- i\sigma}{\Lambda r} \Bigg] \ed
\label{HemiFree}
\end{align}
One can check that  $d_{B, N}^+ + d_{B, N-1}^- - d_{F, N}^- = 0$ and $d_{B, N}^- + d_{B, N-1}^+ - d_{F, N}^+ = 1$, whence
\es{HemiFreeAgain}{
	F_{HS^3} = \sum_{N=0}^\infty  \log \frac{N + \frac{1 + \abs{b}}{2}- i\sigma}{\Lambda r} 
	= -\frac{d}{ds} \left[ (\Lambda r)^s \zeta\left(s,  \frac{1 + \abs{b}}{2} - i \sigma\right) \right] \bigg|_{s=0} \,,
}
where we have used zeta function regularization to evaluate the divergent sum.  From \eqref{HemiFreeAgain}, using $\zeta(0,q)=\frac{1}{2}-q$ and $\frac{d\zeta(s,q)}{ds}\Big|_{s=0}=\log\frac{\Gamma(q)}{\sqrt{2\pi}}$, we then extract the regularized value of the hemisphere partition function $Z_{HS^3} = e^{-F_{HS^3}}$:
\es{GotZHS}{
	Z_{HS^3} = \frac{1}{(\Lambda r)^{\frac{\abs{b}}{2} - i \sigma}} \frac{\Gamma(\frac{1 + \abs{b}}{2} - i\sigma)}{\sqrt{2 \pi}} \,.
}
In a general abelian theory of the form described in Section \ref{abeliantheories}, the $HS^3$ wavefunction with a twisted monopole operator of charge $\vec{b}\in\Gamma_m$ at the tip generalizes to
\begin{align}
Z_{HS^3} &= \prod_{I=1}^{N_h}\frac{1}{ (\Lambda r)^{\frac{\abs{\vec{q}_I\cdot\vec{b}}}{2} - i \vec{q}_I\cdot\vec{\sigma}} }\frac{\Gamma\left(\frac{1+|\vec{q}_I\cdot\vec{b}|}{2}-i\vec{q}_I\cdot\vec{\sigma}\right)}{\sqrt{2\pi}}   \ed \label{GotZHSgen}
\end{align}
The cutoff dependence $(\Lambda r)^{i\vec{q}_I\cdot\vec{\sigma}}$ in \eqref{GotZHSgen} can be interpreted as the logarithmic running of the FI term induced on the 2D boundary of $HS^3$ by the $I^\textrm{th}$ bulk hypermultiplet. The dependence on $(\Lambda r)^{\frac{\vec{q}_I\cdot\vec{b}}{2}}$ arises because the monopole operator acquires conformal dimension $\sum_I\abs{\vec{q}_I\cdot\vec{b}}/2$. This power of $\Lambda$ can be removed by formally renormalizing the monopole operator itself. 

\subsubsection{Reproducing Two-Point Function from Gluing} \label{matchsection}

Armed with the gluing measure \eqref{mudef} and the $HS^3$ partition function \eqref{GotZHS} corresponding to $\Psi_{\pm}$ through \eqref{psipb} and \eqref{psipm}, we can now reproduce from gluing the two-point function of $\cM^{\pm\vec{b}}$ on $S^3$ computed in Section \ref{S3loc}. In particular, the $S^3$ result $Z(\vec{\sigma},\vec{b})$, prior to the $\vec{\sigma}$ integration in \eqref{2moncor}, is written in \eqref{detRatioGen}. From the point of view of this section, $Z(\vec{\sigma},\vec{b})$ should be reproduced by the $\vec{\sigma}$ integrand of the gluing formula \eqref{Glue_F}. Indeed,
\begin{align}
Z(\vec{\sigma},\vec{b}) &= \sum_{B\in \Gamma_m}\mu(\vec{\sigma},\vec{B}) \Psi_{+}(\vec{\sigma},\vec{B};\cM^{\vec{b}}) \Psi_{-}(\vec{\sigma},\vec{B};\cM^{-\vec{b}})  \notag\\
&= \prod_{I=1}^{N_h} \frac{(-1)^{\frac{|\vec{q}_I\cdot \vec{b}|-\vec{q}_I\cdot \vec{b}}{2}}}{2\pi(\Lambda r)^{|\vec{q}_I\cdot \vec{b}|}}\Gamma\bigg(\frac{1+|\vec{q}_I\cdot \vec{b}|}{2}-i\vec{q}_I\cdot \vec{\sigma}\bigg)\Gamma\bigg(\frac{1+|\vec{q}_I\cdot \vec{b}|}{2}+i\vec{q}_I\cdot \vec{\sigma}\bigg) \label{match}
\end{align}
precisely as in \eqref{detRatioGen}, including all numerical factors!

The match exhibited in \eqref{match} is a strong consistency check on the technical details of the gluing procedure that we have developed. In particular, it is pleasing that the cutoff dependence due to the logarithmic running of the 2D FI term, which appears in the gluing measure $\mu(\vec{\sigma},\vec{B})$ as well as in the wavefunctions $\Psi_{\pm}$, precisely cancels in the gluing. Indeed, no such running should arise on $S^3$. Moreover, the $\vec{b}$-dependent sign that was conjectured in Section \ref{S3loc} based on general considerations is reproduced in the gluing computation, coming entirely from the gluing measure.

\subsection{Bilinear Form and Conjugation}\label{sec:conj}

So far, we have described the gluing procedure as the composition of a state vector $|\Psi_+\rangle$ and a covector $\langle\Psi_-|$. The wavefunctions appearing in the gluing formula can be thought of as
\begin{align}
\Psi_-(\sigma, B) &= \langle \Psi_-|\sigma, B\rangle\ec\cr
\Psi_+(\sigma, B) &= \langle\sigma, B|\Psi_+\rangle\ec
\end{align}
where $\langle\sigma,B|$ represents the boundary condition \eqref{locloc2d} imposed at the boundary of the upper hemisphere $HS^3_+$ and $|\sigma,B\rangle$ represents the same boundary condition applied to the lower hemisphere $HS^3_-$. We have assumed that the upper hemisphere path integral prepares a vector, while that of the lower hemisphere prepares a covector. This formally follows from the fact that gluing requires the boundaries of the two hemispheres to have opposite orientations.

Can we ``glue'' two vectors? The answer is obviously yes, since the physical Hilbert space is always equipped with a sesquilinear inner product that can be used to compose two states into a number. Here, however, we would like to define a different bilinear form that is natural to our construction. To do so, we turn one of the state vectors into a covector and then compose it with another state vector. There exists a natural operation, a simple reflection across the equator, which flips the upper and lower hemispheres and thereby turns a vector into a covector. In our fibration coordinates, it can be written as:
\begin{align}
\theta \to \theta,\quad \varphi \to - \varphi,\quad \tau \to \tau.
\end{align}
On one hand, this is simply a coordinate change that leaves boundary conditions unaffected.  Hence the wavefunction $\Psi(\sigma, B)$ stays unchanged.  On the other hand, it can be thought of as a reflection of the hemisphere. Such an operation flips magnetic charges, so it turns the upper hemisphere with a monopole insertion into the lower hemisphere with an antimonopole insertion. If we also assign orientations properly, then we are in the situation where the gluing procedure works and we can simply apply the gluing formula.

Thus if we are given two vectors $|\Psi_1\rangle, |\Psi_2\rangle$, then we can apply reflection to one of them, say $|\Psi_1\rangle$, thereby obtaining a covector $\widetilde{\langle \Psi_1|}$ with the property that $\widetilde{\langle \Psi_1}|\sigma,B\rangle = \langle \sigma, B|\Psi_1\rangle$. Using the gluing formula, we then arrive at the definition of a bilinear form on $\cH$:
\begin{equation}
\label{innerPr}
( \Psi_1, \Psi_2 ) = \sum_{\vec{B}} \int d\vec{\sigma}\, \mu(\vec{\sigma},\vec{B}) \Psi_1(\vec{\sigma},\vec{B}) \Psi_2(\vec{\sigma}, \vec{B}),\quad \Psi_1, \Psi_2 \in \cH.
\end{equation}

Notice that if we have a monopole inserted on $HS^3_+$ close to the North pole of $S^2= \partial HS^3_+$, then after applying the reflection, it turns into a monopole of the opposite charge inserted close to the North pole of $S^2$ but from the $HS^3_-$ side. We can move it slightly upward without affecting the answer for the glued correlator, so that it crosses the boundary and enters $HS^3_+$. Now it is again inserted on $HS^3_+$, except that its charge has flipped. If we represent the insertion of a monopole of charge $\vec{b}$ on $HS^3_+$ through the North pole of the boundary $S^2$ by an operator $\cM^{\vec b}_N$, then this statement can be written as
\begin{equation}
(\cM^{\vec b}_N\Psi_1, \Psi_2)=(\Psi_1, \cM^{-\vec b}_N\Psi_2),
\end{equation}
i.e., the following conjugation property should hold with respect to the bilinear form \eqref{innerPr}:
\begin{equation}
\label{Nconj}
(\cM^{\vec b}_N)^\dagger = \cM^{-\vec b}_N.
\end{equation}
To derive a similar statement for the analogous South pole operator, we would have to move it through the South pole. Recall from Section \ref{periSubtle}, however, that monopoles of half-integral R-charge are antiperiodic on $S^1$ and therefore defined with respect to a branch point at the South pole of $S^2$ ($\varphi = \pm\pi$).  For a monopole operator of charge $\vec{b}$, the periodicity is determined by the sign $\prod_I (-1)^{\vec{q}_I\cdot\vec{b}}$. As a consequence, the conjugation rule for South pole operators is slightly different:
\begin{equation}
\label{Sconj}
(\cM^{\vec b}_S)^\dagger = \cM^{-\vec b}_S \prod_{I=1}^{N_h}(-1)^{\vec{q}_I\cdot\vec{b}}.
\end{equation}
Later, when we derive explicit expressions for these operators, it will be instructive to check that \eqref{Nconj} and \eqref{Sconj} hold.

\section{Correlators with Multiple Insertions}
\label{multipleinsertions}

In this section, we derive a general expression for correlators of arbitrarily many twisted CBOs inserted anywhere along the great circle in $S^3$. In particular, we will represent these insertions by certain shift operators acting on the $HS^3$ partition function. As described in Section \ref{SGLUE}, two $HS^3$ partition functions with insertions can then be glued to obtain correlators on $S^3$. Furthermore, we will show that our results can be reproduced by dimensional reduction of the 4D $\cN=2$ Schur index with line defects.

\subsection{Shift Operators}

Let us first study the abelian theories defined in Section \ref{abeliantheories} with $\widehat{m}=\widehat{\zeta}=0$, deferring a discussion of nonzero mass and FI parameters to Section \ref{massFI}. Consider a general correlator of twisted CBOs in such a theory:
\begin{align}
\langle \cO^{\vec{b}_1}(\varphi_1)\cdots \cO^{\vec{b}_n}(\varphi_n)\rangle_{S^3}  \ec \label{gencor}
\end{align}
where the $\cO^{\vec{b}_i}(\varphi_i)$ ($i=1,\ldots,n$) carry monopole charges $\vec{b}_i\in\Gamma_m$ and are ordered on the circle as $-\pi<\varphi_1<\cdots<\varphi_n\leq\pi$. When $\widehat{\zeta}=0$, the twisted translation \eqref{twistedTR} is $\cQ^C_\beta$-exact, so the correlator \eqref{gencor} only depends on the order of the insertions on the circle. In particular, one can translate all the operators to the tip of $HS^3_+$ (i.e., the point $(\theta,\varphi)=(\pi/2,\pi/2)$) while maintaining their order, without changing the value of \eqref{gencor}. Then, by using the OPE at the tip, the above correlator can be represented by a one-point function
\begin{align}
\langle\cO^{\vec{b}_1}(\varphi_1)\cdots\cO^{\vec{b}_n}(\varphi_n)\rangle_{S^3}=\langle\cO^{\vec{b}}(\pi/2)\rangle_{S^3} \label{gencorO}
\end{align} 
where $\cO^{\vec{b}}(\pi/2)$ is a twisted CBO of charge $\vec{b}=\sum_{i=1}^n\vec{b}_i$ defined by
\begin{align}
\cO^{\vec{b}}(\pi/2) \equiv  \lim_{\substack{\varphi_i\to \pi/2 \\ \varphi_1<\cdots<\varphi_n}} \cO^{\vec{b}_1}(\varphi_1)\cdots\cO^{\vec{b}_n}(\varphi_n) \label{Ohat}
\end{align}
(of course, the correlator \eqref{gencorO} vanishes unless $\vec{b}=0$). In \eqref{Ohat}, the $\varphi_i\to \pi/2$ limit is taken in a way that maintains the order of the $\cO^{\vec{b}_i}(\varphi_i)$ on the circle. The topological property of the 1D theory then implies that $\cO^{\vec{b}}(\pi/2)$ is some position-independent linear combination of twisted CBOs defined by the OPE, up to $\cQ^C_\beta$-exact terms that do not affect our correlation functions.\footnote{Upon passing to the cohomology of $\cQ^C_\beta$, the order-preserving OPE defined in \eqref{Ohat} is simply the non-commutative star product of \cite{Beem:2016cbd}.} In Section \ref{SGLUE}, it was shown how to obtain an $S^3$ correlator of twisted CBOs at the tips of $HS^3_{\pm}$ by gluing the $HS^3_{\pm}$ partition functions along their $\partial HS^3_{\pm}=S^2$ boundary. In what follows, without loss of generality, we will only consider the representation \eqref{gencorO} of twisted correlators, in which there is an insertion at the tip of $HS^3_+$ and none at $HS^3_-$.

In this case, the properties of the $HS^3$ partition functions and of the gluing formula in Section \ref{SGLUE} are simple to describe. 
The $HS^3$ wavefunctions with insertions at the tip only depend on a finite-dimensional set of boundary conditions. These boundary conditions are parametrized by $\vec{\sigma}\in\mathbb{R}^r$ and the topological charge $\vec{B}\in\Gamma_m$ measuring the number of units of magnetic flux through the boundary $S^2$.  Due to this simplicity, the wavefunction corresponding to an insertion of a twisted CBO $\cO^{\vec{b}}$ of charge $\vec{b}\in\Gamma_m$ and dimension $\Delta$ at the tip can be evaluated explicitly: it takes the form
\begin{align}
\Psi(\vec{\sigma},\vec{B}; \cO^{\vec{b}}) &= \delta_{\vec{B},\vec{b}}\frac{P(\vec{\sigma},\vec{b})}{r^{\Delta}}\prod_{I=1}^{N_h}\frac{1}{\sqrt{2\pi} }\Gamma\left(\frac{1+|\vec{q}_I\cdot\vec{b}|}{2}-i\vec{q}_I\cdot\vec{\sigma}\right)  \equiv \delta_{\vec{B},\vec{b}}\, \psi(\vec{\sigma},\vec{b}; \cO^{\vec{b}}) \ec \label{HS3pPsi}
\end{align}
where $P(\vec{\sigma},\vec{b})$ is some polynomial. In the final equality of \eqref{HS3pPsi}, we have factored out the trivial dependence of the wavefunction on $\vec{B}$. 

For example, the insertion of a bare twisted monopole operator $\cM^{\vec{b}}$ of charge $\vec{b}\in\Gamma_m$ is represented by a wavefunction \eqref{HS3pPsi} with $P=1$:
\begin{align}
\Psi(\vec{\sigma},\vec{B}; \cM^{\vec{b}}) &= \delta_{\vec{B},\vec{b}}\prod_{I=1}^{N_h}\frac{1}{\sqrt{2\pi} r^{\frac{\abs{\vec{q}_I\cdot\vec{b}}}{2}}}\Gamma\left(\frac{1+|\vec{q}_I\cdot\vec{b}|}{2}-i\vec{q}_I\cdot\vec{\sigma}\right)  \ed \label{psimon}
\end{align}
The ``vacuum wavefunction'' is defined by inserting the identity at the tip, and is given by simply setting $\vec{b}=0$ in \eqref{psimon}:
\begin{align}
\Psi(\vec{\sigma},\vec{B};1) &= \delta_{\vec{B},\vec{0}}\prod_{I=1}^{N_h}\frac{1}{\sqrt{2\pi}} \Gamma\left(\frac{1}{2}-i\vec{q}_I\cdot\vec{\sigma}\right) \equiv \delta_{\vec{B},\vec{0}}\,\psi_0(\vec{\sigma}) \label{psi0} \ed
\end{align}
Unlike in previous sections, we work with renormalized quantities in what follows, thus removing the explicit dependence on the UV cutoff. This requires formally renormalizing the monopole operators by powers of the cutoff. Moreover, in the $HS^3$ partition function \eqref{GotZHSgen} and gluing measure \eqref{mudef}, we set the renormalized 2D FI coupling to zero at the scale at which we are working. This is done to avoid notational clutter, and will have no effect on the final results. In particular, as we saw in \eqref{match}, the running 2D FI terms cancel anyway after gluing, as they must.\footnote{In other words, we define our shift operators without the factors of $(\Lambda r)^{i\vec{q}_I\cdot\vec{\sigma}}$ in \eqref{GotZHSgen}, which are independent of the monopole charges, by absorbing them into the factors of $(\Lambda r)^{-2i\vec{q}_I\cdot\vec{\sigma}}$ in \eqref{mudef}.  This definition is consistent because FI parameters, whether in 3D or 2D, should not affect the definition of twisted CBOs (or the shift operators that create them) in an essential way: see \eqref{QCsqr}.} 

The $S^3$ correlator \eqref{gencorO} is given by gluing the appropriate wavefunction \eqref{HS3pPsi} to the vacuum \eqref{psi0} using the gluing formula \eqref{Glue_F}, resulting in
\begin{gather}
\langle\cO^{\vec{b}_1}(\varphi_1)\cdots\cO^{\vec{b}_n}(\varphi_n)\rangle_{S^3} =\langle\cO^{\vec{b}}(\pi/2)\rangle_{S^3} 
= \frac{1}{Z_{S^3}}\sum_{B\in \Gamma_m}
\int d^r \sigma\, \mu(\vec{\sigma},\vec{B})\Psi(\vec{\sigma},\vec{B};1)\Psi(\vec{\sigma},\vec{B}; \cO^{\vec{b}}) \notag \\
=\frac{\delta_{\vec{b}, \vec{0}}}{Z_{S^3}}\int d^r \sigma\, (\psi_0(\vec{\sigma}))^* \psi(\vec{\sigma},\vec{b};\cO^{\vec{b}}) \ed \label{gluing}
\end{gather}
In \eqref{gluing}, $Z_{S^3}$ is the $S^3$ partition function, and in the last line, we have evaluated the sum over $\vec{B}$ while noting that $\mu(\vec{\sigma},\vec{B})$, defined in \eqref{mudef}, satisfies $\mu(\vec{\sigma},0)\psi_0(\vec{\sigma})=(\psi_0(\vec{\sigma}))^\ast$. The normalization in \eqref{gluing} is such that $\langle\hat{1}\rangle=1$. Indeed, assuming that $\langle\hat{1}\rangle=1$ and substituting the explicit form \eqref{psi0} of the vacuum wavefunction into \eqref{gluing}, we find that
\begin{align}
Z_{S^3} = Z_{S^3}\langle\hat{1}\rangle 
=  \int d^r \sigma\, \prod_{I=1}^{N_h}\frac{1}{2\cosh(\pi\vec{q}_I\cdot\vec{\sigma})} \ec \label{ZS3}
\end{align}
which is the correct $S^3$ partition function of our theory.

\subsubsection{Twisted CBOs as Shift Operators}

Let us now argue that insertions of twisted CBOs at the tip of $HS^3_+$ can be realized by differential operators acting on the wavefunctions \eqref{HS3pPsi}. The operation of inserting a twisted CBO along the $R=1$ semicircle of $HS^3_+$ and moving it to the tip can be viewed as the action of an operator on the Hilbert space of the 3D theory on $S^2$. In particular, such operators act on the subspace of the Hilbert space containing states whose $HS^3_+$ wavefunctions are \eqref{HS3pPsi}. On such states, these operators are represented by differential operators in $\vec{\sigma}$ and $\vec{B}$ acting on \eqref{HS3pPsi}: they turn out to be simple shift operators.\footnote{Order operators are usually represented by finite-order differential operators. For instance, we will see that insertions of $\vec{\Phi}(\varphi)$ are represented by differential operators of order zero---that is, simply by multiplication by a function. On the other hand, disorder operators such as monopoles are represented by differential operators of infinite order. Operators of this type, such as $e^{a \partial_x}$, will be called shift operators because, e.g., $e^{a \partial_x}f(x)=f(x+a)$. We will employ terminology in which we refer to all of the operators that we use as shift operators.} The goal of this section is to construct these shift operators for the CBOs corresponding to the generators of the Coulomb branch chiral ring. 

In fact, in our case, there are two isomorphic sets of such shift operators. We define $\cO_N$ as the shift operator implementing the insertion of the twisted CBO $\cO(\varphi)$ near the North pole $(R,\varphi)=(1,0)$ of $\partial HS^3_+=S^2$ and translating it to the tip, while $\cO_S(\varphi)$ is defined by the same operation but starting from the South pole $(R,\varphi)=(1,\pm \pi)$ (when the insertion through the South pole is in the upper hemisphere, we should take $\vphi = \pi - \epsilon$, and when this insertion is in the lower hemisphere, we should take $\vphi = -\pi + \epsilon$, with $\epsilon>0$).  The wavefunctions $\cO_N\Psi(\vec{\sigma},\vec{B}; \cO')$ and $\cO_S\Psi(\vec{\sigma},\vec{B}; \cO')$ are generally distinct, because it is not possible to move $\cO(\varphi)$ from the North pole to the South pole along the semicircle of $HS^3_+$ without crossing $\cO'(\frac{\pi}{2})$ at the tip. Therefore, these two wavefunctions lie in different $\cQ^C_\beta$-cohomology classes, corresponding to taking the OPE of $\cO(\varphi)$ and $\cO'(\varphi)$ at the tip in different orders on the semicircle. It follows that in general, the operators $\cO_N$ and $\cO_S$ should also be different.

A wavefunction corresponding to multiple insertions of twisted CBOs can be represented in several equivalent ways by acting on the vacuum wavefunction with the $\cO_{N,S}$ in different orders. For example, consider an $HS^3_+$ wavefunction  $\Psi(\vec{\sigma},\vec{B};\cO^{\vec{b}})$ representing the insertion of two twisted CBOs $\cO^{\vec{b}_1}(\varphi_1)$ and $\cO^{\vec{b}_2}(\varphi_2)$, which are translated to the tip while keeping $\varphi_1<\varphi_2$ and fused into $\cO^{\vec{b}}(\pi/2)$ with $\vec{b}=\vec{b}_1+\vec{b}_2$. This wavefunction can be obtained in three different ways by acting on the vacuum wavefunction \eqref{psi0} as
\begin{align}
\Psi(\vec{\sigma},\vec{B};\cO^{\vec{b}}) = \cO_N^{\vec{b}_2}\cO_N^{\vec{b}_1}\Psi(\vec{\sigma},\vec{B};1) = \cO_S^{\vec{b}_1}\cO_S^{\vec{b}_2}\Psi(\vec{\sigma},\vec{B};1) = \cO_N^{\vec{b}_2}\cO_S^{\vec{b}_1}\Psi(\vec{\sigma},\vec{B};1)\ed \label{2insertions}
\end{align}
An important consequence of the definition of these differential operators is that for any two twisted CBOs $\cO(\varphi)$ and $\cO'(\varphi)$, we have
\begin{align}
[\cO_N, \cO'_S] = 0 \ed\label{NScom}
\end{align}
The commutativity \eqref{NScom} expresses the fact that in bringing two operators to the tip from opposite sides, they never intersect, regardless of which operator is brought to the tip first. This is also related to associativity of the operator algebra: it does not matter whether we first fuse $\cO_N$ with whatever was already at the tip and then fuse the result with $\cO'_S$, or whether we first fuse $\cO'_S$ with the operator at the tip and then fuse the result with $\cO_N$.  

The shift operators $\cO_{N,S}$ corresponding to insertions of twisted monopole operators and vector multiplet scalars can be uniquely fixed from the explicit computations we have done so far. Let us start by determining those corresponding to the twisted CBOs $\vec{\Phi}(\varphi)$ defined in \eqref{HatPhiDef}, which are constructed from the vector multiplet scalars. As shown in Section \ref{S3loc}, for any configuration of twisted monopole operators, $\vec{\Phi}(\varphi)$ localizes to $\vec{\Phi}_{\rm loc}(\varphi)$ as defined in \eqref{HatPhiMon} and \eqref{PHiBdy}. In particular, in the presence of a twisted operator $\cO^{\vec{b}}$ with topological charge $\vec{b}\in\Gamma_m$ at the tip of $HS^3_+$, we find:
\begin{align}
\vec{\Phi}_{\rm loc}(\varphi) = \frac{1}{r}\left[\vec{\sigma} + \frac{i}{2}\vec{b} \sgn\left(\cos\varphi\right)\right] \ed \label{HatPhiLocTip}
\end{align}
The action of the operators $\vec{\Phi}_N$ and $\vec{\Phi}_S$ on the wavefunction $\Psi(\vec{\sigma},\vec{B};\cO^{\vec{b}})$, defined in \eqref{psimon}, is obtained by evaluating \eqref{HatPhiLocTip} either in the segment  $0<\varphi<\frac{\pi}{2}$ connecting the tip to the North pole or in the segment $\frac{\pi}{2}<\varphi<\pi$ connecting the tip to the South pole. The result is
\begin{align}
\vec{\Phi}_N\Psi(\vec{\sigma},\vec{B};\cO^{\vec{b}}) &= \frac{1}{r}\left(\vec{\sigma} + \frac{i}{2}\vec{b}\right)\Psi(\vec{\sigma},\vec{B};\cO^{\vec{b}}) \ec\notag\\ \vec{\Phi}_S\Psi(\vec{\sigma},\vec{B};\cO^{\vec{b}}) &= \frac{1}{r}\left(\vec{\sigma} - \frac{i}{2}\vec{b}\right)\Psi(\vec{\sigma},\vec{B};\cO^{\vec{b}}) \ed \label{hatPhiNSonState}
\end{align}
From their action \eqref{hatPhiNSonState} on the $HS^3_+$ wavefunctions \eqref{HS3pPsi}, $\Phi_{N,S}$ can easily be re-expressed as operators in the variables  $\vec{\sigma}$ and $\vec{B}$. In particular, from the factorized form $\Psi(\vec{\sigma},\vec{B};\cO^{\vec{b}}) = \delta_{\vec{B},\vec{b}}\psi(\vec{\sigma},\vec{b};\cO^{\vec{b}})$ of the wavefunctions \eqref{HS3pPsi}, one can reproduce the action \eqref{hatPhiNSonState} by setting
\begin{align}
\vec{\Phi}_N = \frac{1}{r}\left(\vec{\sigma} + \frac{i}{2}\vec{B}\right) \ecq\,\, \vec{\Phi}_S = \frac{1}{r}\left(\vec{\sigma} - \frac{i}{2}\vec{B}\right) \ed \label{hatPhiNS}
\end{align}
The construction of the shift operators $\cM^{\vec{b}}_{N,S}$ corresponding to a twisted bare monopole operator $\cM^{\vec{b}}(\varphi)$ of charge $\vec{b}\in\Gamma_m$ requires slightly more elaborate reasoning. Clearly, by acting with $\cM^{\vec{b}}_{N,S}$ on any wavefunction $\Psi(\vec{\sigma},\vec{B};\cO^{\vec{b}'})$ of topological charge $\vec{b}'$, one obtains a new wavefunction of the form  \eqref{HS3pPsi} of topological charge $\vec{b}+\vec{b}'$. This fact, together with the $\delta_{\vec{B},\vec{b}}$-dependence of the wavefunctions \eqref{HS3pPsi} mentioned above, implies that
\begin{align}
\cM^{\vec{b}}_{N,S} = v^{\vec{b}}_{N,S}(\vec{\sigma},\vec{B})e^{-\vec{b}\cdot\partial_{\vec{B}}}\label{MNS1}
\end{align}
where $v^{\vec{b}}_{N,S}(\vec{\sigma},\vec{B})$ are some differential operators in $\vec{\sigma}$ with only polynomial dependence on $\vec{B}$. The operators $v^{\vec{b}}_{N,S}(\vec{\sigma},\vec{B})$ can be constrained by using the commutativity property \eqref{NScom} of North and South operators. In particular, demanding
\begin{align}
[\cM^{\vec{b}}_{N}, P(\Phi_S)] = [\cM^{\vec{b}}_{S}, P(\Phi_N)] = 0
\end{align}
for any polynomial $P(x)$ and using the definitions \eqref{hatPhiNS}, \eqref{MNS1} implies that
\begin{align}
v^{\vec{b}}_N(\vec{\sigma},\vec{B}) = w_N(\vec{\sigma},\vec{B})e^{-\frac{i}{2}\vec{b}\cdot\partial_{\vec{\sigma}}} \ecq\,  v^{\vec{b}}_S(\vec{\sigma},\vec{B}) = w_S(\vec{\sigma},\vec{B})e^{\frac{i}{2}\vec{b}\cdot\partial_{\vec{\sigma}}} \ec \label{vNS1}
\end{align}
where $w_{N,S}(\vec{\sigma},\vec{B})$ are simply polynomials in $\vec{\sigma}$ and $\vec{B}$. Moreover, imposing $[\cM^{\vec{b}}_N, \cM^{\vec{b}'}_S] = 0$ and using \eqref{MNS1}, \eqref{vNS1} further restricts the dependence of $w_{N,S}(\vec{\sigma},\vec{B})$ on $\vec{\sigma}$ and $\vec{B}$ to be
\begin{align}
w_N(\vec{\sigma},\vec{B}) &= w_N\left(\vec{\sigma}+\frac{i}{2}\vec{B}\right) = w_N(r\widehat{\Phi}_N) \ec\notag\\
w_S(\vec{\sigma},\vec{B}) &= w_S\left(\vec{\sigma}-\frac{i}{2}\vec{B}\right) = w_S(r\widehat{\Phi}_S) \label{vNS2} \ed
\end{align}
In summary, $\cM^{\vec{b}}_{N,S}$ must take the form
\begin{align}
\cM^{\vec{b}}_N = w_N(r\vec{\Phi}_N) e^{\vec{b}\cdot\left(-\frac{i}{2}\partial_{\vec{\sigma}} - \partial_{\vec{B}}\right)} \ecq \,\, \cM^{\vec{b}}_S = w_S(r\vec{\Phi}_S) e^{-\vec{b}\cdot\left(-\frac{i}{2}\partial_{\vec{\sigma}} + \partial_{\vec{B}}\right)} \ec
\end{align}
for some polynomials $w_{N,S}(x)$. To determine these polynomials, we demand that when the operators $\cM^{\vec{b}}_{N,S}$ act on the vacuum wavefunction \eqref{psi0}, they give rise to the wavefunction \eqref{psimon} with $\cM^{\vec{b}}(\varphi)$ inserted at the tip, i.e.,
\begin{align}
\cM^{\vec{b}}_{N,S}\Psi(\vec{\sigma},\vec{B};1) = \Psi(\vec{\sigma},\vec{B}; \cM^{\vec{b}}) \ed
\end{align}
The above equation uniquely determines the polynomials $w_{N,S}(x)$, giving the final results
\begin{align}
\cM_N^{\vec{b}} &=  \left[\prod_{I=1}^{N_h}  \frac{ (-1)^{(\vec{q}_I\cdot\vec{b})_+}}{  r^{\frac{\abs{\vec{q}_I\cdot\vec{b}}}{2}}}  \left(\frac{1}{2} + i r\vec{q}_I\cdot\vec{\Phi}_N  \right)_{(\vec{q}_I\cdot\vec{b})_+}\right] e^{-\vec{b}\cdot\left(\frac{i}{2}\partial_{\vec{\sigma}}+\partial_{\vec{B}}\right)} \ec \label{MN}\\
\cM_S^{\vec{b}} &= \left[\prod_{I=1}^{N_h}   \frac{(-1)^{(-\vec{q}_I\cdot\vec{b})_+}}{ r^{\frac{\abs{\vec{q}_I\cdot\vec{b}}}{2}}}  \left(\frac{1}{2} + i r\vec{q}_I\cdot\vec{\Phi}_S \right)_{(-\vec{q}_I\cdot\vec{b})_+}\right] e^{\vec{b}\cdot\left(\frac{i}{2}\partial_{\vec{\sigma}}-\partial_{\vec{B}}\right)} \ec \label{MS}
\end{align}
where $(x)_+\equiv \max(x,0)$, $r\vec{\Phi}_N = \vec{\sigma} +\frac{i}{2}\vec{B}$, and $r\vec{\Phi}_S = \vec{\sigma} -\frac{i}{2}\vec{B}$. Note that Dirac quantization implies that $(\pm \vec{q}_I\cdot\vec{b})_+$ is a non-negative integer, and therefore that the Pochhammer symbols\footnote{We use $(x)_n = \Gamma(x+n)/\Gamma(x)$, which equals $x(x+1)(x+2)\dots (x+n-1)$ if $n$ is a positive integer.} in \eqref{MN} and \eqref{MS} are polynomials in $\vec{\Phi}_{N,S}$. The twisted CBOs $\vec{\Phi}(\varphi)$ and $\cM^{\vec{b}}(\varphi)$ correspond to the Coulomb branch chiral operators of lowest dimension within their respective topological classes (defined by their magnetic charges). In particular, all other twisted CBOs/chiral operators are generated from their products.\footnote{The generators $\smash{\vec{\Phi}}$ and $\smash{\cM^{\vec{b}}}$ for any $\smash{\vec{b}}\in\Gamma_m$ are, of course, not all independent.} It follows that all the corresponding shift operators are generated from the products of the fundamental ones \eqref{hatPhiNS}, \eqref{MN}, and \eqref{MS} that we have already found. We conclude that any correlator of twisted CBOs can be obtained by acting on the vacuum wavefunction \eqref{psi0} with the shift operators \eqref{hatPhiNS}, \eqref{MN}, and \eqref{MS} in the right order and gluing the result using \eqref{gluing}.\footnote{A subtlety in defining higher-dimensional CBOs as products of the generators is the phenomenon of operator mixing for CFTs on $S^3$. In particular, on $S^3$, operators can mix with lower-dimensional ones, as described in \cite{Gerchkovitz:2016gxx,Dedushenko:2016jxl}. In our case, this mixing can always be resolved by diagonalizing the matrix of two-point functions of twisted CBOs.}

Note that while the $N$ and $S$ operators clearly commute with each other, the algebras of ``all $N$'' or ``all $S$'' operators are complicated by the fact that different $U(1)$ factors of the gauge group can be coupled through mutually charged hypers.  In particular, the shift operators associated to individual $U(1)$ factors do not, in general, commute with each other.

Finally, we stress that above, we have only determined the shift operators implementing insertions of twisted CBOs on the upper hemisphere $HS^3_+$. One could equivalently determine the corresponding operators representing insertions at the tip of the lower hemisphere $HS^3_-$. These operators can be obtained by taking the adjoints of the $HS^3_+$ operators written above with respect to the $(\ec)$ bilinear form \eqref{innerPr} that implements the gluing. Using the explicit expression for the North and South operators, it can be verified that their conjugates are as predicted in \eqref{Nconj} and \eqref{Sconj}.\footnote{In verifying these facts, it is helpful to use the property
\begin{equation}
\mu(\vec{\sigma}\pm i\vec{b}/2, \vec{B} + \vec{b}) = \mu(\vec{\sigma}, \vec{B})\prod_{I=1}^{N_h}\left[\prod_{\ell_I=0}^{|\vec{q}_I\cdot\vec{b}|-1} (\operatorname{sgn}(\vec{q}_I\cdot\vec{b})(1/2 + \ell_I) + \vec{q}_I\cdot\vec{B}/2\mp i\vec{q}_I\cdot\vec{\sigma})\right]^{\mp\operatorname{sgn}(\vec{q}_I\cdot\vec{b})}
\end{equation}
of the gluing measure \eqref{mudef}.  Note that conjugation with respect to the bilinear form \eqref{innerPr} does not involve complex conjugation.}

\subsubsection{Including Mass and FI Parameters}
\label{massFI}

The above results can be generalized to account for real mass $\widehat{m} = \vec{m}\cdot\vec{t}_H\in\mathfrak{t}_H$ and FI $\widehat{\zeta}=\vec{\zeta}\cdot\vec{t}_C\in\mathfrak{t}_C$ deformations where the $t_{H,C}$ are Cartan generators, $\vec{m}\in\mathbb{R}^{N_h-r}$, and $\vec{\zeta}\in\mathbb{R}^r$. We begin by describing the modification from turning on nonzero real masses. The real mass that couples to the $I^\textrm{th}$ hypermultiplet of $G_H$-weight $\vec{Q}_I\in \mathbb{Z}^{N_h-r}$ is given by $\vec{Q}_I \cdot \vec{m}$. To include it, one should simply shift $\vec{q}_I\cdot\vec{\sigma} \to \vec{q}_I\cdot\vec{\sigma} + r \vec{Q}_I \cdot \vec{m}$ in all of the appropriate formulas, except in the expressions \eqref{hatPhiNS} for $\Phi_{N,S}$, which remain unchanged. In particular, the vacuum wavefunction \eqref{psi0} becomes 
\begin{align}
\Psi_{\widehat{m}}(\vec{\sigma},\vec{B};1) = \delta_{\vec{B},\vec{0}}\prod_{I=1}^{N_h}\frac{1}{\sqrt{2\pi}}\Gamma\left(\frac{1}{2}-i\vec{q}_I\cdot\vec{\sigma} - i r \vec{Q}_I\cdot\vec{m}\right) \equiv\delta_{\vec{B},\vec{0}}\psi_0^{\widehat{m}}(\vec{\sigma}) \label{psim0} \ec
\end{align}
and the monopole shift operators \eqref{MN} and \eqref{MS} become
\begin{align}
\cM_N^{\vec{b}} &= \left[\prod_{I=1}^{N_h}\frac{(-1)^{(\vec{q}_I\cdot\vec{b})_+}}{r^{\frac{\abs{\vec{q}_I\cdot b}}{2}}}\left(\frac{1}{2} + i r\vec{q}_I\cdot\vec{\Phi}_N + i r \vec{Q}_I\cdot\vec{m}\right)_{(\vec{q}_I\cdot \vec{b})_+}\right] e^{-\vec{b}\cdot\left(\frac{i}{2}\partial_{\vec{\sigma}}+\partial_{\vec{B}}\right)} \ec \label{MNm}\\
\cM_S^{\vec{b}} &= \left[\prod_{I=1}^{N_h}\frac{(-1)^{(-\vec{q}_I\cdot\vec{b})_+}}{r^{\frac{\abs{\vec{q}_I\cdot\vec{b}}}{2}}}\left(\frac{1}{2} + i r\vec{q}_I\cdot\vec{\Phi}_S + i r \vec{Q}_I\cdot\vec{m}\right)_{(-\vec{q}_I\cdot\vec{b})_+}\right] e^{\vec{b}\cdot\left(\frac{i}{2}\partial_{\vec{\sigma}}-\partial_{\vec{B}}\right)} \ed \label{MSm}
\end{align}
Including FI parameters is slightly more subtle because when they are nonzero, the twisted translation $\widehat{P}_{\varphi}^C$ in \eqref{twistedTR} is no longer $\cQ^C$-exact. In particular, correlators of twisted CBOs acquire position dependence. Nevertheless, because $\widehat{P}_{\varphi}^C + ir\widehat{\zeta}$ is $\cQ^C$-exact, it is a simple matter to infer the position dependence of correlators with  $\widehat{\zeta}\neq 0$ from the known topological correlators with $\widehat{\zeta}=0$. Explicitly, if we modify all twisted CBOs as $\cO^{\vec{b}}(\varphi)\to e^{r(\vec{\zeta}\cdot\vec{b})\varphi}\cO^{\vec{b}}(\varphi)$, then these new operators have topological correlators. Another modification arises because the FI action \eqref{FIAction} localizes to $S_{\rm FI}[\cV] \to 8\pi^2 i  r \vec{\zeta}\cdot\vec{\sigma}$, which can be thought of as a part of gluing measure \eqref{gluing}.\footnote{Here, for notational convenience, we are making a \emph{choice} to regard the factor of $\smash{e^{-8\pi^2 ir\vec{\zeta}\cdot\vec{\sigma}}}$ as part of the gluing measure rather than dressing each wavefunction by a factor of $\smash{e^{-4\pi^2 ir\vec{\zeta}\cdot\vec{\sigma}}}$.  If taking the latter approach, one would also need to modify the shift operators as $\mathcal{O}\to \smash{e^{-4\pi^2 ir\vec{\zeta}\cdot\vec{\sigma}}\mathcal{O}e^{4\pi^2 ir\vec{\zeta}\cdot\vec{\sigma}}}$.  Note also that the shift operators behave differently under conjugation with respect to the FI-deformed gluing measure, so it is important that in \eqref{generalcorrzeta}, all of the shift operators act on a single hemisphere.} A general correlator of $n$ twisted CBOs $\cO^{\vec{b}_i}(\varphi_i)$ of topological charge $\vec{b}_i\in\Gamma_m$ ($i=1,\ldots,n$) can then be written in the matrix model as
\begin{gather}
\langle \cO^{\vec{b}_1}(\varphi_1)\cdots \cO^{\vec{b}_n}(\varphi_n) \rangle_{S^3}^{\widehat{m},\widehat{\zeta}} \notag\\
= \delta_{\sum_{k=1}^n \vec{b}_k,\vec{0}}\frac{e^{-r\sum_{k=1}^n(\vec{\zeta}\cdot\vec{b}_k)\varphi_k}}{Z_{S^3}^{\widehat{m},\widehat{\zeta}}} \int d^r \sigma\, e^{-8\pi^2 i  r \vec{\zeta}\cdot\vec{\sigma}} (\psi_0^{\widehat{m}}(\vec{\sigma}))^* \cO^{\vec{b}_1}_N\cdots\cO^{\vec{b}_n}_N\psi_0^{\widehat{m}}(\vec{\sigma}) \ec \label{generalcorrzeta}
\end{gather}
where we have assumed that $-\pi<\varphi_1<\varphi_2<\cdots<\varphi_n<\pi$ and the vacuum wavefunction $\psi_0^{\widehat{m}}$ and shift operators $\widehat{\cO}^{\vec{b}_i}_{N}$ are modified according to \eqref{psim0} and \eqref{MNm}, respectively. A similar statement holds for the $S$ operators, but with $-\pi<\varphi_n<\varphi_{n-1}<\cdots<\varphi_1<\pi$ (see Figure \ref{fig:sphere}).  As before, the correlator \eqref{generalcorrzeta} can be represented in different ways by replacing some or all of the North operators with South operators $\cO^{\vec{b}_i}_{S}$, modified according to \eqref{MSm} to accommodate the real mass deformations. Finally, note that the $S^3$ partition function $Z_{S^3}^{\widehat{m},\widehat{\zeta}}$ which appears in the normalization of our correlators \eqref{generalcorrzeta} is given by
\begin{align}
Z_{S^3}^{\widehat{m},\widehat{\zeta}} = \int d^r \sigma\, e^{-8\pi^2 i  r \vec{\zeta}\cdot\vec{\sigma}} \prod_{I=1}^{N_h}\frac{1}{2\cosh(\pi(\vec{q}_I\cdot\vec{\sigma}+ r \vec{Q}_I \cdot \vec{m}))} \ec
\end{align}
so that $\langle 1 \rangle_{S^3}^{\widehat{m},\widehat{\zeta}}=1$.

\subsection{Reduction of Schur Index}

Local monopole operators in 3D field theories are related to 't~Hooft loops wrapping $S^1$ in 4D through a dimensional reduction of the 4D theory on $S^1$. In this section, we present a related correspondence between twisted CBOs in our 3D $\cN=4$ theories and certain line operators in 4D $\cN=2$ theories. More specifically, we consider the Schur limit of the superconformal index of 4D $\cN=2$ theories, which can be realized through a path integral on $S^3\times S^1$. As described in \cite{Dimofte:2011py,Cordova:2016uwk}, the Schur index can be decorated by certain 't~Hooft-Wilson loops wrapping $S^1$, which, to preserve supersymmetry, can only be inserted at points along a great circle of $S^3$.\footnote{See \cite{Gang:2012yr} for a localization computation of the index with line operators.} We will argue that upon dimensional reduction on $S^1$,  the Schur index with such line defects reduces to a correlator of twisted CBOs on $S^3$. 

\subsubsection{The Line Defect Schur Index}
Let us start with a brief review of the Schur index of 4D $\cN=2$ abelian gauge theories and its refinements by line defects. The reader is referred to \cite{Gadde:2011uv,Dimofte:2011py,Cordova:2016uwk} for more details. 
The Schur index can be defined as a trace over the Hilbert space $\cH_{S^3}$ of the 4D $\cN=2$ theory on $S^3$, which is given by
\begin{align}
\cI^{(S)}(p, u_1,\ldots,u_{r_f}) \equiv \trace_{\cH_{S^3}}\left[(-1)^F p^{E-R} \prod_{a=1}^{r_f} u_a^{f_a}\right] \ed \label{IS}
\end{align}
In \eqref{IS}, $F$ is the fermion number, $E$ is the energy, $R$ is the $\mathfrak{su}(2)_R$ spin, and $f_a$ ($a=1,\ldots,r_f$) are the Cartan generators of the rank-$r_f$ flavor symmetry algebra. In our conventions, $(-1)^F = e^{2\pi i (j_1+j_2)}$ where $j_{1,2}$ are the spins of the $\mathfrak{su}(2)_1\oplus\mathfrak{su}(2)_2$ isometry of $S^3$. The Schur index only receives contributions from states satisfying $E=2R+j_1+j_2$ and $j_2-j_1-r=0$, where $r$ is the $U(1)_r$ R-symmetry charge.

For example, the index of a hypermultiplet coupled to a background $U(1)$ vector multiplet with corresponding holonomy $u$ is given by
\begin{align}
\cI^{(S)}_{\rm hyper}(p,u) = \prod_{n=0}^{\infty}\frac{1}{(1-u p^{n+\frac{1}{2}})(1-u^{-1}p^{n+\frac{1}{2}})} = \frac{1}{(\sqrt{p} u; p)(\sqrt{p}u^{-1};p)}\ec \label{IShyper}
\end{align}
where we introduced the $q$-Pochhammer symbol $(z;q)\equiv \prod_{k=0}^{\infty}(1-z q^k)$. In order to gauge the $U(1)$ symmetry, one has to project out gauge non-invariant states, which is achieved by integrating \eqref{IShyper} as $\oint_{|u|=1}\frac{du}{2\pi i u}\cI^{(S)}_{\rm hyper}(p,u)$. The index of an arbitrary abelian gauge theory can be constructed simply by taking products of free hypermultiplet indices and gauging flavor symmetries, as described above.

The Schur index can be reconstructed by gluing two copies of the ``half-index'' on $HS^3\times S^1$ along their $S^2\times S^1$ boundary. This is the 4D analog of the 3D setup that have we considered throughout this paper, and which was discussed in \cite{Dimofte:2011py,Cordova:2016uwk}. It is instructive to go through the details of this gluing procedure for the free hypermultiplet. In that case, there are two boundary conditions on $S^2\times S^1$ which preserve 3D $\cN=2$ supersymmetry, resulting in half-indices $\Pi^{\pm}(p,u)$ on $HS^3\times S^1$ given by
\begin{align}
\Pi^{\pm}(p,u) = \frac{1}{(\sqrt{p}u^{\mp 1};p)} \ed \label{Pihyper}
\end{align}
In \eqref{Pihyper}, the half-indices $\Pi^{\pm}$ correspond to fixing a 3D $\cN=2$ chiral multiplet of $U(1)$ flavor charge $\pm 1$ at the $S^2\times S^1$ boundary. The corresponding gluing measure is then simply the $S^2\times S^1$ superconformal index $\cI^{3D}_{\pm}$ (see \cite{Bhattacharya:2008zy,Kim:2009wb,Imamura:2011su,Kapustin:2011jm}) of a 3D $\cN=2$ chiral multiplet of unit $R$-charge and flavor charge $\pm 1$, given by 
\begin{align}
\cI^{3D}_{\pm}(p,u) = \frac{(u^{\mp 1}\sqrt{p};p)}{(u^{\pm 1}\sqrt{p};p)} \ed \label{I3d}
\end{align}
Indeed, one finds that \eqref{IShyper} is recovered from gluing two copies of \eqref{Pihyper} with the corresponding measure \eqref{I3d}:
\begin{align}
\cI^{(S)}_{\rm hyper}(p,u) = \cI^{3D}_{\pm}(p,u)(\Pi^{\pm}(p,u))^2 \ed
\end{align}

Let us now describe the refinement of the index by line defects in abelian theories. As explained in \cite{Dimofte:2011py}, in the presence of a (gauge) $U(1)$ BPS 't~Hooft loop of charge $ b\in\mathbb{Z}$  wrapping $S^1$ and inserted at the tip of $HS^3$, the hypermultiplet Schur half-indices \eqref{Pihyper} are modified to
\begin{align}
\Pi^{\pm}_B(p,u;b) = \delta_{B,b}\frac{1}{(p^{\frac{1+|b|}{2}}u^{\mp 1};p)} \ed \label{Pib}
\end{align}
The gluing measure is now given by the generalized $\cN=2$ superconformal index \cite{Kapustin:2011jm} (see also \cite{Dimofte:2011py,Aharony:2013dha,Aharony:2017adm}), with $b$ units of flux through $S^2$, of a chiral multiplet as described above:
\begin{align}
\cI^{3D}_{\pm}(p,u; B) = u^{-\frac{B}{2}}\frac{(p^{\frac{1+B}{2}}u^{\mp 1};p)}{(p^{\frac{1+B}{2}}u^{\pm 1};p)} = (-1)^{\frac{|B|-B}{2}}u^{-\frac{|B|}{2}}\frac{(p^{\frac{1+|B|}{2}}u^{\mp 1};p)}{(p^{\frac{1+|B|}{2}}u^{\pm 1};p)}\ed \label{I3dB}
\end{align} 
The full Schur index of the hypermultiplet with 't~Hooft loops of charge $\pm b$ inserted at antipodal points on $S^3$ is then given by composing two copies of \eqref{Pib} with the gluing measure \eqref{I3dB}, resulting in
\begin{align}
\cI^{(S)}_{\rm hyper}(p,u;b) &= \sum_{B\in\mathbb{Z}}\int\frac{du}{2\pi i u}\cI^{3D}_{\pm}(p,u;B)(\Pi^{\pm}_B(p,u;b))^2 \notag\\
&= (-1)^{\frac{|b|-b}{2}}\int \frac{du}{2\pi i u}u^{-\frac{|b|}{2}} \frac{1}{(p^{\frac{1+|b|}{2}}u;p)(p^{\frac{1+|b|}{2}}u^{-1};p)} \ed \label{IShyperb}
\end{align}
One could also consider 't~Hooft loops in flavor symmetries, in which case the $\delta_{B,b}$ in \eqref{Pib} should be omitted and there is neither a sum over $B$ nor an integration over $u$ in \eqref{IShyperb}, with the measure \eqref{I3dB} simply evaluated at $B=b$.

More general insertions of multiple 't~Hooft loops on the great (semi)circle of $(H)S^3$ can be realized by acting with certain difference operators on the half-indices, again in perfect analogy with our 3D construction. One can also insert BPS Wilson loops in the index. According to \cite{Dimofte:2011py, Cordova:2016uwk}, inserting a Wilson loop of minimal charge corresponds to multiplying the hemisphere indices by
\begin{align}
\hat{x}_N \equiv p^{\frac{B}{2}}u \ecq\,\, \hat{x}_S \equiv p^{-\frac{B}{2}}u \ed \label{wilsonloop}
\end{align}
As in our 3D setup, $\hat{x}_N$ ($\hat{x}_S$) corresponds to inserting the loop through the North (South) pole of $\partial HS^3_+\cong S^2$ and translating it to the tip along the semicircle. 

\subsubsection{Supercharges of Line Defects and Twisted CBOs}

Let us now show that the line defect Schur index preserves supercharges that can be identified with $\cQ^C_{1}$ and $\cQ^C_{2}$, given in \eqref{QC12}. This implies that line defect Schur indices in some 4D $\cN=2$ theory reduce on $S^1$ to correlators on $S^3$ of local operators in the cohomology of $\cQ^C_{1,2}$. These are precisely the correlators of twisted CBOs in the 3D $\cN=4$ theory, which is the dimensional reduction of the original 4D theory.

The line defect Schur index preserves certain supercharges within the 4D $\cN=2$ superconformal algebra $\mathfrak{sl}(4|2)$ of the theory on $S^3\times \mathbb{R}$. We follow the conventions of \cite{Beem:2013sza,Cordova:2016uwk} for $\mathfrak{sl}(4|2)$, with $\{Q^A{}_{\alpha}, \tilde{Q}_{A\dot{\alpha}}, S_A{}^{\alpha}, \tilde{S}^{A\dot{\alpha}}\}$ denoting its odd generators. The $A,B,\ldots=1,2$ indices label the fundamental irrep of the $\mathfrak{su}(2)_R\subset\mathfrak{sl}(4|2)$ R-symmetry, while $\alpha,\beta,\ldots = +,-$ and $\dot{\alpha},\dot{\beta},\ldots = \dot{+},\dot{-}$, label the fundamental irreps of $\mathfrak{su}(2)_1$ and $\mathfrak{su}(2)_2$, respectively, which combine into the isometry algebra $\mathfrak{so}(4)\cong \mathfrak{su}(2)_1\oplus\mathfrak{su}(2)_2\subset\mathfrak{sl}(2|4)$ of $S^3$. In addition, $M_{\alpha\beta}$, $M_{\dot{\alpha}\dot{\beta}}$, and $R^A{}_B$  denote the generators of $\mathfrak{su}(2)_1$, $\mathfrak{su}(2)_2$, and $\mathfrak{su}(2)_R$, respectively, while $D$ is the generator of dilatations. As shown in \cite{Cordova:2016uwk}, the line defect Schur index preserves two supercharges, which, in the above notation, are given by
\begin{align}
G^1{}_- = Q^1{}_- + \tilde{Q}_{2\dot{-}} \ecq\,\, H_1{}^- = S_1{}^- + \tilde{S}^{2\dot{-}} \ed\label{GH} 
\end{align}
The $\mathfrak{su}(2|1)_{\ell}\oplus\mathfrak{su}(2|1)_r$ symmetry algebra of our 3D $\cN=4$ theories on $S^3$ can be identified as a subalgebra of the $\mathfrak{sl}(4|2)$ algebra of the 4D theory on $S^3\times\mathbb{R}$. Indeed, the $\mathfrak{su}(2|1)_{\ell}$ generators $\{Q^{(\ell_-)}_{\alpha}, Q^{(\ell_+)}_{\alpha}, J^{(\ell)}_{\alpha\beta}, R_{\ell}\}$ can be identified with $\{Q^1{}_{\alpha}, S_{1\alpha}, M_{\alpha\beta}, D-2R^1{}_1\}$, and the generators $\{Q^{(r_-)}_{\alpha}, Q^{(r_+)}_{\alpha}, J^{(r)}_{\alpha\beta}, R_{r}\}$ of $\mathfrak{su}(2|1)_{r}$ with $\{\tilde{Q}_{2\dot{\alpha}}, \tilde{S}^2{}_{\dot\alpha}, M_{\dot{\alpha}\dot{\beta}}, D+2R^2{}_2\}$. Using the explicit form of the $\mathfrak{sl}(4|2)$ algebra given in \cite{Cordova:2016uwk}, it is easy to check that the $\mathfrak{su}(2|1)_{\ell}\oplus\mathfrak{su}(2|1)_r$ generators with the above identifications indeed satisfy \eqref{Commutators}. Furthermore, we find that the supercharges \eqref{GH} preserved by the index lie within $\mathfrak{su}(2|1)_{\ell}\oplus\mathfrak{su}(2|1)_r$, and can be written as
\begin{align}
G^1{}_- = \cQ^{(\ell_-)}_2 + \cQ^{(r_-)}_2 = \cQ^C_2 \ecq \,\, H_1{}^- = \cQ^{(\ell_+)}_1 + \cQ^{(r_+)}_1 = - \cQ^C_1 \ec \label{GHQC}
\end{align}
where we used the definitions \eqref{QC12} in the final equality of \eqref{GHQC}. The identification \eqref{GHQC} is what we wanted to prove. Note that the analysis leading to \eqref{GHQC} is completely general and applies to all 4D $\cN=2$ / 3D $\cN=4$ theories. In particular, it applies to theories with non-abelian gauge groups. 

\subsubsection{Reduction on $S^1$}

In this subsection, we explicitly construct the map between the line defects in the 4D Schur index and our twisted CBOs on $S^3$ in abelian gauge theories. For simplicity, we will focus on the 4D/3D theory of a single hypermultiplet coupled to a $U(1)$ vector multiplet. Restricting to this theory is sufficient to make our point, because all other abelian theories can be constructed by taking products of the free hypermultiplet theory and gauging flavor symmetries. Furthermore, taking products and gauging are simple operations at the level of the index as well as in the matrix model for correlators of twisted CBOs.

To reduce the index on $S^1$, we closely follow \cite{Aharony:2013dha}. We set 
\begin{align}
p = e^{-\beta} \ecq u = p^{i\sigma} \label{pured}
\end{align}
where $\beta = 2\pi r_1/ r_3$, with $r_1$ and $r_3$ being the radii of $S^1$ and $S^3$, respectively. The reduction is obtained by taking the $\beta\to 0$ ($p\to 1$) limit. To determine this limit, note that the $HS^3$ indices \eqref{Pib} can be written as
\begin{align}
\Pi^{\pm}_B(p,u;b) = \Pi^{\pm}_B(p,p^{i\sigma};b) = \delta_{B,b} \frac{\Gamma_p\left(\frac{1+|b|}{2}\mp i \sigma\right)}{(1-p)^{\frac{1-|b|}{2}\pm i\sigma}(p;p)} \ec \label{PibGamma}
\end{align}
where $\Gamma_q(x)$ is the $q$-Gamma function satisfying $\Gamma_q(x) \to \Gamma(x)$ as $q\to 1$. 

In taking the $\beta\to 0$ ($p\to 1$) limit in \eqref{PibGamma}, one encounters divergences from the den\-ominator, which we now analyze. First, it is useful to introduce the Dedekind $\eta$-function:
\begin{align}
(p;p) = p^{-\frac{\pi i \tau}{12}}\eta(\tau) \ecq \qquad p\equiv e^{2\pi i \tau} = e^{-\beta} \ed
\end{align}
Using its $S$-transformation 
\begin{align}
\eta(\tau) = \frac{1}{\sqrt{-i\tau}}\eta\left(-\frac{1}{\tau}\right)  = \sqrt{\frac{2\pi}{\beta}} e^{-\frac{\pi^2}{6\beta}}\prod_{n=1}^{\infty}\left(1-e^{-\frac{4\pi^2 n}{\beta}}\right)\ec
\end{align}
a short calculation gives
\begin{align}
\frac{1}{(1-p)^{\frac{1-|b|}{2}\pm i\sigma}(p;p)_{\infty}} &= \frac{1}{(1-e^{-\beta})^{\frac{1-|b|}{2}\pm i \sigma}e^{\frac{\beta}{24}}\sqrt{\frac{2\pi}{\beta}} e^{-\frac{\pi^2}{6\beta}}\prod_{n=1}^{\infty}\left(1-e^{-\frac{4\pi^2 n}{\beta}}\right)} \notag\\
&\xrightarrow{\beta\to 0} \frac{1}{\sqrt{2\pi}}\beta^{\frac{|b|}{2}\mp i\sigma} e^{\frac{\pi^2}{6\beta}}(1 + O(\beta)) \ed
\end{align}
We conclude that 
\begin{align}
\lim_{p\to 1}\Pi^{\pm}_B(p,u;b) &= \delta_{B,b}e^{\frac{\pi^2}{6\beta}}\beta^{\frac{|b|}{2} \mp i \sigma} \frac{1}{\sqrt{2\pi}}\Gamma\left(\frac{1+|b|}{2} \mp i\sigma\right) = e^{\frac{\pi^2}{6\beta}} \Psi(\pm\sigma,B; \cM^{b})\biggr|_{\beta=(\Lambda r)^{-1}}\ec \label{Pibplim} 
\end{align}
where we have set the arbitrary scale $\beta$ to $(\Lambda r)^{-1}$ in order to match our 3D conventions. After matching those scales, and up to the prefactor $e^{\frac{\pi^2}{6\beta}}$, \eqref{Pibplim} shows that $\Pi^+_B$ dimensionally reduces to the hemisphere wavefunction \eqref{GotZHSgen} with an insertion of a charge-$b$ twisted monopole operator at the tip.\footnote{In this subsection, we retain the explicit dependence on the cutoff $\Lambda$, as in \eqref{mudef} and \eqref{GotZHSgen}.} The exponential prefactor precisely matches the Cardy behavior discussed in \cite{DiPietro:2014bca}, and should simply be removed in extracting the $HS^3$ partition function from the reduced index. A similar calculation shows that the 3D index \eqref{I3dB} reduces to the $S^2$ partition function in \eqref{mudef} (for $N_h=1$),
\begin{align}
\lim_{p\to 1}\cI^{3D}_{\pm}(p,p^{i\sigma};B) = (-1)^{\frac{|B|-B}{2}}\beta^{\pm2i\sigma}\frac{\Gamma(\frac{1+|B|}{2}\pm i\sigma)}{\Gamma(\frac{1+|B|}{2}\mp i\sigma)}\ec
\end{align}
after the same matching of scales. The integral over the compact gauged holonomies decompactifies as $\beta\to 0$, becoming an integral $\int_{-\infty}^{\infty}d\sigma$, which is the expected integration measure in the $S^3$ matrix model. Finally, recall that inserting a Wilson loop can be achieved by acting with $\hat{x}_{N,S}$ in \eqref{wilsonloop}, which, upon substituting \eqref{pured}, become
\begin{align}
\hat{x}_N \equiv p^{\frac{B}{2}}u = e^{ -i \beta(\sigma - i\frac{B}{2})} \ecq\,\, \hat{x}_S\equiv p^{-\frac{B}{2}}u = e^{- i \beta(\sigma + i\frac{B}{2})}\ed
\end{align}
Note that the exponents $\sigma \pm i\frac{B}{2}$ in the above equation coincide with $\Phi_{N,S}$, defined in \eqref{hatPhiNS}. To obtain $\Phi_{N,S}$ in the reduced theory, we act on the $HS^3$ half-index with
\begin{align}
\left(i\frac{\hat{x}_{N,S}- (\hat{x}_{N,S})^{-1}}{2\beta}\right) \xrightarrow{\beta\to 0} \left(\sigma \mp i\frac{B}{2}\right) = \Phi_{N,S}\ed
\end{align}
We have therefore found a one-to-one correspondence between BPS Wilson loops in the Schur index and the twisted CBO $\Phi(\varphi)$. 

To conclude, we have essentially recovered the ingredients that are used to calculate correlators of twisted CBOs on $S^3$ in abelian $\cN=4$ gauge theories from the reduction of the defect Schur index of 4D $\cN=2$ theories. While we have presented the results for a single hypermultiplet, the generalization to an arbitrary abelian theory is straightforward. It would be interesting to apply this logic to non-abelian gauge theories, where the ``monopole bubbling'' phenomenon \cite{Kapustin:2006pk} plays an important role. We hope to return to this problem in future work.

\section{Applications} \label{applications}

We have seen in the previous section how shift operators can be used to compute arbitrary correlators of twisted CBOs in general abelian theories and how these calculations are modified in the presence of mass and FI parameters.  In this section, we give explicit examples of such calculations, and we match the results obtained to those of the corresponding calculations in the 1D Higgs branch sector of the mirror dual theories.  These matches yield more refined tests of 3D mirror symmetry \cite{Intriligator:1996ex} than have been described in the literature. 

In the following, we work with renormalized monopole operators and the corresponding renormalized shift operators \eqref{MN} and \eqref{MS}, which we quote here for convenience:
 \es{MRen}{
 {\cM}_N^{\vec{b}} &=  \left[\prod_{I=1}^{N_h}  \frac{ (-1)^{(\vec{b} \cdot \vec{q}_I)_+}}{ r^{\frac{\abs{\vec{b} \cdot \vec{q}_I}}{2}}} \left(\frac{1}{2} + i r \vec{\Phi}_N \cdot \vec{q}_I \right)_{(\vec{b} \cdot \vec{q}_I)_+}\right] e^{\vec{b}\cdot \left(-\frac{i}{2}\partial_{\vec{\sigma}}-\partial_{\vec{B}}\right)} \,, \\
 {\cM}_S^{\vec{b}} &= \left[\prod_{I=1}^{N_h}   \frac{(-1)^{(-\vec{b} \cdot \vec{q}_I)_+}}{ r^{\frac{\abs{\vec{b} \cdot \vec{q}_I}}{2} }}  \left(\frac{1}{2} + i r \vec{\Phi}_S \cdot \vec{q}_I\right)_{(-\vec{b} \cdot \vec{q}_I)_+}\right] e^{-\vec{b} \cdot \left(-\frac{i}{2}\partial_{\vec{\sigma}}+\partial_{\vec{B}}\right)} \,,
 }
where $r \vec{\Phi}_N = \vec{\sigma} + i\vec{B}/2$ and $r \vec{\Phi}_S = \vec{\sigma} - i \vec{B}/2$ as in \eqref{hatPhiNS}.

\subsection{Chiral Ring Relations}
\label{CHIRALRING}

We first explain how our formalism reproduces the chiral ring relations obeyed by Coulomb branch operators.  As mentioned already, the moduli space of vacua of the theories that we are considering (${\cal N} = 4$ gauge theories with matter) contains a Coulomb branch, which receives quantum corrections and which is a hyperk\"ahler cone.\footnote{In this section, we assume that mass and FI parameters have been set to zero.  For an application of our formalism to non-conformal QFTs, see Section \ref{massandFI}.}  Functions on the Coulomb branch are in one-to-one correspondence with the Coulomb branch operators of these theories.  For instance, the operators $C_{\dot a_1 \ldots \dot a_{2 j_C}}$, which form a spin-$j_C$ multiplet of $SU(2)_C$, correspond to an $SU(2)_C$ multiplet of functions which we may denote as ${\tilde C_{\dot a_1 \ldots \dot a_{2 j_C}}}$.  With respect to a particular complex structure parametrized by an $SU(2)_C$ polarization $v^{\dot a}$, one can identify the holomorphic component $\tilde C = v^{a_1} \cdots v^{a_{2 j_C}} \tilde C_{\dot a_1 \ldots \dot a_{2 j_C}}$ of the multiplet of functions.  Correspondingly, one can regard the operator $C = v^{a_1} \cdots v^{a_{2 j_C}} C_{\dot a_1 \ldots \dot a_{2 j_C}}$ as chiral.  It follows that the algebra of twisted Coulomb branch operators $C(0)$ defined in \eqref{twistedC}, inserted at $\vphi = 0$, is isomorphic to the algebra ${\cal A}$ of holomorphic functions $\tilde C$ or to the algebra of chiral operators, i.e., the chiral ring.  This algebra carries a commutative product structure inherited from the ordinary product of holomorphic functions, as well as a Poisson bracket.

This information (and more) is captured by our 1D topological theory and can be read off from the rules for computing correlation functions presented thus far.  In fact, the algebra ${\cal A}$ admits a non-commutative star product $\star: {\cal A} \times {\cal A} \to {\cal A}$ with a parameter identified as $1/r$ that measures the degree of non-commutativity.  When $1/r$ is taken to zero, the star product reduces to the ordinary product of holomorphic functions, while the terms of order $1/r$ in the star product correspond to the Poisson bracket of these functions (terms of higher order in $1/r$, fixed by deformation quantization, are necessary to ensure associativity).  This star product, which in general takes the form 
 \es{starDef}{
   {\cal O}_i \star  {\cal O}_j = \sum_k c_{ij}^k  {\cal O}_k
 }
for some coefficients $c_{ij}^k$, is simply a shorthand for the OPE
 \es{1dOPE}{
  {\cal O}_i(0) {\cal O}_j(\vphi) \approx \sum_k c_{ij}^k {\cal O}_k(0) \,, \quad  \text{as $\vphi \to 0$ with $\vphi > 0$} \,.
 }
One can thus extract the OPE coefficients from \eqref{1dOPE} to determine \eqref{starDef}.

In general, the chiral ring is not freely generated due to the existence of chiral ring relations.  The chiral ring relations are simply relations obeyed by the regular multiplication of functions and can thus be read off from the $r$-independent term in \eqref{starDef}.  For operators represented by fields, they are sometimes trivial to see:  for instance, products of polynomials in $\vec{\Phi}$ can be trivially related to higher-degree polynomials.\footnote{As we will see, in the 1D Higgs branch theory, the chiral ring relations are sometimes less obvious because one must use the D-term relations.}  What will be nontrivial for us are the chiral ring relations involving monopole operators, for which we will need to use our definitions for the corresponding shift operators.  

To derive the chiral ring relations obeyed by the monopole operators, let us work with the North shift operators for convenience.  We notice that to leading order in $1/r$,
 \es{MNApprox}{
  {\cM}_N^{\vec{b}} = \left[
  \prod_{I=1}^{N_h} \frac{ (-i r \vec{\Phi}_N \cdot \vec{q}_I )^{\frac{\vec{b} \cdot \vec{q}_I + \abs{\vec{b} \cdot \vec{q}_I}}{2}} }{r^{\frac{\abs{\vec{b} \cdot \vec{q}_I}}{2}}  }
  \right] e^{-\vec{b}\cdot\left(\frac{i}{2}\partial_{\vec{\sigma}}+\partial_{\vec{B}}\right)}  + \cdots \,,
 }
which implies that 
 \es{MSProd}{
  {\cM}_N^{\vec{a}}{\cM}_N^{\vec{b}}
   &= \left[\prod_{I=1}^{N_h}  (-i \vec{\Phi}_N \cdot \vec{q}_I )^{\frac{\abs{\vec{a} \cdot \vec{q}_I} + \abs{\vec{b} \cdot \vec{q}_I} - \abs{(\vec{a}+ \vec{b}) \cdot \vec{q}_I}}{2} } \right] {\cM}_N^{\vec{a} + \vec{b}} + \cdots \,.
 }
From \eqref{MSProd}, one can extract the leading term in the OPE of ${\cM}^{\vec{a}}$ and ${\cM}^{\vec{b}}$, which (by \eqref{starDef} and \eqref{1dOPE}) fixes the leading term in the star product:
 \es{RegProductFunctions}{
  {\cM}^{\vec{a}} \star {\cM}^{\vec{b}}
   = \left[\prod_{I=1}^{N_h}  (-i \vec{\Phi} \cdot \vec{q}_I )^{\frac{\abs{\vec{a} \cdot \vec{q}_I} + \abs{\vec{b} \cdot \vec{q}_I} - \abs{(\vec{a} + \vec{b}) \cdot \vec{q}_I}}{2}} \right] {\cM}^{\vec{a} + \vec{b}} + O(1/r) \,.
 }
After taking the limit $r\to \infty$, this equation can be interpreted as a chiral ring relation.  This is precisely the chiral ring relation obtained in \cite{Bullimore:2015lsa}.

Interestingly, in the chiral ring, the product of two monopole operators of charges $\vec{a}$ and $\vec{b}$ is  equal to a monopole operator of charge $\vec{a} + \vec{b}$ that is in general dressed by the vector multiplet scalars.  No dressing is required precisely when $\sgn(\vec{a} \cdot \vec{q}_I)\sgn(\vec{b} \cdot \vec{q}_I)\geq 0$ for all $I$.  Another interesting case is when $\vec{b} = -\vec{a}$, where we see that the chiral ring product between a monopole of charge $\vec{a}$ and its antimonopole can be expressed solely in terms of $\Phi$:
 \es{RegProductFunctionsMA}{
  {\cM}^{\vec{a}} \star {\cM}^{-\vec{a}}
   = \left[\prod_{I=1}^{N_h}  (-i \vec{\Phi} \cdot \vec{q}_I )^{\abs{\vec{a} \cdot \vec{q}_I} } \right] + O(1/r) \,.
 }
Since the operator ${\vec{\Phi}}$ has scaling dimension $1$, this expression provides another derivation of the fact that the monopole operator ${\cal M}^{\vec a}$ has scaling dimension $\sum_{I=1}^{N_h} \abs{\vec{a} \cdot \vec{q}_I}/2$.

\subsection{Mirror Symmetry:  SQED$_N$ and $N$-Node Necklace Quiver}

As a second application, let us show how our results are consistent with 3D mirror symmetry.  The mirror dual of a 3D $\mathcal{N} = 4$ abelian gauge theory built from vector multiplets and hypermultiplets is a theory of the same type (here, we are not being careful to distinguish a theory containing only ordinary multiplets from a theory containing only twisted multiplets).  At a formal level, the duality was proven in  \cite{Kapustin:1999ha}, and a concrete map between the operators of a given such theory and its mirror dual can be found, for instance, in \cite{Bullimore:2015lsa}.  Our construction allows us to go beyond the operator map and show that the correlation functions, or equivalently the star product, match precisely across the mirror duality.  We will do so in a few simple examples.\footnote{For an outline of a strategy for matching all twisted HBO/CBO correlators in arbitrary abelian mirror pairs, see Appendix \ref{moreonmatching}.}

\begin {figure} [!t]
  \centering
  \newcommand {\svgwidth} {0.5\textwidth}

\begingroup
  \makeatletter
  \providecommand\color[2][]{%
    \errmessage{(Inkscape) Color is used for the text in Inkscape, but the package 'color.sty' is not loaded}
    \renewcommand\color[2][]{}%
  }
  \providecommand\transparent[1]{%
    \errmessage{(Inkscape) Transparency is used (non-zero) for the text in Inkscape, but the package 'transparent.sty' is not loaded}
    \renewcommand\transparent[1]{}%
  }
  \providecommand\rotatebox[2]{#2}
  \ifx\svgwidth\undefined
    \setlength{\unitlength}{350.33000488pt}
  \else
    \setlength{\unitlength}{\svgwidth}
  \fi
  \global\let\svgwidth\undefined
  \makeatother
  \begin{picture}(1,0.99670505)%
    \put(0,0){\includegraphics[width=\unitlength]{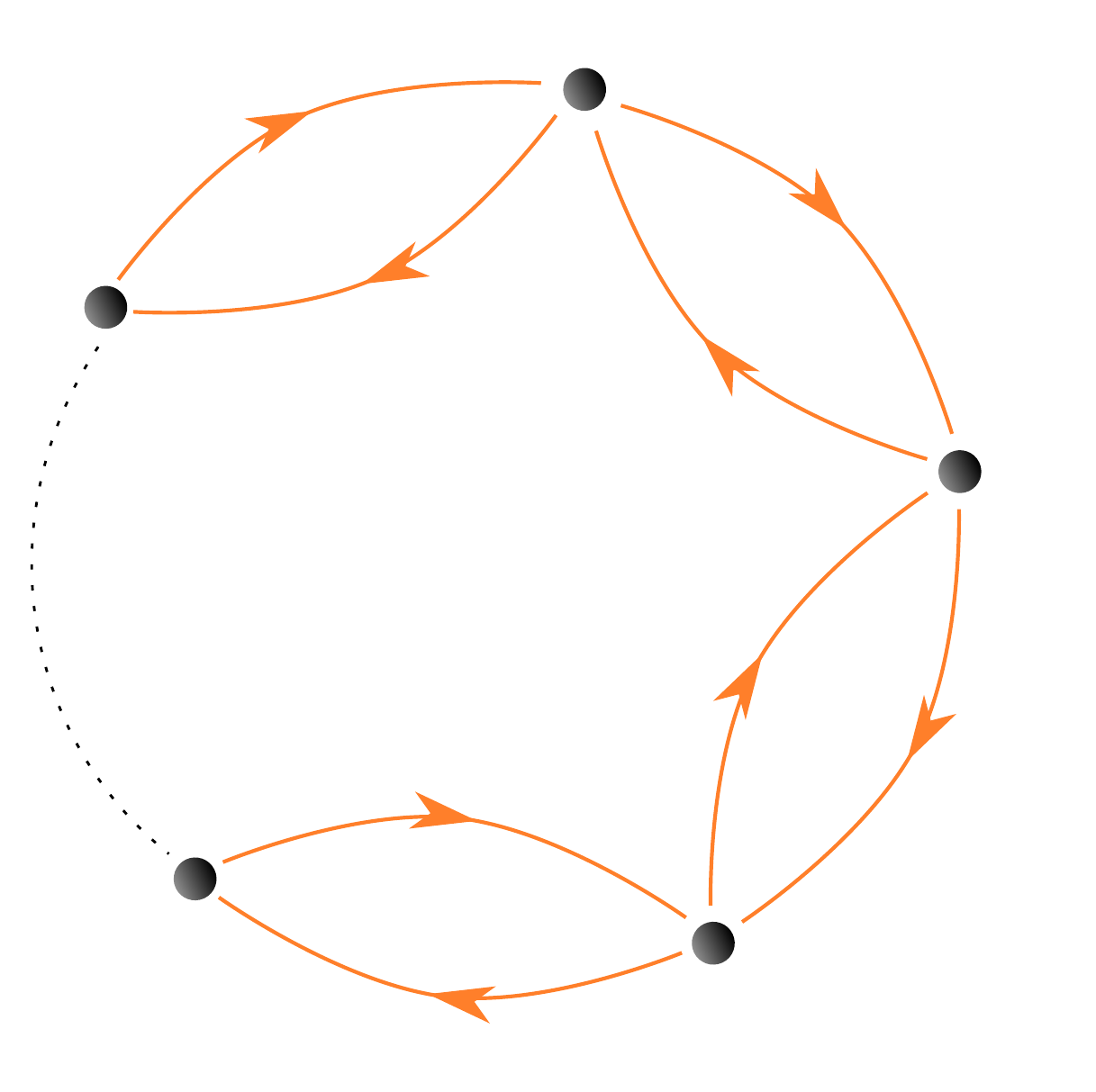}}%
    \put(0.14680037,0.90354507){\color[rgb]{0,0,0}\makebox(0,0)[lb]{\smash{$Q_N$}}}%
    \put(0.72421507,0.87418506){\color[rgb]{0,0,0}\makebox(0,0)[lb]{\smash{$Q_1$}}}%
    \put(0.85470431,0.26414798){\color[rgb]{0,0,0}\makebox(0,0)[lb]{\smash{$Q_2$}}}%
    \put(0.33274746,0.00969401){\color[rgb]{0,0,0}\makebox(0,0)[lb]{\smash{$Q_3$}}}%
    \put(0.56436585,0.61973106){\color[rgb]{0,0,0}\makebox(0,0)[lb]{\smash{$\tilde Q_1$}}}%
    \put(0.60025034,0.41094828){\color[rgb]{0,0,0}\makebox(0,0)[lb]{\smash{$\tilde Q_2$}}}%
    \put(0.38494317,0.29350807){\color[rgb]{0,0,0}\makebox(0,0)[lb]{\smash{$\tilde Q_3$}}}%
    \put(0.30338738,0.65561554){\color[rgb]{0,0,0}\makebox(0,0)[lb]{\smash{$\tilde Q_N$}}}%
  \end{picture}%
\endgroup

  \caption{The $U(1)^N/U(1)$ gauge theory that is mirror dual to SQED$_N$. \label{NecklaceFigure}}
\end {figure}
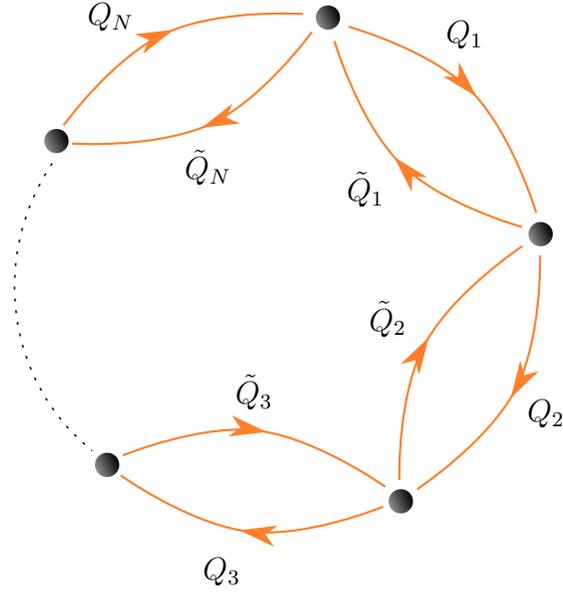

One of the simplest examples of mirror symmetry \cite{Intriligator:1996ex} is the duality between SQED$_N$ (a $U(1)$ gauge theory with $N$ hypermultiplets of unit charge) and the necklace quiver gauge theory with gauge group $U(1)^N/U(1)$ depicted in Figure~\ref{NecklaceFigure}.  In the necklace quiver, there are $N$ $U(1)$ gauge groups and $N$ bifundamental hypermultiplets, the $j^\textrm{th}$ of which has charges $1$ and $-1$ under the $(j-1)^\textrm{st}$ and $j^\textrm{th}$ gauge groups, respectively.  Nothing is charged under the diagonal $U(1)$, so we may regard the gauge group as $U(1)^N / U(1)$.  Each of these two theories has a Higgs branch that is mapped under the mirror duality to the Coulomb branch of the other theory.

\subsubsection{Higgs Branch Topological Sector}

Before we demonstrate how the mirror map works in detail at the level of the corresponding 1D topological sectors, let us briefly review the description given in \cite{Dedushenko:2016jxl} for the Higgs branch topological sector.  For a theory with gauge group $G$ and a hypermultiplet whose scalar fields transform in the representation $R \oplus \overline{R}$ of $G$, the associated 1D theory that allows for the calculation of $n$-point functions of twisted Higgs branch operators is
 \es{1dTheoryHiggs}{
  Z = \frac{1}{|\cW|}\int_{\mathfrak{t}}\, d\sigma \, \det{}'_\text{adj} (2 \sinh(\pi \sigma) )\, Z_\sigma
 }
where $|\cW|$ is the order of the Weyl group of $G$, $\mathfrak{t}$ is a fixed Cartan subalgebra of $\mathfrak{g}$, and
 \es{ZsigmaDef}{
   Z_\sigma \equiv \int \pD Q\, \pD\tQ\, \exp \left[ 4 \pi r \int d\vphi\, \tQ (\partial_\vphi + \sigma) Q \right] \,.
 }  
Here, the 1D fields $Q$ and $\tQ$ transform in the representation $R$ and its dual $\overline{R}$, respectively.  The $Q$ and $\tQ$ obey antiperiodic boundary conditions on the circle, while the Cartan element $\sigma$ is $\varphi$-independent.  The reality condition on bosons selects a certain middle-dimensional integration cycle in $(Q, \widetilde{Q})$-space, which is implicit in \eqref{ZsigmaDef}.  The operators in the 1D theory are gauge-invariant products of $Q$ and $\tQ$.  Correlation functions of these operators can be computed in two steps.  First, one writes the $n$-point function $\langle {\cal O}_1(\vphi_1) \ldots {\cal O}_n (\vphi_n) \rangle$ as
 \es{nPoint}{
  \langle {\cal O}_1(\vphi_1) \ldots {\cal O}_n (\vphi_n) \rangle = 
   \frac{1}{Z}\cdot \frac{1}{|\cW|} \int_{\mathfrak{t}}\, d\mu(\sigma) \, \langle {\cal O}_1(\vphi_1) \ldots {\cal O}_n (\vphi_n) \rangle_\sigma \,,
 }
where 
 \es{dmuDef}{
   d\mu(\sigma) \equiv d\sigma\, \det{}'_\text{adj} (2 \sinh(\pi \sigma) )\, Z_\sigma
    = d\sigma\, \frac{\det{}'_\text{adj} (2 \sinh(\pi \sigma) )}{\operatorname{det}_R (2 \cosh(\pi \sigma) )}
 }
and $\langle {\cal O}_1(\vphi_1) \ldots {\cal O}_n (\vphi_n) \rangle_\sigma$ is a correlation function at fixed $\sigma$.  Second, one computes this correlation function at fixed $\sigma$ by performing Wick contractions using the propagator
 \es{GDef}{
  \langle Q(\vphi_1) \tQ(\vphi_2) \rangle_\sigma \equiv G_\sigma(\varphi_{12}) \equiv -\frac{\sgn\varphi_{12} + \tanh(\pi\sigma)}{8 \pi r}e^{-\sigma\varphi_{12}} \,, \quad
   \vphi_{12} \equiv \vphi_1 - \vphi_2 \,,
 }
which can be derived from the Gaussian theory \eqref{ZsigmaDef}.\footnote{One might wonder how to define $\sgn\varphi_{12}$ for circle-valued variables.  Taking all $\varphi_i$ to lie in $(\varphi_0, \varphi_0 + 2\pi]$ for some $\varphi_0$, one can show that correlation functions are independent of the fiducial $\varphi_0$.  We use $\varphi_0 = -\pi$.}

When dealing with composite operators, one might also need to perform Wick contractions between elementary operators at coincident points. Such Wick contractions suffer from operator ordering ambiguities.  We make the choice that when $\vphi_1 = \vphi_2 = \vphi$, \eqref{GDef} should be interpreted as
  \es{GDefCoincident}{
  \langle Q(\vphi) \tQ(\vphi) \rangle_\sigma \equiv G_\sigma(0) \equiv -\frac{\tanh(\pi\sigma)}{8 \pi r} \,.
 }
Let us now use this formalism to see precisely how the Higgs (Coulomb) branch of SQED$_N$ is mapped to the Coulomb (Higgs) branch of the necklace quiver gauge theory in Figure~\ref{NecklaceFigure}.

\subsubsection{Matching of Partition Functions}

Before explaining the precise map of operators between the two 1D theories, we point out that the partition functions of the two theories agree.  Indeed, for SQED$_N$, we have
 \es{ZSQED}{
  Z = \int d\sigma\, \frac{1}{\left[2 \cosh(\pi \sigma) \right]^N} = \frac{\Gamma \left( \frac{N}{2} \right) }{2^N \sqrt{\pi} \Gamma\left(\frac{N+1}{2} \right) }  \,.
 }
On the necklace quiver side, we have
 \es{ZNecklace}{
  Z = \int d\mu(\sigma) = \int \left(\prod_{j=1}^N d\sigma_j\right)\delta\left(\frac{1}{N}\sum_{j=1}^N \sigma_j\right)
   \prod_{j=1}^N \frac{1}{2\cosh(\pi\sigma_{j-1, j})} \,,
 }
where $\sigma_{j-1,1}\equiv\sigma_{j-1}-\sigma_j$ and $\sigma_0\equiv \sigma_N$. To evaluate this integral, we appeal to the following trick, which we will also use extensively in the matching of correlation functions.  If $F_j(\sigma)$ are arbitrary functions whose Fourier transforms $\tF_j(\tau)$ are defined by 
 \es{FTF}{
  F_j(\sigma) = \int d\tau\, e^{-2 \pi i \sigma \tau} \tF_j(\tau)\,, \qquad
   \tF_j(\tau) = \int d\sigma\, e^{2 \pi i \sigma \tau} F_j(\sigma) \,, 
 }
then the following cyclic convolution identity holds:
 \es{FTId}{
  \int \left(\prod_{j=1}^N d\sigma_j\right)\delta\left(\frac{1}{N}\sum_{j=1}^N \sigma_j\right)
   \prod_{j=1}^N F_j (\sigma_{j-1, j}) = \int d\tau\, \prod_{j=1}^N \tF_j(\tau) \,.
 }
Using \eqref{FTId} with 
  \es{FjBasic}{
   F_j(\sigma) = \frac{1}{2 \cosh (\pi \sigma)}\,, \qquad 
   \tF_j(\tau) = \frac{1}{2 \cosh (\pi \tau)}
 }
for all $j$ shows that \eqref{ZNecklace} is precisely equal to \eqref{ZSQED}.

\subsection{HBOs in $N$-Node Quiver and CBOs in SQED$_N$} \label{CBOSQED-HBOnecklace}

On one side of the mirror duality, we have the Higgs branch of the $N$-node quiver theory.  It is convenient to represent the $(N-1)$-dimensional integration in \eqref{1dTheoryHiggs} and \eqref{nPoint} as an integration over $N$ variables $\sigma_j$ with a delta function constraint.  In particular, let us take the integration measure in \eqref{nPoint} to be
\begin{equation}
d\mu(\sigma_j) = \left(\prod_{j=1}^N d\sigma_j\right)\delta\left(\frac{1}{N}\sum_{j=1}^N \sigma_j\right)Z_\sigma \,,
\label{necklacemeasure}
\end{equation}
where
\begin{equation}
Z_\sigma = \int \left(\prod_{j=1}^N \pD \tQ_j\, \pD Q_j\right)\exp\left( 4 \pi r \int d\varphi\sum_{j=1}^N \tilde{Q}_j(\partial_\varphi + \sigma_{j-1, j})Q_j\right) = \prod_{j=1}^N \frac{1}{2\cosh(\pi\sigma_{j-1, j})} \,.
\end{equation}
The Higgs branch chiral ring is $\C^2 / \Z_N$.  Its generators are
  \es{GeneratorsHiggsC2ZN}{
   {\cal X} = Q_1 Q_2 \cdots Q_N \,, \qquad
    {\cal Y} = \tQ_1 \tQ_1 \cdots \tQ_N \,, \qquad
     {\cal Z} = \tQ_1 Q_1 = \ldots = \tQ_N Q_N \,,
  } 
which obey the chiral ring relation ${\cal X} \star {\cal Y} = {\cal Z}^N + O(1/r)$ (the equalities in the last equation of \eqref{GeneratorsHiggsC2ZN} are enforced by the D-term relations).  All other gauge-invariant operators in the 1D theory are products of ${\cal X}$, ${\cal Y}$, and ${\cal Z}$.

On the other side of the mirror duality is the Coulomb branch of SQED$_N$.  The gauge group is $U(1)$, so after boundary localization, the hemisphere wavefunction is just a function of two variables, $\Psi(\sigma, B)$, with $\sigma \in \R$ and $B \in \Z$.  The operators in the 1D topological theory are products of the twisted vector multiplet scalar $\Phi$ and the monopole operators of charge $b \in \Z$.  Their insertions through the North pole are represented by the shift operators
 \es{PhiMNorth}{
  \Phi_N = \frac{\sigma}{r} + i \frac{B}{2r} \,, \qquad
   {\cal M}^b_N = \frac{(-1)^{N (b)_+}}{r^{\frac{N \abs{b}}{2} } }\left[  \left(  i \sigma + \frac{1-B}{2} \right)_{b_+} \right]^N
    e^{- b \left( \frac{i}{2} \partial_\sigma + \partial_B \right) } 
 }
acting on $\Psi(\sigma, B)$.  The hemisphere wavefunction with no insertions is 
 \es{Psi0}{
  \Psi_0(\sigma, B) =  \delta_{B, 0} \left[ \frac{1}{\sqrt{2 \pi}}\Gamma\left(\frac 12 - i \sigma \right) \right]^N \,.
 }
The Coulomb branch chiral ring is also isomorphic to $\C^2 / \Z_N$ and is generated by
 \es{XYZCoulomb}{
  {\cal X} = \frac{1}{(4 \pi)^{N/2} } {\cal M}^{-1}\,, \qquad {\cal Y} = \frac{1}{(4 \pi)^{N/2} }  {\cal M}^{1} \,, \qquad
   {\cal Z} = -\frac{i}{4 \pi} \Phi \,,
  }
which, as per \eqref{RegProductFunctionsMA}, obey the relation ${\cal X} \star {\cal Y} = {\cal Z}^N + O(1/r)$.  We have used the same letters ${\cal X}$, ${\cal Y}$, ${\cal Z}$ to denote the operators of the two mirror theories to emphasize that, as we will show, their correlation functions in the two theories are identical.

\subsubsection{The Mirror Map}

To begin mapping the operators between the two sides, let us first explain why the mapping works as stated above for the basic operators ${\cal X}$, ${\cal Y}$, ${\cal Z}$.  In the Coulomb branch of SQED$_N$, we calculate that for $0 < \vphi_1 < \vphi_2  < \pi$, 
 \es{BasicTwo}{
   \langle {\cal Z}(\vphi_1) {\cal Z}(\vphi_2) 
    \rangle &= \frac{1}{Z} \int d\sigma\, \frac{\left(- \frac{i \sigma}{4 \pi r} \right)^2}{\left[ 2 \cosh(\pi \sigma) \right]^N} \,, \\
  \langle {\cal X}(\vphi_1) {\cal Y}(\vphi_2) \rangle 
   &= \frac{1}{Z} \sum_B \int d\sigma\, \Psi_0(\sigma, B)^* {\cal M}_N^{-1} {\cal M}_N^{1} \Psi_0(\sigma, B)
    = \frac{1}{Z} \int d\sigma\, \frac{\left(-i \sigma + \frac 1{2} \right)^N}{\left[ 8 \pi r \cosh(\pi \sigma) \right]^N} \,. 
 }
In the Higgs branch of the necklace theory, using the definition ${\cal Z} = \tQ_1 Q_1$ gives
 \es{CorrNecklace}{
  \langle {\cal Z}(\vphi_1) {\cal Z}(\vphi_2) 
    \rangle &= \frac{1}{Z} \int d\mu(\sigma) \left[ G_{\sigma_{12}} (\vphi_{12})
     G_{\sigma_{12}}(-\vphi_{12}) + G_{\sigma_{12}} (0)^2 \right] 
     = \frac{1}{Z} \int d\tau\, \frac{\left(- \frac{i \tau}{4 \pi r} \right)^2}{\left[2 \cosh (\pi \tau) \right]^N} \,, \\
  \langle {\cal X}(\vphi_1) {\cal Y}(\vphi_2) 
    \rangle &= \frac{1}{Z} \int d\mu(\sigma) \prod_{j=1}^N G_{\sigma_{j-1, j}} (\vphi_{12})
     = \frac{1}{Z}  \int d\tau\, \frac{\left(- i \tau + \frac{1}{2} \right)^N}{\left[8 \pi r \cosh (\pi \tau) \right]^N} \,, 
 }
which agrees precisely with \eqref{BasicTwo}.  In deriving the last equality in the first line of \eqref{CorrNecklace}, we used \eqref{FTId} with $F_1(\sigma) =  \left[ G_{\sigma} (\vphi_{12})  G_{\sigma}(-\vphi_{12}) + G_{\sigma} (0)^2 \right] / (2 \cosh (\pi \sigma))$, whose Fourier transform is $\tF_1(\tau) = \left(- \frac{i \tau}{4 \pi r} \right)^2 / (2 \cosh (\pi \tau))$, and with $F_j, \tF_j$ as in \eqref{FjBasic} for $j \geq 2$.  In deriving the last equality in the second line of \eqref{CorrNecklace}, we used \eqref{FTId} with $F_j(\sigma) = G_{\sigma} (\vphi_{12}) / (2 \cosh (\pi \sigma))$ and $\tF_j(\tau) = (-i \tau + 1/2) / ( 8 \pi r \cosh(\pi \tau))$.

Having mapped the chiral ring generators between the two theories, we can construct the mapping of composite operators using the OPE.  In general, we can define composite operators by point splitting:
 \es{Composite}{
  ({\cal O}_1 {\cal O}_2)_\star (0) \equiv ({\cal O}_1 \star {\cal O}_2)(0) =  \lim_{\substack{\epsilon \to 0^+}} {\cal O}_1(0) {\cal O}_2(\epsilon) \,.
 }
We can use this definition on both sides of the duality to find concrete expressions for composite operators in the two theories.  After doing so, one should still perform two nontrivial checks of the mirror symmetry duality: (1) the one-point functions of the composite operators should match, and (2) the star products of any pair of operators should match.  The matching of the one-point functions then guarantees the matching of higher-point functions, because the one-point functions and the OPE determine all correlation functions.\footnote{Note that we are not working in a basis of operators whose two-point functions are diagonal, so the coefficients in \eqref{1dOPE} are not what one usually thinks of as OPE coefficients.  Nonetheless, matching star products and one-point functions in this basis will also guarantee a match after, e.g., Gram-Schmidt orthogonalization.}

Whenever we define composite operators by point splitting as in \eqref{Composite}, we use a subscript $\star$ to indicate that all multiplications in the corresponding expressions are replaced by star products.  For the 1D topological Higgs branch theory reviewed above, we will also define composite operators by simply multiplying the fields $Q_j$ and $\tQ_j$.

\subsubsection{Star Product and Composite Operators}

Let us demonstrate how this procedure works in detail for a few operators.  The simplest composite operator is $({\cal Z}^2)_\star \equiv {\cal Z}  \star {\cal Z}$.  On the Coulomb branch side, each ${\cal Z}$ is represented by $-i \Phi / (4 \pi)$, and we can easily see from the North pole representation of $\Phi$ in \eqref{PhiMNorth} that $({\cal Z}^2)_\star$ is represented by
 \es{ZSqCoulomb}{
  ({\cal Z}^2)_\star = - \frac{1}{(4 \pi)^2} \Phi^2 \,.
 }
On the Higgs branch side, the calculation is slightly more complicated.  If we represent each factor of ${\cal Z}$ in the product by $\tQ_1 Q_1$, then 
 \es{ZSqHiggs}{
  ({\cal Z}^2)_\star = \tQ_1 Q_1 \star \tQ_1 Q_1
   = \tQ_1^2 Q_1^2  -  \frac{1}{64 \pi^2 r^2} \,.
 }
This equality follows from observing that while all self-contractions in $\tQ_1^2 Q_1^2$ are performed with \eqref{GDefCoincident}, the self-contractions in $ \tQ_1 Q_1 \star \tQ_1 Q_1$ between fields on different sides of the star product are performed with the $\vphi_{12} \to 0$ limit of \eqref{GDef}. Thus the difference $\tQ_1 Q_1 \star \tQ_1 Q_1- \tQ_1^2 Q_1^2$ evaluates to 
 \es{QQstarQQ}{
   \tQ_1 Q_1 \star \tQ_1 Q_1- \tQ_1^2 Q_1^2  = \tQ_1 Q_1(0) \, (\delta G_+ + \delta G_-)
    + \delta G_+  \delta G_- \,,
 }
where we have defined
 \es{deltaG}{
  \delta G_\pm\equiv \lim_{\epsilon \to 0^+} \left( G_\sigma(\pm \epsilon) - G_\sigma(0) \right) = \mp \frac{1}{8 \pi r}
 } 
and used \eqref{GDef} and \eqref{GDefCoincident}.  Substituting \eqref{deltaG} into \eqref{QQstarQQ} gives $-1/(64 \pi^2 r^2)$.

Note that we can represent $({\cal Z}^2)_\star$ in a number of equivalent ways coming from the fact that ${\cal Z}$ itself can be represented as $\tQ_j Q_j$ for any $j$ (no summation).  Thus, if we represented the first ${\cal Z}$ factor by $\tQ_1 Q_1$ and the second factor by $\tQ_2 Q_2$, then we would have
 \es{ZSqHiggs2}{
  ({\cal Z}^2)_\star = \tQ_1 Q_1 \star \tQ_2 Q_2
   = \tQ_1 Q_1 \tQ_2 Q_2\,.
 }
The expressions \eqref{ZSqHiggs} and \eqref{ZSqHiggs2} must be equivalent, and one can indeed check that they give identical correlation functions.

More generally, we have that $({\cal Z}^p)_\star$ is represented in the Coulomb branch theory by
 \es{Zn}{
  ({\cal Z}^p)_\star = \left( - \frac{i}{4 \pi} \right)^p \Phi^p \,.
 }
In the Higgs branch theory, the expression for ${\cal Z}^p$ is more complicated.  When $p \leq N$, we can represent the $j^\textrm{th}$ factor in the product by $\tQ_j Q_j$, and since all factors are distinct, we simply have
 \es{ZnHiggs}{
  ({\cal Z}^p)_\star = \tQ_1 Q_1 \star \cdots \star \tQ_p Q_p = \prod_{j=1}^p \tQ_j Q_j \,, \qquad
   1 \leq p \leq N \,.
 }
When $N < p \leq 2N$, we can write $({\cal Z}^p)_\star = \left[ ({\cal Z}^2)^{p-N} {\cal Z}^{2N - p} \right]_\star$.  We can represent the $j^\textrm{th}$ $({\cal Z}^2)_\star$ factor by $\tQ_j^2 Q_j^2  -  \frac{1}{64 \pi^2 r^2}$ (see \eqref{ZSqHiggs}) and the $k^\textrm{th}$ ${\cal Z}$ factor by $\tQ_{p-N + k} Q_{p-N + k}$, giving
 \es{ZnHiggs2}{
  ({\cal Z}^p)_\star = \prod_{j=1}^{p-N} \left( \tQ_j^2 Q_j^2  -  \frac{1}{64 \pi^2 r^2} \right)
   \prod_{j=p-N+1}^N Q_j Q_j \,, \qquad N < p \leq 2N \,.
 }
Similar expressions can be constructed for $p>2N$.

As a test of mirror symmetry, let us calculate the expectation value of $\langle ({\cal Z}^p)_\star \rangle$ on both sides.  On the Coulomb branch side, we have
 \es{ZpExp}{
  \langle ({\cal Z}^p)_\star \rangle = \frac{1}{Z} \left( - \frac{i}{4 \pi r} \right)^p  \int d\sigma\, \frac{\sigma^p}{\left[ 2 \cosh (\pi \sigma) \right]^N} \,.
 }
On the Higgs branch side, when $p \leq N$, we have 
 \es{ZpExpHiggs}{
  \langle ({\cal Z}^p)_\star \rangle =\frac{1}{Z} \int d\mu(\sigma) \prod_{j=1}^p G_{\sigma_{j-1, j}}(0)
   =  \frac{1}{Z} \left( - \frac{i}{4 \pi r} \right)^p  \int d\tau\, \frac{\tau^p}{\left[ 2 \cosh (\pi \tau) \right]^N} \,,
 }
in agreement with \eqref{ZpExp}.  In deriving the last equality in \eqref{ZpExpHiggs}, we used \eqref{FTId} with $F_j(\sigma) = G_{\sigma}(0) / (2 \cosh (\pi \sigma))$ and $\tF_j(\tau) = (- i \tau) / (8 \pi r \cosh (\pi \tau) )$ for $j \leq p$ and with \eqref{FjBasic} for $j>p$.  When $N < p \leq 2N$, we can use \eqref{ZnHiggs2} and a similar calculation to show that the same result \eqref{ZpExpHiggs} holds.  We expect a similar result to hold for $p>2N$.

With these definitions for the composite operators $({\cal Z}^p)_\star$, we can make another consistency check.  Let us compare the star product ${\cal X} \star {\cal Y}$ in both theories.  In SQED$_N$, we use the definitions \eqref{XYZCoulomb} and \eqref{PhiMNorth} in terms of North shift operators to deduce that 
 \es{XYOPE}{
  {\cal X} \star {\cal Y}  = \left[  \left(  {\cal Z} + \frac 1{8 \pi r} \right)^N \right]_\star \,.
 }
In the necklace quiver theory, we use the definitions \eqref{GeneratorsHiggsC2ZN} to write
 \es{XYOPEHiggs}{
  {\cal X} \star {\cal Y} = \prod_{j=1}^N Q_j \star \prod_{j=1}^N \tQ_j
   = \prod_{j=1}^N \left( \tQ_j Q_j + \frac 1{8 \pi r} \right)=  \left[  \left(  {\cal Z} + \frac 1{8 \pi r} \right)^N \right]_\star \,,
 }
which agrees precisely with \eqref{XYOPE} derived in SQED$_N$.

Other composite operators that we can define are powers of ${\cal X}$ and ${\cal Y}$.  In SQED$_N$, we can use again \eqref{XYZCoulomb} to represent
 \es{XpYp}{
  ({\cal X}^p)_\star = \frac{1}{(4 \pi)^{N p/2}} \left[ ({\cal M}^{-1})^p \right]_\star \,, \qquad
   ({\cal Y}^p)_\star = \frac{1}{(4 \pi)^{N p/2}} \left[ ({\cal M}^{1})^p \right]_\star \,.
 }
The star product $\left[ ({\cal M}^{-1})^p \right]_\star$ is easy to compute using the North shift operators \eqref{PhiMNorth} due to the simple form of ${\cal M}_N^{b}$ for $b<0$, namely ${\cal M}_N^{b} = \frac{1}{r^{-b N /2}} e^{-b \left( \frac{i}{2} \partial_\sigma + \partial_B \right) }  $.  This gives
 \es{Xp}{
  ({\cal X}^p)_\star = \frac{1}{(4 \pi)^{N p/2}}  {\cal M}^{-p} \,.
 }
The star product $\left[ ({\cal M}^{1})^p \right]_\star$ is easier to compute using the South shift operators, for which ${\cal M}_S^{b} = \frac{1}{r^{b N /2}} e^{-b \left(- \frac{i}{2} \partial_\sigma + \partial_B \right) }  $ when $b>0$.  This gives
 \es{Yp}{
  ({\cal Y}^p)_\star = \frac{1}{(4 \pi)^{N p/2}}  {\cal M}^{p} \,.
 }
The same expression can, of course, be obtained using North shift operators.  In the necklace quiver, there are no ordering ambiguities in raising $\cal X$ and $\cal Y$ from \eqref{GeneratorsHiggsC2ZN} to the power of $p$, so we can simply define
 \es{XYSQED}{
  ({\cal X}^p)_\star = Q_1^p \cdots Q_N^p \,, \qquad
   ({\cal Y}^p)_\star = \tQ_1^p \cdots \tQ_N^p \,.
 }
We can now perform another check of the mirror symmetry duality by computing $({\cal X}^2)_\star \star ({\cal Y}^2)_\star$ on both sides.  In SQED$_N$, from \eqref{XYZCoulomb} and \eqref{PhiMNorth}, we see that 
 \es{X2Y2star}{
  ({\cal X}^2)_\star \star ({\cal Y}^2)_\star = \frac{1}{(4 \pi r)^{2N}} \left[ \left(  i \sigma - \frac{B}{2} - \frac 12 \right)_2 \right]^N
   =  \left[ \left({\cal Z} + \frac{1}{8 \pi r} \right)^N \left( {\cal Z} + \frac{3}{8 \pi r}\right)^N \right]_\star \,.
 }
To compute $({\cal X}^2)_\star \star ({\cal Y}^2)_\star$ in the necklace quiver, first note that
 \es{QsqQtsq}{
  Q_1^2 \star \tQ_1^2 = Q_1^2 \tQ_1^2 + \frac{1}{2 \pi r} Q_1 \tQ_1 + \frac{1}{32 \pi^2 r^2} 
    = \left[ \left({\cal Z} + \frac{1}{8 \pi r} \right) \left( {\cal Z} + \frac{3}{8 \pi r}\right) \right]_\star \,,
 }
by the definition of ${\cal Z} = \tQ_1 Q_1$ and the definition of $({\cal Z}^2)_\star$ in \eqref{ZSqHiggs}.  Then we see that
 \es{X2Y2star2}{
  ({\cal X}^2)_\star \star ({\cal Y}^2)_\star
   =  \prod_{j=1}^N Q_j^2 \star \prod_{j=1}^N \tQ_j^2 =  \left[ \left({\cal Z} + \frac{1}{8 \pi r} \right)^N \left( {\cal Z} + \frac{3}{8 \pi r}\right)^N \right]_\star \,,
 } 
in agreement with \eqref{X2Y2star}.  Similar checks can be performed by computing $({\cal X}^p)_\star \star ({\cal Y}^p)_\star$.

\subsection{HBOs in SQED$_N$ and CBOs in $N$-Node Quiver} \label{HBOSQED-CBOnecklace}

Let us now turn our attention to the mirror duality between the Higgs branch of SQED$_N$ and the Coulomb branch of the necklace quiver gauge theory in Figure~\ref{NecklaceFigure}.  On the SQED$_N$ side, the 1D Higgs branch theory is described as follows.  Since the gauge group is abelian, we have only one integration variable $\sigma$ and $N$ pairs of 1D fields $(Q_J, \tQ^J)$.  The integration measure in \eqref{nPoint} is simply $d\mu(\sigma) = d\sigma\, Z_\sigma$, with 
 \es{ZSQEDHiggs}{
  Z_\sigma = \int \left(\prod_{j=1}^N \pD\tQ^J\, \pD Q_J \right) \exp
   \left(4 \pi r \int d\vphi\, \sum_{j=1}^N \tQ^J  (\partial_\vphi + \sigma) Q_J  \right) 
    = \frac{1}{\left[2 \cosh(\pi \sigma) \right]^N} \,.
 } 
The 1D theory has an $SU(N)$ flavor symmetry under which the $Q_J$ transform as a fundamental vector and the $\tQ^J$ transform in the antifundamental representation.  The Higgs branch is a minimal nilpotent orbit of the complexified Lie algebra $\mathfrak{su}(N)$.  The Higgs branch chiral ring is generated by the quadratic operators
 \es{JIJDef}{
  {\cal J}_I{}^J = Q_I \tQ^J \,, 
 }
which are traceless ($\sum_{I=1}^N {\cal J}_I{}^I = \sum_{I=1}^N Q_I \tQ^I = 0$) due to the D-term relations.  These operators are also subject to the nilpotency constraint ${\cal J}_I{}^J \star {\cal J}_J{}^K = O(1/r)$, which holds because to leading order in $1/r$, we can treat multiplication of operators as regular multiplication of functions, and we can use the D-term relations (below, we will present the full expression for ${\cal J}_I{}^J \star {\cal J}_J{}^K$).  The operators ${\cal J}_I{}^J$ transform in the adjoint representation of $SU(N)$.  All other operators in the 1D theory can be obtained from products of the ${\cal J}_I{}^J$ and transform in irreducible representations of $SU(N)$.  Due to the nilpotency of ${\cal J}_I{}^J$, the only irreps of $SU(N)$ that appear are those with Dynkin labels $[n0\cdots 0n]$ for positive integer $n$.

The description of the Coulomb branch of the necklace quiver gauge theory is more subtle because the $SU(N)$ symmetry acting on it is not manifest.  It is an emergent symmetry, with only its maximal torus $U(1)^{N-1}$ being visible in the UV\@.  Since the gauge group of the 3D theory is $U(1)^N / U(1)$, the hemisphere partition function after boundary localization must be a function of $N-1$ continuous ``$\sigma$'' variables and $N-1$ discrete ``$B$'' variables.  As in the previous subsection, it is convenient to represent the $N-1$ vector multiplets as $N$ vector multiplets ${\cal V}_I$ obeying the constraint $\sum_{I=1}^N {\cal V}_I = 0$.  Thus both $\vec{\sigma}$ and $\vec{B}$ are $N$-dimensional vectors obeying the constraints $\sum_{I=1}^N \sigma_I  = 0$ and $\sum_{I=1}^N B_I = 0$. These constraints are implemented by supplementing the gluing measure by a factor of 
 \es{deltafunc}{
  \delta \left(\frac{1}{N} \sum_{I=1}^N \sigma_I \right)  \delta_{0, \sum_{I=1}^N B_I} \,.
 }
The lattice in which $\vec{B}$ is valued is determined by the Dirac quantization condition, which implies that $\vec{q}_I \cdot \vec{B} = B_{I-1} - B_{I} \in \Z$ for all $I$.  Here, $\vec{q}_I = (0, \ldots, 0, 1, -1, 0, \ldots, 0)$ are the hypermultiplet gauge charges (the $I^\textrm{th}$ hypermultiplet has charge $+1$ under ${\cal V}_{I-1}$ and charge $-1$ under ${\cal V}_{I}$).

The operators in the 1D Coulomb branch theory are products of the twisted vector multiplet scalars $\Phi_I$ (obeying $\sum_{I=1}^N \Phi_I = 0$) as well as monopole operators of charge $\vec{b}$.  Their insertions through the North pole are represented by the operators
 \es{InsertionsCoulomb}{
  \Phi_{IN} = \frac{\sigma_I}{r} + i \frac{B_I}{2r} \,, \quad
   {\cal M}_N^{\vec{b}} = \prod_{I=1}^N \frac{(-1)^{\left( b_{I-1, I}\right)_+}}{r^{\frac{\abs{b_{I-1, I}}}{2}}}
    \left(i \sigma_{I-1, I} + \frac{1 - B_{I-1, I}}{2}   \right)_{(b_{I-1, I})_+} e^{- \vec{b} \cdot  \left( \frac{i}{2} \partial_{\vec{\sigma}} + \partial_{\vec{B}} \right) } \,,
 }
where $\sigma_{I-1, I} \equiv \sigma_{I-1} - \sigma_{I}$ as in the previous subsection and $b_{I-1, I} \equiv b_{I-1} - b_{I}$.  The operators \eqref{InsertionsCoulomb} act on the wavefunction $\Psi(\vec{\sigma}, \vec{B})$, which takes the form
 \es{wavefnNoIns}{
  \Psi_0(\vec{\sigma}, \vec{B}) = \delta_{\vec{B}, 0} \prod_{I=1}^N \left[\frac{1}{\sqrt{2 \pi}} \Gamma \left(\frac 12 - i \sigma_{I-1, I} \right) \right]
 }
in the absence of insertions.

\subsubsection{The Mirror Map} \label{ourmirror}

Identifying which operators in the necklace quiver gauge theory correspond to the generators ${\cal J}_I{}^J$ of the Higgs branch chiral ring of SQED$_N$ is aided by symmetries.  The necklace quiver has a $U(1)^{N-1}$ topological symmetry generated by the currents $j_{I}^\mu = \frac{1}{4\pi} \epsilon^{\mu\nu\rho} (F_{I-1, \nu \rho} - F_{I, \nu\rho})$ that should be identified with the Cartan of $SU(N)$.  The twisted vector multiplet scalars $\Phi_{I-1, I} = \Phi_{I-1} - \Phi_{I}$ should thus be identified with the Cartan elements ${\cal J}_I{}^I$.  We take
 \es{JDiag}{
  {\cal J}_I{}^I = \frac{i\Phi_{I-1, I}}{4 \pi} \quad \text{(no summation over $I$)}\,.
 }
We have $\sum_{I=1}^N {\cal J}_I{}^I = 0$, just as for the corresponding operators \eqref{JIJDef} in SQED$_N$.  The off-diagonal ${\cal J}_I{}^J$, with $I \neq J$, are monopole operators because they carry charges $+1$ and $-1$ under $j_I^\mu$ and $j_J^\mu$, respectively, and are uncharged under all other $j_K^\mu$ with $K \neq I, J$.  They are thus given by 
 \es{JOffDiag}{
  \qquad {\cal J}_I{}^J = -\frac{{\cal M}^{\vec{b}_I{}^J}}{4 \pi} \,, \quad 
   \vec{b}_I{}^J &\equiv (\underbrace{0, \ldots, 0}_{I-1}, \underbrace{1, \ldots, 1}_{J-I}, \underbrace{0, \ldots, 0}_{N-J+1}) 
   - \frac{J-I}{N} (1, 1, \ldots, 1) \,, \\
 \qquad {\cal J}_J{}^I = -\frac{{\cal M}^{\vec{b}_J{}^I}}{4 \pi} \,, \quad 
   \vec{b}_J{}^I &\equiv (\underbrace{1, \ldots, 1}_{I-1}, \underbrace{0, \ldots, 0}_{J-I}, \underbrace{1, \ldots, 1}_{N-J+1}) 
   - \frac{N+I-J}{N} (1, 1, \ldots, 1) \,, 
 }
where $J>I$. The expressions for $\vec{b}_I{}^J$ can be determined from the conditions that $\vec{b}_I{}^J \cdot \vec{q}_I  = - \vec{b}_I{}^J \cdot \vec{q}_J  = 1$, $\vec{b}_I{}^J \cdot \vec{q}_K = 0$ if $K \neq I, J$, and $\sum_{K=1}^N (b_I{}^J)_K = 0$.  The overall factors in the expressions for ${\cal J}_I{}^J$ are found empirically by matching the two- and three-point functions of these operators across the mirror symmetry duality.  We see that there are $2\binom{N}{2} = N(N - 1)$ independent monopoles with charges $\vec{b}_I{}^J$ and $\vec{b}_J{}^I = -\vec{b}_I{}^J$.

The mapping \eqref{JDiag}--\eqref{JOffDiag}, which relies on a description of the mirror theory to SQED$_N$ as a circular quiver, should be compared to that in \cite{Assel:2017hck}.  An alternate but equivalent presentation of the mirror map, which represents the mirror to SQED$_N$ as a linear quiver, is given in \cite{Bullimore:2015lsa}. In particular, note that our description of the necklace quiver as a $U(1)^N/U(1)$ gauge theory involves fractional monopole charges. This is only because we find it convenient to embed the $\Z^{N-1}$ charge lattice in $\R^N$.

A consistency check of the identification \eqref{JDiag}--\eqref{JOffDiag} comes from the chiral ring.  Indeed, given \eqref{RegProductFunctions}, we have 
 \es{ProdMon}{
  {\cal J}_I{}^J \star {\cal J}_J{}^I
   &=
   -   \Phi_{I-1, I} \Phi_{J-1, J}  + O(1/r) \,, \\
    {\cal J}_I{}^J \star {\cal J}_J{}^K
   &= i  \Phi_{J-1, J} {\cal J}_I{}^K  + O(1/r) \,,
 }
with no summation over $I$, $J$, $K$.  These equations hold even if $J=I$ or $J=K$.  Then it is easy to see that since $\sum_{J=1}^N \Phi_{J-1, J} = 0$, we have
 \es{RelationCoul}{
  \sum_{J=1}^N  {\cal J}_I{}^J \star {\cal J}_J{}^K  = O(1/r)
 }
for any $I, K$.  This nilpotency constraint matches the constraint obeyed by ${\cal J}_I{}^J$ in SQED$_N$. 

\subsubsection{Star Product and Composite Operators}

Let us provide more evidence for our proposed correspondence between the chiral ring generators, and provide a construction of more complicated operators that are dual on both sides.  We first point out that computing correlators in the Higgs branch topological sector of SQED$_N$ can be done without evaluating any integrals over $\sigma$, for the following reasons.  First, one can compute star products of various operators at fixed $\sigma$, as we did in the previous section for the Higgs branch of the necklace quiver theory.  Second, if we are careful to work with operators transforming in irreps of $SU(N)$, then all such operators have zero expectation value unless they are singlets of $SU(N)$.  The only singlet is the identity operator.

Explicitly, let us compute ${\cal J}_I{}^J \star {\cal J}_K{}^L = Q_I \tQ^J \star Q_K \tQ^L$:
 \es{QQtQQt}{
  Q_I \tQ^J \star Q_K \tQ^L = Q_I Q_K \tQ^J \tQ^L + \delta G_- \delta_I^L Q_K \tQ^J + \delta G_+ \delta_K^J Q_I \tQ^L 
   + \delta_I^L \delta_K^J \delta G_+ \delta G_-\,,
 }
with $\delta G_\pm = \mp \frac{1}{8 \pi r}$ (defined in \eqref{deltaG}) being the difference between the coincident limit of the propagator and the value assigned to the propagator at coincident points.  The operator $Q_I Q_K \tQ^J \tQ^L$ does not transform in an irreducible representation of $SU(N)$:  it is a linear combination of a singlet and an operator transforming in the $[20\cdots 02]$ irrep.  The latter is a traceless symmetric tensor ${\cal J}_{IK}{}^{JL} = Q_I Q_K \tQ^J \tQ^L - (\text{traces})$ that can be written explicitly as
 \es{J2}{
  {\cal J}_{IK}{}^{JL} 
   = Q_I Q_K \tQ^J \tQ^L - \frac{4 Q_M \tQ^M}{N+2} \delta^{(J}_{(I} Q_{K)} \tQ^{L)}  + \frac{2 (Q_M \tQ^M)^2}{(N+1)(N+2)} 
    \delta^{(J}_{(I}  \delta^{L)}_{K)}  \,.
 }
This expression can be simplified using the D-term relation $Q_K \tQ^K = 0$, which implies that $Q_I \tQ^J \star Q_K \tQ^K = 0$.  Then from \eqref{QQtQQt}, we conclude that
 \es{Q4Relation}{
  Q_I  \tQ^J Q_K \tQ^K 
   = \frac{1}{64 \pi^2 r^2} \delta_I^J  \,.
 }
Combining \eqref{QQtQQt}--\eqref{Q4Relation} and doing a bit of algebra gives 
 \es{JJstar}{
  {\cal J}_I{}^J \star {\cal J}_K{}^L = {\cal J}_{IK}{}^{JL} 
   - \frac{1}{8 \pi r} \left(  \delta_K^J {\cal J}_I{}^L 
    -  \delta_I^L {\cal J}_K{}^J \right) - \frac{N}{64 \pi^2 r^2 (N+1)} \left(\delta_I^L \delta_K^J - \frac{1}{N} \delta_I^J \delta_K^L \right)   \,.
 }
Since $\langle  {\cal J}_{IK}{}^{JL} \rangle = \langle {\cal J}_I{}^J \rangle = 0$, we immediately have that
 \es{JJTwoPoint}{
  \langle {\cal J}_I{}^J (\vphi_1)  {\cal J}_K{}^L (\vphi_2)  \rangle
   = - \frac{N}{64 \pi^2 r^2 (N+1)} \left(\delta_I^L \delta_K^J - \frac{1}{N} \delta_I^J \delta_K^L \right) \,.
 }
Seeing as $\langle {\cal J}_{IJ}{}^{KL} \star {\cal J}_M{}^N \rangle= 0$ (because this product does not contain an $SU(N)$ singlet), \eqref{JJstar} also implies the three-point function (for $\vphi_1 < \vphi_2 <  \vphi_3$)
 \es{JJThreePoint}{
  \langle {\cal J}_I{}^J (\vphi_1)  {\cal J}_K{}^L (\vphi_2) {\cal J}_M{}^N (\vphi_3)  \rangle
    &=  \frac{N}{(8 \pi r)^3 (N+1)} \left( \delta^J_K \delta_I^N \delta_M^L - \delta^L_I \delta_K^N \delta_M^J \right)  \,.
 }
We can now write the nilpotency condition mentioned above more precisely:  setting $K=J$ in \eqref{JJstar} and summing over $J$, we have
 \es{nilpotency}{
   {\cal J}_I{}^J \star {\cal J}_J{}^L =
    -\frac{N}{8 \pi r}  {\cal J}_I{}^L - \frac{N - 1}{64 \pi^2 r^2}  \delta_I^L  \,.
 }
This expression is $O(1/r)$, as mentioned above.

Let us now reproduce these formulas from the Coulomb branch 1D sector of the necklace quiver gauge theory.  First, based on the definitions \eqref{JDiag}--\eqref{JOffDiag}, we represent ${\cal J}_I{}^J$ by the North shift operators
 \es{JNorth}{
  ({\cal J}_I{}^I)_N = \frac{1}{4 \pi r} \left(i \sigma_{I-1, I} - \frac{B_{I-1, I}}{2} \right) \,, \quad
   ({\cal J}_I{}^J)_N = \frac{1}{4 \pi r}
    \left(i \sigma_{I-1, I} + \frac{1 - B_{I-1, I}}{2}   \right) e^{- \vec{b}_I{}^J \cdot  \left( \frac{i}{2} \partial_{\vec{\sigma}} + \partial_{\vec{B}} \right) } \,,
 } 
where $I \neq J$ (no summation over $I$).  Using \eqref{JNorth} and \eqref{wavefnNoIns}, we can then show that the two-point functions of ${\cal J}_I{}^I$ agree with \eqref{JJTwoPoint}:
\begin{align}
  \langle {\cal J}_I{}^I(\vphi_1) {\cal J}_I{}^I(\vphi_2) \rangle 
   &= \frac{1}{Z} \int d\mu(\sigma)  \left(\frac{i \sigma_{I-1, I}}{4 \pi r} \right)^2 
   = \frac{1}{Z} \int d\tau\, \frac{G_\tau(\vphi_{12})G_\tau(-\vphi_{12}) + G_\tau(0)^2}{\left[ 2 \cosh(\pi \tau) \right]^N} \nonumber \\
   &= - \frac{N-1}{64 \pi^2 r^2 (N+1)} \,,
\label{JIITwo}
\end{align}
where in the second equality, we used \eqref{FTId} with $F_I(\sigma) = \left(\frac{i \sigma}{4 \pi r} \right)^2 / (2 \cosh (\pi \sigma))$ and $\tF_I(\tau) = \left[G_\tau(\vphi_{12})G_\tau(-\vphi_{12}) + G_\tau(0)^2 \right]/(2 \cosh(\pi \tau))$ and $F_j, \tF_j$ given in \eqref{FjBasic} for $j \neq I$.  We can also show that the two-point functions of the off-diagonal ${\cal J}_I{}^J$ agree with \eqref{JJTwoPoint}.  For $\vphi_1 < \vphi_2$, 
\begin{align}
  \langle {\cal J}_I{}^J(\vphi_1) {\cal J}_J{}^I(\vphi_2) \rangle 
   &= \frac{1}{Z} \int d\mu(\sigma) \, \frac{\left( i \sigma_{I-1, I} + \frac12 \right)\left( i \sigma_{J-1, J} - \frac12 \right) }{16 \pi^2 r^2} \nonumber \\
    &= \frac{1}{Z} \int d\tau\, \frac{G_\tau(\vphi_{12})G_\tau(-\vphi_{12})}{\left[ 2 \cosh(\pi \tau)  \right]^N} 
   = - \frac{N}{64 \pi^2 r^2 (N+1)} \,,
\label{JIJTwo}
\end{align}
where in the second equality, we used \eqref{FTId} with $F_I(\sigma) = \frac{i \sigma + \frac 12}{8 \pi r \cosh(\pi \sigma)}$,  $F_J(\sigma) =  \frac{i \sigma - \frac 12}{8 \pi r \cosh(\pi \sigma)}$, $\tF_I(\tau) = \left[ 1 - \tanh(\pi \tau) \right]/(16 \pi r \cosh(\pi \tau))$, $\tF_J(\tau) = \left[ -1 - \tanh(\pi \tau) \right]/(16 \pi r \cosh(\pi \tau))$, and $F_j, \tF_j$ given in \eqref{FjBasic} for $j \neq I$ and $j \neq J$.  One can similarly check that all other two-point functions of the ${\cal J}_I{}^J$ vanish, thus reproducing \eqref{JJTwoPoint}.

Lastly, we check that the three-point functions agree with \eqref{JJThreePoint}.  For instance, we have (with no summation over distinct $I$, $J$, $K$, and $\vphi_1 < \vphi_2 < \vphi_3$)
\begin{align}
  \langle {\cal J}_I{}^J (\vphi_1) {\cal J}_J{}^K (\vphi_2) {\cal J}_K{}^I (\vphi_3) \rangle &=
   \frac{1}{Z} \int d\mu(\sigma) \, \frac{\left( i \sigma_{I-1, I} + \frac12 \right)\left( i \sigma_{J-1, J} - \frac12 \right) \left( i \sigma_{K-1, K} - \frac12 \right) }{64 \pi^3 r^3} \nonumber
    \\
    &= \frac{N}{(8 \pi r)^3 (N+1)} 
\label{ThreePoint1}
\end{align}
and 
\begin{align}
  \langle {\cal J}_I{}^I (\vphi_1) {\cal J}_I{}^J (\vphi_2) {\cal J}_J{}^I (\vphi_3) \rangle &=
   \frac{1}{Z} \int d\mu(\sigma) \, \frac{\left( i \sigma_{I-1, I}  \right)\left( i \sigma_{I-1, I} + \frac12 \right) \left( i \sigma_{J-1, J} - \frac12 \right) }{64 \pi^3 r^3} \nonumber
    \\
    &= \frac{N}{(8 \pi r)^3 (N+1)} \,,
\label{ThreePoint2}
\end{align}
in agreement with \eqref{JJThreePoint}.

We have thus matched all two- and three-point functions of the operators $\mathcal{J}_I{}^J$, which are neatly summarized in the star product \eqref{JJstar}, across the duality.  Note that in SQED$_N$, we can take derivatives with respect to mass parameters to compute correlation functions of the diagonal $\mathcal{J}_I{}^I$, so equality of the partition functions of SQED$_N$ and the $N$-node necklace quiver theory enriched with mass/FI parameters already guarantees matching of correlation functions of $\mathcal{J}_I{}^I$ and $\Phi_{I-1, I}$.  Hence our nontrivial check is of the correspondence between the off-diagonal $\mathcal{J}_I{}^J$ and monopole operators.

Having matched the chiral ring generators ${\cal J}_I{}^J$ between the two sides of the mirror symmetry duality, one can construct composite operators by taking star products of ${\cal J}_I{}^J$.  For example, for fixed $I$ and $J$ (no summation), we can consider on the SQED$_N$ side
 \es{JIIstarJJJ}{
  {\cal J}_I{}^I \star {\cal J}_J{}^J = {\cal J}_{IJ}{}^{IJ} 
    - \frac{N}{64 \pi^2 r^2 (N+1)} \left(\delta_I^J  - \frac{1}{N}\right) \,.
 }
On the necklace side, we have
 \es{JIIstarJJJNeck}{
  {\cal J}_I{}^I \star {\cal J}_J{}^J = - \frac{\Phi_{I-1, I} \Phi_{J-1, J}}{16 \pi^2} \,.
 }
The two expressions must match, so we conclude that
 \es{JIIJJNeck}{
   {\cal J}_{IJ}{}^{IJ}  = - \frac{\Phi_{I-1, I} \Phi_{J-1, J}}{16 \pi^2} + \frac{N}{64 \pi^2 r^2 (N+1)} \left(\delta_I^J  - \frac{1}{N}\right) \,.
 }
As another example, let $I>J>K>L$.  Then on the SQED$_N$ side, we have ${\cal J}_I{}^J \star {\cal J}_K{}^L = {\cal J}_{IJ}{}^{KL}$.  On the necklace side, we have
 \es{JIJKLExample}{
  {\cal J}_{IJ}{}^{KL} = \frac{{\cal M}^{\vec{b}_I{}^J} {\cal M}^{\vec{b}_K{}^L}}{16 \pi^2}
   = \frac{{\cal M}^{\vec{b}_I{}^J +\vec{b}_K{}^L}}{16 \pi^2} \,.
 }
One can construct other composite operators along the same lines.  Note that because all twisted HBOs in SQED$_N$ can be obtained by taking traceless, symmetric products of the $\mathcal{J}_I{}^J$, we expect to be able to construct all bare monopoles in the necklace quiver as polynomials in the basic monopoles \eqref{JOffDiag} and twisted scalars (unlike in generic abelian theories \cite{Bullimore:2015lsa}).

\section{Discussion}

Twisted Higgs and Coulomb branch operators comprise protected 1D topological sectors of 3D $\mathcal{N} = 4$ theories.  Their OPE algebras can be viewed as noncommutative deformations of the Higgs and Coulomb branch chiral rings, and their correlation functions can be calculated exactly.  In this paper, we have studied correlation functions of twisted CBOs using supersymmetric localization.  An arbitrary number of such operators can be inserted anywhere along a great circle of $S^3$ while preserving a common supercharge, with the resulting correlators depending only on their ordering along the circle. These correlators determine the Coulomb branch chiral ring of our theories, and moreover, completely fix the two- and three-point functions of all CBOs at the IR fixed point. In combination with the results of \cite{Dedushenko:2016jxl}, where similar results were obtained for the Higgs branch, we now have a complete story for two- and three-point functions of half-BPS operators in 3D $\cN=4$ abelian gauge theories.  We have leveraged our results to perform new tests of abelian 3D mirror symmetry, amounting to a proof at the level of two- and three-point functions of half-BPS local operators.

Unlike in the Higgs branch case, a challenging aspect of dealing with twisted CBOs is that they include defect monopole operators. As a result, while the Higgs branch 1D TQFTs admit very explicit 1D Lagrangians \cite{Dedushenko:2016jxl}, constructing such  Lagrangians for the Coulomb branch proved to be difficult. Instead, we have devised an alternative approach, in which insertions of twisted CBOs are represented by certain shift operators acting on hemisphere wavefunctions, which in turn can be glued into the desired correlators on $S^3$. The same approach was also used in the context of the line defect Schur index in 4D $\cN=2$ theories \cite{Dimofte:2011py,Gang:2012yr,Cordova:2016uwk}, which we have shown to be related to our 3D computations by dimensional reduction.

The natural next step is to extend this work to non-abelian theories, where the Coulomb branch chiral ring and mirror symmetry are less understood. In those theories, the BPS equations in the presence of monopole operators have ``monopole bubbling'' solutions in which the GNO charge of a singular monopole is screened away from the insertion point \cite{Kapustin:2006pk}. These solutions have to be summed over, which considerably complicates the analysis. Fortunately, this problem has been addressed in some examples in 4D $\cN=2$ theories (see, e.g., \cite{Gomis:2011pf,Ito:2011ea,Gang:2012yr}). Therefore, the 4D/3D relation we have uncovered could  prove to be useful in incorporating the monopole bubbling effect into our 3D localization framework. 

So far, in both the Higgs and Coulomb branch cases, only theories with hypermultiplets and vector multiplets have been studied. It would be interesting to generalize our localization computations to other theories that also include twisted multiplets. One class of examples where the generalization is rather trivial is that of abelian gauge theories with BF couplings \cite{Brooks:1994nn};\footnote{These couplings are simply FI actions that couple background twisted vector multiplets to dynamical vector multiplets. Introducing twisted hypermultiplets coupled to the background twisted vector multiplets, and gauging the latter, produces BF-type theories.} some aspects of these theories are discussed in Appendix \ref{moreonmatching}. There are also theories with Chern-Simons terms for which application of our results is less trivial, such as those of Gaiotto-Witten \cite{Gaiotto:2008ak}, ABJ(M) \cite{Aharony:2008ug, Aharony:2008gk}, and generalizations thereof \cite{Imamura:2008dt, Hosomichi:2008jd}.\footnote{For recent progress on combining supersymmetric localization results with the conformal bootstrap in the maximally supersymmetric case, see \cite{Agmon:2017xes}.} A technical obstruction to applying our formalism to those theories is that only an $\cN=3$ subalgebra of the $\cN=4$ SUSY algebra is realized off shell on their vector multiplet. The supercharge that we wish to use for localization, however, does not reside in this $\cN=3$ subalgebra, and therefore does not close off shell (as required for localization). Nevertheless, it is plausible that this technical difficulty could be overcome by closing off shell only the particular supercharge in which we are interested.

An interesting offshoot of our analysis is the careful treatment of the gluing of hemisphere partition functions into the $S^3$ partition function (with insertions). In particular, in our approach, gluing is performed through supersymmetric localization of the path integral over boundary conditions. It could be interesting to apply this approach to other supersymmetric theories on manifolds with boundaries as studied in, e.g., \cite{Beem:2012mb, Pasquetti:2016dyl, Sugishita:2013jca,Hori:2013ika,Honda:2013uca,Yoshida:2014ssa,Gava:2016oep,LeFloch:2017lbt,Bawane:2017gjf}. 

Finally, another open question, of a somewhat academic nature, is whether the 3D gluing bilinear form has a 1D Hilbert space interpretation. For example, it would be interesting to understand whether the hemisphere wavefunctions can really be thought of as representing states in the 1D TQFT. In particular, in passing to cohomology, one is tempted to view a state in the Hilbert space of the 3D theory on $S^2$ as a state in the product $H_N\otimes H_S$, where $H$ is the Hilbert space of the 1D theory and the two copies correspond to the North and South boundary points of the semicircle. The North and South shift operators that we have constructed are then simply interpreted as operators acting on $H_N$ and $H_S$, respectively. One fantasy is that the answers to these questions could provide an interpretation of the $S^3$ partition function of 3D $\cN=4$ theories as some trace over the Hilbert space of the 1D TQFTs. We hope to address some of the questions raised here in future work.

\section*{Acknowledgements}

We thank Ofer Aharony, Clay C\'ordova, Tudor Dimofte, Bruno Le Floch, Davide Gaiotto, Sergei Gukov, Masazumi Honda, Mikhail Isachenkov, Anton Kapustin, Victor Mikhaylov, Shlomo S. Razamat, Shu-Heng Shao, Genis Torrents, Gustavo J. Turiaci, and Itamar Yaakov for useful discussions. We especially thank Davide Gaiotto for sharing unpublished notes and for a fruitful discussion. While preparing this work, we have benefited from multiple meetings of the Simons Collaboration on the Nonperturbative Bootstrap. The work of MD was supported by the Walter Burke Institute for Theoretical Physics and the U.S. Department of Energy, Office of Science, Office of High Energy Physics, under Award No.\ DE-SC0011632, as well as the Sherman Fairchild Foundation. YF gratefully acknowledges support from the NSF GRFP under Grant No.\ DGE-1656466 and from the Graduate School at Princeton University. The work of SSP was supported in part by the Simons Foundation Grant No.~488651. The work of RY was supported in part by a grant from the Israel Science Foundation Center for Excellence, by the Minerva Foundation with funding from the Federal German Ministry for Education and Research, and by the ISF within the ISF-UGC joint research program framework (grant no.\ 1200/14).

\appendix

\section{Conventions} \label{conventions}

Our conventions largely follow those of \cite{Dedushenko:2016jxl}.  In particular, spacetime indices are denoted by $\mu, \nu, \ldots$, frame indices are denoted by $i, j, \ldots$, and fundamental indices of $SU(2)_H$, $SU(2)_C$, and $SU(2)_\text{rot}$ are denoted by $a, b, \ldots = 1, 2$; $\dot{a}, \dot{b}, \ldots = 1, 2$; and $\alpha, \beta, \ldots = 1, 2$, respectively.  $SU(2)_{H, C, \, \text{rot}}$ indices are all raised and lowered from the left with the antisymmetric tensor, which satisfies $\epsilon^{12} = \epsilon_{21}=1$.  $SU(2)_{H, C}$ indices are typically explicit while spinor ($SU(2)_\text{rot}$) and gauge (color) indices are typically suppressed; spinor contractions are defined by $\psi\chi\equiv \psi^\alpha\chi_\alpha$.  The spinor parameter $\xi$ is always taken to be commuting, so that $\delta_\xi$ is anticommuting.  For any given $SU(2)$ index, we have the Fierz identity
\begin{equation}
x_\alpha y^\beta z_\beta + x_\beta y_\alpha z^\beta + x^\beta y_\beta z_\alpha = 0,
\end{equation}
which holds regardless of whether the objects $x, y, z$ are Grassmann-even or Grassmann-odd, or $c$-numbers or $q$-numbers.

Unless otherwise stated, the gamma matrices in any local frame are the Pauli matrices, which satisfy $\gamma^i\gamma^j = \delta^{ij} + i\epsilon^{ijk}\gamma_k$.  Recall that $\nabla_\mu = \partial_\mu + \frac{i}{4}\omega_{\mu ij}\epsilon^{ijk}\gamma_k$ on spinors.

\subsection{Coordinates} \label{coordinates}

Let us summarize the various coordinate systems on round $S^3$ of radius $r$ used throughout the text.  It is useful to relate all of them to embedding coordinates $(X_1, X_2, X_3, X_4)\in \mathbb{R}^4$.

\begin{itemize}
	\item In the usual fibration coordinates $\theta, \tau, \varphi$ with $\theta\in [0, \pi/2]$ and $\tau, \varphi\in [-\pi, \pi]$, we have
	\begin{equation}
	(X_1, X_2, X_3, X_4) = r(\cos\theta\cos\tau, \cos\theta\sin\tau, \sin\theta\cos\varphi, \sin\theta\sin\varphi).
	\end{equation}
	The metric takes the form
	\begin{equation}
	ds^2 = r^2\cos^2\theta\, d\tau^2 + ds_{D^2}^2, \quad ds_{D^2}^2 = r^2(d\theta^2 + \sin^2\theta\, d\varphi^2).
	\end{equation}
	Operator insertions lie along the $\theta = \pi/2$ circle parametrized by $\varphi$:
	\[
	S^1 : \{(X_1 + iX_2, X_3 + iX_4) = (0, re^{i\varphi})\}.
	\]
	We cut the $S^3$ along an $S^2$ parametrized by $\theta, \tau$ (= $\tau$ circle fibered over a line segment) orthogonal to this $S^1$ that meets this $S^1$ at $\varphi = 0, \pm\pi$:
	\[
	S^2 : \bigcup_\pm \{(X_1 + iX_2, X_3 + iX_4) = (r\cos\theta e^{i\tau}, \pm r\sin\theta)\}.
	\]
	The hemispheres $HS^3_\pm$ bounded by this $S^2$ correspond to $\varphi > 0$ and $\varphi < 0$, respectively.
	\item In stereographic coordinates $x_{1, 2, 3}$, we have
	\begin{equation}
	x_{i = 1, 2} = \frac{2X_i}{1 + X_3/r}, \quad x_3 = \frac{2X_4}{1 + X_3/r}, \quad x^2\equiv x_1^2 + x_2^2 + x_3^2.
	\end{equation}
	The metric takes the simple form
	\begin{equation}
	ds^2 = e^{2\Omega}(dx_1^2 + dx_2^2 + dx_3^2), \quad e^\Omega = \frac{1}{1 + x^2/4r^2} = \frac{1 + X_3/r}{2}.
	\end{equation}
	Stereographic projection maps the insertion circle to the line $(x_1, x_2, x_3) = (0, 0, 2r\tan\frac{\varphi}{2})$ and the boundary $S^2$ to the $(1, 2)$-plane
	\[
	\bigcup_\pm \left\{(x_1, x_2, x_3) = \left(\frac{2r\cos\theta\cos\tau}{1\pm \sin\theta}, \frac{2r\cos\theta\sin\tau}{1\pm \sin\theta}, 0\right)\right\},
	\]
	here written as the union of the interior/exterior of a circle.
	\item In spherical coordinates $\eta, \psi, \tau$ adapted to our two-monopole background (so that the monopole and antimonopole insertions at $\eta = 0, \pi$ correspond to $\varphi = \pm\pi/2$), we have
	\begin{equation}
	(X_1, X_2, X_3, X_4) = r(\sin\eta\sin\psi\cos\tau, \sin\eta\sin\psi\sin\tau, -\sin\eta\cos\psi, \cos\eta)
	\end{equation}
	where $\tau\in [-\pi, \pi]$ and $\eta, \psi\in [0, \pi]$ ($\tau$ is the same as in fibration coordinates).  The metric takes the form
	\begin{equation}
	ds^2 = r^2(d\eta^2 + \sin^2\eta\, ds_{S^2}^2), \quad ds_{S^2}^2 = d\psi^2 + \sin^2\psi\, d\tau^2.
	\end{equation}
	The boundary $S^2$ corresponds to setting $\eta = \pi/2$:
	\[
	(X_1, X_2, X_3, X_4) = \underbrace{r(\cos\theta\cos\tau, \cos\theta\sin\tau, \pm\sin\theta, 0)}_{\text{fibration}} = \underbrace{r(\sin\psi\cos\tau, \sin\psi\sin\tau, -\cos\psi, 0)}_{\text{spherical}}.
	\]
\end{itemize}
The $\pm$ in $\pm\sin\theta$ can be suppressed by assuming that on the boundary, $\theta\in [-\pi, \pi]$.

For fermions, we work mainly in the stereographic or the spherical frame.  The stereographic frame is defined as $(e_\text{st})_\mu^i = e^\Omega\delta_\mu^i$ while the spherical frame is defined as
\begin{equation}
(e_\text{sph})^1 = d\eta, \quad (e_\text{sph})^2 = \sin\eta\, d\psi, \quad (e_\text{sph})^3 = \sin\eta\sin\psi\, d\tau,
\end{equation}
in their respective coordinates.

\subsection{Supersymmetry Transformations}
\label{susytrans}

The supersymmetry transformations used in the main text are as follows.

\subsubsection{3D $\cN=4$}

These transformations are parametrized by the conformal Killing spinors \eqref{killing1} on $S^3$. The transformations of the vector multiplet $\cV$ in \eqref{Vmul} are given by
\begin{align}
\delta_{\xi} A_{\mu} &= \textstyle \frac{i}{2} \xi^{a\dot{b}}\gamma_{\mu}\lambda_{a\dot{b}} \ec \label{Avar}\\
\delta_{\xi} \lambda_{a\dot{b}} &= \textstyle - \frac{i}{2}\varepsilon^{\mu\nu\rho}\gamma_{\rho}\xi_{a\dot{b}}F_{\mu\nu} - D_a{}^c\xi_{c\dot{b}} -i\gamma^{\mu}\xi_a{}^{\dot{c}} \cD_{\mu}\Phi_{\dot{c}\dot{b}} + 2i \Phi_{\dot{b}}{}^{\dot{c}}\xi_{a\dot{c}}' + \frac{i}{2}\xi_{a\dot{d}} [ \Phi_{\dot{b}}{}^{\dot{c}}, \Phi_{\dot{c}}{}^{\dot{d}}] \ec \label{lamvar}\\
\delta_{\xi}\Phi_{\dot{a}\dot{b}} &= \xi^c{}_{(\dot{a}}\lambda_{|c|\dot{b})} \ec \label{phivar}\\
\delta_{\xi} D_{ab} &= -i\cD_{\mu}(\xi_{(a}{}^{\dot c}\gamma^{\mu}\lambda_{b)\dot c}) - 2i\xi'_{(a}{}^{\dot c}\lambda_{b)\dot c} + i [\xi_{(a}{}^{\dot{c}}\lambda_{b)}{}^{\dot{d}}, \Phi_{\dot{c}\dot{d}}] \ed \label{dvar}
\end{align}
The transformations of the hypermultiplet $\cH$ in \eqref{Hmul} are given by
\begin{align}
&\delta_{\xi} q^a = \xi^{a\dot{b}} \psi_{\dot{b}}\ecq \delta_{\xi} \psi_{\dot{a}} = i\gamma^{\mu}\xi_{a\dot{a}} \cD_{\mu}q^a + i\xi'_{a\dot{a}}q^a - i\xi_{a\dot{c}}\Phi^{\dot{c}}{}_{\dot{a}}q^a \ec \label{qpsivar}\\
&\delta_{\xi} \tq^{a} = \xi^{a\dot{b}} \tpsi_{\dot{b}} \ecq \delta_{\xi} \tpsi_{\dot{a}} = i\gamma^{\mu}\xi_{a\dot{a}} \cD_{\mu}\tq^a + i\tq^a\xi'_{a\dot{a}} + i\xi_{a\dot{c}}\tq^a\Phi^{\dot{c}}{}_{\dot{a}} \ed \label{qtpsitvar}
\end{align}
In terms of Poincar\'e and conformal supercharges of $\mathfrak{osp}(4|4)$, the supercharges of primary interest for us are
\begin{align}
\mathcal{Q}_1^H &= \textstyle Q_{11\dot{2}} - \frac{1}{2r}S_{12\dot{2}}, \nonumber \\
\mathcal{Q}_2^H &= \textstyle Q_{21\dot{1}} + \frac{1}{2r}S_{22\dot{1}}, \nonumber \\
\mathcal{Q}_1^C &= \textstyle \frac{1}{2}\left(Q_{11\dot{2}} + iQ_{12\dot{2}} + Q_{11\dot{1}} + iQ_{12\dot{1}} + \frac{i}{2r}S_{11\dot{2}} - \frac{1}{2r}S_{12\dot{2}} - \frac{i}{2r}S_{11\dot{1}} + \frac{1}{2r}S_{12\dot{1}}\right), \nonumber \\
\mathcal{Q}_2^C &= \textstyle \frac{1}{2}\left(Q_{21\dot{1}} - iQ_{22\dot{1}} + Q_{21\dot{2}} - iQ_{22\dot{2}} + \frac{i}{2r}S_{21\dot{1}} + \frac{1}{2r}S_{22\dot{1}} - \frac{i}{2r}S_{21\dot{2}} - \frac{1}{2r}S_{22\dot{2}}\right),
\end{align}
from which $\mathcal{Q}_\beta^H$ and $\mathcal{Q}_\beta^C$ follow.  To derive the corresponding Killing spinors $\xi_\beta^H$ and $\xi_\beta^C$, we use that in $\mathbb{R}^3$, the action of supersymmetries is
\begin{equation}
\xi_{a\dot{a}} = \epsilon_{a\dot{a}} + x^i\gamma_i\eta_{a\dot{a}}, \mbox{ } \xi'_{a\dot{a}} = \eta_{a\dot{a}} \implies \delta_\xi\mathcal{O} = \frac{i}{2}[\epsilon^{\alpha a\dot{a}}Q_{\alpha a\dot{a}} + \eta^{\alpha a\dot{a}}S_{\alpha a\dot{a}}, \mathcal{O}]
\end{equation}
(for the explicit action of the generators of $\mathfrak{osp}(4|4)$ on fields, see Appendix C of \cite{Dedushenko:2016jxl}).  Expressions for $(\xi_\beta^H)_{\alpha a\dot{a}}$ and $(\xi_\beta^C)_{\alpha a\dot{a}}$ are given in (5.5) of \cite{Dedushenko:2016jxl} and \eqref{xiCbeta}, respectively.

\subsubsection{2D $\cN=(2,2)$}

These transformations are parametrized by a pair of Killing spinors $\epsilon$ and $\bar\epsilon$ on $S^2$ satisfying
\begin{align}
\label{KillingS2}
\nabla_\mu \epsilon &= -\frac{i}{2r}\gamma_\mu \epsilon\ec\quad
\nabla_\mu \bar{\epsilon} = \frac{i}{2r}\gamma_\mu\bar{\epsilon}\ec
\end{align}
where $\mu=\theta,\tau$ is restricted to the directions along $S^2$. We define the 2D gamma matrices
\begin{equation}
\Gamma_\theta = i\sigma_3 \gamma_\theta\ec \quad \Gamma_\tau = i\sigma_3 \gamma_\tau \ec
\end{equation}
in terms of which the 2D Killing spinor equations \eqref{KillingS2} become
\begin{align}
\label{KillingS2_mod}
\nabla_\mu \epsilon &= \frac{1}{2r}\Gamma_\mu \sigma_3 \epsilon\ec\quad
\nabla_\mu \bar{\epsilon} = -\frac{1}{2r}\Gamma_\mu\sigma_3\bar{\epsilon}\ec
\end{align}
precisely matching those of \cite{Doroud:2012xw}. 

The spinors parametrizing the $\cN=(2,2)$ supercharges \eqref{Qp22} and \eqref{Qm22} are given by
\begin{align}
Q_1^+:\ \ &\bar\epsilon_{\alpha}=0,\, \epsilon_\alpha=\left(\begin{matrix}-\frac{e^{-i\tau}\cos\theta}{\sqrt{2+2\sin\theta}}\cr -\frac{i\sqrt{1+\sin\theta}}{\sqrt{2}}\end{matrix}\right),\cr
Q_2^+:\ \ &\bar\epsilon_{\alpha}=0,\, \epsilon_\alpha=\left(\begin{matrix}-\frac{i\sqrt{1+\sin\theta}}{\sqrt{2}}\cr -\frac{ e^{i\tau}\cos\theta}{\sqrt{2+2\sin\theta}}\end{matrix}\right),\cr
Q_1^-:\ \ &\epsilon_{\alpha}=0,\, \bar\epsilon_\alpha=\left(\begin{matrix}-\frac{e^{-i\tau}\cos\theta}{\sqrt{2+2\sin\theta}}\cr \frac{i\sqrt{1+\sin\theta}}{\sqrt{2}}\end{matrix}\right),\cr
Q_2^-:\ \ &\epsilon_{\alpha}=0,\, \bar\epsilon_\alpha=\left(\begin{matrix}-\frac{i\sqrt{1+\sin\theta}}{\sqrt{2}}\cr \frac{e^{i\tau}\cos\theta}{\sqrt{2+2\sin\theta}}\end{matrix}\right)\ed
\end{align}
These can be derived by demanding that the corresponding Killing vectors contain no $\partial_\varphi$ terms on the boundary $S^2$ (see Appendix A of \cite{Dedushenko:2016jxl}).

The SUSY transformations of the 2D $\cN=(2,2)$ chiral multiplets $\Phi^{(2d)}$ in \eqref{PH2d} are
\begin{align}
\delta \phi &= \bar{\epsilon}\chi\ec\cr
\delta \widetilde{\phi}&= \epsilon\widetilde{\chi}\ec\cr
\delta\chi &= \textstyle i( \Gamma^\mu\cD_\mu \phi + s_1 \phi - is_2 \phi\sigma_3 + \frac{1}{2r}\phi\sigma_3 )\epsilon + f\bar{\epsilon}\ec\cr
\delta\widetilde\chi &= \textstyle i( \Gamma^\mu\cD_\mu \widetilde{\phi} + s_1 \widetilde{\phi} + is_2 \widetilde{\phi}\sigma_3 - \frac{1}{2r}\widetilde{\phi}\sigma_3 )\bar\epsilon + \widetilde{f}\epsilon\ec\cr
\delta f &= \textstyle i(\cD_\mu\chi \Gamma^\mu + s_1\chi - is_2\chi\sigma_3 + \phi\lambda +\frac1{2r} \chi\sigma_3 )\epsilon\ec\cr
\delta \widetilde{f} &= \textstyle i(\cD_\mu\widetilde{\chi}\Gamma^\mu + s_1\widetilde\chi + is_2\widetilde\chi\sigma_3 - \widetilde{\phi}\bar\lambda -\frac1{2r} \widetilde\chi\sigma_3 )\bar\epsilon\ed
\end{align}
The SUSY transformations of the vector multiplets $V^{(2d)}$ in \eqref{V2d} are
\begin{align}
\delta a_\mu &= \textstyle -\frac{i}{2} ( \bar{\epsilon}\Gamma_\mu\lambda + \epsilon\Gamma_\mu\widetilde{\lambda})\ec\cr
\delta s_1 &= \textstyle \frac{1}{2} ( \bar{\epsilon}\lambda - \epsilon\widetilde{\lambda})\ec\cr
\delta s_2 &= \textstyle -\frac{i}{2}( \bar{\epsilon}\sigma_3\lambda - \epsilon\sigma_3\widetilde{\lambda} )\ec\cr
\delta\lambda&= ( i V^\mu \Gamma_\mu + i V^3\sigma_3 - D_{2d} )\epsilon\ec\cr
\delta\widetilde{\lambda}&= ( i\bar{V}^\mu \Gamma_\mu + i\bar{V}^i \sigma_3 + D_{2d} )\bar{\epsilon}\ec\cr
\delta D_{2d} &= \textstyle -\frac{i}{2}\bar{\epsilon}( \Gamma^\mu\cD_\mu\lambda +[s_1,\lambda] -i[s_2,\sigma_3\lambda]) + \frac{i}{2}\epsilon( \Gamma^\mu\cD_\mu\widetilde{\lambda} - [s_1,\widetilde{\lambda}] - i[s_2,\sigma_3\widetilde{\lambda}])\ed
\end{align}
In the SUSY variations of $\lambda$ and $\widetilde{\lambda}$, we have used the following combinations:
\begin{align}
V^\mu &= \varepsilon^{\mu\nu}\cD_\nu s_2 + \cD^\mu s_1\ec\quad V^3 = \frac12 \varepsilon^{\mu\nu}F_{\mu\nu} +i[s_1,s_2]+\frac1{r} s_1\ec \nonumber \\[5 pt]
\bar{V}^\mu &= \varepsilon^{\mu\nu}\cD_\nu s_2 - \cD^\mu s_1\ec\quad \bar{V}^3 = \frac12 \varepsilon^{\mu\nu}F_{\mu\nu} -i[s_1,s_2]+\frac1{r} s_1\ec
\end{align}
where the 2D $\varepsilon$-symbol is induced from the 3D orientation:
\begin{equation}
\varepsilon^{\theta\tau}=-\frac{1}{r^2\cos\theta}.
\end{equation}
These results are in complete agreement with the SUSY variations from \cite{Doroud:2012xw}, up to a minor change of notation.\footnote{There is only a sign difference in the variations of the auxiliary fields $f$ and $\widetilde{f}$, as compared to \cite{Doroud:2012xw}. The reason is that our SUSY parameters $\xi$ in 3D, and consequently the $\epsilon$ in 2D, are commuting, whereas their $\epsilon$ are anticommuting.}

Finally, we comment on two issues regarding how these transformations are verified when $\Phi^{(2d)}$ and $V^{(2d)}$ are identified with the boundary values of the 3D $\cN=4$ fields according to \eqref{xi}, \eqref{phif}, and \eqref{A2d}--\eqref{D2d}. First, to obtain the variations of $f$ and $\widetilde{f}$, one must use the equations of motion of the hypermultiplet fermions $\rho$ and $\widetilde{\rho}$ defined in \eqref{rho}. This is related to the fact that the 3D $\cN=4$ algebra, and consequently, its $\mathfrak{su}(2|1)$ subalgebra that we are using, are not closed off shell.\footnote{While the 3D $\cN=4$ algebra cannot be completely closed off shell, it can be done for the $\mathfrak{su}(2|1)$ sub\-algebra by introducing auxiliary fields. For our purposes, there is no need to perform this exercise explicitly.}  A similar subtlety does not arise in the computation of $\delta D_{2d}$, which is related to the 3D $\cN=4$ vector multiplet being closed off shell.

The second issue is related to the partial gauge-fixing condition $A_\perp\big|=0$ in \eqref{Aperp}. The SUSY variation breaks this gauge, so to fix this, we must supplement it by some gauge transformation with parameter $\kappa$. A convenient way to do this, which does not affect any of the other boundary conditions, is to find a $\kappa$ that vanishes at the boundary, and such that $(A_\perp + \partial_\perp \kappa)\big| =0$. Because $\kappa\big|=0$, it does not affect the boundary values of any fields except for $A_\perp$, whose gauge transformation at the boundary becomes $A_\perp\big| \to A_\perp\big| + \partial_\perp\kappa\big| =0$. Note that this is true even in non-abelian theories, simply because $\kappa\big|=0 \Rightarrow [A_\perp, \kappa]\big|=0$.

\section{More on Monopoles}

\subsection{Global Symmetries and Defects}\label{DefectsSymm}

A local \emph{order} operator $\cO[\phi]$ in quantum field theory is constructed as a functional of local fields $\phi$, and a symmetry transformation acts on it by transforming the argument:
\begin{equation}
U \cO[\phi] U^{-1} = \cO[U\phi U^{-1}].
\end{equation}
On the other hand, given a local \emph{disorder} operator $M[b]$ defined by imposing some boundary condition $b$ close to its insertion point, the action of a symmetry transformation can formally be written as
\begin{equation}
\label{defectTRANSF}
U M[b] U^{-1} \approx M[U^{-1}b],
\end{equation}
where the notation ``$\approx$'' means ``up to normalization'' and accounts for the fact that the normalization of the defect operator $M[b]$ might not be fixed by the boundary condition $b$ alone (as is the case for monopoles). In other words, to act with a global symmetry $U$ on a defect operator, one must act with $U^{-1}$ on the boundary condition that was used to define it, and extra care should be taken to determine normalization.

Let us prove this statement by deriving the Ward identities in the path integral formulation separately for order and disorder operators. The results will differ by a sign.

Consider a symmetry transformation which also acts on boundary conditions:
\begin{equation}
\phi' = \phi + \delta\phi\ec \quad b'=b+\delta b\ed
\end{equation}
Here, $\phi$ stands for all fields in the theory, and the transformation of a boundary condition is simply given by restricting the transformation of $\phi$ to the boundary. The fact that it is a symmetry means that
\begin{equation}
\pD\phi'\, e^{-S[\phi']}=\pD\phi\, e^{-S[\phi]}.
\end{equation}
This transformation takes $\cO[\phi]$ to $\cO[\phi']$ and $M[b]$ to $\widetilde{M}[b']$, where the tilde represents the fact that the normalization (e.g., the phase) of the defect $M[b]$ might change in a way not fixed by $b$. Let us define
\begin{equation}
\delta\cO[\phi]\equiv\cO[\phi']-\cO[\phi]\ec\quad \delta M[b]\equiv\widetilde{M}[b']-M[b]\ed
\end{equation}
Now consider the change of variables
\begin{equation}
\label{change}
\phi'=\phi + \rho(x)\delta\phi,
\end{equation}
where $\rho(x)$ is a smooth function supported in a small neighborhood $U(x_0)$ of the insertion point $x_0$ of the operator of interest and equal to 1 in a compact $V(x_0)\subset U(x_0)$. Since $\rho$ is non-constant, this transformation is no longer a symmetry: instead,
\begin{equation}
\pD\phi'\, e^{-S[\phi']}=\pD\phi\, e^{-S[\phi]}\left(1 - \int d^n x\, \partial_\mu\rho(x) j^\mu(x)\right)
\end{equation}
where $j^\mu$ is the conserved current.

First suppose that the local operator inserted at $x_0$ is of type $\cO[\phi]$. The trivial identity
\begin{equation}
\label{inv}
\int \pD\phi'\, e^{-S[\phi']} \cO[\phi']({\cdots}) = \int \pD\phi\, e^{-S[\phi]} \cO[\phi]({\cdots}),
\end{equation}
where $({\cdots})$ represents insertions outside of $U(x_0)$, implies that:
\begin{equation}
\label{W1}
\left\langle\left(\delta\cO[\phi(x_0)] - \int d^n x\, \partial_\mu\rho(x) j^\mu(x)\cO[\phi(x_0)]\right)({\cdots})\right\rangle=0.
\end{equation}
Now suppose that the operator inserted at $x_0$ is a defect. In this case, we should proceed slightly differently: instead of \eqref{inv}, we start with $\langle \delta M[b]({\cdots})\rangle = \langle \widetilde{M}[b']({\cdots})\rangle - \langle M[b]({\cdots})\rangle $, which is equivalent to
\begin{equation}
\langle \delta M[b]({\cdots})\rangle = \int_{b'} \pD\phi'\, e^{-S[\phi']}({\cdots}) - \int_{b} \pD\phi\, e^{-S[\phi]}({\cdots}),
\end{equation}
where the notation $\int_b$ means that we compute the path integral with boundary conditions $b$. We also assume that the path integral with boundary conditions $b'$ is properly normalized so as to precisely represent the defect operator $\widetilde{M}[b']$. Let us perform the coordinate change \eqref{change} in the first integral. Close to the point $x_0$, we have $\rho(x)=1$, so the coordinate change is simply $\phi'=\phi + \delta\phi$ there; it transforms the boundary condition $b'$ into $b$ and the operator $\widetilde{M}[b']$ into $M[b]$. As a result, we obtain
\begin{align}
\langle \delta M[b]({\cdots})\rangle &= \int_{b} \pD\phi\, e^{-S[\phi]}\left(1 - \int d^n x\, \partial_\mu\rho(x) j^\mu(x)\right)({\cdots}) - \int_{b} \pD\phi\, e^{-S[\phi]}({\cdots}) \nonumber \\
&= -\int_{b} \pD\phi\, e^{-S[\phi]}\int d^n x\, \partial_\mu\rho(x) j^\mu(x)({\cdots}),
\end{align}
or simply:
\begin{equation}
\label{W2}
\left\langle\left(\delta M[b] + \int d^n x\, \partial_\mu\rho(x) j^\mu(x) M[b] \right)({\cdots})\right\rangle=0.
\end{equation}
Notice that \eqref{W1} and \eqref{W2} differ by the sign, which is what we wanted to show.

In the situation where the boundary condition $b$ determines $M[b]$ only up to normalization, there exist symmetries of the theory that act nontrivially in the bulk without changing $b$. Such symmetries multiply $M[b]$ by a number.\footnote{For example, in the case of half-BPS monopole operators in 3D $\cN=4$ theories, $b$ represents a monopole singularity, and it remains invariant under $U(1)_C\subset SU(2)_C$ transformations preserved by this singularity. However, such transformations act nontrivially on $M[b]$: they multiply it by a phase.}  Therefore, we could choose to consider a different bulk symmetry,
\begin{equation}
\label{symmMOD}
\widetilde\phi' = \phi + \widetilde\delta\phi\ec
\end{equation}
which restricts to \emph{the same} transformation of the defect singularity $b'=b+\delta b$. Following the steps above, we obtain the same equation \eqref{W2}, except that the current $j_\mu(x)$ is replaced by the current $\widetilde{j}_\mu(x)$ for the symmetry \eqref{symmMOD}. The difference between $j_\mu(x)$ and $\widetilde{j}_\mu(x)$ is a symmetry that multiplies $M[b]$ by a number. This is why the finite transformation of the defect operator in \eqref{defectTRANSF} is written only up to normalization.

To have a precise equality, one must also determine whether $U$ changes the normalization of $M[b]$. However, it might be impossible to pick a consistent normalization of $M[b]$ for all possible $b$. An example of this kind was explained in the main text: it is impossible to pick a normalization of the monopole operator for all possible singular boundary conditions, as it would require choosing a global section of the Hopf fibration. It is possible, nevertheless, to pick a normalization of $M[b]$ for some subset $\cB$ of possible boundary conditions. In this situation, one can write $M[b]$ only for $b\in\cB$, and the transformation becomes
\begin{equation}
U M[b] U^{-1} = \lambda(U)M[U^{-1}b],\quad \lambda(U)\in \C^*,
\end{equation}
where $\lambda(U)$ encodes the change of normalization. Moreover, we are only allowed to consider those $U$ for which $U^{-1}b\in \cB$.

\subsection{Boundary Terms and Monopole Counterterms}\label{CounterSigma}

Writing \eqref{Shyper} as $S_\text{hyper}[\cH, \cV] = \int d^3 x\, \sqrt{g}\cL_\text{hyper}[\cH, \cV]$, the boundary term in the SUSY variation of the gauged hypermultiplet Lagrangian for $\xi\in \mathfrak{osp}(4|4)$ is
\begin{align}
\label{hyperBNDRY}
\delta_\xi\cL_\text{hyper}[\cH, \cV] &= D_\mu\bigg((\xi_{a\dot{a}}\tilde{\psi}^{\dot{a}})D^\mu q^a + \tilde{q}^a(D^\mu\xi_{a\dot{a}}\psi^{\dot{a}}) + i\epsilon^{\mu\nu\rho}\tilde{q}^a D_\nu(\xi_{a\dot{a}}\gamma_\rho\psi^{\dot{a}}) \nonumber \\
&\hspace{1.5 cm} + \tilde{q}^a\Phi^{\dot{a}\dot{b}}(\xi_{a\dot{b}}\gamma^\mu\psi_{\dot{a}}) + \frac{1}{2}\tilde{q}^a(\xi_a{}^{\dot{b}}\gamma^\mu\lambda^c{}_{\dot{b}})q_c + \frac{1}{2}\tilde{q}^a(\xi^{c\dot{b}}\gamma^\mu\lambda_{a\dot{b}})q_c\bigg).
\end{align}
Writing \eqref{SYM} as $S_\text{YM}[\cV] = g_\text{YM}^{-2}\int d^3 x\, \sqrt{g}\cL_\text{YM}[\cV]$, the boundary term in the SUSY variation of the Yang-Mills Lagrangian for $\xi\in \mathfrak{su}(2|1)_\ell\oplus \mathfrak{su}(2|1)_r$ is
\begin{align}
\label{vectorBNDRY}
\delta_\xi\cL_\text{YM}[\cV] &= D_\mu\Tr\bigg(i\xi^{a\dot{a}}\gamma_\nu\lambda_{a\dot{a}}F^{\mu\nu} - \frac{1}{2}\epsilon^{\mu\nu\rho}\xi^{a\dot{a}}\lambda_{a\dot{a}}F_{\nu\rho} + iD_a{}^c\xi_{c\dot{a}}\gamma^\mu\lambda^{a\dot{a}} + \frac{1}{2}\xi_a{}^{\dot{c}}\gamma^\mu\lambda^{a\dot{a}}[\Phi_{\dot{a}}{}^{\dot{b}}, \Phi_{\dot{b}\dot{c}}] \nonumber \\
&\hspace{-1 cm} - \xi_{a\dot{b}}\lambda^{a\dot{a}}D^\mu\Phi_{\dot{a}}{}^{\dot{b}} - i\epsilon^{\mu\nu\rho}\xi^{a\dot{a}}\gamma_\rho\lambda_a{}^{\dot{b}}D_\nu\Phi_{\dot{a}\dot{b}} - 2\xi'_{a\dot{b}}\gamma^\mu\lambda^{a\dot{a}}\Phi_{\dot{a}}{}^{\dot{b}} - \frac{i}{r}h^{ab}\bar{h}^{\dot{a}\dot{b}}\xi_a{}^{\dot{c}}\gamma^\mu\lambda_{b\dot{c}}\Phi_{\dot{a}\dot{b}}\bigg).
\end{align}
For the monopole counterterm, it suffices to consider the abelian case, for which the SUSY transformations are obtained by omitting the terms involving commutators in \eqref{lamvar} and \eqref{dvar}.  We need not include fermionic terms in the monopole counterterm because fermions are set to zero in BPS configurations.  Letting
\begin{equation}
V = h^{ab}\bar{h}^{\dot{a}\dot{b}}\left(\frac{1}{2}\lambda_{a\dot{a}}\lambda_{b\dot{b}} - D_{ab}\Phi_{\dot{a}\dot{b}}\right),
\end{equation}
we compute that for arbitrary Killing spinors $\xi, \tilde{\xi}$ in $\mathfrak{su}(2|1)_\ell\oplus \mathfrak{su}(2|1)_r$,
\begin{align}
\delta_\xi\delta_{\tilde{\xi}}V|_\text{bos} &= h^{ab}\bar{h}^{\dot{a}\dot{b}}\bigg(\frac{1}{2}\tilde{\xi}_{a\dot{a}}\xi_{b\dot{b}}F_{\mu\nu}F^{\mu\nu} - \frac{1}{2}\tilde{\xi}_{a\dot{a}}\xi_{b\dot{b}}D^{cd}D_{cd} - \frac{1}{2}\tilde{\xi}_{a\dot{a}}\xi_{b\dot{b}}\partial^\mu\Phi^{\dot{c}\dot{d}}\partial_\mu\Phi_{\dot{c}\dot{d}} \nonumber \\
&\hspace{0.5 cm} - \tilde{\xi}'_{a\dot{a}}\xi'_{b\dot{b}}\Phi^{\dot{c}\dot{d}}\Phi_{\dot{c}\dot{d}} - \frac{3}{4r^2}\tilde{\xi}_{a\dot{a}}\xi_{b\dot{b}}\Phi^{\dot{c}\dot{d}}\Phi_{\dot{c}\dot{d}} - 2i(\tilde{\xi}_{b\dot{a}}\xi_{c\dot{c}}' + \tilde{\xi}'_{b\dot{a}}\xi_{c\dot{c}})D_a{}^c\Phi_{\dot{b}}{}^{\dot{c}}\bigg) + \nabla_\mu\Sigma^\mu
\end{align}
where
\begin{align}
\Sigma^\mu &= h^{ab}\bar{h}^{\dot{a}\dot{b}}\bigg(\tilde{\xi}_a{}^{\dot{c}}\xi_b{}^{\dot{d}}\Phi_{\dot{a}\dot{b}}\partial^\mu\Phi_{\dot{c}\dot{d}} + i\epsilon^{\mu\nu\rho}\tilde{\xi}_a{}^{\dot{c}}\gamma_\rho\xi_b{}^{\dot{d}}\Phi_{\dot{a}\dot{c}}\partial_\nu\Phi_{\dot{b}\dot{d}} + 2\tilde{\xi}_a{}^{\dot{c}}\gamma^\mu\xi'_b{}^{\dot{d}}\Phi_{\dot{a}\dot{c}}\Phi_{\dot{b}\dot{d}} \nonumber \\
&\hspace{1.5 cm} + \frac{1}{2}\epsilon^{\mu\nu\rho}(\tilde{\xi}_{a\dot{a}}\xi_b{}^{\dot{c}} + \tilde{\xi}_b{}^{\dot{c}}\xi_{a\dot{a}})F_{\nu\rho}\Phi_{\dot{c}\dot{b}} - i\tilde{\xi}_a{}^{\dot{c}}\gamma_\nu\xi_{b\dot{c}}F^{\mu\nu}\Phi_{\dot{a}\dot{b}} - i\tilde{\xi}_a{}^{\dot{c}}\gamma^\mu\xi_{c\dot{c}}D_b{}^c\Phi_{\dot{a}\dot{b}}\bigg).
\end{align}
Specializing to $\xi = \xi_\beta^C$, $\tilde{\xi} = \xi_{-\beta}^C$, $h_a{}^b = -(\sigma^2)_a{}^b$, $\bar{h}^{\dot{a}}{}_{\dot{b}} = -(\sigma^3)^{\dot{a}}{}_{\dot{b}}$, we have
\begin{equation}
h^{ab}\bar{h}^{\dot{a}\dot{b}}\tilde{\xi}_{a\dot{a}}\xi_{b\dot{b}} = 8i\beta, \quad h^{ab}\bar{h}^{\dot{a}\dot{b}}\tilde{\xi}'_{a\dot{a}}\xi'_{b\dot{b}} = -\frac{2i\beta}{r^2}, \quad h^{ab}\bar{h}^{\dot{a}\dot{b}}(\tilde{\xi}_{b\dot{a}}\xi_{c\dot{c}}' + \tilde{\xi}'_{b\dot{a}}\xi_{c\dot{c}}) = -\frac{2\beta}{r}h_c{}^a\bar{h}_{\dot{c}}{}^{\dot{b}}
\end{equation}
(the explicit form of $\xi_\beta^C$ is given in \eqref{xiCbeta}), and substituting these results gives
\begin{equation}
\delta_\xi\delta_{\tilde{\xi}}V = \delta_\xi\delta_{\tilde{\xi}}V|_\text{bos} + \delta_\xi\delta_{\tilde{\xi}}V|_\text{fer} = 4i\beta\mathcal{L}_\text{Maxwell} + \nabla_\mu(\Sigma^\mu + \Sigma_f^\mu)
\end{equation}
where
\begin{equation}
\begin{aligned}
\mathcal{L}_\text{Maxwell} &= F^{\mu\nu}F_{\mu\nu} - \partial^\mu\Phi^{\dot{c}\dot{d}}\partial_\mu\Phi_{\dot{c}\dot{d}} + i\lambda^{a\dot{a}}\slashed{\nabla}\lambda_{a\dot{a}} - D^{cd}D_{cd} \\
&\hspace{1 cm} - \frac{1}{2r}h^{ab}\bar{h}^{\dot{a}\dot{b}}\lambda_{a\dot{a}}\lambda_{b\dot{b}} + \frac{1}{r}(h_a{}^b D_b{}^a)(\bar{h}^{\dot{a}}{}_{\dot{b}}\Phi^{\dot{b}}{}_{\dot{a}}) - \frac{1}{r^2}\Phi^{\dot{c}\dot{d}}\Phi_{\dot{c}\dot{d}}
\end{aligned}
\end{equation}
and $\Sigma_f^\mu$ are fermionic terms that are irrelevant for our purposes.  The monopole counterterm $\Sigma^\mu$ can likewise be simplified using the explicit forms of $\xi, \tilde{\xi}, h_a{}^b, \bar{h}^{\dot{a}}{}_{\dot{b}}$.  In stereographic coordinates, we obtain
\begin{align}
\Sigma^\mu &= 8i\beta(\partial^\mu(\Phi_{\dot{1}\dot{1}} + \Phi_{\dot{2}\dot{2}} - 2\Phi_{\dot{1}\dot{2}}) - e^\Omega(2 + ix_3/r)\partial^\mu\Phi_{\dot{1}\dot{1}} - e^\Omega(2 - ix_3/r)\partial^\mu\Phi_{\dot{2}\dot{2}})\Phi_{\dot{1}\dot{2}} \nonumber \\
&\phantom{==} + 4i\beta\epsilon^{\mu\nu\rho}F_{\nu\rho}(\Phi_{\dot{1}\dot{1}} - \Phi_{\dot{2}\dot{2}} - e^\Omega(2 + ix_3/r)\Phi_{\dot{1}\dot{1}} + e^\Omega(2 - ix_3/r)\Phi_{\dot{2}\dot{2}}) \nonumber \\
&\phantom{==} + 4\epsilon^{\mu\nu\rho}U_\nu(\Phi_{\dot{1}\dot{1}}\partial_\rho\Phi_{\dot{2}\dot{2}} - \Phi_{\dot{2}\dot{2}}\partial_\rho\Phi_{\dot{1}\dot{1}}) + 16U_\nu F^{\mu\nu}\Phi_{\dot{1}\dot{2}} + V^\mu
\end{align}
where $U_\mu = \beta e^{2\Omega}(\delta^1_\mu x_2 - \delta^2_\mu x_1)/r$ and
\begin{align*}
V^{i = 1, 2} &= \frac{4i\beta e^{2\Omega}x_i}{r^2}\left[\left(2 + \frac{ix_3}{r}\right)\Phi_{\dot{1}\dot{1}} + \left(2 - \frac{ix_3}{r}\right)\Phi_{\dot{2}\dot{2}}\right]\Phi_{\dot{1}\dot{2}}, \\
V^3 &= \frac{2\beta e^{2\Omega}}{r}\left[\frac{x_1^2 + x_2^2}{r^2}(\Phi_{\dot{1}\dot{1}} - \Phi_{\dot{2}\dot{2}}) + \left(2 + \frac{ix_3}{r}\right)^2\Phi_{\dot{1}\dot{1}} - \left(2 - \frac{ix_3}{r}\right)^2\Phi_{\dot{2}\dot{2}}\right]\Phi_{\dot{1}\dot{2}}.
\end{align*}
Strictly speaking, our $\Sigma = \Sigma_\mu dx^\mu$ is actually the Hodge dual of the $\Sigma$ defined in \eqref{moncounterterm}.

\section{General BPS Monopole Backgrounds}
\label{moremonopoles}

\subsection{Singular Solutions to BPS Equations}

In this section, we construct the singular solutions to \eqref{PHeq1}--\eqref{BogoEq} that describe insertions of multiple twisted-translated monopole operators anywhere on the $R=1$ great circle of $S^3$. Consider $n$ such operator insertions at angles $-\pi<\varphi_1\leq\varphi_2\leq\cdots\leq\varphi_n\leq \pi$. Let the monopole at $\varphi=\varphi_k$ have charge $b_k\in\Gamma_m$ ($k=1,\ldots,n$). Because $S^3$ is compact, the charges must satisfy $\sum_{k=1}^n b_k=0$. Our task is to solve \eqref{PHeq1}--\eqref{BogoEq} on $S^3$ with punctures at $(R,\varphi)=(1,\varphi_k)$ such that the fields near the $k^\textrm{th}$ puncture approach a charge-$b_k$ monopole singularity, as prescribed in \eqref{MonSing}.

To define the gauge bundle on the punctured $S^3$, we cover it with patches $D^{(i)}$ given by
\begin{align}
D^{(i)} = \{0\leq R< 1\}\cup\{ R = 1 \,, \varphi_i<\varphi<\varphi_{i+1}\} \ec \label{patches}
\end{align}
where it is understood that $D^{(n)} \supset\{ R = 1 \,, \varphi_n<\varphi\leq\pi\}\cup\{R=1\,,-\pi< \varphi<\varphi_1\}$. On each patch $D^{(i)}$, the gauge connection $A^{(i)}$ is a well-defined one-form, and $A^{(i)}-A^{(j)}$ is a valid gauge transformation. In abelian gauge theories, the other fields in the vector multiplet are neutral, so they must be globally defined functions on the punctured $S^3$.

An important consequence of \eqref{PHeq1}--\eqref{BogoEq} is that the gauge field is related to $\Phi_i$. Indeed, by combining \eqref{BogoEq} with \eqref{PHeq1} and \eqref{PHeq2}, it is straightforward to see that\footnote{First, the equation $F_{R\varphi}=0$ implies that we can set $A_R= A_{\varphi}=0$, since there are no nontrivial flat connections on the disk with punctures at its boundary. The other equations for $A_{\tau}$  can be written in each patch as $\partial_{\varphi}(A^{(i)}_{\tau} - r \Phi_i)=\partial_{R}(A^{(i)}_{\tau} - r \Phi_i)=0$, which can be integrated to \eqref{AiPhi}.}
\begin{align}
A^{(i)} = (r\Phi_i + c^{(i)})d\tau \ec \label{AiPhi} 
\end{align}
where the $c^{(i)}$ are constants. For $A$ to be well-defined, $c^{(i)}-c^{(j)}$ must be integrally quantized for all $1\leq i,j\leq n$, and moreover, 
\begin{align}
c^{(i)} = -r\Phi_i\biggr|_{R=1} \label{ciDef}
\end{align}
because the $\tau$-circle shrinks at the boundary of the disk.  We conclude that $\Phi_i$ must be a piecewise constant function on the $R=1$ circle.

Let us now show that $\Phi_i$ is uniquely determined by its value at $R=1$. First, combining \eqref{PHeq1} and \eqref{PHeq2}, we find that $\Phi_i$ must satisfy the second-order differential equation
\begin{align}
(R(1-R^2)\partial_R(R\partial_R) + \partial_{\varphi}^2) \Phi_i(R,\varphi) = 0 \ed \label{PHiEq}
\end{align}
The general solution to \eqref{PHiEq}, which is smooth in the interior of the disk, can be found using separation of variables: 
\begin{align}
\Phi_i = \sum_{n=-\infty}^{\infty} a_n \frac{\Gamma(|n|/2+1)^2}{\Gamma(|n|+1)} e^{in\varphi} R^{|n|} {}_2F_1\left(\frac{|n|}{2},\frac{|n|}{2},|n|+1,R^2\right) \ed \label{PHiExp}
\end{align}
In particular, at $R=1$, we find
\begin{align}
\Phi_i(R=1,\varphi) = \sum_{n=-\infty}^{\infty} a_n e^{in\varphi} \ec
\end{align}
from which the coefficients $a_n$ are uniquely determined. As argued around \eqref{PHirLim}, the field $\Phi_{i}$ must vanish at $R=0$, which implies that $a_0 = 0$. We show below that $\Phi_i(R=1,\varphi)$ is completely fixed by this requirement and the boundary conditions at the punctures.

Once $\Phi_i$ is fixed, $\Phi_r$ can be obtained simply by integrating the  BPS equations \eqref{PHeq1} and \eqref{PHeq2}. In particular, integrating \eqref{PHiExp} term-by-term, we find that
\begin{align}
\Phi_r = i\sum_{n\neq 0} \sgn(n)a_n e^{i n \varphi} R^{|n|}{}_2F_1\left(\frac{|n|}{2},\frac{|n|+2}{2},|n|+1,R^2\right) \ec \label{PHrExp}
\end{align}
where the constant mode has been set to zero, as before.  Note that \eqref{PHrExp} is an expansion of $\Phi_r$ in $\tau$-independent solutions of the Laplace equation on $S^3$. That $\nabla^2\Phi_r=0$ is satisfied follows directly from the Bogomolny equation \eqref{BogoEq}, and also by combining \eqref{PHeq1} and \eqref{PHeq2} into a second-order equation for $\Phi_r$. The linear equations \eqref{PHeq1}, \eqref{PHeq2} provide the relation between the mode expansions of $\Phi_i$ and $\Phi_r$, as shown explicitly in \eqref{PHiExp} and \eqref{PHrExp}.

To summarize, the solutions of the BPS equations \eqref{PHeq1}--\eqref{BogoEq} on the punctured $S^3$ are uniquely determined by $\Phi_i(R=1,\varphi)$, which, according to \eqref{ciDef}, must be a piecewise constant periodic function of $\varphi$. Furthermore, $\Phi_i(R=1,\varphi)$ must not have a zero mode, i.e., $\int_{-\pi}^{\pi} d\varphi\, \Phi_i(R=1,\varphi)=0$. Let us finally spell out the connection between the above construction and monopole operators. In Appendix \ref{matchingsingularities}, we show that the singular monopole boundary conditions fix $\Phi_i(R=1,\varphi)$ up to an overall constant:
\begin{align}
\Phi_i(R=1,\varphi) = -\frac{1}{2r}\sum_{k=1}^nb_k\sgn\left(\cos\frac{\varphi}{2}\sin\frac{\varphi-\varphi_k}{2}\right)  +\mathrm{constant}\ed \label{PHiBdyTmp}
\end{align}
The undetermined constant in \eqref{PHiBdyTmp} is fixed by imposing that $\Phi_i(R=1,\varphi)$ have no zero mode, resulting in the final expression\footnote{If we restrict to the range $\varphi\in(-\pi,\pi]$, then $\operatorname{sgn}(\cos\frac{\varphi}{2}\sin\frac{\varphi - \varphi_k}{2})$ can be replaced by $\operatorname{sgn}(\varphi - \varphi_k)$.}
\begin{align}
\Phi_i(R=1,\varphi) = -\frac{1}{2r}\sum_{k=1}^nb_k\left[\sgn\left(\cos\frac{\varphi}{2}\sin\frac{\varphi-\varphi_k}{2}\right) + \frac{\varphi_k}{\pi}\right] \ed \label{PHiBdyAP}
\end{align}
This concludes our description of the solution for the background corresponding to $n$ twisted-translated monopole operators.

\subsection{Relation to Monopole Singularities}
\label{matchingsingularities}

Let us now derive \eqref{PHiBdyTmp} by showing how the piecewise constant function $\Phi_i(R=1,\varphi)$ is determined by the monopole singularities \eqref{MonSing}. In stereographic coordinates $x_{\mu}$, the insertions lie along the line $x_1=x_2=0$, and the monopole background is given by
\begin{align}
\ast F \sim \left(1+\frac{x^2}{4r^2}\right)\sum_{k=1}^n b_k \frac{x_{\mu}-y_{\mu}^k}{|\vec{x}-\vec{y}^k|^3} \ec \label{Fmon}
\end{align}
where $\vec{y}^k = (0,0,2r\tan\frac{\varphi_k}{2})$. The $\sim$ sign in \eqref{Fmon} implies equality up to non-singular terms. We will use this notation throughout this section.

Because BPS configurations are functions on the $(R,\varphi)$ disk, it will be more convenient to use the $(R,\varphi,\tau)$ coordinates. In these coordinates, the insertions are located at angles $\varphi_k$ on the $R=1$ boundary of the disk, and \eqref{Fmon} takes a more complicated form:
\begin{align}
\ast F &\sim -\frac{1}{r}\sum_{k=1}^n b_k\left(\frac{\cos\varphi+\tan\frac{\varphi_k}{2}\sin\varphi}{(1-R^2 + \tan^2\frac{\varphi_k}{2}\left(R(\sin\varphi-\cos\varphi) -1\right)^2)^{\frac{3}{2}}}\, dR\right. \notag\\
&\hspace{4 cm} + \left.\frac{\cos^2\frac{\varphi_k}{2}\left(R\sin\frac{\varphi_k}{2} - \sin\left(\varphi-\frac{\varphi_k}{2}\right)\right)}{\left((1+R\cos\varphi)(1-R\cos(\varphi-\varphi_k))\right)^{\frac{3}{2}}}R\, d\varphi\right)\ed \label{FRmon} 
\end{align}
The gauge connection that reproduces the magnetic field \eqref{FRmon} is given by
\begin{align}
A^{(i)} \sim -\frac{1}{2}\left(\sum_{k=1}^n b_k\frac{R\sin\left(\varphi-\frac{\varphi_k}{2}\right) - \sin\frac{\varphi_k}{2}}{\sqrt{(1+R\cos\varphi)(1-R\cos(\varphi-\varphi_k))}}- \sum_{k=1}^i b_k + \sum_{k=i+1}^n b_k \right)d\tau \ec \label{AiR}
\end{align}
where $A^{(i)}$ is defined in the patch $D^{(i)}$ defined in \eqref{patches}. The constant terms in \eqref{AiR} are chosen such that $A^{(i)}$ vanishes at $R=1$, making it a well-defined one-form on $D^{(i)}$. Moreover, in $D^{(i)}\cap D^{(j)}$, we have that $A^{(i)} - A^{(j)}$ is a well-defined gauge transformation.

Up to regular terms, the scalars $\Phi_{r,i}$ are determined by \eqref{PHeq2} and \eqref{BogoEq} to be
\begin{align}
\Phi_r &\sim \frac{1}{4r}\sum_{k=1}^n \frac{b_k}{\cos\left(\varphi-\frac{\varphi_k}{2}\right)}\left[\frac{\cos\varphi-(1+2R\cos\varphi)\cos(\varphi-\varphi_k)}{\sqrt{(1+R\cos\varphi)(1-R\cos(\varphi-\varphi_k))}} + \cos(\varphi-\varphi_k) - \cos\varphi\right] \ec \label{PHrSing}\\
\Phi_i &\sim -\frac{1}{2r}\sum_{k=1}^n b_k\left[\frac{R\sin\left(\varphi-\frac{\varphi_k}{2}\right)-\sin\frac{\varphi_k}{2}}{\sqrt{(1+R\cos\varphi)(1-R\cos(\varphi-\varphi_k))}} + \sin\frac{\varphi_k}{2}\right] \ed \label{PHiSing}
\end{align}
At $R=1$, the singular part of $\Phi_i$, given in \eqref{PHiSing}, becomes a piecewise constant function:
\begin{align}
\lim_{R\to 1}\Phi_i(R,\varphi)\biggr|_{\mathrm{singular}} = -\frac{1}{2r}\sum_{k=1}^nb_k\left[\sgn\left(\sin\frac{\varphi-\varphi_k}{2}\right) + \sin\frac{\varphi_k}{2}\right] \ed
\end{align}
Any contribution to $\Phi_i$ at $R=1$ from the terms suppressed in \eqref{PHiSing} must be a regular periodic function $f(\varphi)$. However, as argued around \eqref{ciDef}, on the BPS locus, $\Phi_i$ must be piecewise constant at $R=1$. We conclude that regular terms can only contribute $f(\varphi)=\text{constant}$, so that the expression for $\Phi_i(R=1,\varphi)$ is as in \eqref{PHiBdyTmp}.

\section{Hypermultiplet One-Loop Determinant on $S^3$}
\label{1loopdetails}

In this section, we calculate the hypermultiplet determinant \eqref{Zsig} on $S^3$ in the two-monopole background \eqref{PHspherical}, \eqref{Aspherical}. For simplicity, we consider a $U(1)$ gauge theory with a single hypermultiplet of unit charge. Moreover, to simplify notation slightly, we define $q = b/2$ and set $r=1$ throughout this section.

\subsection{Bosonic Spectrum}

The eigenvalue problem for the bosonic part $\int d^3 x\, \sqrt{g}\tilde{q}^a(\mathcal{D}_B)_a{}^b q_b$ of the action \eqref{Shyper} is
\begin{align}
\cD_B\cdot\vec{f}\equiv\left[\delta_a{}^b\left(-D^{\mu}D_{\mu} + \frac{3}{4} - \frac{1}{2}\Phi^{\dot{a}\dot b}\Phi_{\dot{a}\dot b}\right) + i D_a{}^b\right]f_b = \lambda_B f_a \ed
\end{align}
Diagonalizing the 2-by-2 $R$-symmetry matrix $\cD_B$ and using the solution to the BPS equations leads to the equation
\begin{align}
\left[-D^2 + \frac{3}{4} + \sigma^2 - \Phi_{\dot{1}\dot 1}\Phi_{\dot{2}\dot 2} \pm i(\sigma +i\mathrm{Re}D_{11}) - \lambda_B^{\pm} \right] f_{\pm} = 0 \ed
\end{align}
For the specific configuration of (anti-)monopole at $\eta=0$ ($\eta=\pi$), $\mathrm{Re}D_{11}=0$ and we can write
\begin{align}
\left[-D^2 + \frac{q^2}{\sin^2\eta}-\tilde{\lambda}_B^{\pm}\right]f_{\pm} = \left[-\partial_{\eta}^2 - 2\cot\eta\partial_{\eta}  - \frac{1}{\sin^2\eta}D_{S^2\ec q}^2 + \frac{q^2}{\sin^2\eta} - \tilde{\lambda}^{\pm}_B\right]f_{\pm}=0\ec
\end{align}
where $D^2_{S^2\ec q}$ is the gauge-covariant Laplacian on $S^2$ with metric $ds^2 = d\psi^2+\sin^2\psi d\tau^2$ in the charge-$q$ monopole background and we have defined
\begin{align}
\tilde{\lambda}_B^{\pm} \equiv \lambda_B^{\pm} - \left(\frac{3}{4}+\sigma^2\pm i\sigma\right)\ed
\end{align}
The eigenfunctions can be expanded in monopole spherical harmonics, 
\begin{align}
f_{\pm} = h_{\pm}(\eta)Y_{q ; \ell m}(\psi,\tau)\ec
\label{fpm}
\end{align}
which satisfy
\begin{align}
D_{S^2\ec q}^2 Y_{q; \ell m} = -\left(\ell(\ell+1)-q^2\right)Y_{q; \ell m}
\end{align}
with $\ell= |q|, |q|+1, \ldots$ and $m=-\ell,-\ell+1,\ldots, \ell$.  We are left with an ordinary differential equation
\begin{align}
\left[-\partial_{\eta}^2 - 2\cot\eta\partial_{\eta} + \frac{\ell(\ell+1)}{\sin^2\eta} - \tilde{\lambda}^{\pm}_B\right]h_{\pm}=0 \ec
\end{align}
whose solutions are given by
\begin{equation}
h_\pm(\eta) = \frac{1}{(1 - x^2)^{1/4}}\left[c_1 P_{\sqrt{\tilde{\lambda}_B^\pm + 1} - 1/2}^{\ell + 1/2}(x) + c_2 Q_{\sqrt{\tilde{\lambda}_B^\pm + 1} - 1/2}^{\ell + 1/2}(x)\right]
\label{hpm}
\end{equation}
where $x = \cos\eta$ and $P, Q$ are associated Legendre functions.  The solutions are singular at $x=\pm 1$ unless\footnote{$P_L^m(x)$ is regular on $[-1, 1]$ only if $L, m$ are integers with $0\leq m\leq L$, and a similar statement holds for $Q_L^m(x)$ when $L, m$ are half-integers.  If $q$ is an integer, then $\ell$ is an integer and we keep the $Q$ solution; otherwise, we keep the $P$ solution.}
\begin{align}
\sqrt{\tilde{\lambda}^{\pm}_B+1} = \ell+1, \ell+2, \ldots  \ed
\end{align}
Hence the bosonic spectrum on $S^3$ is
\begin{equation}
\lambda_B^\pm = (\ell + n)^2 - \frac{1}{4}\pm i\sigma + \sigma^2, \quad n = 1, 2, \ldots, \quad \ell = |q|, |q| + 1, \ldots,
\end{equation}
with degeneracy $2\ell + 1$ for each sign.  Equivalently, set $N + |q| + 1 = \ell + n$; then
\begin{equation}
\lambda_B^\pm = (N + |q|)(N + |q| + 2) + \frac{3}{4}\pm i\sigma + \sigma^2, \quad N = 0, 1, \ldots
\end{equation}
with $(\ell, n) = (|q|, N + 1), (|q| + 1, N), \ldots, (N + |q|, 1)$ and therefore degeneracy
\begin{equation}
\sum_{\ell=|q|}^{N + |q|} (2\ell + 1) = (N + 1)^2 + 2|q|(N + 1)
\end{equation}
for each sign, as in \eqref{DbSpec}.

\subsection{Fermionic Spectrum}

On $S^3$, we work in the frame
\begin{align}
e^1 = d\eta \ecq e^2 = \sin\eta\, d\psi \ecq e^3 =\sin\eta\sin\psi\, d\tau \ec
\end{align}
in which the nonvanishing components of the spin connection are given by
\begin{align}
\omega_{\psi}^{21}=- \omega_{\psi}^{12}=\cos\eta \ecq \omega_{\tau}^{31} = -\omega_{\tau}^{13} = \cos\eta\sin\psi \ecq \omega_{\tau}^{32} = -\omega_{\tau}^{23} = \cos\psi \ed
\end{align}
On $S^2$, we work in the frame
\begin{align}
\tilde{e}^1=d\psi \ecq \tilde{e}^2=\sin\psi\, d\tau \ec
\end{align}
and choose the associated 2D gamma matrices to be $\tilde{\gamma}_1=\sigma_1$ and $\tilde{\gamma}_2=\sigma_2$.\footnote{We have chosen these conventions in light of \eqref{boundaryconditions}, to make the fermionic analysis on $HS^3$ more natural.} The nonvan\-ishing components of the spin connection on $S^2$ are then
\begin{align}
\tilde{\omega}_{\tau}^{21}=-\tilde{\omega}_{\tau}^{12}=\cos\psi \ed
\end{align}
Using the above conventions, we can decompose the $S^3$ and $S^2$ covariant Dirac operators in the monopole background as
\begin{equation}
\slashed{D}_{S^3, q} = \sigma_3 d_\eta + \frac{1}{\sin\eta}\slashed{D}_{S^2, q}, \quad \slashed{D}_{S^2, q} = \sigma_1\left(D_\psi + \frac{1}{2}\cot\psi\right) + \sigma_2\frac{1}{\sin\psi}D_\tau,
\end{equation}
where $d_\eta\equiv \partial_\eta + \cot\eta$.  The latter is diagonalized by monopole spinor harmonics, which are two-component spinors that satisfy
\begin{equation}
i\slashed{D}_{S^2, q}Y_{q, \ell m}^\pm = \Delta_{q, \ell}^\pm Y_{q, \ell m}^\pm, \quad \Delta_{q, \ell}^\pm = \pm\sqrt{(\ell + 1/2)^2 - q^2}
\end{equation}
for $\ell = |q| + 1/2, |q| + 3/2, \ldots$ and $m = -\ell, -\ell + 1, \ldots, \ell$. For $q\neq 0$, there also exist zero modes
\begin{equation}
i\slashed{D}_{S^2, q}Y_{q, \ell m}^0 = 0
\end{equation}
with $\ell = |q| - 1/2$. We will make use of the properties
\begin{align}
\sigma_3 Y^{\pm}_{q, \ell m} = Y^{\mp}_{q, \ell m} \ecq \sigma_3 Y^0_{q,\ell m} = \sgn(q)Y^0_{q, \ell m} \ed
\end{align}

\subsubsection{Eigenvalue Problem}

The eigenvalue problem for the fermionic part $-\int d^3 x\, \sqrt{g}\tilde{\psi}^{\dot{a}}(\mathcal{D}_F)_{\dot{a}}{}^{\dot{b}}\psi_{\dot{b}}$ of the action \eqref{Shyper} is
\begin{align}
-i\slashed{D}\psi_{\dot a} -i\Phi_{\dot a}{}^{\dot b}\psi_{\dot b} = \lambda_F\psi_{\dot a} \ed
\end{align}
Substituting the background $\Phi_{\dot{1}}{}^{\dot{1}} = -\Phi_{\dot{2}}{}^{\dot{2}} = \sigma$ and $\Phi_{\dot{1}}{}^{\dot{2}} = -\Phi_{\dot{2}}{}^{\dot{1}} = -iq/\sin\eta$ for the scalar fields, the operator that we wish to diagonalize can be written as
\begin{align}
\cD_F = \begin{pmatrix}
i\slashed D + i \sigma  & \frac{q }{\sin\eta}\mathbb{1}_2 \\ -\frac{q }{\sin\eta}\mathbb{1}_2 & i\slashed{D} - i\sigma 
\end{pmatrix} \ed \label{DF}
\end{align}
Let us start by making some manipulations to eliminate $\sigma$ from the problem. Since we are only interested in the determinant of $\cD_F$, we can instead solve the eigenvalue problem for
\begin{align}
\tilde{\cD}_F\equiv \cD_F(\sigma_3\otimes \mathbb{1}_2) = \begin{pmatrix}
i\slashed D + i \sigma  & -\frac{q}{\sin\eta}\mathbb{1}_2 \\ -\frac{q}{\sin\eta}\mathbb{1}_2 & -i\slashed{D} + i\sigma \label{tDF}
\end{pmatrix} \ed
\end{align}
We can now absorb $\sigma$ into the eigenvalues, i.e., instead of $\tilde{\cD}_F$, we will diagonalize
\begin{align}
\hat{\cD}_F\equiv \begin{pmatrix}
i\slashed D  & -\frac{q}{\sin\eta}\mathbb{1}_2 \\ -\frac{q}{\sin\eta}\mathbb{1}_2 & -i\slashed{D} 
\end{pmatrix} \ec \label{hatDF}
\end{align}
whose eigenvalues are related to those of $\tilde{\cD}_F$ by $\tilde{\lambda}=\hat{\lambda}+i\sigma$.  Using the spherical symmetry of the background, the eigenspinors $\Psi=(\psi_1, \psi_2)^T$ of $\hat{\cD}_F$ can be decomposed into monopole spinor harmonics.  To do so, we consider separately the cases $\ell\geq |q|+\frac{1}{2}$ and $\ell=|q|-\frac{1}{2}$.

\subsubsection{$\ell\geq |q|+\frac{1}{2}$}

In this case, we write
\begin{align}
\psi_{1} = f_{1+}(\eta) Y^+_{q, \ell m} + f_{1-}(\eta) Y^-_{q, \ell m} \ec\\
\psi_{2} = f_{2+}(\eta) Y^+_{q, \ell m} + f_{2-}(\eta) Y^-_{q, \ell m} \ed
\end{align}
In terms of the above decomposition, we have
\begin{align}
\hat{\cD}_F\Psi = 
\begin{pmatrix}
i\sigma_3 d_{\eta} + \frac{1}{\sin\eta}i\slashed{D}_{S^2,q} & -\frac{q}{\sin\eta}\mathbb{1}_2 \\ -\frac{q}{\sin\eta}\mathbb{1}_2 & -i\sigma_3 d_{\eta} - \frac{1}{\sin\eta}i\slashed{D}_{S^2,q} 
\end{pmatrix}\begin{pmatrix}
f_{1+}(\eta) Y^+_{q, \ell m} + f_{1-}(\eta) Y^-_{q, \ell m} \\ f_{2+}(\eta) Y^+_{q, \ell m} + f_{2-}(\eta) Y^-_{q, \ell m}
\end{pmatrix} = \hat{\lambda}\Psi\ed
\end{align}
Using the property $\sigma_3 Y_{q, \ell m}^\pm = Y_{q, \ell m}^\mp$ and linear independence of $Y^\pm$, this is equivalent to
\begin{align}
\cM\cdot\vec{f}\equiv
\begin{pmatrix}
i\sigma_1 d_{\eta} + \frac{\Delta_{q,\ell}^+}{\sin\eta}\sigma_3 & -\frac{q}{\sin\eta}\mathbb{1}_2 \\ -\frac{q}{\sin\eta}\mathbb{1}_2 & -i\sigma_1 d_{\eta} - \frac{\Delta_{q,\ell}^+}{\sin\eta}\sigma_3
\end{pmatrix}\begin{pmatrix}
f_{1+}(\eta) \\ f_{1-}(\eta) \\ f_{2+}(\eta) \\ f_{2-}(\eta) 
\end{pmatrix} = \hat{\lambda}\begin{pmatrix}
f_{1+}(\eta) \\ f_{1-}(\eta) \\ f_{2+}(\eta) \\ f_{2-}(\eta) 
\end{pmatrix} \ed
\end{align}
This system of four coupled equations can be decoupled into a pair of two coupled equations by making a unitary transformation: in terms of
\begin{equation}
\cM_U = U^{-1}\cM U, \quad \vec{f}_U = U^{-1}\vec{f}, \quad U = e^{i\theta(\sigma_2\otimes\sigma_3)}, \quad \tan(2\theta) = \frac{q}{\Delta^+_{q,\ell}},
\label{theta}
\end{equation}
it becomes $\cM_U\cdot \vec{f}_U = \hat{\lambda}\vec{f}_U$ where
\begin{align}
\cM_U=\begin{pmatrix}
\tilde{\cM}_U & 0 \\ 0 & -\tilde{\cM}_U
\end{pmatrix}, \quad \tilde{\cM}_U\equiv i\sigma_1(\partial_{\eta}+\cot\eta) + \frac{\ell+\frac{1}{2}}{\sin\eta}\sigma_3. \label{MtU}
\end{align}
Now let us make a further rotation on \eqref{MtU} and consider the eigenvalue problem 
\begin{align} \label{ranproblem}
\left(i\sigma_3(\partial_{\eta}+\cot\eta) - \frac{\ell+\frac{1}{2}}{\sin\eta}\sigma_1\right)\cdot\vec{h} = \begin{pmatrix}
i(\partial_{\eta}+\cot\eta) & -\frac{\ell+\frac{1}{2}}{\sin\eta} \\ -\frac{\ell+\frac{1}{2}}{\sin\eta} & -i(\partial_{\eta}+\cot\eta)
\end{pmatrix}\begin{pmatrix}
h_1 \\h_2
\end{pmatrix} = \hat{\lambda} \begin{pmatrix}
h_1\\ h_2
\end{pmatrix} \ed
\end{align}
The solutions are given by
\begin{align}
h_1 &= c_1(x^2)^{\frac{1 - \hat{\lambda}}{2}}(1 - x^2)^{\ell - 1/2}{}_2 F_1\left(\ell + \frac{1}{2}, 1 + \ell - \hat{\lambda}, \frac{1}{2} - \hat{\lambda}, x^2\right) \nonumber \\
&\phantom{==} + c_2(x^2)^{\frac{2 + \hat{\lambda}}{2}}(1 - x^2)^{\ell - 1/2}{}_2 F_1\left(\ell + \frac{3}{2}, 1 + \ell + \hat{\lambda}, \frac{3}{2} + \hat{\lambda}, x^2\right), \label{h1} \\
h_2 &= ic_1\left(\frac{2\ell + 1}{2\hat{\lambda} - 1}\right)(x^2)^{\frac{2 - \hat{\lambda}}{2}}(1 - x^2)^{\ell - 1/2}{}_2 F_1\left(\ell + \frac{3}{2}, 1 + \ell - \hat{\lambda}, \frac{3}{2} - \hat{\lambda}, x^2\right) \nonumber \\
&\phantom{==} - ic_2\left(\frac{2\smash{\hat{\lambda}} + 1}{2\ell + 1}\right)(x^2)^{\frac{1 + \hat{\lambda}}{2}}(1 - x^2)^{\ell - 1/2}{}_2 F_1\left(\ell + \frac{1}{2}, 1 + \ell + \hat{\lambda}, \frac{1}{2} + \hat{\lambda}, x^2\right), \label{h2}
\end{align}
where $x=e^{i\eta}$.  The hypergeometric function ${}_2F_1(a,b,c;z)$ is regular on the unit circle if $c=0,1,\ldots$ and $\mathrm{Re}(c-a-b)>0$, or if either $a$ or $b$ are non-positive integers (in which case the hypergeometric series terminates). The first condition is always violated in the above solution, so we conclude that there exists a regular solution if
\begin{align}
\hat{\lambda} = \pm( n+\ell+1) \ecq n=0,1,\ldots \ecq \ell=|q|+\frac{1}{2}, |q|+\frac{3}{2},\ldots \ed   \label{hatlambda}
\end{align}
The degeneracy of each eigenvalue above, considered as an eigenvalue of $\cM_U$ (rather than of $\tilde{\cM}_U$) in \eqref{MtU} and hence of $\hat{\cD}_F$, is $2(2\ell+1)$.

\subsubsection{$\ell=|q|-\frac{1}{2}$}

In this case, we work directly with \eqref{hatDF} and expand $\Psi=(\psi_1,\psi_2)^T$ in zero modes as
\begin{align}
\psi_1 = h_1(\eta) Y^0_{q,\ell m} \ecq \psi_2 = h_2(\eta) Y^0_{q,\ell m} \ed
\end{align}
Using the property $\sigma_3 Y_{q, \ell m}^0 = \sgn(q)Y_{q, \ell m}^0$, the eigenvalue problem $\hat{\cD}_F\Psi = \hat{\lambda}\Psi$ becomes
\begin{equation}
\begin{pmatrix}
i(\partial_\eta + \cot\eta) & -\frac{|q|}{\sin\eta} \\ -\frac{|q|}{\sin\eta} & -i(\partial_\eta + \cot\eta)
\end{pmatrix}\begin{pmatrix}
h_1 \\ h_2
\end{pmatrix} = \sgn(q)\hat{\lambda}\begin{pmatrix}
h_1 \\ h_2
\end{pmatrix}\ec
\end{equation}
where it is understood that $\ell=|q|-\frac{1}{2}$.  This is precisely \eqref{ranproblem}, with $\hat{\lambda}\to \sgn(q)\hat{\lambda}$.  The corresponding solutions \eqref{h1} and \eqref{h2} are regular when $1 + \ell\pm \sgn(q)\hat{\lambda}\leq 0$, so regardless of $\sgn(q)$, the eigenvalues are the same as for the non-zero modes.  However, the degeneracies are halved relative to that case.  Namely, the eigenvalues are given by
\begin{align}
\hat{\lambda} = \pm\sgn(q)\left(n+|q|+\frac{1}{2}\right) \ecq n=0,1,\ldots \ec
\end{align}
with degeneracy $2|q|$.

\subsubsection{Summary}

The eigenvalues of $\tilde{\cD}_F$ in \eqref{tDF} are
\begin{equation}
\pm(n + \ell + 1) + i\sigma, \quad n = 0, 1, \ldots, \quad \ell = |q| + 1/2, |q| + 3/2, \ldots
\end{equation}
with degeneracy $2(2\ell + 1)$ for each sign and
\begin{equation}
\pm(n + \ell + 1) + i\sigma, \quad n = 0, 1, \ldots, \quad \ell = |q| - 1/2
\end{equation}
with degeneracy $2\ell + 1$ for each sign.  Equivalently, set $N + |q| = n + \ell + 1/2$ ($N = 0, 1, \ldots$); then the eigenvalues are $\pm(N + |q| + 1/2) + i\sigma$ with degeneracy
\begin{equation}
2|q| + \sum_{\ell = |q| + 1/2}^{N + |q| - 1/2} 2(2\ell + 1) = 2N(N + 1) + 2|q|(2N + 1)
\end{equation}
for each sign, as in \eqref{DfSpec} and \eqref{Df0Spec}.

\section{Hypermultiplet One-Loop Determinant on $HS^3$} \label{hemisphere-details}

In this section, we perform the $HS^3$ counterpart of the calculation in the previous section, using the same conventions throughout.  To implement the boundary conditions \eqref{boundaryconditions}, it will be necessary to keep careful track of the relevant eigenvectors and eigenspinors.

\subsection{Bosonic Spectrum}

Recall that the bosonic $R$-symmetry matrix and its eigenvectors are
\begin{equation}
(\mathcal{D}_B)_a{}^b = \begin{pmatrix} -D^2 + \frac{3}{4} + \sigma^2 + \frac{q^2}{\sin^2\eta} & -\sigma \\ \sigma & -D^2 + \frac{3}{4} + \sigma^2 + \frac{q^2}{\sin^2\eta} \end{pmatrix}, \quad \begin{pmatrix} q_1 \\ q_2 \end{pmatrix} = \begin{pmatrix} f_\pm \\ \mp if_\pm \end{pmatrix}
\end{equation}
with corresponding eigenvalues $\lambda_B^\pm$, where $f_\pm$ can be written in terms of monopole spherical harmonics as in \eqref{fpm} and \eqref{hpm}.

On $HS^3$, we have two cases:
\begin{enumerate}
\item The eigenvectors with eigenvalues $\lambda_B^+$ have $q_+ = 2f_+$ and $q_- = 0$, so the boundary conditions reduce to
\[
q_+| = 0 \Longleftrightarrow f_+| = 0.
\]
By linear independence of the $Y_{q; \ell m}$, this is equivalent to $h_+(\pi/2) = 0$.  Both $P_L^m(0) = 0$ and $Q_L^m(0) = 0$ when $L - m$ is an odd integer, so allowed eigenfunctions have $n$ even.  This means that we sum over only those $\ell$ with $N + |q| - \ell$ odd (i.e., those $\ell' = \ell - |q|$ with $N - \ell'$ odd).  Hence the degeneracies are modified to
\begin{equation}
\frac{N(N + 1)}{2} + |q|N \text{ ($N$ even)}, \quad \frac{N(N + 1)}{2} + |q|(N + 1) \text{ ($N$ odd)}
\label{bosdegen1}
\end{equation}
for the ``$+$'' sign.
\item The eigenvectors with eigenvalues $\lambda_B^-$ have $q_+ = 0$ and $q_- = 2f_-$, so the boundary conditions reduce to
\[
\partial_\perp q_-| = 0 \Longleftrightarrow \partial_\eta f_-| = 0 \Longleftrightarrow \partial_\eta h_-(\pi/2) = 0.
\]
If $q$ is an integer, then we keep only the $Q$ solution in $h$ and
\[
\partial_\eta h_-(\pi/2)\propto \left(\sqrt{\tilde{\lambda}_B^- + 1} - \ell\right)Q_{\sqrt{\tilde{\lambda}_B^- + 1} + 1/2}^{\ell + 1/2}(0),
\]
which vanishes when $(\tilde{\lambda}_B^- + 1)^{1/2} - \ell$ is an odd integer (it is never zero).  Similarly, if $q$ is a half-integer, then we keep only the $P$ solution in $h$ and
\[
\partial_\eta h_-(\pi/2)\propto \left(\sqrt{\tilde{\lambda}_B^- + 1} - \ell\right)P_{\sqrt{\tilde{\lambda}_B^- + 1} + 1/2}^{\ell + 1/2}(0),
\]
which again vanishes when $(\tilde{\lambda}_B^- + 1)^{1/2} - \ell$ is an odd integer.  Hence in either case, the degeneracies are modified to
\begin{equation}
\frac{(N + 1)(N + 2)}{2} + |q|(N + 2) \text{ ($N$ even)}, \quad \frac{(N + 1)(N + 2)}{2} + |q|(N + 1) \text{ ($N$ odd)}
\label{bosdegen2}
\end{equation}
for the ``$-$'' sign.
\end{enumerate}
Note that in \eqref{bosdegen1} and \eqref{bosdegen2}, $|q|$ is always multiplied by an even integer.  Combining these results gives \eqref{bosdegen}.

\subsection{Fermionic Spectrum}

Let $\cD_F$ denote the fermionic $R$-symmetry matrix \eqref{DF}, let $\tilde{\cD}_F = \cD_F(\sigma_3\otimes \mathbb{1}_2)$ as in \eqref{tDF}, and let $\lambda_F$ denote the eigenvalues of $\tilde{\cD}_F$ (not of $\cD_F$).  Our basic approach to evaluating the fermionic functional determinant is as follows.  The space of four-component spinors splits as $V = X\oplus Y$ where spinors in $X$ satisfy the $\psi$ boundary condition and spinors in $Y$ satisfy the $\tilde{\psi}$ boundary condition.  Left multiplication by $\sigma_3$ (in the sense of $R$-symmetry indices) takes the subspaces $X$ and $Y$ to each other: that is, $\chi = (\sigma_3\otimes \mathbb{1}_2)\psi$ and $\tilde{\psi}$ satisfy the same boundary conditions.  Thus the path integral with action $\tilde{\psi}\mathcal{D}_F\psi$ computes the determinant of $\tilde{\cD}_F$, restricted to the subspace $Y$.  As we will see, however, $Y$ is not an invariant subspace of $\tilde{\cD}_F$.  Hence one cannot simply diagonalize $\tilde{\cD}_F$ in $Y$.  Rather, for a linear operator $M$ and a subspace $S$, we define the determinant of $M$ ``restricted to $S$'' as ${\det}_S M = \exp(\tr_S\log M)$, regardless of whether the operator $M|_S$ makes sense.

To begin, we know that the eigenvalue problem
\begin{equation}
\left(i\sigma_1(\partial_\eta + \cot\eta) + \frac{\ell + 1/2}{\sin\eta}\sigma_3\right)\left(\begin{array}{c} h_+(\eta) \\ h_-(\eta) \end{array}\right) = \lambda\left(\begin{array}{c} h_+(\eta) \\ h_-(\eta) \end{array}\right)
\end{equation}
has the following solutions for the eigenvalues:
\begin{equation}
\lambda = \pm(n + \ell + 1), \quad n = 0, 1, \ldots
\end{equation}
with degeneracy $2\ell + 1$ for each sign.\footnote{Use $g = \frac{1}{\sqrt{2}}\left(\begin{smallmatrix} 1 & 1 \\ -1 & 1 \end{smallmatrix}\right)\in SU(2) \implies g(\sigma_1, \sigma_2, \sigma_3)g^{-1} = (\sigma_3, \sigma_2, -\sigma_1)$ to change basis to \eqref{ranproblem}.}  The corresponding eigenspinors are given by
\begin{equation}
h_+(\eta) = \frac{1}{\sqrt{2}}(h_1(\eta) - h_2(\eta)), \quad h_-(\eta) = \frac{1}{\sqrt{2}}(h_1(\eta) + h_2(\eta))
\end{equation}
where for $\lambda = +(n + \ell + 1)$, we substitute
\begin{equation}
\left(\begin{array}{c} h_1 \\ h_2 \end{array}\right) = \left(\begin{array}{c} F_{n, \ell}^{(1)}(x) \\ \frac{i(2\ell + 1)}{2(n + \ell) + 1}F_{n, \ell}^{(2)}(x) \end{array}\right)
\end{equation}
and for $\lambda = -(n + \ell + 1)$, we substitute
\begin{equation}
\left(\begin{array}{c} h_1 \\ h_2 \end{array}\right) = \left(\begin{array}{c} F_{n, \ell}^{(2)}(x) \\ \frac{i(2(n + \ell) + 1)}{2\ell + 1}F_{n, \ell}^{(1)}(x) \end{array}\right),
\end{equation}
with
\begin{align}
F_{n, \ell}^{(1)}(x) &\equiv x^{-n - \ell}(1 - x^2)^{\ell - 1/2}{}_2 F_1(\ell + 1/2, -n, -n - \ell - 1/2, x^2), \\
F_{n, \ell}^{(2)}(x) &\equiv x^{1 - n - \ell}(1 - x^2)^{\ell - 1/2}{}_2 F_1(\ell + 3/2, -n, 1/2 - n - \ell, x^2),
\end{align}
and $x = e^{i\eta}$.

From the previous paragraph and the manipulations of the previous section, we deduce that the eigenspinors $\chi$ of $\tilde{\cD}_F$ are as follows.  First consider the non-zero ($\pm$) modes, with $\ell\geq |q| + 1/2$.  Define
\begin{align*}
s_{q, n\ell m+}(\theta) &\equiv \textstyle \frac{1}{\sqrt{2}}(\cos\theta(F_{n, \ell}^{(1)}(x) - \frac{i(2\ell + 1)}{2(n + \ell) + 1}F_{n, \ell}^{(2)}(x)) + \sin\theta(F_{n, \ell}^{(2)}(x) - \frac{i(2(n + \ell) + 1)}{2\ell + 1}F_{n, \ell}^{(1)}(x)))Y_{q, \ell m}^+ \\
&+ \textstyle \frac{1}{\sqrt{2}}(\cos\theta(F_{n, \ell}^{(1)}(x) + \frac{i(2\ell + 1)}{2(n + \ell) + 1}F_{n, \ell}^{(2)}(x)) - \sin\theta(F_{n, \ell}^{(2)}(x) + \frac{i(2(n + \ell) + 1)}{2\ell + 1}F_{n, \ell}^{(1)}(x)))Y_{q, \ell m}^-, \\
s_{q, n\ell m-}(\theta) &\equiv \textstyle \frac{1}{\sqrt{2}}(\cos\theta(F_{n, \ell}^{(2)}(x) - \frac{i(2(n + \ell) + 1)}{2\ell + 1}F_{n, \ell}^{(1)}(x)) + \sin\theta(F_{n, \ell}^{(1)}(x) - \frac{i(2\ell + 1)}{2(n + \ell) + 1}F_{n, \ell}^{(2)}(x)))Y_{q, \ell m}^+ \\
&+ \textstyle \frac{1}{\sqrt{2}}(\cos\theta(F_{n, \ell}^{(2)}(x) + \frac{i(2(n + \ell) + 1)}{2\ell + 1}F_{n, \ell}^{(1)}(x)) - \sin\theta(F_{n, \ell}^{(1)}(x) + \frac{i(2\ell + 1)}{2(n + \ell) + 1}F_{n, \ell}^{(2)}(x)))Y_{q, \ell m}^-.
\end{align*}
For $\lambda_F = +(n + \ell + 1) + i\sigma$, we have
\begin{equation}
\chi_{\dot{1}} = \sum_m a_m s_{q, n\ell m+}(\theta), \quad \chi_{\dot{2}} = \sum_m a_m' s_{q, n\ell m-}(-\theta).
\end{equation}
For $\lambda_F = -(n + \ell + 1) + i\sigma$, we have
\begin{equation}
\chi_{\dot{1}} = \sum_m b_m s_{q, n\ell m-}(\theta), \quad \chi_{\dot{2}} = \sum_m b_m' s_{q, n\ell m+}(-\theta).
\end{equation}
Now consider the zero modes, with $\ell = |q| - 1/2$.  For $\lambda_F = \sgn(q)(n + \ell + 1) + i\sigma$, we have
\begin{equation}
\left(\begin{array}{c} \chi_{\dot{1}} \\ \chi_{\dot{2}} \end{array}\right) = \sum_m a_m\left(\begin{array}{c} F_{n, \ell}^{(1)}(x)Y_{q, \ell m}^0 \\ \frac{i(2\ell + 1)}{2(n + \ell) + 1}F_{n, \ell}^{(2)}(x)Y_{q, \ell m}^0 \end{array}\right).
\end{equation}
For $\lambda_F = -\sgn(q)(n + \ell + 1) + i\sigma$, we have
\begin{equation}
\left(\begin{array}{c} \chi_{\dot{1}} \\ \chi_{\dot{2}} \end{array}\right) = \sum_m b_m\left(\begin{array}{c} F_{n, \ell}^{(2)}(x)Y_{q, \ell m}^0 \\ \frac{i(2(n + \ell) + 1)}{2\ell + 1}F_{n, \ell}^{(1)}(x)Y_{q, \ell m}^0 \end{array}\right).
\end{equation}
The coefficients $a, b, a', b'$ parametrize linear combinations of degenerate eigenspinors.

Specializing to the hemisphere with boundary $S^2$ at $\eta = \pi/2$ means restricting to those spinors $\chi$ satisfying $\chi|_{\dot{1}} = -\sigma_3\chi|_{\dot{2}}$.  Clearly, among non-zero modes, the allowed spinors reduce at the boundary to linear combinations of
\begin{equation}
\left(\begin{array}{c} Y_{q, \ell m}^+ \\ -Y_{q, \ell m}^- \end{array}\right), \quad \left(\begin{array}{c} Y_{q, \ell m}^- \\ -Y_{q, \ell m}^+ \end{array}\right),
\label{allowedspinors}
\end{equation}
which span a $2(2\ell + 1)$-dimensional subspace of the $4(2\ell + 1)$-dimensional subspace of spinors with fixed $n, \ell$.  Using the property
\begin{equation}
\frac{{}_2 F_1(\ell + 1/2, -n, -n - \ell - 1/2, -1)}{{}_2 F_1(\ell + 3/2, -n, 1/2 - n - \ell, -1)} = \frac{(-1)^n(2\ell + 1)}{2(n + \ell) + 1} \Longleftrightarrow \frac{F_{n, \ell}^{(1)}(i)}{F_{n, \ell}^{(2)}(i)} = \frac{(-1)^n(2\ell + 1)}{i(2(n + \ell) + 1)}
\end{equation}
allows us to write $Y^\pm$ as linear combinations of $s_\pm|$: namely, for fixed $n, \ell$, we have up to an $m$-independent constant that
\begin{align}
Y_{q, \ell m}^+ &\propto c_+^+(\theta)s_{q, n\ell m+}(\theta)| + c_-^+(\theta)s_{q, n\ell m-}(\theta)|, \label{YasS1} \\
Y_{q, \ell m}^- &\propto c_+^-(\theta)s_{q, n\ell m+}(\theta)| + c_-^-(\theta)s_{q, n\ell m-}(\theta)|, \label{YasS2}
\end{align}
where
\begin{align}
c_+^+(\theta) &= i(1 + (-1)^n)(2(n + \ell) + 1)\cos\theta + (1 - (-1)^n)(2\ell + 1)\sin\theta, \\
c_-^+(\theta) &= (1 - (-1)^n)(2\ell + 1)\cos\theta + i(1 + (-1)^n)(2(n + \ell) + 1)\sin\theta, \\
c_+^-(\theta) &= -i(1 - (-1)^n)(2(n + \ell) + 1)\cos\theta - (1 + (-1)^n)(2\ell + 1)\sin\theta, \\
c_-^-(\theta) &= (1 + (-1)^n)(2\ell + 1)\cos\theta + i(1 - (-1)^n)(2(n + \ell) + 1)\sin\theta.
\end{align}
We see that none of the eigenspinors of $\tilde{\cD}_F$ survive the boundary conditions, and moreover, that $\tilde{\cD}_F$ does not act in a simple way on the subspace of spinors that do (it is neither an invariant subspace nor mapped to its orthogonal complement).  Therefore, to compute the desired determinant of $\tilde{\cD}_F$, we exponentiate the trace of $\log\tilde{\cD}_F$ in the subspace $Y$ of allowed spinors.  In view of \eqref{allowedspinors}, \eqref{YasS1}, \eqref{YasS2}, an orthonormal basis for this subspace is given by
\begin{align*}
s_{1, m} &\equiv \frac{1}{\sqrt{\mathcal{N}}}\left(\begin{array}{c}
c_+^+(\theta)s_{q, n\ell m+}(\theta) + c_-^+(\theta)s_{q, n\ell m-}(\theta) \\
-c_+^-(-\theta)s_{q, n\ell m+}(-\theta) - c_-^-(-\theta)s_{q, n\ell m-}(-\theta)
\end{array}\right), \\
s_{2, m} &\equiv \frac{1}{\sqrt{\mathcal{N}}}\left(\begin{array}{c}
c_+^-(\theta)s_{q, n\ell m+}(\theta) + c_-^-(\theta)s_{q, n\ell m-}(\theta) \\
-c_+^+(-\theta)s_{q, n\ell m+}(-\theta) - c_-^+(-\theta)s_{q, n\ell m-}(-\theta)
\end{array}\right),
\end{align*}
where the normalization constant is $\mathcal{N} = 4((2(n + \ell) + 1)^2 + (2\ell + 1)^2)$ under the assumption that $\smash{s_{q, n\ell m\epsilon}^\dag}\cdot s_{q, n\ell m\epsilon'} = \delta_{\epsilon\epsilon'}$ for some suitably defined inner product.\footnote{This assumption is justified because $\tilde{\cD}_F - i\sigma\mathbb{1}_4$ is Hermitian.}  We compute that
\begin{align*}
s_{1, m}^\dag(\log\tilde{\cD}_F)s_{1, m} &= \sum_\pm \frac{1\pm (-1)^n\cos 2\theta}{2}\log(\pm(n + \ell + 1) + i\sigma), \\
s_{2, m}^\dag(\log\tilde{\cD}_F)s_{2, m} &= \sum_\pm \frac{1\mp (-1)^n\cos 2\theta}{2}\log(\pm(n + \ell + 1) + i\sigma),
\end{align*}
whereupon
\begin{equation}
\tr_Y\log\tilde{\cD}_F = \sum_{i, m} s_{i, m}^\dag(\log\tilde{\cD}_F)s_{i, m} = (2\ell + 1)\sum_\pm \log(\pm(n + \ell + 1) + i\sigma).
\end{equation}
Hence the degeneracies of the $\pm$ eigenmodes are halved on the hemisphere.  We now turn to the zero modes with $\ell = |q| - 1/2$:
\begin{itemize}
\item For $\lambda_F = \sgn(q)(n + \ell + 1) + i\sigma$, we have
\begin{equation}
\left(\begin{array}{c} \chi|_{\dot{1}} \\ \chi|_{\dot{2}} \end{array}\right) = F_{n, \ell}^{(1)}(i)\sum_m a_m\left(\begin{array}{c} Y_{q, \ell m}^0 \\ (-1)^{n+1}Y_{q, \ell m}^0 \end{array}\right),
\end{equation}
so the boundary condition reduces to $1 = (-1)^n\sgn(q)$.
\item For $\lambda_F = -\sgn(q)(n + \ell + 1) + i\sigma$, we have
\begin{equation}
\left(\begin{array}{c} \chi|_{\dot{1}} \\ \chi|_{\dot{2}} \end{array}\right) = F_{n, \ell}^{(2)}(i)\sum_m b_m\left(\begin{array}{c} Y_{q, \ell m}^0 \\ (-1)^n Y_{q, \ell m}^0 \end{array}\right),
\end{equation}
so the boundary condition reduces to $1 = (-1)^{n+1}\sgn(q)$.
\end{itemize}
In other words, regardless of $\sgn(q)$, we must have $n$ even for $\lambda_F = +(n + \ell + 1) + i\sigma$ and $n$ odd for $\lambda_F = -(n + \ell + 1) + i\sigma$ when $\ell = |q| - 1/2$.  Hence on the hemisphere, the eigenvalues $+(N + |q| + 1/2) + i\sigma$ (resp.\ $-(N + |q| + 1/2) + i\sigma$) have degeneracies
\begin{align}
2|q| + \sum_{\ell = |q| + 1/2}^{N + |q| - 1/2} (2\ell + 1) &= N(N + 1) + 2|q|(N + 1) \quad \text{($N$ even, resp.\ odd)}, \\
\sum_{\ell = |q| + 1/2}^{N + |q| - 1/2} (2\ell + 1) &= N(N + 1) + 2|q|N \quad \text{($N$ odd, resp.\ even)}.
\end{align}
This completes the derivation of \eqref{ferdegen}.

\subsection{Monopole Spinor Harmonics}

The explicit forms of the $Y_{q, \ell m}^{\pm, 0}$, while not needed here due to our judicious conventions, can be obtained from \cite{Benna:2009xd}.  Set $y = \cos\psi$.  Matching to our conventions, let
\begin{equation}
\Omega_{q, \ell m}^\pm = N_{q, \ell m}(\tau)\begin{pmatrix}
\mp e^{\mp i\pi/4}\omega_{q, \ell m}^\uparrow(y) \\
\pm e^{\pm i\pi/4}\omega_{q, \ell m}^\downarrow(y)
\end{pmatrix}
\end{equation}
where
\begin{align}
N_{q, \ell m}(\tau) &= \frac{(-1)^{\ell - m}(i/2)^{\ell + 1/2}(\ell + 1/2)}{\sqrt{\Gamma(\ell + 3/2 - q)\Gamma(\ell + 3/2 + q)}}\smash{\sqrt{\frac{(\ell - m)!}{(\ell + m)!}}}\frac{e^{i(m + q)\tau}}{\sqrt{2\pi}}, \\
\omega_{q, \ell m}^\uparrow(y) &= (1 - y)^{(m - 1/2 + q)/2}(1 + y)^{(m + 1/2 - q)/2}\frac{d^{\ell + m}}{dy^{\ell + m}}((1 - y)^{\ell + 1/2 - q}(1 + y)^{\ell - 1/2 + q}), \\
\omega_{q, \ell m}^\downarrow(y) &= (1 - y)^{(m + 1/2 + q)/2}(1 + y)^{(m - 1/2 - q)/2}\frac{d^{\ell + m}}{dy^{\ell + m}}((1 - y)^{\ell - 1/2 - q}(1 + y)^{\ell + 1/2 + q}),
\end{align}
and then set
\begin{align}
Y_{q, \ell m}^+ &= \sqrt{\frac{1 + r_{q, \ell}}{2}}e^{-i\pi/4}\Omega_{q, \ell m}^+ + \sgn(q)\sqrt{\frac{1 - r_{q, \ell}}{2}}e^{i\pi/4}\Omega_{q, \ell m}^-, \\
Y_{q, \ell m}^- &= \sgn(q)\sqrt{\frac{1 - r_{q, \ell}}{2}}e^{-i\pi/4}\Omega_{q, \ell m}^+ + \sqrt{\frac{1 + r_{q, \ell}}{2}}e^{i\pi/4}\Omega_{q, \ell m}^-
\end{align}
where $r_{q, \ell} = \sqrt{1 - q^2/(\ell + 1/2)^2}$.  Note that
\begin{equation}
\cos\theta = \sqrt{\frac{1 + r_{q, \ell}}{2}}, \quad \sin\theta = \sgn(q)\sqrt{\frac{1 - r_{q, \ell}}{2}},
\end{equation}
with $\theta$ defined in \eqref{theta}.  Further define, for $\ell = |q| - 1/2$,
\begin{equation}
Y_{q, \ell m}^0 = \frac{1}{\sqrt{2}}(Y_{q, \ell m}^+ + \sgn(q)Y_{q, \ell m}^-).
\end{equation}
The desired properties $\sigma_3 Y_{q, \ell m}^\pm = Y_{q, \ell m}^\mp$ and $\sigma_3 Y_{q, \ell m}^0 = \sgn(q)Y_{q, \ell m}^0$ are satisfied by virtue of $\sigma_3\Omega_{q, \ell m}^\pm = \pm i\Omega_{q, \ell m}^\mp$.  The $\smash{Y_{q, \ell m}^\pm}$ and $\smash{Y_{q, \ell m}^0}$ are eigenmodes of $i\slashed{D}_{S^2, q}$ where
\begin{equation}
\slashed{D}_{S^2, q} = \sigma_1\left(D_\psi + \frac{1}{2}\cot\psi\right) + \sigma_2\frac{1}{\sin\psi}D_\tau = \sigma_1\left(\partial_\psi + \frac{1}{2}\cot\psi\right) + \sigma_2\frac{1}{\sin\psi}(\partial_\tau - iA_\tau)
\end{equation}
and the gauge field is defined in the coordinate patches $0 < \psi < \pi/2$ and $\pi/2 < \psi < \pi$ as $A_\tau^\pm = -q(\cos\psi\mp 1)$, respectively.  The formulas above are suited to the patch $0 < \psi < \pi/2$.

\section{More on Matching} \label{moreonmatching}

In this appendix, we elaborate on several aspects of the matching of twisted correlators across mirror symmetry.  Throughout this section, for notational convenience, we leave all correlators unnormalized (i.e., we omit an overall factor of $1/Z$) and set $r = 1$.

The mirror dual of any 3D $\mathcal{N} = 4$ abelian gauge theory consisting of only ordinary or twisted multiplets is known: therefore, the 1D topological theory for twisted HBOs in such a theory gives a completely general prescription for computing correlators of twisted CBOs in its mirror dual.  On the other hand, shift operators provide a completely general prescription for computing correlators of twisted CBOs in any such theory directly.  To show that these two prescriptions give identical results for all correlators consists of two steps:
\begin{enumerate}
\item Prove this statement for the fundamental abelian mirror symmetry: namely, an arbitrary twisted HBO correlator in the free massive hyper is equal to the corresponding twisted CBO correlator in SQED$_1$ with matching FI parameter.
\item Show how to obtain twisted CBO correlators in a general abelian theory from those of the free hyper/SQED$_1$, namely as sums of products of two-point functions, integrated over appropriate subsets of mass/FI parameters.
\end{enumerate}
We carry out the first step in Appendix \ref{proof1} by proving that all twisted correlators match across the basic duality between a free hyper with mass $m$ and SQED$_1$ with FI parameter $m$.  We then illustrate the second step in Appendix \ref{proofn} by proving that all twisted CBO correlators in SQED$_N$ match the corresponding twisted HBO correlators in the $N$-node abelian necklace quiver.  In this case, the map between CBOs and HBOs is very simple, and we derive explicit formulas for all correlators.  In principle, our arguments can be extended to match correlators of twisted HBOs and CBOs in arbitrary abelian mirror pairs using the general mirror map between chiral ring generators presented in \cite{Bullimore:2015lsa}.

\subsection{Mass and FI Parameters} \label{massandFI}

Before embarking on this program, we first review how the shift operator prescription works in the presence of nonzero mass and FI parameters.  As explained in Section \ref{massFI}, real masses modify the vacuum wavefunctions, the gluing measure, and the multiplicative factors in the monopole shift operators via $\sigma\to \sigma + m$.  On the other hand, FI parameters modify the gluing measure by a factor of $e^{-8\pi^2 i\zeta\sigma}$ for each $U(1)$ factor in the gauge group.  Moreover, in the non-conformal case, correlators take the form of topological correlators dressed with simple position-dependent factors.  The latter are fixed by symmetry, and the shift operator prescription allows us to compute the topological parts, which we denote by $\langle\rangle_\text{top}$.  In particular, mass (FI) parameters leave the topological nature of CBO (HBO) correlators unchanged while making HBO (CBO) correlators non-topological.  For an $n$-point function of twisted Higgs/Coulomb branch operators, each global (flavor/topological) $U(1)$ symmetry contributes a factor of $e^{-\zeta\sum_{i=1}^n q_i\varphi_i}$ where $q_i$ is the charge of the $i^\textrm{th}$ operator in the correlation function and $\zeta$ is the associated mass/FI parameter.\footnote{Strictly speaking, our conventions require an extra factor in the map between mass and FI parameters: $m\leftrightarrow -4\pi\zeta$.}

Let us demonstrate how these rules work in practice in the case of the SQED$_N$/abelian necklace quiver duality by matching the three-point function of a monopole $\mathcal{X}^q$, antimonopole $\mathcal{Y}^q$, and (composite) product of twisted scalars $(\mathcal{Z}^p)_\star$.  This correlator will be a useful base case in the arguments to follow.

\subsubsection*{Masses in SQED$_N$/FI Parameters in $N$-Node Quiver}

FI parameters in the abelian necklace quiver correspond to real masses for the Cartan of the $SU(N)$ flavor symmetry in SQED$_N$.  For massive SQED$_N$, we use
\begin{equation}
\mu(\sigma, 0) = \prod_{I=1}^N \frac{\Gamma(1/2 + i(\sigma + m_I))}{\Gamma(1/2 - i(\sigma + m_I))}, \quad \Psi_0(\sigma, B) = \delta_{B, 0}\prod_{I=1}^N \frac{\Gamma(1/2 - i(\sigma + m_I))}{\sqrt{2\pi}},
\end{equation}
with the mass parameters $m_I$ satisfying $\sum_{I=1}^N m_I = 0$.  Using a slightly more natural convention for the Coulomb branch chiral ring generators than in the main text, namely
\begin{equation}
\cX = \frac{1}{(-4\pi)^{N/2}}\cM^1, \quad \cY = \frac{1}{(-4\pi)^{N/2}}\cM^{-1}, \quad \cZ = \frac{i}{4\pi}\Phi,
\label{naturaldef}
\end{equation}
the corresponding North shift operators (appropriately modified by $m_I$) are
\begin{equation}
\mathcal{M}_N^1 = \left[\prod_{I=1}^N \left(\frac{B - 1}{2} - i(\sigma + m_I)\right)\right]e^{-\frac{i}{2}\partial_\sigma - \partial_B}, \quad \mathcal{M}_N^{-1} = e^{\frac{i}{2}\partial_\sigma + \partial_B}, \quad \Phi_N = \sigma + \frac{iB}{2}.
\label{massiveNorthshift}
\end{equation}
Using \eqref{naturaldef} and \eqref{massiveNorthshift}, we compute that for $\varphi_1 > \varphi_2 > \varphi_3$,
\begin{equation}
\langle(\cZ^p)_\star(\varphi_1)\cX^q(\varphi_2)\cY^q(\varphi_3)\rangle = \int \frac{d\sigma\, (i\sigma)^p}{(4\pi)^{qN + p}}\prod_{I=1}^N \left[\frac{\prod_{\ell=1}^q (i(\sigma + m_I) - \ell + 1/2)}{2\cosh(\pi(\sigma + m_I))}\right]
\label{zxywithmasses}
\end{equation}
in SQED$_N$.  On the necklace quiver side, we write the $N$ FI parameters (of which $N - 1$ are independent) as $\zeta_j = \omega_{j-1} - \omega_j$ subject to the condition $\sum_j \omega_j = 0$.  We now define
\begin{equation}
\mathcal{X} = Q_1\cdots Q_N, \quad \mathcal{Y} = \tilde{Q}_1\cdots \tilde{Q}_N, \quad (\mathcal{Z}^p)_\star = \prod_{j=1}^p (\tilde{Q}_j Q_j + i\omega_j),
\label{withFI}
\end{equation}
assuming for simplicity that $p\leq N$.  The definition of $(\mathcal{Z}^p)_\star$ is the natural one from the point of view of the D-term relations (the parameters $\omega_j$ resolve the geometry of the Higgs branch).  The integration measure \eqref{necklacemeasure} is modified as
\begin{equation}
Z_\sigma = \prod_{j=1}^N \frac{e^{8\pi^2 i\omega_j\sigma_{j, j+1}}}{2\cosh(\pi\sigma_{j, j+1})},
\end{equation}
while the 1D propagator \eqref{GDef} (which is sensitive to mass parameters) remains unchanged.  Counting Wick contractions carefully yields the basic three-point function
\begin{equation}
\begin{gathered}
\langle(\mathcal{Z}^p)_\star(\varphi_1)\mathcal{X}^q(\varphi_2)\mathcal{Y}^q(\varphi_3)\rangle = (q!)^N\int d\mu(\sigma_j)\prod_{j = p + 1}^N G_{\sigma_{j, j+1}}(\varphi_{23})^q \\
\times \prod_{a=1}^p (G_{\sigma_{a, a+1}}(0)G_{\sigma_{a, a+1}}(\varphi_{23}) + qG_{\sigma_{a, a+1}}(\varphi_{21})G_{\sigma_{a, a+1}}(\varphi_{13}))G_{\sigma_{a, a+1}}(\varphi_{23})^{q-1}.
\end{gathered}
\label{intermediate}
\end{equation}
Assuming that $\varphi_1 > \varphi_2 > \varphi_3$, we may use \eqref{FTId}, the identity
\begin{equation}
\label{intbyparts}
\frac{(\sgn\varphi_{12} + \tanh(\pi\sigma))^m}{2\cosh(\pi\sigma)} = \frac{1}{m!}\left[\prod_{j=1}^m \left((2j - 1)\sgn\varphi_{12} - \frac{1}{\pi}\frac{d}{d\sigma}\right)\right]\frac{1}{2\cosh(\pi\sigma)},
\end{equation}
integration by parts, and $\frac{1}{2\cosh(\pi\sigma)} = \int d\tau\, \frac{e^{2\pi i\sigma\tau}}{2\cosh(\pi\tau)}$ to simplify \eqref{intermediate} to
\begin{equation}
\begin{split}
\langle(\mathcal{Z}^p)_\star(\varphi_1)\mathcal{X}^q(\varphi_2)\mathcal{Y}^q(\varphi_3)\rangle = \int \frac{d\tau\, (i\tau)^p}{(4\pi)^{qN + p}}\prod_{I=1}^N \left[\frac{\prod_{j=1}^q (i(\tau - 4\pi\omega_I) - j + 1/2)}{2\cosh(\pi(\tau - 4\pi\omega_I))}\right].
\end{split}
\label{zxywithFI}
\end{equation}
This matches the SQED$_N$ result if we identify $m_I\leftrightarrow -4\pi\omega_I$.

\subsubsection*{FI Parameters in SQED$_N$/Masses in $N$-Node Quiver}

Mass parameters in the abelian necklace quiver correspond to FI parameters in SQED$_N$.  Consider adding a real mass associated to the $U(1)$ flavor symmetry of the necklace quiver under which $Q_i, \tilde{Q}_i$ carry charge $\pm 1/N$.  In practice, this means replacing all instances of $\sigma_{j, j+1}$ by $\sigma_{j, j+1} + m/N$ in the 1D theory computations.  Using the identity
\begin{equation}
\int \left(\prod_{j=1}^N d\sigma_j\right)\delta\left(\frac{1}{N}\sum_{j=1}^N \sigma_j\right)\prod_{j=1}^N F_j(\sigma_{j, j + 1} + m/N) = \int d\tau\, e^{2\pi im\tau}\prod_{j=1}^N \tF_j(\tau),
\end{equation}
which is the appropriate modification of \eqref{FTId}, we obtain (with $\varphi_1 > \varphi_2 > \varphi_3$)
\begin{equation}
\langle(\mathcal{Z}^p)_\star(\varphi_1)\mathcal{X}^q(\varphi_2)\mathcal{Y}^q(\varphi_3)\rangle_\text{top} = \int \frac{d\tau\, e^{2\pi im\tau}}{(4\pi)^{qN + p}}\frac{(i\tau)^p}{(2\cosh(\pi\tau))^N}\prod_{j=1}^q (i\tau - j + 1/2)^N.
\end{equation}
This matches the expression
\begin{equation}
\langle(\mathcal{Z}^p)_\star(\varphi_1)\mathcal{X}^q(\varphi_2)\mathcal{Y}^q(\varphi_3)\rangle_\text{top} = \int \frac{(-i)^p\, d\sigma\, e^{2\pi im\sigma}}{(-4\pi)^{qN + p}}\mu(\sigma, 0)\Psi_0(\sigma, 0)[\mathcal{M}_N^{-q}\mathcal{M}_N^q\Phi_N^p\Psi_0(\sigma, B)]|_{B=0}
\end{equation}
on the SQED$_N$ side.

\subsection{Proof: Basic Mirror Duality} \label{proof1}

With this warmup complete, we now match all twisted correlators in SQED$_1$ with FI parameter $\zeta$ and a free hyper of mass $m = -4\pi\zeta$.  In the free hyper theory, correlation functions of $\mathcal{X} = Q$, $\mathcal{Y} = \tilde{Q}$, $\mathcal{Z} = Q\tilde{Q}$ are computed using the measure
\begin{equation}
d\mu(\sigma) = \frac{d\sigma\, \delta(\sigma)}{2\cosh(\pi m)},
\end{equation}
and Wick contractions are performed using the $\sigma$-independent Green's function
\begin{equation}
G(\varphi_{12}) = \langle Q(\varphi_1)\tilde{Q}(\varphi_2)\rangle = -\frac{\sgn\varphi_{12} + \tanh(\pi m)}{8\pi}e^{-m\varphi_{12}}.
\end{equation}
Correlators are no longer topological due to the factor of $e^{-m\varphi_{12}}$.

In matching all correlators, let us focus only on the topological parts (as the position-dependent parts match trivially).  We wish to show that
\begin{equation}
\langle\cS\rangle_\text{top, SQED$_1$} \stackrel{!}{=} \langle\cS\rangle_\text{top, free hyper}
\end{equation}
where $\mathcal{S}$ is some operator string in $\mathcal{X}, \mathcal{Y}, \mathcal{Z}$ and operators appearing in correlation functions are understood to be in \emph{descending} order by insertion point (i.e., $\varphi_1 > \cdots > \varphi_n$).\footnote{In SQED$_N$, when restricting our attention to the operators $\cX$ and $\cY$, it suffices to consider correlators of the form $\langle\cX^{a_1}\cY^{b_1}\cX^{a_2}\cY^{b_2}\cdots \cX^{a_n}\cY^{b_n}\rangle$ for $a_i, b_i\in \mathbb{Z}_{> 0}$ for two reasons.  First, if $\cX^{p_j}(\varphi_j)\cX^{p_{j+1}}(\varphi_{j+1})$ appears somewhere in the operator string, then we may replace it by $\cX^{p_j + p_{j+1}}$, and similarly for $\cY$: this is obvious from composition of shift operators, and also from the mirror 1D theory because Wick contractions depend only on the ordering between $\cX$ and $\cY$.  Second, correlators on the circle simply change by signs under cyclic permutations of the insertions: for example, $\langle\cX^m\cY^{m+n}\cX^n\rangle = (-1)^{Nn}\langle\cX^{m+n}\cY^{m+n}\rangle$; this property is clear from moving shift operators past the branch point but harder to see from the 1D theory.}  Shift operators in SQED$_1$ with FI parameter $\zeta$ give
\begin{equation}
\langle\cO_1^{p_1}\cdots \cO_n^{p_n}\rangle_\text{top} = \int d\sigma\, e^{-8\pi^2 i\zeta\sigma}\mu(\sigma, 0)\Psi_0(\sigma, 0)[(\cO_n)_N^{p_n}\cdots (\cO_1)_N^{p_1}\Psi_0(\sigma, B)]|_{B=0}
\label{shiftresultgeneral}
\end{equation}
where $\cO_i\in \{\cX, \cY, \cZ\}$ and
\begin{equation}
\cX_N = \left(\frac{B - 1}{2} - i\sigma\right)\frac{e^{-\frac{i}{2}\partial_\sigma - \partial_B}}{(-4\pi)^{1/2}}, \quad \cY_N = \frac{e^{\frac{i}{2}\partial_\sigma + \partial_B}}{(-4\pi)^{1/2}}, \quad \cZ_N = \frac{i}{4\pi}\left(\sigma + \frac{iB}{2}\right).
\label{shiftsqed1}
\end{equation}
Here, the notation $\cZ^p$ is understood to mean $p$ adjacent insertions of $\cZ$ at separated points, which is equivalent to a single insertion of the composite operator $(\cZ^p)_\star$.  On the other hand, the 1D theory for the free hyper with mass $m$ gives
\begin{equation}
\langle\mathcal{O}_1^{p_1}\cdots \mathcal{O}_n^{p_n}\rangle_\text{top} = \int d\tau\, e^{2\pi im\tau}\int d\sigma\, \frac{e^{-2\pi i\tau\sigma}}{2\cosh(\pi\sigma)}w(\mathcal{O}_1^{p_1}\cdots \mathcal{O}_n^{p_n})
\label{1Dresultgeneral}
\end{equation}
where $w(s)$ denotes the sum of all full Wick contractions of the operator string $s$ and Wick contractions are performed using the ``topological'' propagators
\begin{equation}
G_\pm = -\frac{\pm 1 + \tanh(\pi\sigma)}{8\pi}, \quad G_0 = -\frac{\tanh(\pi\sigma)}{8\pi}.
\end{equation}
We proceed by induction.  In the previous subsection, we established the base case
\begin{equation}
\langle\cZ^p\cX^q\cY^q\rangle_\text{top, SQED$_1$} = \langle\mathcal{Z}^p\mathcal{X}^q\mathcal{Y}^q\rangle_\text{top, free hyper}.
\end{equation}
Now fix some $\cS$ and suppose we have established that $\langle\cS\rangle_\text{top, SQED$_1$} = \langle\mathcal{S}\rangle_\text{top, free hyper}$, as well as a similar statement for all operator strings containing fewer operators than $\cS$.  Consider swapping two adjacent operators in $\cS$ to form a new operator string $\cS'$.  Starting from the basic string $\cZ^p\cX^q\cY^q$, one can obtain any other string by performing three types of swaps (below, let $\cS_{L, R}$ denote substrings of $\cS$):
\begin{enumerate}[leftmargin=20pt]
\item Let $\cS\equiv \cS_L \cX\cY\cS_R$, $\cS'\equiv \cS_L \cY\cX\cS_R$, and $\cS_0\equiv \cS_L \cS_R$.
\item Let $\cS\equiv \cS_L \cZ\cX\cS_R$, $\cS'\equiv \cS_L \cX\cZ\cS_R$, and $\cS_0\equiv \cS_L \cX\cS_R$.
\item Let $\cS\equiv \cS_L \cZ\cY\cS_R$, $\cS'\equiv \cS_L \cY\cZ\cS_R$, and $\cS_0\equiv \cS_L \cY\cS_R$.
\end{enumerate}
In all three cases, the Wick contractions of the strings so defined are related in a simple way, implying relations between the corresponding correlators \eqref{1Dresultgeneral} in the free hyper theory:
\begin{enumerate}[leftmargin=20pt]
\item $w(\mathcal{S}') = w(\mathcal{S}) + (G_- - G_+)w(\mathcal{S}_0) = w(\mathcal{S}) + \frac{1}{4\pi}w(\mathcal{S}_0) \implies \langle\mathcal{S}'\rangle_\text{top} = \langle\mathcal{S}\rangle_\text{top} + \frac{1}{4\pi}\langle\mathcal{S}_0\rangle_\text{top}$.
\item $w(\mathcal{S}') = w(\mathcal{S}) + (G_+ - G_-)w(\mathcal{S}_0) = w(\mathcal{S}) - \frac{1}{4\pi}w(\mathcal{S}_0) \implies \langle\mathcal{S}'\rangle_\text{top} = \langle\mathcal{S}\rangle_\text{top} - \frac{1}{4\pi}\langle\mathcal{S}_0\rangle_\text{top}$.
\item Same as in case (1).
\end{enumerate}
On the other hand, the shift operators \eqref{shiftsqed1} for SQED$_1$ satisfy the commutation relations
\begin{equation}
[\cX_N, \cY_N] = \frac{1}{4\pi}, \quad [\cX_N, \cZ_N] = \frac{1}{4\pi}\cX_N, \quad [\cY_N, \cZ_N] = -\frac{1}{4\pi}\cY_N,
\end{equation}
implying that the correlators \eqref{shiftresultgeneral} in SQED$_1$ satisfy identical relations in the three cases:
\begin{enumerate}[leftmargin=20pt]
\item $\langle \cS'\rangle_\text{top} = \langle \cS\rangle_\text{top} + \frac{1}{4\pi}\langle \cS_0\rangle_\text{top}$.
\item $\langle \cS'\rangle_\text{top} = \langle \cS\rangle_\text{top} - \frac{1}{4\pi}\langle \cS_0\rangle_\text{top}$.
\item Same as in case (1).
\end{enumerate}
By the induction hypothesis, $\langle\mathcal{S}\rangle_\text{top}$ and $\langle\mathcal{S}_0\rangle_\text{top}$ both match in SQED$_1$ and the free hyper, which immediately implies that $\langle \cS'\rangle_\text{top, SQED$_1$} = \langle\mathcal{S}'\rangle_\text{top, free hyper}$, as desired.

\subsection{Proof: HBOs in $N$-Node Quiver and CBOs in SQED$_N$} \label{proofn}

All correlation functions of twisted CBOs in SQED$_N$ can be written very explicitly with the aid of the shift operators
\begin{equation}
\cX_N = \left(\frac{B - 1}{2} - i\sigma\right)^N\frac{e^{-\frac{i}{2}\partial_\sigma - \partial_B}}{(-4\pi)^{N/2}}, \quad \cY_N = \frac{e^{\frac{i}{2}\partial_\sigma + \partial_B}}{(-4\pi)^{N/2}}, \quad \cZ_N = \frac{i}{4\pi}\left(\sigma + \frac{iB}{2}\right),
\label{shiftsqedn}
\end{equation}
which are the appropriate generalizations of \eqref{shiftsqed1}.  Namely, consider a correlator with $n$ operators, drawn from $\cX, \cY, \cZ$, having positive integer powers $p_1, \ldots, p_n$ and labeled by signs $\epsilon_1, \ldots, \epsilon_n\in \{0, \pm 1\}$ indicating whether the operator is $\cX$ ($\epsilon = +1$), $\cY$ ($\epsilon = -1$), or $\cZ$ ($\epsilon = 0$).  We assume that the charges sum to zero, so that the correlator is nontrivial: $\sum_{i=1}^n \epsilon_i p_i = 0$.  For arbitrary $f(\sigma, B)$, we have that
\begin{equation}
\cX_N^p f(\sigma, B) = \frac{(-1)^{pN}}{(-4\pi)^{pN/2}}\prod_{\ell=1}^p \left(\ell - \frac{1}{2} - \frac{B}{2} + i\sigma\right)^N f(\sigma - ip/2, B - p)
\end{equation}
while $\cY_N^p f(\sigma, B) = (-4\pi)^{-pN/2}f(\sigma + ip/2, B + p)$.  Hence we obtain, using \eqref{shiftsqedn},
\begin{equation}
\begin{aligned}
\langle\mathcal{O}_1^{p_1}\cdots \mathcal{O}_n^{p_n}\rangle &= \frac{(-1)^{N\sum_j p_j\epsilon_j^2/2}}{(-4\pi)^{N\sum_j p_j\epsilon_j^2/2 + \sum_j p_j(1 - \epsilon_j^2)}}\int \frac{d\sigma}{(2\cosh(\pi\sigma))^N} \\
&\times \prod_{j=1}^n \Bigg[\left(\sum_{k=1}^j \epsilon_k p_k - i\sigma \right)^{p_j(1 - \epsilon_j)}\prod_{\ell=1}^{p_j} \left(\ell - \frac{1}{2} + i\sigma - \sum_{k=1}^j \epsilon_k p_k\right)^{N\epsilon_j/2}\Bigg]^{(1 + \epsilon_j)}
\end{aligned}
\label{shift-n}
\end{equation}
where the insertion points of the $\cO_i^{p_i}$ satisfy $\varphi_1 > \cdots > \varphi_n$.  Note that $\sum_j p_j\epsilon_j^2/2$ is always an integer, by the $(\text{mod 2})$-version of the zero-charge condition $\sum_{i=1}^n \epsilon_i p_i = 0$.  The formula \eqref{shift-n} encodes all possible correlators of twisted CBOs in SQED$_N$.  One can check that \eqref{shift-n} includes \eqref{zxywithmasses} (without mass parameters) as a special case.  The shift operator approach to twisted CBOs in SQED$_N$ is significantly simpler than the mirror approach to twisted HBOs in the necklace quiver using the Higgs branch topological theory: reproducing \eqref{shift-n} in full generality using the latter approach is so laborious as to be intractable.  Nonetheless, we now present a proof that all twisted HBO/CBO correlators match across this duality.\footnote{Mirror symmetry seems to entail a principle of ``conservation of effort'': for twisted CBO correlators in the necklace quiver, using the mirror 1D theory is simpler in practice than using shift operators (unlike for twisted CBO correlators in SQED$_N$).}

Let us use the result of the previous subsection to match all shift operator results for SQED$_N$ to the mirror correlators computed using the 1D theory in the necklace quiver.  Our argument relies on the procedure of building mirror pairs from (copies of) the basic mirror duality and gauging subsets of mass/FI parameters.  The basic ingredients are as follows.  The 1D theories for the free hyper with mass parameter $m$ associated to the $U(1)$ flavor symmetry under which $Q, \tilde{Q}$ have charge $\pm 1$ and for SQED$_1$ with FI parameter $\zeta$ are
\begin{align}
Z_\text{free}(m) &= \int \pD\tilde{Q}\, \pD Q\, e^{4\pi\int d\varphi\, \tilde{Q}(\partial_\varphi + m)Q}, \\
Z_\text{SQED$_1$}(\zeta) &= \int d\sigma\, e^{-8\pi^2 i\zeta\sigma}\int D\tilde{Q}\, DQ\, e^{4\pi\int d\varphi\, \tilde{Q}(\partial_\varphi + \sigma)Q}.
\end{align}
The basic examples of gauging/ungauging are
\begin{equation}
\int dm\, Z_\text{free}(m) = Z_\text{SQED$_1$}, \quad \int d\zeta\, Z_\text{SQED$_1$}(-\zeta/4\pi) = Z_\text{free},
\end{equation}
where $Z\equiv Z(0)$ (compare these operations to the $S$-transformation in \cite{Witten:2003ya}).

In our case, the 1D theory for SQED$_N$ is obtained by taking $N$ copies of $Z_\text{free}$ and gauging the diagonal $U(1)$ subgroup:
\begin{equation}
Z_\text{SQED$_N$} = \int d\sigma\int \left(\prod_{j=1}^N D\tilde{Q}_j\, DQ_j\right)e^{4\pi\int d\varphi\sum_{j=1}^N \tilde{Q}_j(\partial_\varphi + \sigma)Q_j} = \int d\zeta\, Z_\text{free}(\zeta)^N.
\end{equation}
The 1D theory for the necklace quiver is obtained by writing
\begin{equation}
Z_{U(1)^N/U(1)} = \int \left(\prod_{j=1}^N d\sigma_j\right)\delta\left(\frac{1}{N}\sum_{j=1}^N \sigma_j\right)\prod_{j=1}^N Z_\text{free}(\sigma_{j, j+1}) = \int d\tau\, \widetilde{Z}_\text{free}(\tau)^N,
\end{equation}
where we have used \eqref{FTId}.  Any correlator of the form $\langle\mathcal{O}_1^{p_1}(\varphi_1)\cdots \mathcal{O}_n^{p_n}(\varphi_n)\rangle$ in the necklace quiver where $\mathcal{O}_i\in \{\mathcal{X}, \mathcal{Y}\}$ ($\mathcal{X}\equiv Q_1\cdots Q_N$ and $\mathcal{Y}\equiv \tilde{Q}_1\cdots \tilde{Q}_N$) can be written as
\begin{align}
\langle\mathcal{O}_1^{p_1}(\varphi_1)\cdots \mathcal{O}_n^{p_n}(\varphi_n)\rangle &= \int d\tau\left(\int d\zeta\, e^{-2\pi i\tau\zeta}Z_\text{free}(\zeta)[o_1^{p_1}(\varphi_1)\cdots o_n^{p_n}(\varphi_n)]\right)^N, \label{combinatorial} \\
Z_\text{free}(\zeta)[o_1^{p_1}(\varphi_1)\cdots o_n^{p_n}(\varphi_n)] &\equiv \int \pD\tilde{Q}\, \pD Q\, e^{4\pi\int d\varphi\, \tilde{Q}(\partial_\varphi + \zeta)Q}o_1^{p_1}(\varphi_1)\cdots o_n^{p_n}(\varphi_n),
\end{align}
with $o_i\in \{Q, \tilde{Q}\}$.  There remains a correspondence between operator insertions in $Z_{U(1)^N/U(1)}$ and operator insertions in $Z_\text{free}(\zeta)$ when the operators include $\mathcal{Z}$: letting $\mathcal{O}_i\in \{\mathcal{X}, \mathcal{Y}, \mathcal{Z}\}$ be specified by signs $\epsilon_i$, we have in the necklace quiver that $\langle\mathcal{O}_1^{p_1}(\varphi_1)\cdots \mathcal{O}_n^{p_n}(\varphi_n)\rangle$ is given by
\begin{gather}
Z_{U(1)^N/U(1)}[\mathcal{O}_1^{p_1}(\varphi_1)\cdots \mathcal{O}_n^{p_n}(\varphi_n)] = \int \left(\prod_{j=1}^N d\sigma_j\right)\delta\left(\frac{1}{N}\sum_{j=1}^N \sigma_j\right)\prod_{j=1}^N Z_\text{free}(\sigma_{j, j+1})[\Pi_j], \nonumber \\
\Pi_j\equiv \prod_{k=1}^n (Q(\varphi_k)^{\epsilon_k(1 + \epsilon_k)/2}\tilde{Q}(\varphi_k)^{-\epsilon_k(1 - \epsilon_k)/2})^{p_k}Q\tilde{Q}(\varphi_k)^{\theta(p_k - j)(1 - \epsilon_k)(1 + \epsilon_k)},
\end{gather}
where the Heaviside step function $\theta$ is defined so that $\theta(0) = 1$; here, we have assumed that if $\epsilon_i = 0$, then the corresponding $p_i\leq N$.  By \eqref{FTId}, this can be written as
\begin{equation}
Z_{U(1)^N/U(1)}[\mathcal{O}_1^{p_1}(\varphi_1)\cdots \mathcal{O}_n^{p_n}(\varphi_n)] = \int d\tau\left(\prod_{j=1}^N\int d\sigma_j\, e^{-2\pi i\tau\sigma_j}Z_\text{free}(\sigma_j)[\Pi_j]_\text{top}\right)
\label{useful}
\end{equation}
where
\begin{equation}
Z_\text{free}(\sigma_j)[\Pi_j] = e^{-\sigma_j\sum_{k=1}^n p_k\epsilon_k\varphi_k}Z_\text{free}(\sigma_j)[\Pi_j]_\text{top}
\end{equation}
and we have shifted the $\tau$ contour to replace $Z_\text{free}(\sigma_j)[\Pi_j]$ by $Z_\text{free}(\sigma_j)[\Pi_j]_\text{top}$ in \eqref{useful}.  Using the shift operator formula
\begin{gather}
\langle\mathcal{O}_1^{p_1}\cdots \mathcal{O}_n^{p_n}\rangle_\text{top} = \frac{(-1)^{\sum_j p_j\epsilon_j^2/2}}{(-4\pi)^{\sum_j p_j\epsilon_j^2/2 + \sum_j p_j(1 - \epsilon_j^2)}}\int d\tau\, \frac{e^{2\pi i\zeta\tau}}{2\cosh(\pi\tau)} \nonumber \\
\times \prod_{j=1}^n \Bigg[\left(\sum_{k=1}^j \epsilon_k p_k - i\tau\right)^{p_j(1 - \epsilon_j)(1 + \epsilon_j)}\prod_{\ell=1}^{p_j} \left(\ell - \frac{1}{2} + i\tau - \sum_{k=1}^j \epsilon_k p_k\right)^{\epsilon_j(1 + \epsilon_j)/2}\Bigg]
\end{gather}
for SQED$_1$ with FI parameter $-\zeta/4\pi$ (the $N = 1$ case of \eqref{shift-n}, with an extra insertion of $e^{2\pi i\zeta\sigma}$) and the result of the previous subsection, we have in the free hyper theory that
\begin{gather}
Z_\text{free}(\sigma_j)[\Pi_j]_\text{top} = \frac{(-1)^{\sum_k p_k\epsilon_k^2/2}}{(-4\pi)^{\sum_k p_k\epsilon_k^2/2 + \sum_k \theta(p_k - j)(1 - \epsilon_k^2)}}\int d\tau_j\, \frac{e^{2\pi i\sigma_j\tau_j}}{2\cosh(\pi\tau_j)} \nonumber \\
\times \prod_{k=1}^n \Bigg[\left(\sum_{\ell=1}^k \epsilon_\ell p_\ell - i\tau_j\right)^{\theta(p_k - j)(1 - \epsilon_k)(1 + \epsilon_k)}\prod_{m=1}^{p_k} \left(m - \frac{1}{2} + i\tau_j - \sum_{\ell=1}^k \epsilon_\ell p_\ell\right)^{\epsilon_k(1 + \epsilon_k)/2}\Bigg]. \label{shiftuseful}
\end{gather}
Substituting \eqref{shiftuseful} into \eqref{useful} and simplifying shows that
\begin{equation}
Z_{U(1)^N/U(1)}[\mathcal{O}_1^{p_1}(\varphi_1)\cdots \mathcal{O}_n^{p_n}(\varphi_n)] = \langle\mathcal{O}_1^{p_1}(\varphi_1)\cdots \mathcal{O}_n^{p_n}(\varphi_n)\rangle
\end{equation}
where the right-hand side is given precisely by the shift operator formula \eqref{shift-n} for SQED$_N$.  This completes our proof of matching for the SQED$_N$/necklace quiver duality.

\subsection{$BF$ Theories: An Appetizer}

In some cases, it is possible to test mirror symmetry at the level of 1D topological sectors by working purely on the Higgs branch.  This observation dovetails with another application of our formalism, namely to $BF$ theories.

So far, the 1D formalism for HBOs has been applied to theories containing only ordinary or only twisted $\mathcal{N} = 4$ multiplets.  There are some situations in which it can describe theories containing both ordinary and twisted multiplets.  Namely, one can couple ordinary and twisted abelian vector multiplets through a $BF$ (mixed Chern-Simons) term that preserves $\mathcal{N} = 4$ supersymmetry \cite{Kapustin:1999ha}.  In addition, one can couple the vector multiplet to hypermultiplets and the twisted vector multiplet to twisted hypermultiplets.  Call such abelian $\mathcal{N} = 4$ CSM theories, which have only mixed ordinary-twisted $BF$ terms, ``of $BF$ type.''

As an example, consider the $\mathcal{N} = 4$ CSM theories of Jafferis-Yin \cite{Jafferis:2008em}.  These are special cases of their model II$(N_f)_k$, which is defined (in $\mathcal{N} = 3$ notation) as a $U(1)_k\times U(1)_{-k}$ theory with $N_f - 1$ hypermultiplets $(X_i, \smash{\tilde{X}_i})$ of charge $((+1, +1), (-1, -1))$ and one hypermultiplet $(Y, \smash{\tilde{Y}})$ of charge $((+1, -1), (-1, +1))$ where $X_i, \smash{\tilde{X}_i}, Y, \smash{\tilde{Y}}$ are $\mathcal{N} = 2$ chiral multiplets.  The II$(N_f)_k$ theory is of $BF$ type: in $\mathcal{N} = 4$ language, it consists of one vector coupled to $N_f - 1$ hypers $(X_i, \smash{\tilde{X}_i})$, one twisted vector coupled to one twisted hyper $(Y, \smash{\tilde{Y}})$, and (after a simple change of variables) a mixed $BF$ term at level $k$.  The classical moduli space has two Higgs branches $\mathcal{M}_X$ and $\mathcal{M}_Y$, of complex dimension $2(N_f - 1)$ and 2, respectively.  These are parametrized by $X_i, \smash{\tilde{X}_i}$ and $Y, \smash{\tilde{Y}}$ (modulo constant gauge transformations), respectively.  An important feature of $\mathcal{N} = 4$ CSM theories is that their Higgs branches can receive quantum corrections \cite{Gaiotto:2007qi}.  Assuming that $k$ is even, the quantum-corrected Higgs branches are
\begin{equation}
\mathcal{M}_X = \mathbb{C}^{2N_f}///U(1), \quad \mathcal{M}_Y = \mathbb{C}^2/\mathbb{Z}_{k/2 + N_f - 1}
\end{equation}
where, in the first case, the action of $U(1)$ on the coordinates $(X_i, \smash{\tilde{X}_i}, X', \smash{\tilde{X}'})$ of $\mathbb{C}^{2N_f}$ is
\begin{equation}
\label{action}
X_i\to e^{2i\theta/k}X_i, \quad \tilde{X}_i\to e^{-2i\theta/k}\tilde{X}_i, \quad X'\to e^{i\theta}X', \quad \tilde{X}'\to e^{-i\theta}\tilde{X}'
\end{equation}
(we have introduced extra variables $X', \smash{\tilde{X}'}$, whose charges we have swapped relative to those of \cite{Jafferis:2008em}).  Concretely, $\mathcal{M}_X$ can be described by the equations
\begin{equation}
\label{quotient}
\sum_{i=1}^{N_f - 1} (|X_i|^2 - |\tilde{X}_i|^2) + \frac{k}{2}(|X'|^2 - |\tilde{X}'|^2) = 0, \quad \sum_{i=1}^{N_f - 1} X_i\tilde{X}_i + \frac{k}{2}X'\tilde{X}' = 0
\end{equation}
modulo the action \eqref{action}.  The theory II$(N_f)_{k = 2}$ is argued to describe the same IR fixed point as $\mathcal{N} = 4$ SQED$_{N_f}$.  Indeed, SQED$_{N_f}$ has Coulomb branch $\mathbb{C}^2/\mathbb{Z}_{N_f}$ and Higgs branch equal to the hyperk\"ahler quotient \eqref{quotient} with $k = 2$.

Let us write down, and qualitatively discuss, the 1D theory for the Jafferis-Yin theory II$(N_f)_k$.  Let $\sigma$ and $\tau$ denote the scalar components of the ordinary and twisted abelian vector multiplets, respectively; let $Q_i, \smash{\tilde{Q}_i}$ denote the twisted scalars of the $N_f - 1$ hypermultiplets $(X_i, \smash{\tilde{X}_i})$, and let $R, \smash{\tilde{R}}$ denote the twisted scalars of the twisted hypermultiplet $(Y, \smash{\tilde{Y}})$ (hopefully, confusion will not arise between the two senses of ``twisted'').  Motivated by the identification with SQED$_{N_f}$, let us interpret $SU(2)_L$ as $SU(2)_H$ (acting on the Higgs branch) and $SU(2)_R$ as $SU(2)_C$ (acting on the would-be Coulomb branch).  The $\mathcal{N} = 4$ Yang-Mills term is both $\cQ_\beta^H$- and $\cQ_\beta^C$-exact, so we may use it to localize with respect to either supercharge.  If we localize with respect to $\cQ_\beta^H$, then we obtain a 1D theory for $Q_i, \smash{\tilde{Q}_i}$ with a determinant contribution from the twisted part:
\begin{equation}
\label{QHloc}
\mathcal{Z} = \int d\sigma\, d\tau\, \frac{e^{-ik\pi\sigma\tau}}{2\cosh(\pi\tau)}\int \pD\tilde{Q}_i\, \pD Q_i\, e^{4\pi\int d\varphi\, \tilde{Q}_i(\partial_\varphi + \sigma)Q_i}.
\end{equation}
If we localize with respect to $\cQ_\beta^C$, then we obtain a 1D theory for $R, \smash{\tilde{R}}$ with a determinant contribution from the untwisted part:
\begin{equation}
\label{QCloc}
\mathcal{Z} = \int d\sigma\, d\tau\, \frac{e^{-ik\pi\sigma\tau}}{(2\cosh(\pi\sigma))^{N_f - 1}}\int \pD\tilde{R}\, \pD R\, e^{4\pi\int d\varphi\, \tilde{R}(\partial_\varphi + \tau)R}.
\end{equation}
These two representations are equivalent, and they can be summarized by writing a 1D theory for both ordinary and twisted fields as follows:
\begin{equation}
\label{loc}
\mathcal{Z} = \int d\sigma\, d\tau\, e^{-ik\pi\sigma\tau}\int \pD\tilde{Q}_i\, \pD Q_i\, \pD\tilde{R}\, \pD R\, e^{4\pi\int d\varphi\, (\tilde{Q}_i(\partial_\varphi + \sigma)Q_i + \tilde{R}(\partial_\varphi + \tau)R)}.
\end{equation}
Integrating out $R, \smash{\tilde{R}}$ in \eqref{loc} reproduces \eqref{QHloc}, and integrating out $Q_i, \smash{\tilde{Q}_i}$ in \eqref{loc} reproduces \eqref{QCloc}.  Operators in the cohomology of $\cQ_\beta^H$ are $Q_i, \smash{\tilde{Q}_i}$ and monopoles for the twisted $U(1)$.  Operators in the cohomology of $\cQ_\beta^C$ are $R, \smash{\tilde{R}}$ and monopoles for the untwisted $U(1)$.  The Coulomb branch chiral ring of this theory is simple to describe (again, by ``Coulomb branch,'' we mean the Higgs branch $\mathcal{M}_Y$ that would be interpreted as the Coulomb branch of SQED$_{N_f}$ when $k = 2$).  Let $\mathcal{M}, \smash{\overline{\mathcal{M}}}$ denote the basic monopole/antimonopole of the untwisted $U(1)$.  Due to the mixed Chern-Simons term, $\mathcal{M}$ and $\smash{\overline{\mathcal{M}}}$ are charged under the twisted $U(1)$ (the hypermultiplets do not contribute to the monopole charges).  Given the explicit description of $\mathcal{M}_Y$ as the hyperk\"ahler cone $\mathbb{C}^2/\mathbb{Z}_{k/2 + N_f - 1}$, we expect that the Coulomb branch chiral ring is generated by three gauge-invariant twisted CBOs $X, Y, Z$, modulo the relation $XY = \smash{Z^{k/2 + N_f - 1}}$.  The natural candidates for these operators are
\begin{equation}
X\sim R^{k/2}\mathcal{M}, \quad Y\sim \tilde{R}^{k/2}\overline{\mathcal{M}}, \quad Z\sim R\tilde{R}.
\end{equation}
In particular, when $k = 2$, we may identify the dressed monopoles $R\mathcal{M}, \smash{\tilde{R}\overline{\mathcal{M}}}$ of the CSM theory with the (gauge-neutral) bare monopoles of SQED$_{N_f}$, which satisfy the chiral ring relation $(R\mathcal{M})(\smash{\tilde{R}\overline{\mathcal{M}}}) = Z^{N_f}$.  The Higgs branch chiral ring (i.e., that of $\mathcal{M}_X$) is more complicated.  If, in addition to $Q_i, \smash{\tilde{Q}_i}$, one introduces twisted scalars $Q', \smash{\tilde{Q}'}$ for the $(X', \smash{\tilde{X}'})$ in \eqref{action} and \eqref{quotient}, then one can construct the generators from the gauge-invariant combinations
\begin{equation}
Q_i^{k/2}\tilde{Q}', \quad \tilde{Q}_i^{k/2}Q'.
\end{equation}
Let us simply observe, using the basic Fourier transform identity for $(2\cosh(\pi\sigma))^{-1}$, that the Higgs branch representation of the 1D theory \eqref{QHloc} can be written as follows:
\begin{align}
\mathcal{Z} &= \int \frac{d\sigma}{2\cosh(k\pi\sigma/2)}\int \pD\tilde{Q}_i\, \pD Q_i\, e^{4\pi\int d\varphi\, \tilde{Q}_i(\partial_\varphi + \sigma)Q_i} \\
&= \int d\sigma\int \pD\tilde{Q}_i\, \pD Q_i\, \pD\tilde{Q}'\, \pD Q'\, e^{4\pi\int d\varphi\, (\tilde{Q}_i\partial_\varphi Q_i + \tilde{Q}'\partial_\varphi Q' + \sigma(Q_i\tilde{Q}_i + \frac{k}{2}Q'\tilde{Q}'))}. \label{rewriting}
\end{align}
Hence $\sigma$ can be interpreted as a Lagrange multiplier enforcing the constraint
\begin{equation}
\sum_{i=1}^{N_f - 1} Q_i\tilde{Q}_i + \frac{k}{2}Q'\tilde{Q}' = 0,
\end{equation}
which corresponds to the second of the defining conditions \eqref{quotient} for the Higgs branch.  When $k = 2$, we obtain the usual D-term relation in SQED$_{N_f}$.  This is a consistency check of the CSM description of SQED$_{N_f}$ from the point of view of the 1D theory.\footnote{Note that in going from \eqref{QHloc} to \eqref{rewriting}, we are really using the equivalence of SQED$_1$ to a free hyper.  Thus the new fields $Q'$ and $\tilde{Q}'$ correspond to the monopole operators for $\tau$ in the theory \eqref{QHloc}.} 

\section{Supergravity Background}\label{SUGRA_BG}

In this section, we briefly show how to obtain our non-conformal rigid $\mathcal{N} = 4$ supersymmetry algebra on $S^3$, namely $\mathfrak{su}(2|1)_\ell\oplus \mathfrak{su}(2|1)_r$, from a supergravity background (analogous constructions are known in the 2D $\mathcal{N} = (2, 2)$ context, which is similar to 3D $\mathcal{N} = 4$ in terms of how mirror symmetry acts on $R$-symmetries; see, e.g., \cite{Closset:2014pda}).

We use the off-shell formulation of 3D $\mathcal{N} = 4$ conformal supergravity presented in \cite{Banerjee:2015uee}, which dimensionally reduces off-shell 4D $\mathcal{N} = 2$ SUGRA to off-shell 3D $\mathcal{N} = 4$ SUGRA.  In the process, the 4D $R$-symmetry group $(SU(2)\times U(1))/\mathbb{Z}_2$ is enhanced to the 3D $R$-symmetry group $(SU(2)\times SU(2))/\mathbb{Z}_2\cong SO(4)$.  The 4D Weyl multiplet decomposes into a 3D Weyl multiplet and a 3D Kaluza-Klein vector multiplet.  In 4D and 3D, matter multiplets are defined in a superconformal background of 4D or 3D Weyl multiplet fields, according to the superconformal method for constructing matter-coupled Poincar\'e supergravity.  There is a direct correspondence between 4D and 3D matter multiplets, namely vector multiplets, tensor multiplets, and hypermultiplets (i.e., these multiplets are irreducible under reduction).

A note on conventions: \cite{Banerjee:2015uee} uses indices $i, j$ (our $a, b$) for the fundamental of $SU(2)_H$ and $p, q$ (our $\dot{a}, \dot{b}$) for the fundamental of $SU(2)_C$.  The spinor parameters of $Q$- and $S$-supersymmetry are $\epsilon^{ip}$ and $\eta^{ip}$, which have Weyl weights $-1/2$ and $1/2$, respectively (the former should not be confused with the Levi-Civita symbol, for which $\epsilon^{12} = 1$).  Below, spinor indices are suppressed.

The 3D background multiplets are as follows:
\begin{itemize}
\item The 3D Weyl multiplet consists of fields $e_\mu{}^a, \psi_\mu{}^{ip}, b_\mu, \mathcal{V}_\mu{}^i{}_j, \mathcal{A}_\mu{}^p{}_q, C, \chi^{ip}, D$ (vielbein, gravitino, dilatation gauge field, $SU(2)_H$ $R$-symmetry gauge field, $SU(2)_C$ $R$-symmetry gauge field, and auxiliary fields) with Weyl weights $-1, -1/2, 0, 0, 0, 1, 3/2, 2$, respectively.  Its transformation rules are given by (3.1) of \cite{Banerjee:2015uee}.  The BPS conditions require that
\begin{align}
\delta\psi_\mu{}^{ip} &= 2\mathcal{D}_\mu\epsilon^{ip} - \gamma_\mu\eta^{ip} = 0, \label{bps1} \\
\delta\chi^{ip} &= 2\slashed{D}C\epsilon^{ip} + D\epsilon^{ip} + \frac{1}{2}R(\mathcal{A})_{ab}{}^p{}_q\gamma^{ab}\epsilon^{iq} - \frac{1}{2}R(\mathcal{V})_{ab}{}^i{}_j\gamma^{ab}\epsilon^{jp} + 2C\eta^{ip} = 0. \label{bps2}
\end{align}
\item The 3D Kaluza-Klein (compensator) vector multiplet consists of a scalar triplet $(L^0)^p{}_q$ (antihermitian), a spinor $\psi^{ip}$, a gauge field $B_\mu$, and an auxiliary scalar triplet $(Y^0)^i{}_j$ (Hermitian), with Weyl weights $1, 0, 3/2, 2$, respectively.  Its transformation rules are given by (2.40) and (2.41) of \cite{Banerjee:2015uee}.  The BPS conditions require that
\begin{equation}
\label{bps3}
\delta\psi^{ip} = \slashed{D}(L^0)^p{}_q\epsilon^{iq} - \frac{1}{2}F(B)_{ab}\gamma^{ab}\epsilon^{ip} + C(L^0)^p{}_q\epsilon^{iq} + (Y^0)^i{}_j\epsilon^{jp} + (L^0)^p{}_q\eta^{iq} = 0.
\end{equation}
\end{itemize}
In the above, the derivative $\mathcal{D}_\mu$ is covariant with respect to Lorentz, dilatation, and $R$-symmetry transformations, while the derivative $D_\mu$ is covariant with respect to all superconformal symmetries and includes fermionic terms.  The 3D matter multiplets are as follows:
\begin{itemize}
\item The 3D vector multiplet, like the KK vector multiplet, consists of fields $L^p{}_q, W_\mu, \Omega^{ip}, Y^i{}_j$ with Weyl weights $1, 0, 3/2, 2$, respectively.  Its transformation rules are given by (4.6) of \cite{Banerjee:2015uee}.  Setting the background fermions to zero, these are:
\begin{align}
\delta W_\mu &= \bar{\epsilon}_{ip}\gamma_\mu\Omega^{ip}, \\
\delta\Omega^{ip} &= \slashed{D}L^p{}_q\epsilon^{iq} - \frac{1}{2}F(W)_{ab}\gamma^{ab}\epsilon^{ip} + Y^i{}_j\epsilon^{jp} + CL^p{}_q\epsilon^{iq} + L^p{}_q\eta^{iq}, \\
\delta L^p{}_q &= 2\bar{\epsilon}_{iq}\Omega^{ip} - \delta^p{}_q\bar{\epsilon}_{ir}\Omega^{ir}, \\
\delta Y^i{}_j &= 2\bar{\epsilon}_{jp}\slashed{D}\Omega^{ip} - 2C\bar{\epsilon}_{jp}\Omega^{ip} - \bar{\eta}_{jp}\Omega^{ip} - \text{(trace)}.
\end{align}
The transformation rules for the 3D tensor (twisted vector) multiplet are given by (4.17) of \cite{Banerjee:2015uee}; these are similar.
\item The 3D hypermultiplet consists of fields $(A^\alpha)_i, \zeta^\alpha$ with Weyl weights $1/2, 1$, respectively.  Its transformation rules are given by (4.22) of \cite{Banerjee:2015uee}:
\begin{align}
\delta(A^\alpha)_i &= 2\bar{\epsilon}_{ip}(\zeta^\alpha)^p, \\
\delta(\zeta^\alpha)^p &= \slashed{D}(A^\alpha)_i\epsilon^{ip} - \frac{1}{2}C(A^\alpha)_i\epsilon^{ip} + \frac{1}{2}(A^\alpha)_i\eta^{ip}.
\end{align}
Here, $\alpha$ can be thought of as a flavor index (superconformal invariance requires that the hypermultiplet target space be a hyperk\"ahler cone, so that $\alpha$ takes an even number of values).  The transformation rules for the 3D twisted hypermultiplet, which cannot be obtained by dimensional reduction, are given by (4.23) of \cite{Banerjee:2015uee}; these are similar.
\end{itemize}
The desired BPS configuration of the background fields is as follows.  We set
\begin{equation}
\label{bkgd}
b_\mu, \mathcal{V}_\mu{}^i{}_j, \mathcal{A}_\mu{}^p{}_q, C, D = 0, \quad B_\mu = 0
\end{equation}
in the 3D Weyl and KK vector multiplets, thus reducing the BPS conditions \eqref{bps1}, \eqref{bps2}, \eqref{bps3} to
\begin{align}
0 &= 2\nabla_\mu\epsilon^{ip} - \gamma_\mu\eta^{ip}, \label{cond1} \\
0 &= \slashed{\partial}(L^0)^p{}_q\epsilon^{iq} + (Y^0)^i{}_j\epsilon^{jp} + (L^0)^p{}_q\eta^{iq}. \label{cond2}
\end{align}
Keeping in mind the Weyl weights and hermiticity properties of $L^0$ and $Y^0$, if we take
\begin{equation}
\epsilon_{a\dot{a}} = \xi_{a\dot{a}}, \quad \eta_{a\dot{a}} = 2\xi_{a\dot{a}}', \quad (L^0)^{\dot{a}}{}_{\dot{b}} = i\bar{h}^{\dot{a}}{}_{\dot{b}}, \quad (Y^0)_a{}^b = \frac{1}{r}h_a{}^b
\end{equation}
where $h$ and $\bar{h}$ are constant $\mathfrak{su}(2)_{H, C}$ matrices, then the conditions \eqref{cond1}, \eqref{cond2} become
\begin{align}
\nabla_\mu\epsilon_{a\dot{a}} = \frac{1}{2}\gamma_\mu\eta_{a\dot{a}} &\Longleftrightarrow \nabla_\mu\xi_{a\dot{a}} = \gamma_\mu\xi'_{a\dot{a}}, \\
(L^0)^{\dot{b}}{}_{\dot{a}}\eta^{a\dot{a}} = -(Y^0)^a{}_b\epsilon^{b\dot{b}} &\Longleftrightarrow \xi'_{a\dot{a}} = \frac{i}{2r}h_a{}^b\xi_{b\dot{b}}\bar{h}^{\dot{b}}{}_{\dot{a}},
\end{align}
which are precisely the Killing spinor equations \eqref{killing1} and \eqref{killing2}.

We now substitute the background values into the transformation rules for the matter multiplets.  In doing so, we recover only the transformations for abelian vector multiplets and ungauged hypermultiplets rather than the full supersymmetry transformations \eqref{Avar}, \eqref{lamvar}, \eqref{phivar}, \eqref{dvar} for the vector multiplet and \eqref{qpsivar}, \eqref{qtpsitvar} for the hypermultiplet.  This is because \cite{Banerjee:2015uee} considers only abelian 3D vector multiplets and 3D hypermultiplets that are not coupled to vector multiplets (although one can gauge-covariantize the SUSY transformations of the latter by hand).  From \eqref{bkgd}, we have for the vector multiplet that
\begin{align}
\delta W_\mu &= \bar{\epsilon}^{ip}\gamma_\mu\Omega_{ip}, \\
\delta\Omega_{ip} &= -\frac{i}{2}F(W)_{ab}\epsilon^{abc}\gamma_c\epsilon_{ip} - Y_i{}^j\epsilon_{jp} + \gamma^\mu\epsilon_i{}^q\partial_\mu L_{qp} - L_p{}^q\eta_{iq}, \\
\delta L_{pq} &= -(\bar{\epsilon}^i{}_p\Omega_{iq} + \bar{\epsilon}^i{}_q\Omega_{ip}), \\
\delta Y_{ij} &= \bar{\epsilon}_{ip}\slashed{\nabla}\Omega_j{}^p + \bar{\epsilon}_{jp}\slashed{\nabla}\Omega_i{}^p - \frac{1}{2}(\bar{\eta}_{ip}\Omega_j{}^p + \bar{\eta}_{jp}\Omega_i{}^p).
\end{align}
If we now assume that the Dirac conjugates satisfy $\bar{\epsilon} = \frac{i}{2}\epsilon = \frac{i}{2}\xi$ and $\bar{\eta} = \frac{i}{2}\eta = i\xi'$ and identify
\begin{equation}
(W_\mu, \Omega_{a\dot{b}}, L_{\dot{a}\dot{b}}, Y_{ab}) = (A_\mu, \lambda_{a\dot{b}}, -i\Phi_{\dot{a}\dot{b}}, D_{ab}),
\end{equation}
then we reproduce the abelian vector multiplet transformations
\begin{align}
\delta_\xi A_\mu &= \frac{i}{2}\xi^{a\dot{b}}\gamma_\mu\lambda_{a\dot{b}}, \\
\delta_\xi\lambda_{a\dot{b}} &= \textstyle -\frac{i}{2}\epsilon^{\mu\nu\rho}\gamma_\rho\xi_{a\dot{b}}F_{\mu\nu} - D_a{}^c\xi_{c\dot{b}} - i\gamma^\mu\xi_a{}^{\dot{c}}\partial_\mu\Phi_{\dot{c}\dot{b}} + 2i\Phi_{\dot{b}}{}^{\dot{c}}\xi_{a\dot{c}}', \\
\delta_\xi\Phi_{\dot{a}\dot{b}} &= \textstyle \frac{1}{2}(\xi^c{}_{\dot{a}}\lambda_{c\dot{b}} + \xi^c{}_{\dot{b}}\lambda_{c\dot{a}}), \\
\delta_\xi D_{ab} &= -\frac{i}{2}(\xi_a{}^{\dot{c}}\gamma^\mu\nabla_\mu\lambda_{b\dot{c}} + \xi_b{}^{\dot{c}}\gamma^\mu\nabla_\mu\lambda_{a\dot{c}}) + \frac{i}{2}(\xi'_a{}^{\dot{c}}\lambda_{b\dot{c}} + \xi'_b{}^{\dot{c}}\lambda_{a\dot{c}}).
\end{align}
From \eqref{bkgd}, we have for the hypermultiplet that
\begin{align}
\delta(A^\alpha)^a &= -2\bar{\epsilon}^{a\dot{b}}(\zeta^\alpha)_{\dot{b}}, \\
\delta(\zeta^\alpha)_{\dot{a}} &= -\slashed{\partial}(A^\alpha)^a\epsilon_{a\dot{a}} - \frac{1}{2}(A^\alpha)^a\eta_{a\dot{a}},
\end{align}
so that if we take $\alpha = 1, 2$ and identify
\begin{equation}
((A^1)_i, (\zeta^1)_p)\propto (q_a, i\psi_{\dot{a}}), \quad ((A^2)_i, (\zeta^2)_p)\propto (\tilde{q}_a, i\tilde{\psi}_{\dot{a}}),
\end{equation}
then we reproduce the ungauged hypermultiplet transformations
\begin{align}
\delta_\xi q^a = \xi^{a\dot{b}}\psi_{\dot{b}}, \quad \delta_\xi\psi_{\dot{a}} = i\gamma^\mu\xi_{a\dot{a}}\partial_\mu q^a + i\xi'_{a\dot{a}}q^a, \\
\delta_\xi\tilde{q}^a = \xi^{a\dot{b}}\tilde{\psi}_{\dot{b}}, \quad \delta_\xi\tilde{\psi}_{\dot{a}} = i\gamma^\mu\xi_{a\dot{a}}\partial_\mu\tilde{q}^a + i\tilde{q}^a\xi'_{a\dot{a}}.
\end{align}

\bibliographystyle{utphys} 
\bibliography{CoulombAbel}

\end{document}